\documentclass[11pt, a4paper, logo, copyright]{googledeepmind}

\pdftrailerid{redacted}

\makeatletter
\providecommand\bibentry[1]{\nocite{#1}\fullcite{#1}}
\makeatother

\usepackage{kantlipsum, lipsum}
\usepackage{dsfont}
\usepackage{gdm-colors}
\usepackage[utf8]{inputenc}   
\usepackage{newunicodechar}   
\usepackage{amssymb}          
\usepackage{CJKutf8}
\usepackage{tcolorbox}
\tcbuselibrary{breakable}
\usepackage{enumitem}
\usepackage{arydshln}
\usepackage[flushleft]{threeparttable}
\usepackage{wasysym,marvosym}

\definecolor{catBlue}{HTML}{2A6FB5}
\definecolor{catPurple}{HTML}{6E3FA3}
\definecolor{catTeal}{HTML}{1F6F6B}
\definecolor{catRed}{HTML}{C0392B}
\definecolor{cellGreen}{HTML}{6CC180}
\definecolor{cellGray}{HTML}{ECECEC}

\newcommand{\yes}{\textcolor{white}{\ding{51}}}
\newcommand{\no}{\textcolor{gray!70}{\ding{55}}}
\newcommand{\cyes}{\cellcolor{cellGreen}\yes}
\newcommand{\cno}{\cellcolor{cellGray}\no}

\usepackage{graphicx}
\newcommand{\catlabel}[3]{%
  \multirow{#1}{*}{\rotatebox[origin=c]{90}{\textcolor{#2}{\textbf{#3}}}}}

\usepackage{caption}
\usepackage{xcolor}
\usepackage{soul}

\soulregister\cite7
\soulregister\ref7
\soulregister\cref7

\usepackage[
  style=authoryear-comp,
  sorting=ynt,
  natbib=true,
  backend=bibtex,
  maxcitenames=2,
  maxbibnames=99,
  uniquelist=false,
  uniquename=false,
  giveninits=true,
  dashed=false
]{biblatex}
\let\cite\textcite

\DeclareFieldFormat{citehyperref}{%
  \DeclareFieldAlias{bibhyperref}{noformat}%
  \bibhyperref{#1}}
\DeclareFieldFormat{textcitehyperref}{%
  \DeclareFieldAlias{bibhyperref}{noformat}%
  \bibhyperref{%
    #1%
    \ifbool{cbx:parens}
      {\bibcloseparen\global\boolfalse{cbx:parens}}
      {}}}
\savebibmacro{cite}
\savebibmacro{textcite}
\renewbibmacro*{cite}{%
  \printtext[citehyperref]{%
    \restorebibmacro{cite}%
    \usebibmacro{cite}}}
\renewbibmacro*{textcite}{%
  \ifboolexpr{
    ( not test {\iffieldundef{prenote}} and
      test {\ifnumequal{\value{citecount}}{1}} )
    or
    ( not test {\iffieldundef{postnote}} and
      test {\ifnumequal{\value{citecount}}{\value{citetotal}}} )
  }
    {\DeclareFieldAlias{textcitehyperref}{noformat}}
    {}%
  \printtext[textcitehyperref]{%
    \restorebibmacro{textcite}%
    \usebibmacro{textcite}}}

\let\Cross\undefined

\usepackage{bbding}
\usepackage[utf8]{inputenc} 
\usepackage[T1]{fontenc}    
\usepackage{hyperref}       
\usepackage{url}            
\usepackage{booktabs}       
\usepackage{nicefrac}       
\usepackage{microtype}      
\usepackage{amsmath}
\usepackage{graphicx}
\usepackage{tablefootnote}
\usepackage{multicol}
\usepackage[nameinlink]{cleveref}
\usepackage{bbm}
\usepackage{multirow}
\usepackage{soul}
\usepackage{float}
\usepackage{wrapfig}
\usepackage{blindtext}
\usepackage{tablefootnote}
\usepackage{amsfonts}
\usepackage{colortbl}
\usepackage{mathtools,amssymb}
\usepackage{bm}
\usepackage{makecell}
\usepackage{caption}
\usepackage{capt-of}
\usepackage{array}
\usepackage{calc}      
\usepackage{caption}   
\usepackage{subcaption}  
\usepackage{xcolor,colortbl}
\usepackage[bottom]{footmisc}

\usepackage{xcolor}   
\usepackage{soul}     

\sethlcolor{yellow!20}

\definecolor{mydeepgreen}{RGB}{0, 100, 0}

\newcommand{\benchname}{OneReason-Bench}

\usepackage{xspace}

\newcommand{\cdashlinerow}[2]{%
  \cdashline{#1}%
  \noalign{\global\let\CT@row@color\relax\vskip0pt}%
  \rowcolor{#2}%
}

\newcommand{\eat}[1]{}

\crefformat{section}{\S#2#1#3}

\graphicspath{{figures/}}

\title{OneReason Technical Report}

\renewcommand{\today}{}

\author{\large OneRec Team}

\begin{abstract}
Generative recommendation models in the OneRec family have been widely deployed in many real-world services, such as short-video, live-streaming, advertising, and e-commerce.
However, these generative models can only benefit from the scaling advantage, while their reasoning ability is hard to activate, since we cannot construct meaningful Chain-of-Thought (CoT) sequences consisting of itemic tokens only.
Inspired by the success of the reasoning-style ``think before answer'' paradigm in the LLM field, we conduct preliminary studies (i.e., OneRec-Think, OpenOneRec) to explore reasoning capability in generative recommendation. 
Nevertheless, we notice an unexpected phenomenon: the thinking mode does not show advantages over the non-thinking mode.
Drawing insights from recent findings on CoT robustness in multi-modal language models, we argue that effective reasoning in recommendation rests on two factors: \emph{perception}, the ability to ground itemic tokens in their underlying language semantics, and \emph{cognition}, the ability to reorganize a user's behavior sequence into coherent latent interest points.
We therefore propose OneReason, which includes: (1) strong itemic token perception in pre-training, (2) a three-level cognition-enhanced CoT format for recommendation tasks in SFT, and (3) a specialize-then-unify training recipe in RL to enhance the thinking ability.
Equipped with the above techniques, OneReason successfully achieves a thinking mode that outperforms the non-thinking mode across multiple real-world business benchmarks at Kuaishou.
Moreover, we observe an interesting phenomenon: replacing unCoT recommendation data with CoT-supervised data can improve non-thinking inference in several domains. This suggests that some benefits of CoT supervision may transfer to direct decoding, although the current evidence is behavioral and does not by itself identify whether the effect comes from compression, reasoning, or their interaction.
Finally, we systematically present a series of reasoning benchmarks and design insights, and we will open-source the OneReason-8B and OneReason-0.8B models to facilitate research on generative recommendation models.
\end{abstract}

\begin{document}

\maketitle

\begin{figure*}[th!]
\centering
\includegraphics[width=\linewidth]{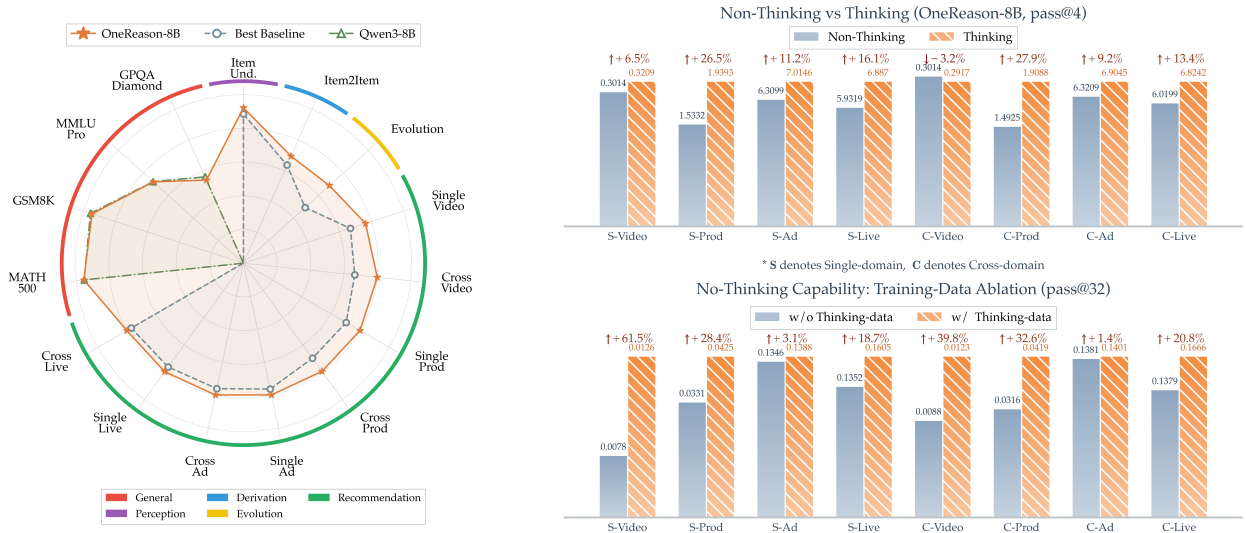}
\caption{Performance overview: (1) compared with Qwen3-8B, OneReason-8B maintains its general capabilities; (2) OneReason-8B exhibits SOTA performance across real-world recommendation benchmarks; (3) OneReason-8B showing promising ``thinking'' advantage to recommendation tasks.}
\label{fig:first}
\end{figure*}

\newpage
\setcounter{tocdepth}{2} 

\tableofcontents 

\newpage
\section{Introduction}

In recent years, Transformer-based large language models (LLMs) have pushed the frontier of machine intelligence to unprecedented levels.
Driven by massive compute scale and advanced training methodologies, LLMs have come to exhibit two exciting properties that guide us to build more powerful models:
\begin{itemize}
    \item \textbf{Scaling property in pre-training}: When model size, training data, and compute resources are scaled up, the convergence loss follows a regular, extrapolable decay trend \citep{kaplan2020scaling, hoffmann2022trainingcomputeoptimallargelanguage}.
    As scaling proceeds, the evaluation scores across multiple benchmarks improve simultaneously, demonstrating the models' remarkable semantic understanding and generalization capability.

    \item \textbf{Reasoning property in post-training}: Following pre-training, the post-training stage further unlocks and refines the model's advanced capabilities via high-quality Supervised Fine-Tuning (SFT) and Reinforcement Learning (RL) \citep{ouyang2022traininglanguagemodelsfollow, shao2024deepseekmath, zheng2025groupsequencepolicyoptimization, yu2025dapoopensourcellmreinforcement}.
    Driven by these post-training techniques, LLMs learn to follow CoT-style procedures \citep{wei2023chainofthoughtpromptingelicitsreasoning, kojima2023largelanguagemodelszeroshot}, forming a ``think-before-answer'' reasoning paradigm.
    This paradigm significantly enhances their capabilities to solve complex, in-depth problems, as exemplified by the OpenAI o1 \citep{jaech2024openai} and DeepSeek-R1 \citep{guo2025deepseek} series.

\end{itemize}

Inspired by the \textbf{Scaling} and \textbf{Reasoning} properties of LLMs, we are deeply curious about whether these two properties can be further leveraged to push the capability of industrial recommender systems.
Driven by this motivation, we have started a series of exploration efforts.
For the Scaling property, our OneRec V1/V2  series of generative models \citep{zhou2025onerectechnicalreport, zhou2025onerecv2technicalreport} have validated \textbf{the scaling property in industrial recommendation scenarios}, contributing significant business gains at Kuaishou.
For the Reasoning property, we acknowledge that the recommendation-data-driven OneRec model cannot unlock the reasoning ability on its own.
Since the OneRec model is trained exclusively on pure itemic sequential data, it has only learned from flat transition patterns without any underlying logical thinking traces.
To elicit the reasoning capabilities, our recent models, OneRec-Think \citep{liu2025onerecthinkintextreasoninggenerative} and OpenOneRec \citep{zhou2026openonerectechnicalreport}, have explored introducing interleaved itemic-text recommendation data and general-domain reasoning data. 
In these two models, we have successfully observed that \textbf{the ``think-before-answer'' pattern can be generalized to recommendation tasks}, allowing us to inspect the underlying logic of the model and to understand the user's interests as well as the rationale behind the next-item prediction.
However, we also notice an unexpected phenomenon: \textbf{the thinking mode does not show significant advantages on recommendation benchmarks compared with the non-thinking mode.}

\begin{figure}[t!]\centering 
    \includegraphics[width=\textwidth]{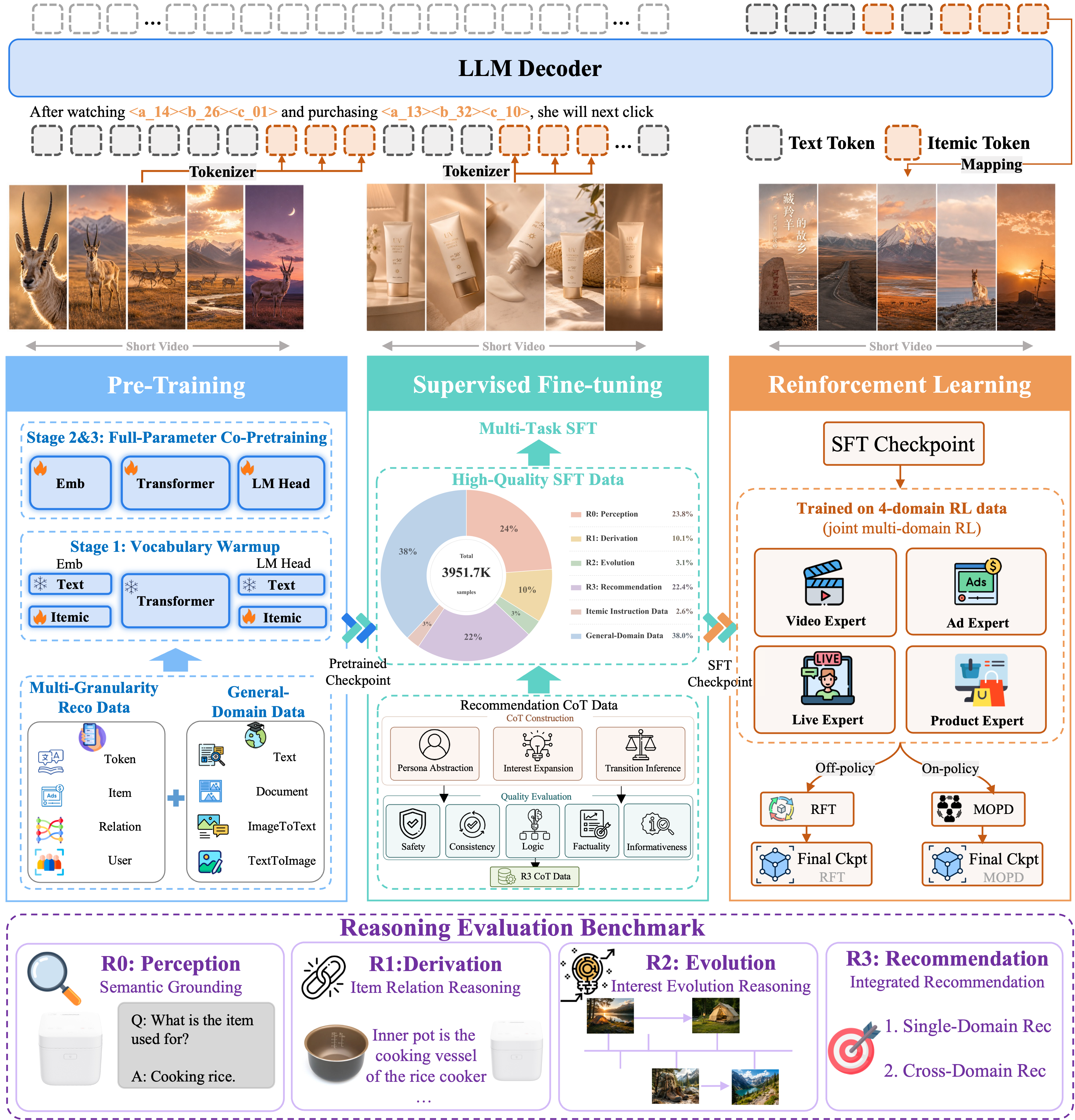}
    \caption{The pre-training, SFT, RL, and benchmarking pipeline of OneReason.}\label{fig:main}
\end{figure}

\begin{figure}[t]
  \centering
  \begin{minipage}[c]{0.32\textwidth}
    \centering
    \begin{tcolorbox}[
      colback=gray!5!white,
      colframe=gray!50!black,
      boxrule=0.4pt,
      arc=3pt,
      left=4pt, right=4pt, top=3pt, bottom=3pt,
      boxsep=0pt,
      before=\centering,
      after skip=6pt,
    ]
      \textbf{Question: How many complete fruits are there?}
    \end{tcolorbox}
    \includegraphics[width=\linewidth]{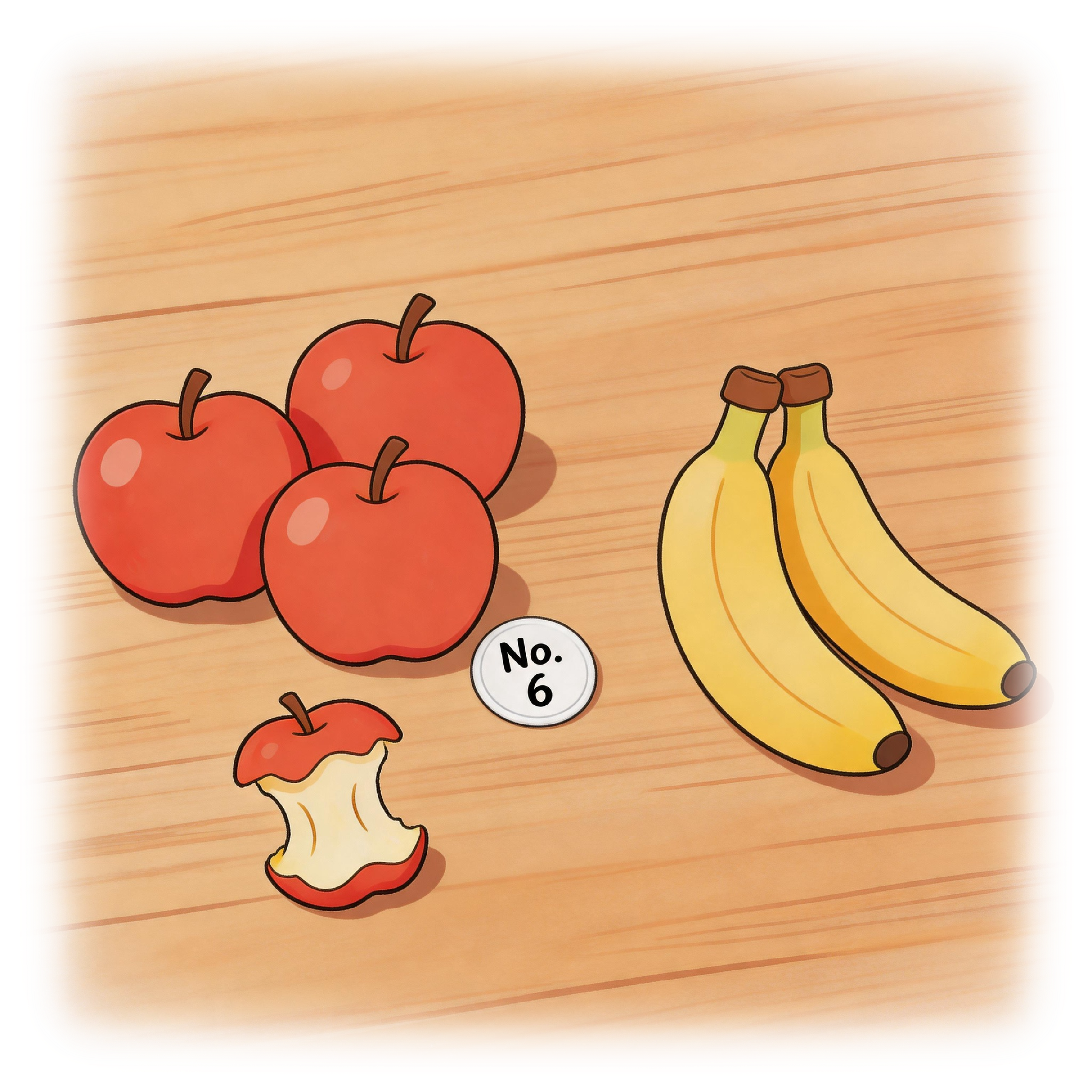}
  \end{minipage}%
  \hfill
  \begin{minipage}[c]{0.65\textwidth}
    \raggedright
    \begin{tcolorbox}[
      colback=red!5!white,
      colframe=red!70!black,
      title=\footnotesize\bfseries (a) Alignment Failure,
      fonttitle=\footnotesize\bfseries,
      before upper={\textcolor{red!70!black}{\textbf{\ding{55} Answer: 6}}\\[2pt]},
      left=4pt, right=4pt, top=3pt, bottom=3pt,
      after skip=4pt
    ]
      \footnotesize
      \texttt{"I see 4 apples and 2 bananas. So answer is 6."} \\[3pt]
      \colorbox{red!15}{\footnotesize\textbf{Cause:} Misperceived objects; weak visual grounding}
    \end{tcolorbox}

    \begin{tcolorbox}[
      colback=red!5!white,
      colframe=red!70!black,
      title=\footnotesize\bfseries (b) CoT Quality Failure,
      fonttitle=\footnotesize\bfseries,
      before upper={\textcolor{red!70!black}{\textbf{\ding{55} Answer: 6}}\\[2pt]},
      left=4pt, right=4pt, top=3pt, bottom=3pt,
      after skip=4pt
    ]
      \footnotesize
      \texttt{"3 apples + 2 bananas = 5. But apple core \& No.~6 suggest 6 originally. Answer: 6."} \\[3pt]
      \colorbox{red!15}{\footnotesize\textbf{Cause:} Perception OK but faulty reasoning chain}
    \end{tcolorbox}

    \begin{tcolorbox}[
      colback=green!5!white,
      colframe=green!60!black,
      title=\footnotesize\bfseries (c) Robust Reasoning,
      fonttitle=\footnotesize\bfseries,
      before upper={\textcolor{green!50!black}{\textbf{\ding{51} Answer: 5}}\\[2pt]},
      left=4pt, right=4pt, top=3pt, bottom=3pt,
      after skip=0pt
    ]
      \footnotesize
      \texttt{"3 apples + 2 bananas = 5. Core \& tag are irrelevant."} \\[3pt]
      \colorbox{green!15}{\footnotesize\textbf{Key:} Deep alignment + logical CoT}
    \end{tcolorbox}
  \end{minipage}

  \caption{Three failure/success modes of MLLM chain-of-thought reasoning given the same visual input.
           (a) Alignment failure: the model miscounts due to weak visual grounding.
           (b) CoT quality failure: correct perception but faulty reasoning drift.
           (c) Robust reasoning: deep alignment and logical CoT lead to the correct answer.}
  \label{fig:motivation}
\end{figure}

To explore the root cause of this unexpected phenomenon, we turn to the multi-modal LLM (MLLM) literature, where a strikingly similar reasoning-mode brittleness has been observed.
\citep{sun2026readingthinkingunderstandingbridging} reveals that when textual and visual modalities are insufficiently aligned, models tend to mechanically ``read'' surface-level visual text rather than genuinely reason about the underlying visual semantics, providing a compelling indication that deep cross-modal alignment is a prerequisite for authentic thinking.
Extending this insight, \citep{zhou2025perceptioncognitionsurveyvisionlanguage} systematically surveys the perception-to-cognition pathway in MLLMs and concludes that advanced interactive reasoning cannot emerge without first establishing robust perceptual alignment between modalities.
Building on this alignment foundation, \cite{jiang2025corvidimprovingmultimodallarge} further demonstrates that, even with aligned modalities, reasoning robustness remains fragile unless the CoT traces themselves are carefully structured, specifically by adopting logically coherent, coarse-to-fine reasoning formats that prevent hallucinations and over-thinking.

Taken together, these works converge on two essential and complementary pillars for unlocking genuine reasoning capability: \textbf{(i) modality alignment in perception}, which grounds perception and prevents the model from operating on hollow surface signals; and \textbf{(ii) CoT quality in cognition}, which provides coherent coarse-to-fine reasoning trajectories and enables a faithful \textit{think-before-answer} process once alignment is in place.

For generative recommendation, as illustrated in Figure~\ref{fig:motivation}, these insights carry a clear implication: enabling a model to truly \textit{think} about user interests demands both
(1) \textbf{deep alignment in perception} between itemic modality and natural-language modality, so that itemic tokens become referable, composable semantic units rather than opaque identifiers; and
(2) \textbf{recommendation-specific and logically fluent CoT structures in cognition} that are designed to scaffold deliberate reasoning over user histories and item properties on top of this alignment.
These two dimensions serve as the cornerstones for building a powerful reasoning model, where we explicitly strengthen text-item alignment and devise tailored coarse-to-fine reasoning formats to make the model genuinely capable of recommendation-oriented thinking.

In light of the above, we propose OneReason, a recommendation foundation model with genuine reasoning capability.
Specifically, for better \textbf{item perception ability}, we collect complex and diverse coarse-to-fine alignment pre-training corpora to align the newly inserted itemic tokens with text tokens in the same semantic space.
For better \textbf{reasoning cognition ability}, we elaborate a standard thinking structure to ensure CoT quality and build the SFT corpora.
Moreover, building on the model after pre-training and SFT, we find that the thinking mode still underperforms the non-thinking mode under multi-domain mixed RL, yet consistently surpasses it when RL is restricted to a single domain.
This contrast motivates us to adopt a ``specialize-then-unify'' RL strategy - we first conduct domain-focused RL to fully unlock thinking-mode benefits within each domain, and then apply rejection sampling fine-tuning \citep{yuan2023scaling} or Multi-Teacher On-Policy Distillation \citep{xiao2026mimo} for cross-domain balancing and refinement.
The overall training process of OneReason is visualized in Figure~\ref{fig:main}.
Meanwhile, we also make an interesting observation: under an equal number of training tokens, adding recommendation CoT supervision can improve non-thinking-mode performance compared with training on pure unCoT data in several domains (see the bottom right panel of Figure~\ref{fig:first}). We treat this as behavioral evidence that some supervision signals from CoT traces may transfer to direct decoding, rather than as proof that CoT is universally necessary or that the mechanism has been identified.

To evaluate the reasoning ability of our model in recommendation scenarios, we introduce OneReason-Bench, a curated collection of reasoning-oriented recommendation benchmarks.
Organized along multiple axes, this benchmark is designed to assess reasoning abilities in a comprehensive and structured manner, and has played a central role in guiding and validating our model development.
As illustrated in Figure~\ref{fig:first}, our OneReason model achieves superior performance across this benchmark, demonstrating its robust reasoning capabilities.
To facilitate research on generative recommendation models, we will open-source the OneReason-8B and OneReason-0.8B models and other fine-tuning-related materials to the public.
%
%

\section{Reasoning Design Philosophy}
Before describing the architecture and training recipes in detail, we first outline the design principle of recommendation reasoning: what is a good CoT trace in recommendation?
In classical LLM reasoning tasks involving math, code, or symbolic logic, it is usually assumed that there exist intermediate steps that maintain task constraints and apply relevant world knowledge to logically answer the question step by step.
Unlike classical reasoning tasks with a single correct answer, recommendation admits multiple plausible items at once, and the underlying user intent is never directly observable; it must be inferred from a long sequence of historical itemic tokens (where each complete sequence represents some item).
Therefore, recommendation reasoning is abductive rather than deductive: it requires hypothesizing latent interest points, modeling how they evolve over time, and using them to justify why specific candidates fit the current context.
In other words, a recommendation CoT should select relevant behaviors as hypothesized interest points, compress them into an interpretable preference, model the temporal transitions between these interests, and finally associate the inferred preference state with the recommended item.
This distinction motivates our training and evaluation design. We decompose recommendation reasoning into four diagnostic capabilities along two dimensions: (1) \textbf{Perception (R0)}, which grounds itemic tokens in their semantic content; and (2) \textbf{Cognition (R1--R3)}, which builds progressively deeper reasoning on top of solid perception.
\begin{itemize}
    \item \textbf{R0: Perception.} Interpreting itemic patterns via their explicit semantic content is the foundational ability; without it, user behavior is uninterpretable.
    \item \textbf{R1: Derivation.} Reasoning from individual item semantics to item-to-item relations via common-sense or knowledge-based associations; without it, latent interests cannot be extracted from noisy interaction histories.
    \item \textbf{R2: Evolution.} Reasoning over items of the same latent interest as a temporal process; without it, long-term, short-term, and periodic preferences cannot be modeled.
    \item \textbf{R3: Recommendation.} Reasoning coherently to produce high-quality decisions across all served domains (e.g., Short-Video, Live-Streaming).
\end{itemize}

The immediately following section operationalizes this hierarchy as an evaluation instrument, so that the trajectory from pre-training (Section~\ref{sec:pre-training}) to post-training (Sections~\ref{sec:sft} and~\ref{sec:rl}) to deployment (Section~\ref{sec:deployment}) can be measured at every step against the same diagnostic axes.

\section{Benchmark Design for Reasoning Recommendation}\label{sec:Bench}

Placing the benchmark before the model description is deliberate: the four-granularity pre-training (Section~\ref{sec:pre-training}), the compression-then-reasoning SFT (Section~\ref{sec:sft}), and the specialize-then-unify RL (Section~\ref{sec:rl}) are each motivated by specific gaps revealed at the R0--R3 layers of \benchname{}. The benchmark therefore functions less as a leaderboard and more as a measurement protocol that justifies, monitors, and ultimately validates every downstream design choice.

While RecIF-Bench in OpenOneRec~\citep{zhou2026openonerectechnicalreport} broadens the evaluation of recommendation foundation models, its reasoning evaluation is still coarse-grained and not sufficiently diagnostic. To address this, we develop \benchname{}, extending RecIF-Bench toward multi-level reasoning.

\subsection{Task Taxonomy}

\begin{figure*}[t!]
\centering
\includegraphics[width=\textwidth]{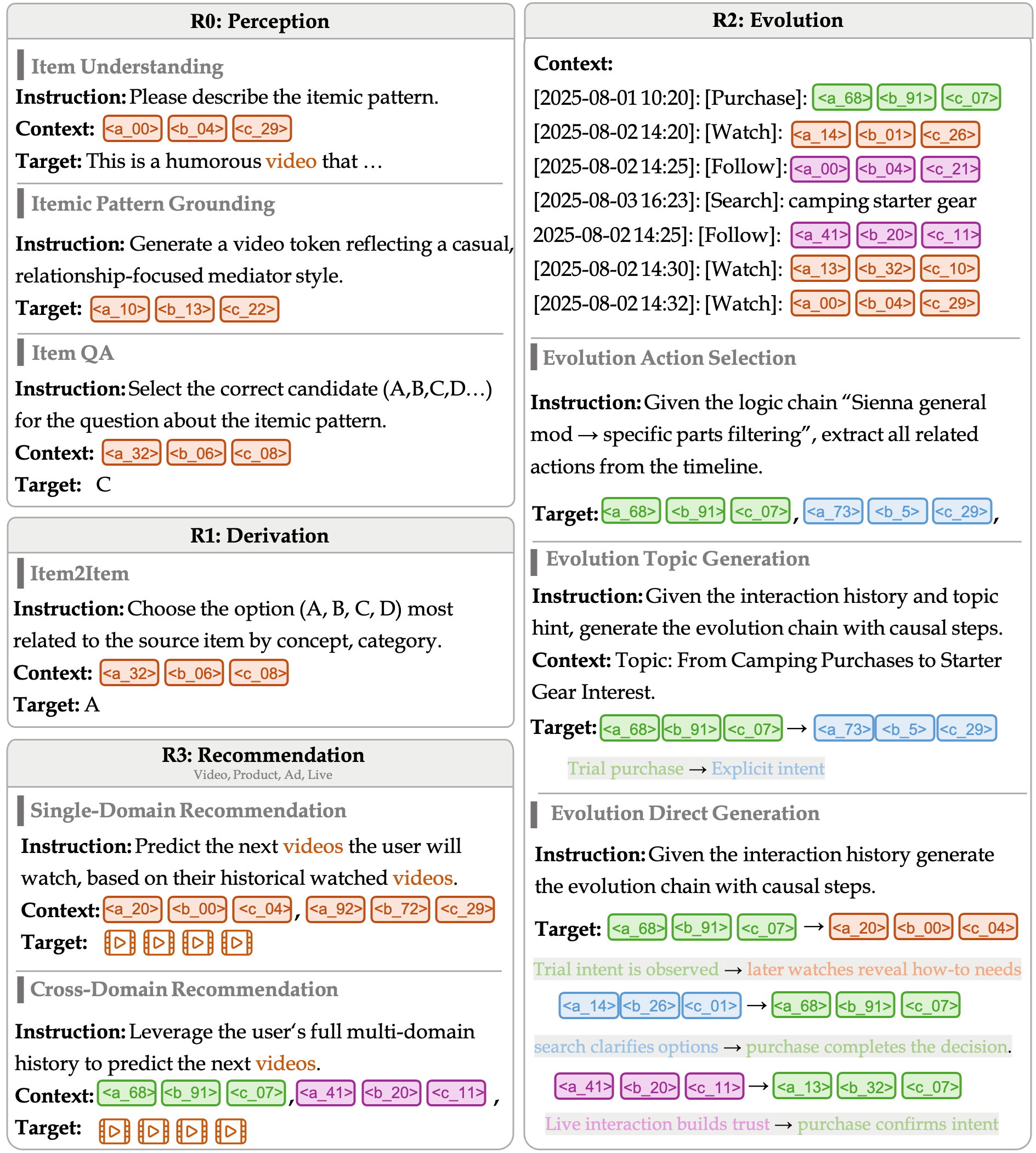}
\caption{Illustration of the task taxonomy. Each task is specified with its instruction, context, and target format.}
\label{fig:taxonomy}
\end{figure*}

\benchname{} organizes reasoning-oriented benchmark tasks into a progressive hierarchy: Perception (R0), Derivation (R1), Evolution (R2), and Recommendation (R3). These reasoning layers progress from semantic grounding to relational and temporal reasoning, and finally to integrated recommendation tasks. Table~\ref{tab:taxonomy} summarizes the taxonomy, and examples are shown in Figure~\ref{fig:taxonomy}.

\medskip
\noindent\textbf{Unified Task Formulation.}
All tasks are formulated as sequence generation $Y=\mathcal{F}(X)$, where $X$ combines a task instruction $I$ with a context $C$ such as itemic patterns, user portraits, or interaction histories. The target $Y$ can be an itemic pattern, an answer option, a natural-language response, or a structured evolution chain, allowing R0--R3 tasks to share a common generative evaluation protocol.

\begin{table*}[htbp]
\centering
\caption{Task taxonomy of \benchname{} organized across four reasoning layers.}
\label{tab:taxonomy}
\resizebox{\textwidth}{!}{%
\begin{tabular}{@{}lllll@{}}
\toprule
\textbf{Layer} & \textbf{Task} & \textbf{Input ($X$)} & \textbf{Target ($Y$)} & \textbf{Metric} \\
\midrule
\multirow{3}{*}{\makecell[l]{\textbf{R0: Perception}}}
  & Item Understanding      & Item $i$                                          & Item description                & LLM-as-a-Judge \\
  & Itemic Pattern Grounding     & Item description                                  & Item $i$                        & \makecell[l]{Pass@K, Recall@K} \\
  & Item QA                 & Item $i$ + Answer options $\mathcal{O}^{a}$       & Correct option $o^{a}$          & Accuracy \\
\midrule
\multirow{1}{*}{\makecell[l]{\textbf{R1: Derivation}}}
  & Item2Item               & Source item $i$ + Item options $\mathcal{O}^{item}$   & Correct option $o^{item}$   & Accuracy \\
\midrule
\multirow{3}{*}{\makecell[l]{\textbf{R2: Evolution}}}
  & Evolution Action Selection      & \makecell[l]{History $\mathcal{H}$ + Topic $t$} & Related actions $\mathcal{A}$   & F1  \\
  & Evolution Topic Gen.  & History $\mathcal{H}$ + Topic $t$                                  & Evolution chain $\mathcal{E}_t$     & Action--Logic Score \\
  & Evolution Direct Gen. & History $\mathcal{H}$                                              & Evolution chain(s) $\{\mathcal{E}\}$  & Multi-Chain Action--Logic Score \\
\midrule
\multirow{2}{*}{\makecell[l]{\textbf{R3: Recommendation}}}
  & \makecell[l]{Single-Domain Rec.\\(Video / Product / Ad / Live)}    & \makecell[l]{Portrait $\mathcal{P}$ + History $\mathcal{H}^d$}             & Next items $\{i\}$                & \makecell[l]{Pass@K, Recall@K} \\
  & \makecell[l]{Cross-Domain Rec.\\(Cross-Video / Cross-Product / \\Cross-Ad / Cross-Live)}     & \makecell[l]{Portrait $\mathcal{P}$ + History $\mathcal{H}$} & Next items $\{i\}$      & \makecell[l]{Pass@K, Recall@K} \\
\bottomrule
\end{tabular}%
}
\end{table*}

\subsubsection{R0: Perception}
Perception grounds itemic patterns in explicit semantic content, providing the semantic basis for higher-level reasoning and recommendation~\citep{zheng2024adapting,zhou2026openonerectechnicalreport}. We instantiate this layer with three task types:

\begin{itemize}
    \item \textbf{Item Understanding}: given an item $i$, the model generates its natural-language description, decoding item semantics from itemic patterns.
    \item \textbf{Itemic Pattern Grounding}: given a natural-language description, the model grounds it to the corresponding item.
    \item \textbf{Item QA}: given an item $i$ and content-grounded multiple-choice options $\mathcal{O}^{a}$, the model selects the correct option, assessing whether it can use item semantics to answer targeted questions.
\end{itemize}

\subsubsection{R1: Derivation}
Building upon Perception, Derivation evaluates whether the model can reason from individual item semantics to item-to-item relationships under common-sense or knowledge-based associations.
\begin{itemize}
    \item \textbf{Item2Item}: given a source item $i$ and a set of candidate items $\mathcal{O}^{item}$, the model selects the item that best matches the intended item-to-item relation.
\end{itemize}

\subsubsection{R2: Evolution}
Evolution evaluates whether the model can reason over user interests as a temporal process rather than a static preference profile. We design three task variants with increasing autonomy, from extracting topic-relevant actions to generating topic-guided or fully autonomous evolution chains:
\begin{itemize}
    \item \textbf{Evolution Action Selection}: given a history $\mathcal{H}$ and a specified interest topic $t$, the model identifies the historical actions $\mathcal{A}$ logically connected to that topic.
    \item \textbf{Evolution Topic Generation}: given a history $\mathcal{H}$ and a specified interest topic $t$, the model generates the evolution chain $\mathcal{E}_t$ describing how this interest developed.
    \item \textbf{Evolution Direct Generation}: given only the history $\mathcal{H}$, the model autonomously generates valid interest-evolution chains $\{\mathcal{E}\}$ without topic guidance.
\end{itemize}

This layer evaluates multi-hop, temporally structured reasoning about user behavior, providing a bridge from item-level reasoning to context-aware recommendation.

\subsubsection{R3: Recommendation}
While the preceding tasks focus on specific R0--R2 capabilities, real-world recommendation scenarios often require models to flexibly combine perception, derivation, and evolution abilities. We therefore introduce integrated recommendation tasks to evaluate the model's overall reasoning capability in practical recommendation settings. We assess this across two settings:
\begin{itemize}
    \item \textbf{Single-Domain Recommendation}: given a user portrait $\mathcal{P}$ and the user's interaction history $\mathcal{H}^d$ within a target domain (short video, advertising, e-commerce, or live-streaming\footnote{For brevity, we denote short video as Video, advertising as Ad, e-commerce as Product, and live-streaming as Live.}), the model predicts the next items the user will engage with.
    \item \textbf{Cross-Domain Recommendation}: given a user portrait $\mathcal{P}$, the user's multi-domain interaction history $\mathcal{H}$, and a specified target domain $d$, the model predicts the next items in that domain.
\end{itemize}

\noindent Detailed construction procedures, R3 data statistics, and a comparison with existing recommendation benchmarks are provided in Appendix~\ref{appendix:benchmark_construction_and_comparison} and \ref{appendix:r3_statistics}.

\subsection{Evaluation Protocols}

We employ a set of task-specific metrics to cover recommendation accuracy, QA correctness, interest-evolution reasoning, and open-ended generation quality.

\paragraph{Text Generation Metrics.}
For the Item Understanding task, we reuse the LLM-as-a-Judge protocol introduced in OpenOneRec~\citep{zhou2026openonerectechnicalreport} without modifying its metric definition.

\paragraph{Choice and Selection Metrics.}
For single-answer choice tasks such as Item QA and Item2Item, we report \textbf{Accuracy}. For multi-answer selection tasks, including Evolution Action Selection, we report \textbf{F1} computed over the predicted and ground-truth sets.

\paragraph{Evolution Generation Metrics.}
Evolution Topic Generation and Evolution Direct Generation are evaluated by the \textbf{Action--Logic Score} family. This metric family combines action alignment between the generated and ground-truth evolution chains with logic alignment between the corresponding reasoning statements. The detailed implementation of these metrics is provided in Appendix~\ref{appendix:action_logic_score}.

\paragraph{Recommendation Metrics.}
For recommendation and itemic pattern grounding tasks, we report \textbf{Pass@$K$} and \textbf{Recall@$K$}. Pass@$K$ indicates whether at least one ground-truth target appears in the top-$K$ generated outputs, while Recall@$K$ measures the fraction of ground-truth targets recovered within the top-$K$ list. The generated itemic patterns are first decoded into item IDs, and metrics are then computed at the item granularity.

\subsection{General-Intelligence Sanity Check}
Following OpenOneRec~\citep{zhou2026openonerectechnicalreport}, we also retain a general-intelligence sanity-check suite, covering general-purpose tasks such as MMLU-Pro~\citep{wang2024mmlu}, to ensure that recommendation specialization does not come at the cost of general reasoning or instruction-following ability.

With the diagnostic axes (R0--R3) and the general-intelligence guard rail now in place, we are ready to describe the training stack that \benchname{} is designed to evaluate. The next section opens with the perception foundation, itemic tokenization and four-granularity pre-training data, that R0 directly probes and on which every higher-order capability is built.

\section{Pre-Training Pipeline}
\label{sec:pre-training}

This section details our pre-training methodology. We begin by describing the itemic tokenizer in Section~\ref{sec:pre-training-tokenizer}. Section~\ref{sec:pre-training-data} then introduces our core contribution: a four-granularity data system that hierarchically integrates recommendation corpora with general-domain data to achieve deep modality alignment. Section~\ref{sec:pre-training-ablation} systematically ablates each granularity to quantify its specific contribution to the model's overall performance. Finally, Section~\ref{sec:pre-training-recipe} outlines the three-stage training recipe that operationalizes this data design, from vocabulary warm-up to full-scale training.

\subsection{Itemic Tokenizer}
\label{sec:pre-training-tokenizer}

To obtain compact yet semantically grounded item embeddings, we jointly train a multimodal encoder and a decoder LLM on item-understanding tasks.
The encoder integrates a Vision Transformer (ViT)~\citep{radford2021learning} and an LLM initialized from the open-source Qwen3-VL~\citep{yang2025qwen3technicalreport}, together with an audio encoder~\citep{vyas2026pushing}.
It distills rich content signals---including cover images, video frames, text descriptions, and audio---into a compact dense embedding.
To optimize the item embedding in an end-to-end manner, we prepend the embedding as a soft prefix to a separate decoder LLM, followed by the item's text descriptions.

We quantize these embeddings via RQ-KMeans~\citep{luo2025qarm}.
We use a three-layer codebook, with each layer containing 8192 codes.
Each item is thus represented by an itemic pattern consisting of one domain-aware begin token and three sub-tokens.
For instance, an itemic pattern is formatted as:
$$
\text{<|}\texttt{domain}\text{\_begin|><a\_5028><b\_6733><c\_2559>},
$$
where \texttt{domain} is one of \texttt{video} (video), \texttt{prod} (product), \texttt{ad} (ad), or \texttt{living} (live) for our four recommendation scenarios, or \texttt{sid} for general-domain multimodal data.
Unlike OpenOneRec~\citep{zhou2026openonerectechnicalreport}, we drop the trailing end token.
This reduces the overall context length consumed by itemic patterns, leaving more capacity for the reasoning traces that are essential to our thinking paradigm.

\subsection{Pre-Training Data}
\label{sec:pre-training-data}

Prior generative recommendation works, including OneRec-Think and OpenOneRec, organize pre-training data by task type---item captioning, user behavior sequences, and persona-text interleaving---and mix them for joint training. While this establishes a basic mapping between itemic patterns and natural language, it leaves three structural gaps. First, the semantic expression of items and users tends to be homogeneous, limiting the linguistic diversity the model encounters and weakening representational robustness. Second, the data organization stays at the task level without explicitly modeling the fine-grained semantic hierarchy inside itemic tokens or the relational logic across items. Third, user behavior modeling relies on a narrow conditioning paradigm in which a full user profile maps to a complete behavioral sequence, restricting the model's ability to generalize across diverse recommendation contexts.

To address these gaps, we restructure the pre-training corpora into four progressive granularities, moving from microscopic token semantics to macroscopic user behavior. This hierarchical design systematically expands the alignment between the textual and itemic modalities along three axes: broader expression diversity; explicit structural reasoning from sub-token composition to inter-item relations; and richer conditioning patterns for user modeling. In the remainder of this section, we detail the construction of both the recommendation corpora and the general-domain corpora that jointly realize this vision.

\subsubsection{Recommendation Corpora}

\begin{figure}[t]
    \centering
    \includegraphics[width=0.95\textwidth,height=0.47\textheight,keepaspectratio]{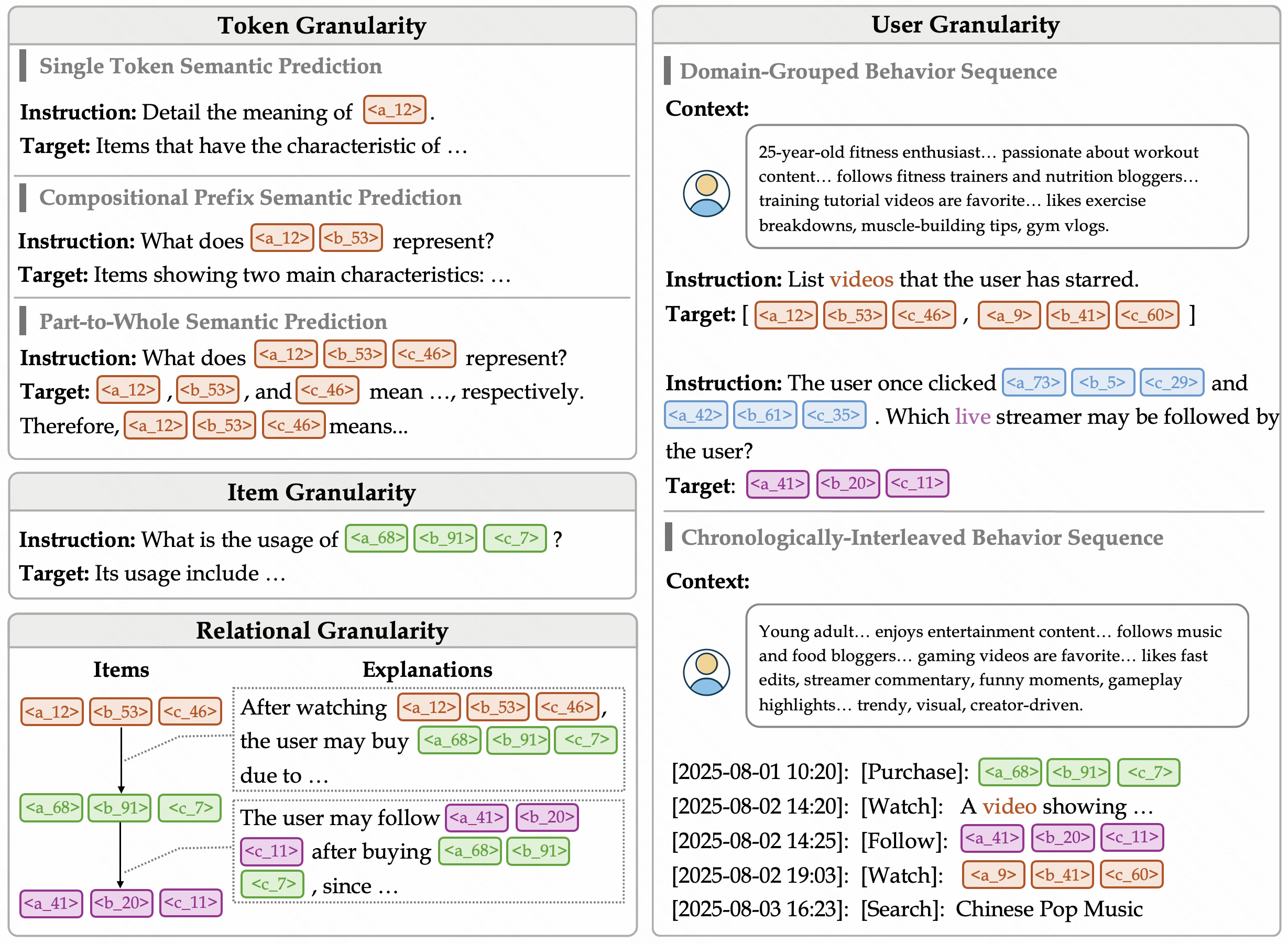}
    \caption{Overview of our four-granularity pre-training recommendation corpora, ordered from fine-grained token-level semantics to coarse-grained user behavior modeling.}
    \label{fig:pre-training_data}
\end{figure}

To achieve a deep and seamless fusion between the textual and itemic modality, as shown in Figure~\ref{fig:pre-training_data}, we structure our pre-training corpora into four progressive granularities:

\begin{itemize}
    \item \textbf{Token Granularity} data focuses on individual or compositional sub-itemic-token semantics and achieves modality alignment at the finest granularity.
    \item \textbf{Item Granularity} data aligns itemic pattern with the textual modality. It pushes cross-modal fusion to the full item level and leverages multi-view tasks and diversified item sources to match complete itemic patterns with comprehensive textual descriptions.
    \item \textbf{Relational Granularity} data explains the connections between items via natural language. It summarizes the interest-transfer logic of associated items and enables collaborative signals to be encoded into text representations and model parameters.
    \item \textbf{User Granularity} data accomplishes thorough cross-modal fusion by enabling the model to understand and generate multi-modal content under unified, personalized user contexts.
\end{itemize}

\begin{figure}[htbp]
    \centering
    \includegraphics[width=\linewidth]{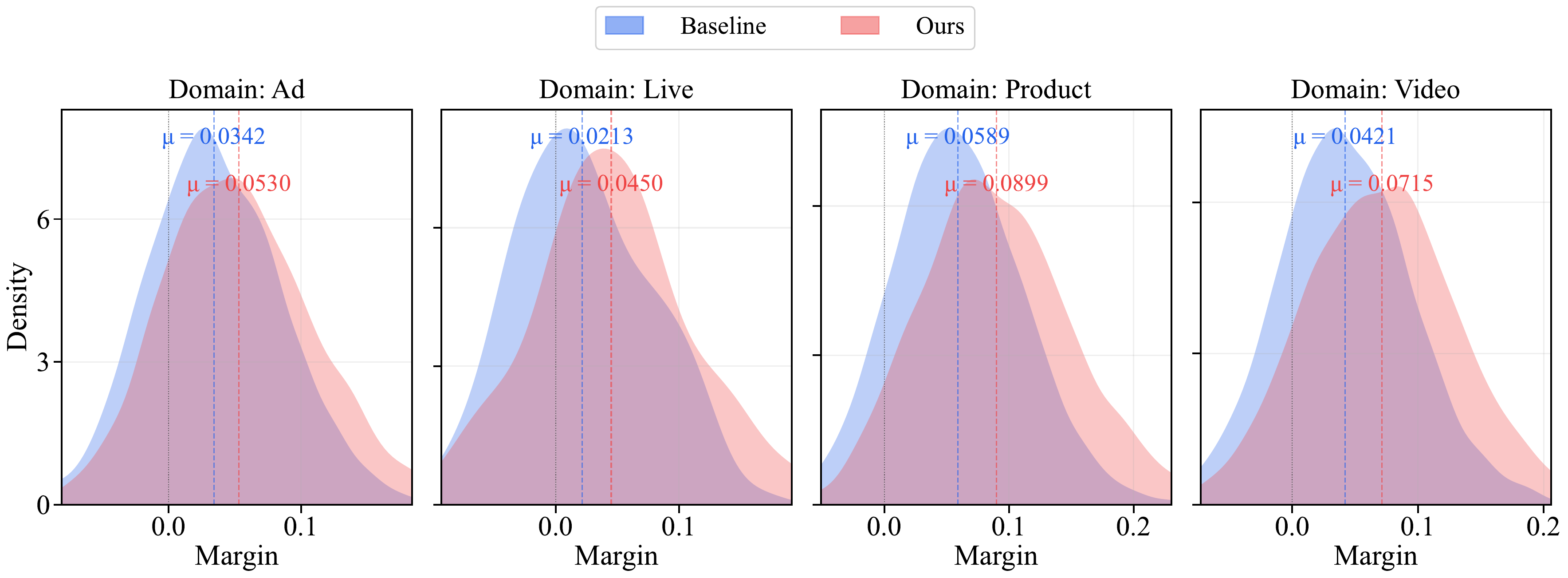}

    \caption{Distribution of the \textit{positive--negative} similarity margin across the four itemic domains. A rightward shift indicates stronger cross-modal alignment. Our corpora consistently outperform the reproduced OpenOneRec baseline.}
    \label{fig:pos-neg-delta-embed}
\end{figure}

To empirically validate the enhanced modality alignment afforded by our multi-granularity corpora, we randomly sample a set of items and extract embeddings for both their itemic patterns and textual captions. Each item is paired with its own caption to form a positive pair, while a caption randomly drawn from another item serves as the corresponding negative. Whereas prior work typically reports positive and negative pair similarities in isolation~\citep{lee2025on,rusak2024infonce}, we additionally compute their difference, yielding a \textit{positive--negative} similarity margin that more directly quantifies the cross-modal alignment. As shown in Figure~\ref{fig:pos-neg-delta-embed}, replacing the pre-training recommendation corpora of OpenOneRec with our proposed corpora consistently shifts the distribution of this margin to the right across all itemic domains, evidencing stronger cross-modal alignment.
We defer the comprehensive methodology of this evaluation to Appendix~\ref{appendix:pretrain-embedding-comp}.

In the remainder of this section, we describe the detailed construction methodology for the pre-training corpora at each of the four granularities.
The detailed token ratios for each corpus are provided in Appendix~\ref{appendix:pretrain-data}.

\paragraph{Token Granularity.}
Achieving precise alignment between language and itemic patterns requires modeling semantics across a continuous spectrum of granularities.
Traditional methods~\citep{zhou2026openonerectechnicalreport} align each itemic pattern with its holistic caption.
At the opposite end, recent explorations~\citep{feng2026finegrainedsemanticsintegrationlarge} focus on single-token semantic prediction by pairing each individual sub-token with its semantic meaning. However, the crucial compositional process through which the semantics of these isolated sub-tokens hierarchically combine into the semantics of the full itemic pattern remains implicit and is not explicitly modeled.

To bridge this gap, our token-granularity data is designed to explicitly model the compositional process through which sub-token semantics are progressively integrated into complete itemic-pattern semantics. Building upon the basic premise of \textbf{single-token semantic alignment}, we further introduce two novel components that strengthen sub-token-level semantic understanding while explicitly modeling the hierarchical composition of semantics across granularities:

\begin{enumerate}
    \item \textbf{Compositional Prefix Semantic Prediction.}
    We formulate tasks that require the model to predict the combined semantic meaning of a prefix sub-token pair <a\_xxxx><b\_xxxx>.
    The data is constructed by leveraging an LLM to summarize the shared semantics of items that share the given prefix.
    This kind of data explicitly teaches the model how isolated semantics merge and interact at the intermediate sub-token-pair granularity.

    We further construct the reverse direction, \textbf{Prefix Itemic Token Grounding}, where the model must retrieve the correct prefix sub-token pair given a textual description of the combined semantics.
    This bidirectional formulation ensures that the model not only interprets what a prefix means but also maps semantic concepts back to the corresponding sub-token space.
    
    \item \textbf{Part-to-Whole Semantic Prediction.}
    We structure the data to elicit a two-step semantic prediction process.
    The first step predicts the fine-grained semantic explanation for each individual sub-token within the complete sub-token sequence corresponding to a given item, while the second step subsequently synthesizes these discrete semantics into a cohesive, holistic caption for the item.
    The data thus enforces a structured, multi-step semantic prediction paradigm from parts to the whole.
\end{enumerate}

\paragraph{Item Granularity.}
Similar to the \textit{Itemic Dense Caption Data} introduced in OpenOneRec~\cite{zhou2026openonerectechnicalreport}, which comprises a bi-directional mapping between itemic patterns and captions, our item-granularity data serve to connect itemic patterns with natural-language descriptions.
However, our item-granularity data significantly advances this paradigm through two critical enhancements:

\begin{enumerate}

    \item \textbf{Capacity-Aware Caption Coarse-Graining.}
    The captions used as item-granularity alignment targets are often packed with very fine-grained details, yet each item is represented by only three sub-tokens.
    These sub-tokens capture an item's coarse semantics but cannot encode such fine details, so aligning them with an over-detailed caption forces the model to hallucinate and pollutes the semantic space.
    We therefore coarse-grain each caption before alignment, keeping only what three sub-tokens can reliably carry, via three operations.
    (i) We remove instance-specific noise that the item code cannot recover, such as on-screen OCR text, ASR lyrics, dates, and product model numbers.
    (ii) We map fine-grained continuous attributes onto coarse buckets---an exact price becomes a price band and an exact age becomes a semantic label such as \textit{young adult}---and we drop decorative modifiers while keeping the underlying entity (e.g., ``pink top with a brown skirt'' becomes ``top with a skirt'').
    (iii) We preserve the high-discrimination content skeleton, including the item category, brand, IP, and title names, core selling points, target audience, and key material.
    Since the noise distribution differs across business domains, we design a separate set of coarse-graining rules for video, product, live, and ad.
    
    \item \textbf{Multi-Perspective Item QA.}
    While standard captioning builds an initial perception of an item, it often falls short of capturing its nuanced attributes.
    To deepen alignment, we further construct multi-perspective question-answering data that queries the model from various angles, including target-audience preferences, core product attractions, visual styles, and negative-feedback rationales.
    This task design elevates alignment from a singular descriptive mapping to a robust, multi-dimensional semantic interrogation of the itemic pattern.
    
\end{enumerate}

\paragraph{Relational Granularity.}
In recommendation scenarios, semantic alignment remains incomplete without capturing the collaborative connections between items.
Existing paradigms typically rely on implicit collaborative-filtering signals from raw user interaction sequences, leaving item-to-item transitions devoid of explicit and interpretable semantic grounding.
To bridge this gap, our relational-granularity data explicitly aligns these collaborative signals with the textual space by constructing interleaved sequences of itemic patterns and natural-language explanations.
Specifically, each training sample takes the following form:
\begin{equation}
\text{Itemic\_Pattern}_0 \rightarrow \text{Textual\_Explanation}_0 \rightarrow \text{Itemic\_Pattern}_1 \rightarrow \cdots \rightarrow \text{Itemic\_Pattern}_n.
\end{equation}
This data corpus encompasses both \textbf{direct item-to-item associations} (where $n=1$) and \textbf{multi-item interest flows} (where $n>1$).

We use the internal \textit{search-after-play} table to construct
\textbf{direct item-to-item associations}.
Each row in this table links a consumed source video to a specific search result product.
This connection is established when a user issues a search query after a valid playback and positively engages with a product from the search results.
The accompanying natural-language explanation is generated by fusing the captions of the video and the product into predefined templates.

Compared with direct item-to-item associations, the \textbf{multi-item interest flows} not only feature greater lengths and encapsulate richer collaborative signals per sample, but also provide broader modality coverage, ensure inter-item non-triviality, and yield higher explanation quality.
\begin{enumerate}
    \item To expand the \textbf{modality coverage}, we further include two data sources
    apart from the \textit{search-after-play} table.
    The first source is the internal \textit{TagNext CF Relations}, where collaborative video pairs are extracted from positive user behaviors.
    The second involves aggregating co-occurring item pairs within sliding windows over temporally ordered, multi-domain user behavior sequences.
    Specifically, an item pair is retained if a substantial number of users provide positive feedback on two items that appear within a specific window size in their temporal behavior sequence.
\item To improve the \textbf{inter-item non-triviality}, we employ a dedicated process.
    First, we build a cross-user global item graph based on the three aforementioned data sources and randomly sample item chains from it.
    The sampled item chains capture latent relations that may not be prominent within a single user's behavior sequence.
    Since transitions between adjacent items in a sampled chain $[Item_0, Item_1, Item_2, \dots]$ may be somewhat intuitive, we further sample a sub-chain $[Item_0, Item_{\text{interval}}, Item_{2\cdot \text{interval}}, \dots]$ to form the training sample.
    Sub-chains are discarded if the semantic similarity between any two adjacent item nodes within the sub-chain exceeds a certain threshold.

\item To boost the \textbf{explanation quality}, we leverage an LLM to generate natural-language explanations that articulate the underlying interest transitions from $Item_{i\cdot \text{interval}}$ to $Item_{(i+1)\cdot \text{interval}}$.
    To ground these explanations in real-world transition paths, the LLM is also provided with the semantic captions of the intermediate item nodes $[Item_{i\cdot \text{interval}+1}, \dots, Item_{(i+1)\cdot \text{interval}-1}]$.
\end{enumerate}

\paragraph{User Granularity.}

As the final part of our pre-training recommendation corpora, the user-granularity data expands the cross-modal alignment to the complete sequential evolution of user interactions.

A key prerequisite for this alignment is high-quality user-side text. However, raw user profiles assembled from business logs typically suffer from noise, incoherence, and rigid template patterns. We therefore employ an open-source LLM to recaption each profile across diverse aspects and formats, transforming them into fluent and natural narratives while preserving the original factual content. Building on these purified profiles, we design two data formats that seamlessly integrate natural language with itemic patterns to enable deep modality alignment:

\begin{enumerate}
    \item \textbf{Domain-Grouped Behavior Sequences.}
    Each sequence in this format is anchored by a textual user profile, followed by the itemic patterns derived from the user's interaction history.
    The involved itemic patterns are grouped into sub-sequences by their domains.
    Rather than unfolding the domain-grouped subsequences as a flat narrative, we dynamically structure them through multi-turn QA dialogues.
    By using natural language to explicitly query specific domain behaviors---often conditioning the prediction in one domain on the history in another---the text acts as an active control signal.
    The model must comprehend the semantic constraints of the question and map them accurately into the itemic space, thereby solidifying a robust alignment between the textual and itemic modalities.
    \item \textbf{Chronologically Interleaved Behavior Sequence.}
    This data format reconstructs the user behavior sequence strictly by timestamps, naturally interleaving interactions across all domains.
    This timeline-based restructuring enables the model to differentiate short-term behavioral bursts from long-term interest evolution.
    Most importantly, to further enforce cross-modal alignment, we probabilistically replace a subset of itemic patterns within this timeline with their corresponding textual captions.
    This creates an interleaved, hybrid timeline that forces the model to achieve deep semantic integration between continuous natural language and discrete itemic patterns within a unified sequential context.
\end{enumerate}

\subsubsection{General-Domain Corpora}
\paragraph{Text Corpora.}
Beyond the four-granularity recommendation corpora, we maintain a substantial proportion of general-domain text data throughout pre-training, spanning mathematical reasoning, code, scientific literature, medical text, and general instruction-following corpora.
This mixture serves to preserve the model's core capabilities in logical reasoning, structured knowledge understanding, and instruction adherence while it absorbs recommendation-specific knowledge.

\paragraph{Multimodal Corpora.}
Beyond text-only corpora, we further enrich pre-training with diverse multimodal data, which we convert into itemic-text format by tokenizing the images into itemic patterns. These data span four complementary sources: \textbf{image-text understanding pairs}, \textbf{text-image interleaved documents}, \textbf{image-editing triplets} (source image, editing instruction, target image), and \textbf{text-to-image generation data}. Across all four sources, we selectively retain only samples whose visual semantics operate at a coarse granularity---such as overall image style, dominant subject matter, and high-level scene composition---while discarding samples that hinge on fine-grained details that exceed the representational capacity of three sub-token codes. This deliberate diversity in data format, together with the coarse-granularity filter, jointly serves three purposes:
\begin{enumerate}
    \item First, each source brings a distinct modality composition and linguistic style, collectively acting as a regularizer that prevents the model from overfitting to the homogeneous patterns of rule-synthesized recommendation corpora.
    \item Second, these datasets are curated with high-quality semantic annotations, providing clean supervision signals for concept grounding and compositional visual reasoning.
    \item Third, by mapping open-world visual concepts into itemic patterns, they function as external semantic anchors that expand the model's knowledge boundary beyond the closed recommendation domain, thereby maintaining general multimodal-understanding capability throughout intensive recommendation training.
\end{enumerate}

\paragraph{Data Composition Strategy.}
Since we continue pre-training from the instruction-tuned Qwen3 backbone~\citep{yang2025qwen3technicalreport}, a central concern is injecting domain knowledge without eroding its pre-existing instruction-following ability and textual fluency~\citep{parmar2024reusedontretrainrecipe,ovadia2025knowledge}. To this end, and unlike OpenOneRec~\citep{zhou2026openonerectechnicalreport}, we adopt a unified next-token prediction objective and route noisy, template-synthesized business data into the loss-masked context, reserving the loss-bearing target for higher-quality text. Consequently, the effective training targets naturally exhibit a roughly 1:1 ratio of QA-style and free-form data: the former keeps the model aligned with instruction formats, while the latter anchors it in natural text distributions. The loss-masked context further allows the model to condition on domain signals without learning to reproduce their noise.

\subsection{Ablation Study on Recommendation Corpora}
\label{sec:pre-training-ablation}

\subsubsection{Experimental Settings}

We perform the ablation study on the 0.8B model with a fixed budget of 30B tokens, keeping the computational cost manageable while our main results are obtained on the 8B model. The five configurations are defined as follows:

\begin{itemize}
    \item \textbf{Exp1 (Baseline):} the standard OpenOneRec dataset, comprising original item-caption data and user-granularity data (including user-behavior data and user-persona grounding data).
    \item \textbf{Exp2 (Exp1 + Token):} incorporates our token-granularity data into Exp1, proportionally downsampling the original mixture to maintain the budget.
    \item \textbf{Exp3 (Exp2 + Item):} replaces the baseline item-caption data in Exp2 with our item-granularity data.
    \item \textbf{Exp4 (Exp3 + Relational):} integrates our relational-granularity data into Exp3, rebalancing the overall mixture accordingly.
    \item \textbf{Exp5 (Exp4 + User):} substitutes the baseline user-granularity data in Exp4 with our user-granularity data.
\end{itemize}

Notably, the token and relational granularity data introduce novel capabilities absent from the baseline, whereas the item and user granularity data serve as upgraded replacements for their baseline counterparts. This sequential pipeline effectively isolates the marginal contribution of each granularity while ruling out any confounding effects from increased data volume.

\begin{table*}[htbp]
\footnotesize
\centering
\caption{Each experiment builds on the previous one by adding a new granularity or replacing an existing component. 
We report \texttt{pass@64} (\%) for Itemic Pattern Grounding and cross-domain recommendation, and macro-weighted F1 (\%) for the remaining tasks. All values are reported as percentages. The results show a consistent performance progression, with the largest gains in cross-domain recommendation emerging after User-granularity data is introduced.}
\label{tab:ablation_results}
\begin{tabular}{llccccc}
\toprule
\multicolumn{2}{c}{\textbf{Task}} & \textbf{Exp1} & \textbf{Exp2} & \textbf{Exp3} & \textbf{Exp4} & \textbf{Exp5} \\
\midrule
\multirow{4}{*}{\textbf{R0: Itemic Pattern Grounding}}
& ad    & 0.48 & 0.80 & \textbf{2.40} & 0.48 & \underline{1.92} \\
& live  & 0.40 & \underline{1.01} & \textbf{1.21} & \textbf{1.21} & \textbf{1.21} \\
& prod  & 2.42 & \textbf{5.81} & \underline{5.33} & 2.91 & \underline{5.33} \\
& video & 0.35 & 0.35 & 0.87 & \textbf{1.05} & \textbf{1.05} \\
\cmidrule(lr){1-7}
\multirow{4}{*}{\textbf{R0: Item Understanding}}
& ad    & 16.37 & \textbf{37.86} & 31.65 & 29.69 & \underline{32.56} \\
& live  & 35.16 & \textbf{46.16} & \underline{39.27} & 37.39 & 35.52 \\
& prod  & 20.58 & 28.62 & \underline{29.51} & \textbf{29.78} & 28.75 \\
& video & 22.56 & \textbf{36.76} & 30.77 & 28.68 & \underline{31.63} \\
\cmidrule(lr){1-7}
\multirow{1}{*}{\textbf{R1: Item2Item QA}} & & 0.00 & 20.57 & 20.73 & \underline{25.65} & \textbf{29.72} \\
\cmidrule(lr){1-7}
\multirow{1}{*}{\textbf{R2: Evolution Direct Gen}}& mixed & 0.13 & 0.00 & \underline{0.32} & 0.18 & \textbf{0.37} \\
\cmidrule(lr){1-7}
\multirow{4}{*}{\textbf{R3: Cross-Domain Rec.}}
& ad    & 9.06 & 8.75 & \underline{9.54} & 8.58 & \textbf{10.84} \\
& live  & 2.29 & 2.32 & \underline{3.49} & 3.25 & \textbf{8.56} \\
& product  & 1.15 & 0.94 & 1.29 & \underline{1.54} & \textbf{1.65} \\
& video & 0.65 & \textbf{0.74} & 0.72 & \underline{0.73} & 0.66 \\
\bottomrule
\end{tabular}
\end{table*}

\subsubsection{Results}
To systematically demonstrate the effectiveness of our multi-granularity data construction pipeline, we present the performance trajectory across OneReason-Bench tasks (Table~\ref{tab:ablation_results}). The results reveal a clear capability progression, confirming that each granularity level successfully injects its targeted compositional or semantic competencies into the model.

\paragraph{Exp1 $\rightarrow$ Exp2: Token Granularity.}
Adding token-granularity data sharply improves \textbf{R0} retrieval and understanding across all domains (e.g., \texttt{Itemic Pattern Grounding\_prod} 2.42\% $\rightarrow$ 5.81\%; \texttt{Item Understanding\_ad} 16.37\% $\rightarrow$ 37.86\%). \textbf{R1} Item2Item QA surges from 0.00\% to 20.57\%, confirming that token-level QA-style data unlocks the QA format absent in the baseline. \textbf{R2} temporarily drops to 0.00\%, indicating disruption of higher-level reasoning.

\paragraph{Exp2 $\rightarrow$ Exp3: Item Granularity.}
Replacing the baseline's captions with multi-perspective item data pushes several \textbf{R0} retrieval metrics to their peaks. \textbf{R2} recovers to 0.32\%, and \textbf{R3} cross-domain recommendation improves (e.g., Cross-Live 2.32\% $\rightarrow$ 3.49\%). \textbf{R0} Item Understanding regresses on most domains, attributable to residual noise in business-sourced captions (see Table~\ref{tab:item_hallucination_case}).

\begin{table*}[t]
\centering
\caption{A representative Item Understanding case in the \texttt{video} domain. For each model, we show its actual caption (excerpted; omissions are marked by ellipses) together with an English analysis. The baseline (Exp1) invents a drama title and degenerates into repetitive, contentless lines; after the token-granularity data is added (Exp2), the model stays faithful to the genre; once the item-granularity data is further incorporated (Exp3), the model fabricates a concrete, wrong drama title and re-frames the clip as a medical drama. Fabricated or incorrect content is marked in {\color{red}red}.}
\label{tab:item_hallucination_case}
\setlength{\tabcolsep}{3pt}
\renewcommand{\arraystretch}{1.05}
\begin{CJK*}{UTF8}{gbsn}
{\scriptsize
\begin{tabular}{@{}*{4}{p{\dimexpr(\textwidth-6\tabcolsep)/4\relax}}@{}}
\toprule
\multicolumn{4}{@{}p{\textwidth}@{}}{\textbf{Prompt:}~请解读以下视频\,token\,并还原视频内容：~\texttt{<|video\_begin|><a\_8164><b\_6973><c\_7395>}} \\
\midrule
\textbf{Ground Truth} & \textbf{Exp1 (Baseline)} & \textbf{Exp2 (Exp1 + Token)} & \textbf{Exp3 (Exp2 + Item)} \\
{\color{gray}\textbf{[Reference]}} & {\color{red}\textbf{[Repetitive Hallucination]}} & {\color{teal}\textbf{[Faithful Genre]}} & {\color{red}\textbf{[Fabricated Title]}} \\
\midrule
这是一段都市言情剧情短视频\ldots{}女主角池晚作为神医救醒了病危的霍老爷子，霍老爷子为报恩极力撮合池晚与小儿子霍南洲结婚\ldots{}前未婚夫霍司寒决定与池晚退婚、迎娶白月光池娇\ldots{}两份婚书同时送到池家\ldots{}最终池晚选择了霍南洲的婚书。
\par\smallskip
{\tiny\color{gray}\emph{Analysis:} a modern urban-romance drama about an arranged-marriage conflict within a wealthy family.}
&
这段视频是关于一部名为{\color{red}《我爱的他》}的电视剧的\ldots{}通过对话和场景切换展现角色之间复杂的情感纠葛\ldots{}角色之间的对话如{\color{red}``我爱的他''、``我爱的他''}\ldots{}揭示了复杂的情感纠葛和关系变化。
\par\smallskip
{\tiny\color{gray}\emph{Analysis:} invents a drama title and degenerates into repeated, contentless lines, capturing only a vague emotional conflict and nothing specific to the plot.}
&
这些短视频主要属于剧情类短剧与影视解说\ldots{}强剧情冲突与反转：聚焦人物关系的复杂纠葛（如豪门恩怨、复仇、误会）\ldots{}鲜明的人物标签（如``霸道总裁''、``灰姑娘''）\ldots{}面向喜爱现代都市情感、家庭伦理剧的受众\ldots{}
\par\smallskip
{\tiny\color{gray}\emph{Analysis:} correct at the genre level (modern urban romance / family-conflict short drama) and consistent with the ground truth, though generic and not tied to the specific story.}
&
这是一个关于{\color{red}《医祖凌云传》}的故事\ldots{}女性角色身着白色上衣展现专业形象，男性角色穿着西装\ldots{}画面中出现文字{\color{red}``医祖凌云传''}\ldots{}还出现了{\color{red}手术台、医疗器械}等元素\ldots{}兴趣点在于吸引喜爱{\color{red}《医祖凌云传》和关注医疗题材}的观众。
\par\smallskip
{\tiny\color{gray}\emph{Analysis:} after item-caption data is added, the model fabricates a concrete, wrong title and re-frames the clip as a medical drama, contradicting the actual urban-romance plot; the hallucination worsens.}
\\
\bottomrule
\end{tabular}
}
\end{CJK*}
\end{table*}

\paragraph{Exp3 $\rightarrow$ Exp4: Relational Granularity.}
Integrating relational data lifts \textbf{R1} Item2Item QA to 25.65\% and \textbf{R3} Cross-Prod to 1.54\%. Some \textbf{R0} retrieval metrics decline (e.g., \texttt{Itemic Pattern Grounding\_prod} 5.33\% $\rightarrow$ 2.91\%), reflecting a trade-off toward relational reasoning. \textbf{R2} dips slightly (0.32\% $\rightarrow$ 0.18\%), since item-to-item relational data does not directly benefit this user-side task.

\paragraph{Exp4 $\rightarrow$ Exp5: User Granularity.}
Substituting the baseline's user data with our temporally-aware sequences pushes \textbf{R2} to its peak (0.37\%) and drives \textbf{R3} cross-domain recommendation to its highest levels, most notably Cross-Live (3.25\% $\rightarrow$ 8.56\%) and Cross-Ad (8.58\% $\rightarrow$ 10.84\%). \textbf{R1} reaches 29.72\%, and \textbf{R0} retrieval metrics recover to earlier peaks.

\paragraph{Key Insights.}
The ablation reveals three findings: (1) Token- and relational-granularity data expand the model's skill set through augmentation, while item- and user-granularity data refine existing capabilities through replacement. (2) Capability trade-offs emerge during training---gains in relational reasoning may temporarily reduce static retrieval, but subsequent stages recover these capabilities. (3) User granularity acts as the ultimate integrator, contextualizing all prior knowledge within temporal behavior to achieve peak performance across diverse recommendation scenarios.

\subsection{Training Recipe}
\label{sec:pre-training-recipe}

\begin{figure}[htbp]\centering 
    \includegraphics[width=0.75\textwidth]{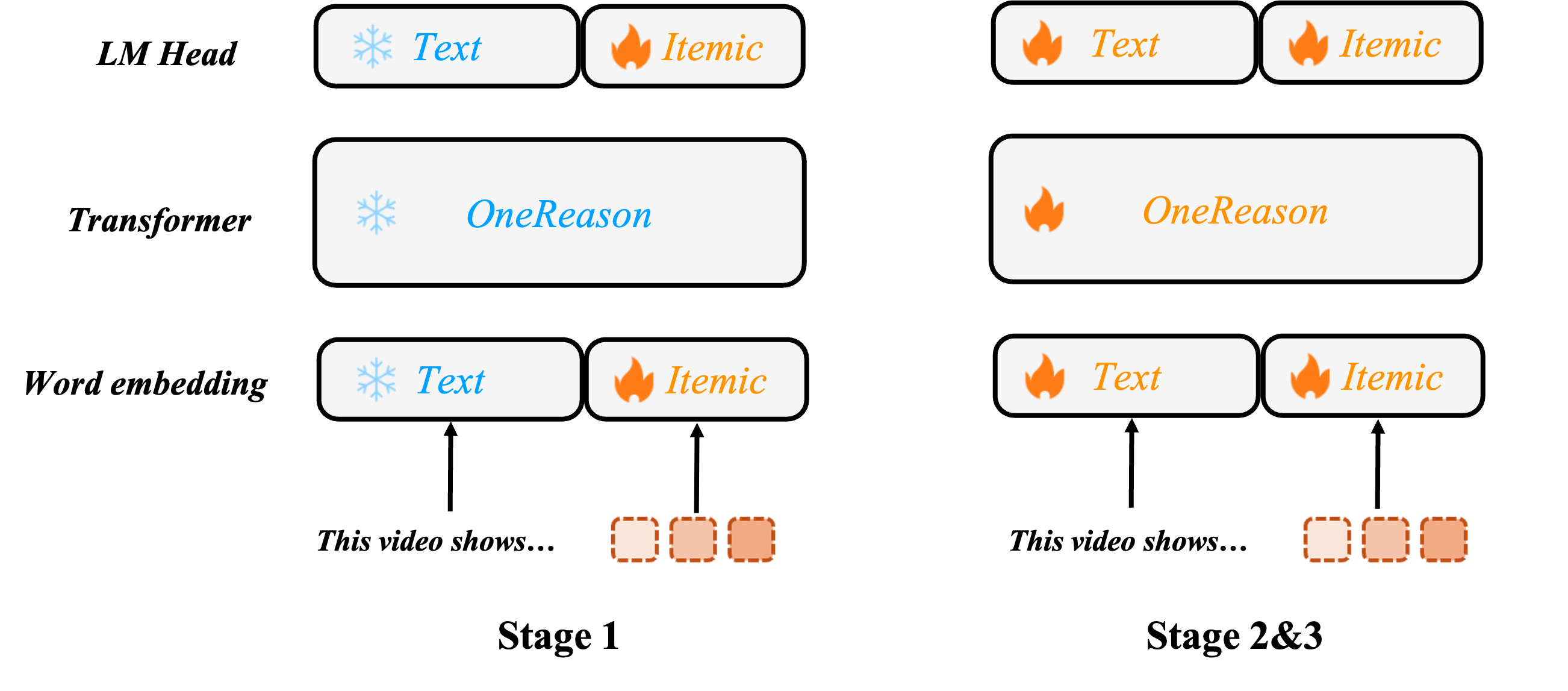}
    \caption{The three-stage pre-training recipe. Stage~1 only tunes the incremental itemic-pattern-related parameters, while Stages~2 and~3 tune all parameters.}
    \label{fig:pretrain_pipeline}
\end{figure}

Our training recipe, illustrated in Figure~\ref{fig:pretrain_pipeline}, unfolds in three stages.
The core hyperparameters of each stage are summarized in Table~\ref{tab:training_recipe}.

\begin{table}[H]
  \centering
  \small
  \caption{Summary of training stages.}
  \label{tab:training_recipe}
  \resizebox{0.82\textwidth}{!}{%
  \begin{tabular}{lccc}
    \toprule
    \textbf{Configuration} & \textbf{Stage 1} & \textbf{Stage 2} & \textbf{Stage 3} \\
    \midrule
    Trainable parameters & Extended vocab. $+$ LM head & All & All \\
    Learning rate & $2\times10^{-4} \rightarrow 1\times10^{-4}$ & $1\times10^{-4} \rightarrow 1\times10^{-5}$  & $1\times10^{-5} \rightarrow 1\times10^{-6}$  \\
    Token budget & 110B & 449B & 19B \\
    \bottomrule
  \end{tabular}%
  }
\end{table}

All stages employ sample packing to improve sequence utilization, producing packed sequences of 48K tokens.
The key distinction across stages lies in the maximum length of individual samples within each packed sequence.
Stage~1 and Stage~2 restrict individual samples to 4K tokens to prioritize training throughput.
In Stage~1, only the newly introduced embeddings and the corresponding LM head weights are trained, while the rest of the backbone remains frozen, allowing the newly introduced itemic-pattern embeddings to settle into the semantic space without disturbing the pre-trained weights.
Stage~2 unfreezes all parameters for full joint training, enabling the model to absorb recommendation knowledge from the four-granularity data mixture.
Stage~3 lifts the per-sample length limit to 32K tokens, allowing the model to train on full user histories with long-range behavioral dependencies.

\section{SFT Pipeline}\label{sec:sft}
\providecommand{\traceStage}[1]{\textbf{\textit{#1}}}
\providecommand{\sftVideoSID}[3]{\texttt{<|video\_begin|>}\allowbreak\texttt{<a\_#1>}\allowbreak\texttt{<b\_#2>}\allowbreak\texttt{<c\_#3>}}
\providecommand{\sftProdSID}[3]{\texttt{<|prod\_begin|>}\allowbreak\texttt{<a\_#1>}\allowbreak\texttt{<b\_#2>}\allowbreak\texttt{<c\_#3>}}
\providecommand{\sftLiveSID}[3]{\texttt{<|living\_begin|>}\allowbreak\texttt{<a\_#1>}\allowbreak\texttt{<b\_#2>}\allowbreak\texttt{<c\_#3>}}

Itemic--text alignment pre-training gives the model a necessary perception foundation: it can associate itemic tokens with natural-language content and has absorbed collaborative signals from user behavior. SFT builds on this foundation by cultivating recommendation cognition, namely the ability to operate over grounded itemic tokens under instruction format. We organize this stage around the same R0--R3 hierarchy, but with two complementary training axes. The \emph{\textbf{compression axis}} teaches the model to reduce long and noisy user histories into typed persona states and compact interest-evolution motifs, so that the later reasoning step can compare a small set of evidence-backed hypotheses rather than a raw behavior log. The \emph{\textbf{reasoning axis}} teaches dynamic operations: R1 derives one-hop item-to-item bridges, R2 tracks temporal interest evolution, and R3 performs transition judgement for next-interaction prediction. Section~\ref{sec:sft-data-r0} introduces R0 perception supervision; Section~\ref{sec:sft-data-r1} and Section~\ref{sec:sft-data-r2} describe R1 derivation and R2 evolution reasoning; Section~\ref{sec:sft-data-r3} presents R3 recommendation supervision, where compression and transition judgement are composed into recommendation CoT; and Section~\ref{sec:sft-data-sid-instruct} and Section~\ref{sec:sft-data-general} cover itemic-instruction and general-domain data for instruction robustness.

\subsection{R0: Perception}
\label{sec:sft-data-r0}

Building on the four-granularity alignment established during pre-training, we further refine the data and carry out high-quality CoT capability construction. R0 provides the perception layer for the SFT stage. Its role is to make itemic tokens readable, controllable, and semantically usable before the model is asked to perform higher-level recommendation reasoning. The caption data covers four major content domains, short videos, e-commerce products, live streamings, and advertising. QA data is narrower: it covers short-video and e-commerce product domains, while live-streaming and advertising contribute caption-style grounding only. R0 therefore contains two complementary task families. The first is \textbf{bidirectional mapping between itemic tokens and natural-language descriptions}, which reinforces item$\leftrightarrow$language grounding under instruction-style supervision. The second is \textbf{question-answering (QA) over platform content}, which expands perception beyond caption reconstruction to attribute recognition, audience understanding, negative-feedback analysis, and open-ended semantic comprehension for the two QA-covered domains.

Since the pre-training stage has already established the hierarchical itemic-token space, R0 does not introduce a new tokenizer or repeat the full token-granularity construction. Instead, it turns this hierarchy into explicit SFT supervision. Table~\ref{tab:sft-r0-pretrain-comparison} shows this difference with the same video itemic tokens: pre-training provides a direct itemic-token-to-caption mapping, whereas R0 asks the model to explain the hierarchical itemic tokens before producing the caption. In this case, the CoT first uses the coarse token to identify a character-centered or ACG-style content type, then uses the middle token to narrow the evidence toward young-female visual presentation, and finally uses the fine token to support the conclusion about personal style and aesthetic display. The key difference is therefore not the target caption itself, but the supervised verbalization of how coarse-to-fine itemic-token semantics support that caption. More bilingual R0 perception examples are provided in Appendix~\ref{appendix:sft-r0-item-perception}.

The caption data includes both thinking and non-thinking variants. In the thinking format, the model first explains how each itemic token contributes to the final item meaning and then produces the final caption. This explicit coarse-to-fine trace teaches itemic-token-to-text grounding rather than recommendation decision-making: it supplies the semantic substrate later reused by R1 derivation, R2 evolution, and R3 recommendation reasoning. The unCoT variant, in parallel, supervises direct itemic-token-to-caption generation, preserving generation quality and efficiency when explicit reasoning is unnecessary.

\begin{table}[H]
\centering
\caption{Representative R0 case comparing item--caption pre-training with instruction-style SFT perception supervision.}
\label{tab:sft-r0-pretrain-comparison}
\setlength{\tabcolsep}{3pt}
\renewcommand{\arraystretch}{1.18}
\footnotesize
\begin{CJK*}{UTF8}{gbsn}
\begin{tabularx}{\linewidth}{@{}>{\raggedright\arraybackslash}X@{}}
\toprule
\textbf{Pre-training itemic-token-to-caption mapping} \\
\textbf{User:} \sftVideoSID{7300}{7894}{4541}
\par\smallskip
\textbf{Assistant:}
视频为一段以二次元角色扮演为主题的短视频，主要展示一位身着黑色紧身战斗服、金发的女性角色在铁轨上的摆拍画面，以及一位身穿黑色吊带裙的女性在停车场奔跑的片段。视频配有“甜妹”“萌妹子”等标签，整体风格偏向二次元美学与个人形象展示。
\par\smallskip
{\scriptsize\color{gray}\emph{Signal:} pre-training provides direct itemic-token-to-caption grounding. It teaches the model to associate the itemic tokens with the caption, but does not explicitly verbalize how the individual itemic tokens support the description.}
\\
\midrule
\textbf{SFT instruction-style perception} \\
\textbf{User:} 视频 token \sftVideoSID{7300}{7894}{4541} 包含哪些视频内容？
\par\smallskip
\textbf{Assistant:} \texttt{<think>}\allowbreak{} 从 {\color{red}\texttt{<a\_7300>}} 来看，该视频属于以角色或人物形象展示为核心的内容类型，常见于二次元文化、Cosplay 或情感化美学展示领域；结合 {\color{red}\texttt{<b\_7894>}}，可以进一步确定主角是年轻女性，内容围绕美女展示、时尚穿搭或 Cosplay 展开；再融合 {\color{red}\texttt{<c\_4541>}}，视频在视觉呈现上具有较强个人风格和审美导向。三层语义交集指向以年轻女性角色形象为核心、融合二次元或 Cosplay 元素、注重视觉美学与情感氛围营造的展示类短视频。\texttt{</think>}\par
视频为一段以二次元角色扮演为主题的短视频，主要展示金发女性角色摆拍、女性奔跑等片段，配合“甜妹”“萌妹子”等标签，吸引喜爱美女欣赏和视觉审美的年轻观众。
\par\smallskip
{\scriptsize\color{gray}\emph{Signal:} the same itemic-token evidence is converted into instruction-followable perception; the CoT makes the coarse-to-fine grounding explicit before the final caption.}
\\
\bottomrule
\end{tabularx}
\end{CJK*}
\end{table}

QA supervision is retained as an instruction-style perception supplement: it extends caption grounding from itemic-token decoding to content-aware question answering over structured attributes and open-ended semantics in the video and product domains. Together, caption and QA data convert the aligned itemic-token perception acquired during pre-training into instruction-followable perception ability. In total, R0 contains approximately \textbf{941K} samples, including about \textbf{682K} caption samples and \textbf{259K} QA samples.

\subsection{R1: Derivation}
\label{sec:sft-data-r1}

R1 targets \textbf{cross-item relation derivation}. It extends R0 from understanding an individual item to explaining why one item may naturally lead to another. The intended capability is not generic item similarity. Instead, R1 teaches the smallest useful \textit{source-to-follow-up bridge}: given a source item during student SFT, the model should generate and justify a plausible follow-up itemic-token answer by identifying the need, clue, scenario, constraint, or refinement on the source side. Source--destination pairs are used during teacher-side data filtering and consistency checking; the student training input remains source-only, and the source prompt tokens are not counted as target loss. This local bridge-building ability later becomes part of R3 recommendation CoT, where the model must connect observed behavior, textual evidence, and target-domain decisions rather than rely on surface textual overlap.

R1 is constructed from the same raw candidate pools used in relational pre-training, including TagNext CF relations and after-play-search data. The difference is the supervision standard. Relational pre-training exposes the model to large-scale collaborative transitions, whereas R1 requires each retained pair to support explicit, one-hop, evidence-grounded reasoning. After deduplication, diversification, and embedding-similarity filtering, a locally deployed flagship LLM assigns an explicit-relevance judgement to each pair: \textit{explicitly related}, \textit{explicitly unrelated}, or \textit{uncertain}. We retain only pairs whose relation can be directly supported by visible item evidence, such as item metadata and dense captions, and remove weak pairs whose connection depends mainly on broad assumptions such as audience overlap, popularity, or platform co-occurrence. The full judgement prompt and its English translation are provided in Appendix~\ref{appendix:sft-r1-explicit-judgement-prompt}.

\begin{table}[H]
\centering
\caption{Representative R1 case comparing message-style relational pre-training with one-hop SFT derivation supervision.}
\label{tab:sft-r1-pretrain-comparison}
\setlength{\tabcolsep}{3pt}
\renewcommand{\arraystretch}{1.18}
\footnotesize
\begin{CJK*}{UTF8}{gbsn}
\begin{tabularx}{\linewidth}{@{}>{\raggedright\arraybackslash}X@{}}
\toprule
\textbf{Pre-training one-hop data}
\par\smallskip
\textbf{User:} 商品\sftProdSID{2094}{4972}{1852}与什么视频内容最相关？
\par\smallskip
\textbf{Assistant:} \sftVideoSID{1776}{5416}{9172}
\\
\midrule
\textbf{SFT one-hop reasoning}
\par\smallskip
\textbf{User:} 请推荐一条与 \sftVideoSID{1303}{7205}{8005} 主题接近、语义相关的视频。
\par\smallskip
\textbf{Assistant:} \texttt{<think>}\allowbreak{} 我注意到源视频虽然主要展示金刚板等建材的价格与材质，但其描述的温馨家居场景和具体的平米计价方式，精准触达了用户对于装修成本控制的深层关注。这种对硬装材料性价比的敏感，往往会自然激发用户对于整体空间软装搭配的联想。当用户开始计算地板或板材预算时，他们已经在构建一个完整的居住画面，此时床品、地毯等软装不仅是装饰，更是需要纳入同一笔账目的重要支出。基于这种由硬到软的推导逻辑，用户接下来的浏览轨迹很可能滑向家居软装搭配、床品选购或整体空间色彩规划内容。\texttt{</think>}\par
\sftVideoSID{0474}{3348}{1453}
\par\smallskip
{\scriptsize\color{gray}\emph{Signal:} SFT turns the pair into an explicit source-to-target derivation. The target itemic tokens are the answer, while the reasoning must justify the bridge from source-side evidence.}
\\
\bottomrule
\end{tabularx}
\end{CJK*}
\end{table}

Table~\ref{tab:sft-r1-pretrain-comparison} contrasts R1 with relational pre-training. Relational pre-training presents the source item and the related target item in a message-style association. R1 instead keeps a single source itemic-token sequence as the instruction input and supervises the answer side with an explicit bridge followed by the target itemic tokens, making the local relation judgement directly learnable.

For high-confidence candidates, the flagship LLM further extracts compact bridge variables, including source-side need, bridge type, abstract bridge, continuation direction, reason seed, and confidence. These variables separate the relation explanation from the destination item's surface content: the destination item is used only for consistency checking, while the generated explanation must be written from the source-side evidence and bridge variables. This design reduces target leakage and prevents the model from learning to copy target-only entities or restate the destination description as the rationale. The generation prompt is shown in Appendix~\ref{appendix:sft-r1-relation-explanation-prompt}, and bilingual sanitized examples are provided in Appendix~\ref{appendix:sft-r1-training-examples}.

This process distills the raw candidate pools into approximately \(358\)K TagNext CF pairs and \(388\)K after-play-search pairs, from which we sample a refined R1 training set of roughly \textbf{400K} examples. Overall, R1 converts implicit item-to-item signals into concise derivation supervision: it preserves the collaborative signal introduced during pre-training, but upgrades it into explicit bridge reasoning that can be reused by R3 when connecting user evidence to the next-interest direction.

\subsection{R2: Evolution}
\label{sec:sft-data-r2}

R2 targets \textbf{user interest evolution modeling}. It extends recommendation cognition from static preference understanding to temporally structured interpretation of behavior. While R0 grounds item semantics and R1 derives explicit relations between item pairs, R2 asks how intent develops over time: which behaviors act as triggers, which later behaviors refine or correct earlier ones, and how multiple events accumulate into an interpretable trajectory. This capability is essential when the recommendation decision depends not only on what the user interacted with, but also on how the user's underlying need became more specific.

We construct R2 from full-domain user behavior timelines, where heterogeneous actions are organized chronologically. Rather than treating a timeline as a flat context, a strong LLM first identifies key evolution nodes that indicate meaningful shifts, refinements, or closures in user intent. The extraction emphasizes coherent progression, topic continuity, and nontrivial cognitive increment, converting raw behavior streams into candidate evolution chains with explicit temporal structure. In this sense, R2 is the SFT counterpart of user-level pre-training: pre-training exposes the model to long behavior contexts, while R2 teaches it to explain which parts of those contexts form valid interest-evolution paths.

To improve reliability, candidate chains are filtered by a separate LLM-as-a-Judge quality-evaluation stage. This stage checks whether each proposed trajectory is supported by the original timeline and whether its transitions form a plausible, self-consistent reasoning path. The diagnostic criteria cover order sensitivity, cognitive increment, trigger-source evidence, strong causal-style exclusivity, evidence closure, and no-mind-reading constraints. In practice, this filtering removes pseudo-logical chains, random-consumption chains, same-category drift, and weakly grounded transitions before downstream task construction. The main extraction prompt and its English translation are provided in Appendix~\ref{appendix:sft-r2-extraction-prompt}; the complete quality-control prompt contains the full set of eleven checks used for second-pass evaluation.

Based on the validated evolution chains, we construct three SFT task families aligned with the R2 benchmark taxonomy:
\begin{itemize}
    \item \textbf{\textit{Evolution Action Selection}}: Given a user history $\mathcal{H}$ and a specified interest topic or candidate chain description, the model identifies the related historical actions $\mathcal{A}$, encouraging discriminative reasoning over temporal structure.
    \item \textbf{\textit{Evolution Topic Generation}}: Given a user history $\mathcal{H}$ and a specified topic $t$, the model generates the corresponding evolution chain $\mathcal{E}_t$, focusing on topic-conditioned temporal reasoning.
    \item \textbf{\textit{Evolution Direct Generation}}: Given only the user history $\mathcal{H}$, the model directly generates valid evolution chains $\{\mathcal{E}\}$ without topic guidance, requiring the highest degree of autonomy.
\end{itemize}

During data construction, these three families are organized as \textit{selection}, \textit{topic-generation}, and \textit{direct-generation} samples, respectively. Some samples include CoT rationales that make the temporal-transition judgement explicit, while direct-answer samples are kept for response efficiency and format robustness. Representative sanitized examples for all three training formats are provided in Appendix~\ref{appendix:sft-r2-multihop-training-examples}.

The final R2 dataset contains approximately \textbf{130K} training instances. Overall, R2 serves as the temporal cognition layer of the SFT mixture. By transforming user timelines into validated evolution supervision under multiple task formats, it teaches the model to interpret user intent as an evolving process rather than a static profile. This temporal scaffold is later reused by R3, where recommendation traces must compress noisy histories, compare candidate interest directions, and decide which direction is best supported by the user's recent and long-term behavioral evidence.

\subsection{R3: Recommendation}
\label{sec:sft-data-r3}

R3 is the recommendation-level composition layer of the SFT mixture. R0--R2 train dedicated capabilities separately, whereas R3 brings them together in the final next-interaction prediction setting. Its contribution is not another standalone reasoning task, but a trace-construction protocol that places R1-style derivation and R2-style temporal-evolution reasoning over a two-axis compressed user state. The R3 block contains both thinking samples and non-thinking direct-answer samples: the former teach structured recommendation cognition, while the latter preserve direct mapping from user context to target itemic tokens and support non-thinking format alignment. This subsection focuses on the construction and evaluation of the CoT traces.

\subsubsection{Recommendation Reasoning Trace Construction}
\label{sec:sft-cot-synthesis}

Each R3 synthesis sample is built from three ingredients: a user profile $\mathcal{P}_u$, a chronological interaction timeline $\mathcal{H}_u$, and the ground-truth subsequent interaction $y_u^{+}$. The target interaction is used only as the supervision answer and is explicitly prohibited from appearing directly in the reasoning trace. The trace must therefore reason from the observed profile and behavior evidence as if the answer were unknown. During data construction, the teacher first writes the trace over natural-language user evidence and marks referenced historical items with their original domain ids, such as video, product, live-author, or ad ids, so that every concrete evidence reference remains machine-matchable. A subsequent post-processing stage joins this raw trace with the itemic-token-converted user context, replaces each matched historical id with its itemic tokens, removes unmatched references, and writes the final assistant message as a \texttt{\detokenize{<think>}} trace followed by the target itemic-token answer. Thus, the training trace seen by the model is not a text-only rationale: it interleaves natural-language reasoning with itemic-token evidence inside the CoT, while keeping the target itemic tokens outside the reasoning span.

The central design principle is to treat recommendation CoT as a \textbf{two-axis compression plus transition judgement} process. The first axis is persona-type compression, which summarizes stable preference priors, demographic or life-stage clues, temporal rhythm, content preference, price sensitivity, interaction depth, and possible shared-device ambiguity. The second axis is interest-evolution-type compression, which identifies recent trajectory motifs such as triggering search, need expansion, parameter narrowing, scenario continuation, saturation-driven substitution, cross-domain echo, or closure from browsing to purchase. These compressed states reduce long-history noise and provide typed variables that the subsequent reasoning stage can manipulate.

Compression in this setting is task-directed understanding, not a generic summary of the observed history. It preserves evidence useful for next-interaction prediction and suppresses routine, accidental, or weakly supported behavior. This design is motivated by the information bottleneck view of representation learning, where useful representations preserve prediction-relevant information while suppressing input noise~\citep{tishby2015deep}; we use this as a design motivation rather than a claim about universal deep-network dynamics~\citep{saxe2018information}. It is also consistent with work connecting language modeling and compression~\citep{deletang2024language} and with the view that reusable abstractions compress long primitive derivations into higher-level concepts~\citep{aksenov2026compression}. In our setting, the practical implication is that a model capable of compressing a noisy user history into an appropriate persona state and evolution motif has already performed a substantial part of recommendation cognition.

After compression, the trace performs transition judgement over the candidate hypotheses. This stage reuses two earlier SFT abilities. From R1, it inherits local bridge construction: the trace should explain how an earlier item, query, or behavior can naturally lead to a follow-up need. From R2, it inherits temporal evolution judgement: the trace should decide whether a candidate direction is supported by order, continuity, refinement, or closure in the behavior timeline. We use \emph{causal-style} only as a data-construction heuristic: the teacher prefers evidence-supported triggers, refinements, and scenario continuations over shallow co-occurrence, without claiming identifiable causal effects from observational logs. This heuristic is aligned with causal recommendation studies that emphasize separating genuine user interest from confounding factors or conformity effects, as well as counterfactual explainable recommendation methods that seek explanations with strong decision influence~\citep{luo2024causal,tan2021counterfactual,zheng2021disentangling}.

Concretely, we adopt a three-stage protocol: \traceStage{Persona Abstraction}, \traceStage{Interest Expansion}, and \traceStage{Transition Inference}. \traceStage{Persona Abstraction} performs the persona-type compression axis. \traceStage{Interest Expansion} starts from interest-evolution-type compression and expands the most informative behavior signals into a small set of evidence-grounded hypotheses. \traceStage{Transition Inference} compares these hypotheses under evidence strength, recency, temporal continuity, persona compatibility, target-domain compatibility, and leakage control, then commits to the most plausible next-interest direction.

\begin{table}[H]
\centering
\small
\renewcommand{\arraystretch}{1.18}
\setlength{\tabcolsep}{5pt}
\caption{Representative persona-abstraction cases for recommendation reasoning. Personas are used only as soft priors; final decisions must be supported by user behavior.}
\label{tab:sft-persona-examples}
\begin{CJK*}{UTF8}{gbsn}
\begin{tabularx}{\linewidth}{@{}>{\raggedright\arraybackslash}p{0.20\linewidth}>{\raggedright\arraybackslash}X@{}}
\toprule
\textbf{Persona type} & \textbf{Chinese reasoning case and English analysis} \\
\midrule
\textbf{Middle-aged homemaker}
&
由【晚饭前连续看家常菜教程、收纳改造、孩子作业辅导】推测用户偏向中年家庭主妇，并且最明显的需求是家人日常饮食、家居整理和孩子学习支持。后续更应关注实用、低门槛、能直接服务家庭场景的内容，而不是泛泛推荐生活方式内容。
\par\smallskip
{\footnotesize\color{gray}\emph{Analysis:} the abstraction compresses several household-oriented signals into a family-consumption prior, but the next recommendation still has to follow the concrete need revealed by recent cooking, storage, or tutoring behaviors.} \\
\midrule
\textbf{Live-shopping enthusiast}
&
由【长时间停留在带货直播间、反复点击购物车、关注限时优惠】推测用户对直播购物偏好较强，并且最明显的兴趣集中在高性价比日用品、食品补货和家居小件。推理时需要优先判断她信任的主播、常购品类和促销节点，避免只把她归为普通购物用户。
\par\smallskip
{\footnotesize\color{gray}\emph{Analysis:} the persona prior highlights promotion sensitivity and trusted live-room categories, so the trace should reason about the next purchase scenario rather than merely predicting generic e-commerce interest.} \\
\midrule
\textbf{Shared-device user}
&
由【上午看少儿动画、晚饭前看家常菜、深夜又集中看体育赛事】推测该账号可能存在多人共用手机。此时画像不能直接合并成一个稳定用户，而要按时段和近期行为拆分，优先识别当前活跃使用者，避免被其他家庭成员的历史行为带偏。
\par\smallskip
{\footnotesize\color{gray}\emph{Analysis:} the abstraction is used as a noise-control mechanism: conflicting behavior clusters are treated as different possible users, and the final transition should rely on the current stable segment rather than the whole mixed history.} \\
\bottomrule
\end{tabularx}
\end{CJK*}
\end{table}

At a conceptual level, the protocol can be written as
\begin{equation}
\begin{split}
\mathcal{C}_u
=
\mathrm{Abstract}(\mathcal{P}_u,\mathcal{H}_u),
\qquad
\mathcal{Z}_u
=
\mathrm{Expand}(\mathcal{C}_u,\mathcal{H}_u),
\end{split}
\label{eq:sft-r3-compression}
\end{equation}
\begin{equation}
\begin{split}
z_u^{\star}
=
\mathrm{Infer}_{\mathrm{trans}}\bigl(\mathcal{Z}_u \mid \mathcal{C}_u,\mathcal{H}_u,d\bigr)
=
\arg\max_{z\in\mathcal{Z}_u}
 s\bigl(z\mid \mathcal{C}_u,\mathcal{H}_u,d\bigr),
\end{split}
\label{eq:sft-r3-transition-inference}
\end{equation}
where $\mathcal{C}_u$ denotes the compressed user state, $\mathcal{Z}_u$ denotes the expanded set of candidate interest hypotheses, $d$ denotes the target domain, $\mathrm{Infer}_{\mathrm{trans}}(\cdot)$ denotes the final transition-judgment step, and $s(\cdot)$ is a conceptual scoring function that jointly considers evidence strength, recency, temporal continuity, persona compatibility, and target-domain compatibility. The final CoT trace is written as a concise realization of this process with text and itemic-token references interleaved, while the target itemic tokens are emitted only after the reasoning trace. Representative sanitized training samples of the final recommendation CoT format are provided in Appendix~\ref{appendix:sft-recommendation-cot-cases}, and the complete bilingual construction prompt is provided in Appendix~\ref{appendix:sft-recommendation-trace-prompt}.

\paragraph{\traceStage{Persona Abstraction}}
\label{sec:sft-persona-abstraction}

The goal of \traceStage{Persona Abstraction} is to compress sparse and noisy behavioral evidence into a compact, interpretable prior. Given the user profile and interaction history, the teacher model identifies the most plausible user archetype, or summarizes a customized profile when the user does not fit any predefined category. The abstraction is a soft, typed user state rather than a deterministic label: it narrows the set of likely interest directions, but the final inference must still be grounded in observed behaviors. This stage reduces the search space for transition judgement by converting a high-dimensional behavior log into a small number of preference factors, such as family consumption needs, game-skill improvement, live-shopping sensitivity, food preference, fitness goals, or shared-device ambiguity. Table~\ref{tab:sft-persona-examples} gives representative Chinese persona-abstraction cases, together with English analyses of how each abstraction constrains recommendation reasoning.

\paragraph{\traceStage{Interest Expansion}}
\label{sec:sft-interest-expansion}

After \traceStage{Persona Abstraction}, the model performs \traceStage{Interest Expansion} so that the trace does not commit too early to a single explanation. This stage begins with interest-evolution-type compression: it identifies typed motifs from recent trajectory evidence, such as a search-triggered need, a refinement from broad category to concrete parameter, a scenario continuation from content to product, or a cross-domain echo between live, product, ad, and video behaviors. These motifs are then opened into a small set of candidate next-interest hypotheses. In this way, \traceStage{Interest Expansion} remains broad enough to represent recommendation uncertainty, but compact enough to prevent the trace from being overwhelmed by raw behavior tokens.

We use four heuristics to keep this expansion broad but evidence-grounded:

\begin{itemize}
    \item \textbf{\textit{Behavior motivation.}} The trace should explain the intent behind informative behaviors, especially search queries, because they often directly reveal the user's current interest and real need.
    \item \textbf{\textit{Temporal sensitivity.}} The trace should respect event order and explain how later behaviors may evolve from earlier ones, rather than treating the history as an unordered set.
    \item \textbf{\textit{Transition coverage.}} When the user appears to shift from one topic to another, the expansion should preserve plausible before--after directions as candidates, while leaving the final judgement to \traceStage{Transition Inference}. If a later behavior abandons all previous conditions without a logical bridge, it should be treated as weakly related or unrelated.
    \item \textbf{\textit{Noise attenuation.}} Routine replenishment, random consumption, repeated exposure, and one-off actions are down-weighted because they are usually weak signals for stable future recommendation intent.
\end{itemize}

Concretely, \traceStage{Interest Expansion} requires the reasoning trace to reach the finest reliable granularity supported by evidence. If the history supports interest in a specific mobile game, the trace should not stop at the generic gaming category; if the user repeatedly engages with a particular opera genre, the trace should mention that genre rather than only ``traditional culture.'' This design keeps the CoT trace fine-grained and evidence-grounded while preserving multiple plausible future directions before the final decision is made.

\begin{figure}[t]
\centering
\includegraphics[width=\linewidth]{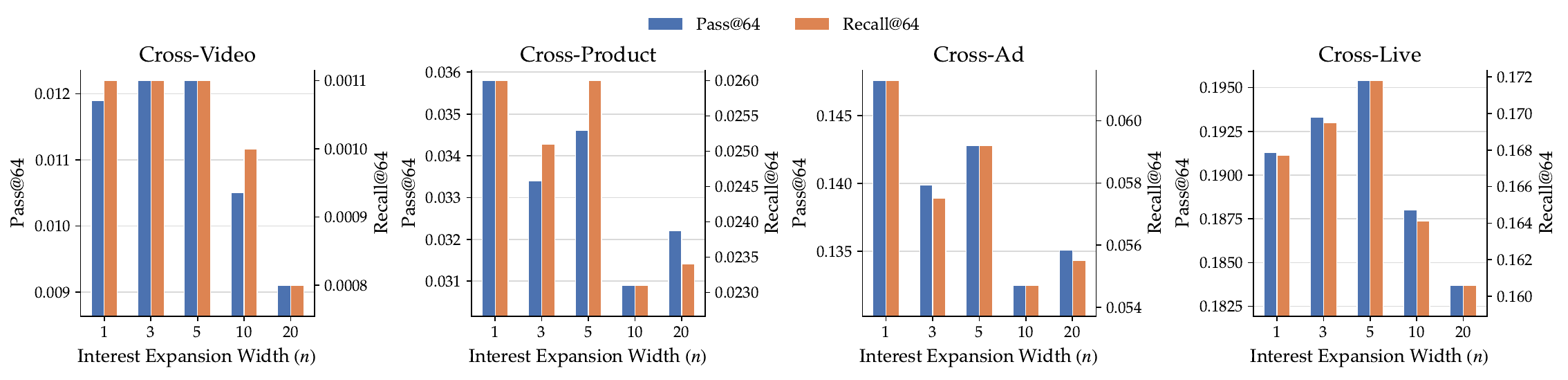}
\caption{Effect of interest expansion width \(n\) during Interest Expansion. Across the cross-video, cross-product, cross-ad, and cross-live domains, compact expansion with \(n\in\{1,3,5\}\) consistently outperforms broader widths such as \(10\) or \(20\) on think-mode Pass@64 and Recall@64.}
\label{fig:sft-interest-width}
\end{figure}

To calibrate how broad this stage should be, we vary the interest expansion width \(n\) during SFT trace construction and evaluate the resulting think-mode recommendation quality across the four domains. Figure~\ref{fig:sft-interest-width} shows a consistent pattern: keeping the expansion compact (\(n\in\{1,3,5\}\)) is consistently better than broader widths such as \(10\) or \(20\). This result fits the compression view above. Once the candidate set becomes too large, the trace starts to carry many weak branches, which dilutes the strongest transition signal and makes the later compare-and-commit step noisier. A small hypothesis set is therefore not a loss of reasoning capacity; rather, it is the point where uncertainty is preserved without letting the reasoning path fragment. In practice, we treat Interest Expansion as a controlled reasoning decompression step: it briefly opens the compressed trajectory into a few plausible directions, then hands a still-compact hypothesis space to \traceStage{Transition Inference}.

\begin{table}[H]
\centering
\caption{R3 recommendation CoT case showing Interest Expansion and Transition Inference.}
\label{tab:sft-r3-expansion-inference-case}
\setlength{\tabcolsep}{5pt}
\renewcommand{\arraystretch}{1.15}
\small
\begin{CJK*}{UTF8}{gbsn}
\begin{tabularx}{\linewidth}{@{}>{\raggedright\arraybackslash}X@{}}
\toprule
\textbf{\traceStage{Interest Expansion}} \\
由【中年女性、家庭日用服装购买、关注穿搭/健康/房产家居】推测用户偏向家庭实用消费型用户，近期最明显的兴趣是家居清洁与子女衣着。从近期行为可发散出多个兴趣点：A. 家居清洁维护，表现为多次购买 \sftProdSID{662}{7813}{7791} 等清洁产品，并长播 \sftVideoSID{192}{6421}{6240} 等教程；B. 子女衣着，表现为搜索“青少年加长版的裤子新款男”并购买 \sftProdSID{6266}{6810}{2477}；C. 健康养生，表现为关注 \sftLiveSID{3241}{7205}{1272}、点击 \sftProdSID{6655}{5260}{329} 及长播 \sftVideoSID{5695}{1638}{6707}；D. 零食/食品，表现为搜索“馋嘴小零食”并购买 \sftProdSID{6488}{6398}{7103}；E. 个人衣着与护肤，表现为购买 \sftProdSID{8099}{2904}{503} 并长播 \sftVideoSID{1409}{4885}{5059}；F. 房产与民生资讯，表现为完播民生、时政类内容，并浏览 \sftVideoSID{4946}{5797}{6364}。
\par\smallskip
{\footnotesize\color{gray}\emph{Analysis:} the trace does not collapse the user into one label. It keeps several evidence-backed branches, each tied to concrete itemic-token evidence and a specific intent.} \\
\midrule
\textbf{\traceStage{Transition Inference}} \\
综合权衡，A(家居清洁)和B(子女衣着)是当前最直接、最可能延续的强需求，因为它们同时具备搜索、点击、购买和长播等多种证据，且都服务于家庭场景中的即时问题解决。C(健康养生)和D(零食/食品)也有明确行为支撑，但更像阶段性补充需求，通常会间歇触发；E(个人衣着与护肤)与用户长期画像一致，但近期行动密度弱于 A/B；F(房产与民生资讯)主要是信息浏览，不足以作为最强后续方向。因此，后续推荐应优先围绕家居清洁技巧、清洁产品测评、青少年男装选购以及实用的家庭生活小窍门展开，同时少量保留健康养生和食品补货作为辅助方向。\par
\textbf{Target:} \sftVideoSID{6166}{1595}{1850}
\par\smallskip
{\footnotesize\color{gray}\emph{Analysis:} the final step chooses the branches with both recency and action density, then emits the target itemic tokens only after the reasoning trace.}
\\
\bottomrule
\end{tabularx}
\end{CJK*}
\end{table}

\paragraph{\traceStage{Transition Inference}}
\label{sec:sft-transition-inference}
\label{sec:sft-causal-reasoning}

The final stage compares the expanded hypotheses against the full context and commits to the most likely subsequent interaction direction. It reuses R1 by requiring a local bridge from antecedent evidence to follow-up need, and reuses R2 by checking whether the candidate direction is supported by temporal order, continuity, refinement, or behavioral closure. The teacher weighs each hypothesis by evidence strength, recency, continuity, persona compatibility, target-domain compatibility, and leakage risk. This compare-and-commit step prevents the trace from merely listing possible interests without making a decision useful for recommendation.

The following constraints make the final judgement decision-oriented and leakage-safe:

\begin{itemize}
    \item \textbf{\textit{Evidence priority.}} Candidate directions supported by repeated actions, active searches, or recent dense interactions receive higher weight; those supported only by isolated or accidental behaviors are down-weighted.
    \item \textbf{\textit{Transition bridge.}} The trace should explain the bridge from earlier evidence to later intent. Valid bridges include feedback, bottleneck, cognitive refinement, scenario continuation, parameter narrowing, or need completion; simple topical similarity is insufficient.
    \item \textbf{\textit{Granularity calibration.}} The final judgement should be as specific as the evidence allows. If the evidence supports a concrete game, opera genre, function, product attribute, or content style, the trace should not stop at a coarse category; if the evidence is insufficient, the trace should avoid over-specific claims.
    \item \textbf{\textit{Conflict resolution.}} When multiple directions are plausible, the trace compares their fit with the persona prior, recent behaviors, long-term interests, and target domain, then chooses the most likely direction.
    \item \textbf{\textit{Leakage control.}} Although the target interaction is visible during data construction, the reasoning must be written as if the answer is unknown. It must not copy the target title, expose the target item id or itemic token, or use target-only entities as evidence.
\end{itemize}

A compact training trace in Table~\ref{tab:sft-r3-expansion-inference-case} shows how \traceStage{Interest Expansion} keeps several grounded directions active, while \traceStage{Transition Inference} commits to the strongest continuation.

Through \traceStage{Persona Abstraction}, \traceStage{Interest Expansion}, and \traceStage{Transition Inference}, the synthesized CoT becomes compact, fine-grained, and decision-oriented. The first two stages denoise the behavior history and organize the hypothesis space, while the final stage compares candidate transitions by reusing R1 bridge reasoning and R2 evolution reasoning. This structure allows SFT to learn not only the final target itemic-token prediction, but also the behavioral logic that connects user evidence to recommendation decisions.

\subsubsection{Quality Evaluation of Reasoning Traces}
\label{sec:sft-cot-quality}

The quality of recommendation reasoning traces is evaluated along five complementary dimensions, each highlighting a different aspect of useful recommendation cognition: they assess whether the trace avoids direct answer leakage, aligns with the intended target, demonstrates logically transferable reasoning over user behavior, accurately reflects observed history, and provides actionable guidance for subsequent recommendations.

\begin{itemize}
    \item \textbf{\textit{Safety.}} Checks whether the CoT trace directly exposes target content, such as item id, itemic token, title, or any entity that could uniquely identify the target. Safe traces avoid direct answer leakage while remaining informative.

    \item \textbf{\textit{Consistency.}} Evaluates whether the trace's conclusions, preference direction, and behavioral explanations align with the intended recommendation target. A consistent trace supports correct downstream predictions.

    \item \textbf{\textit{Logic.}} Measures whether the trace demonstrates transferable and generalizable reasoning over user behavior rather than surface-level concatenation, forced associations, or memory of a single instance.

    \item \textbf{\textit{Factuality.}} Ensures that the trace accurately reflects the observed user history. It checks for fabricated behaviors, incorrect attributions, exaggerated trends, temporal confusions, or unsupported statements.

    \item \textbf{\textit{Informativeness.}} Assesses whether the reasoning provides useful and specific recommendation guidance without leaking the target. It should summarize interest directions, scenarios, functionality, style, attributes, or exclusions that are actionable for recommendation.
\end{itemize}

\begin{figure}[H]
\centering
\includegraphics[width=\linewidth]{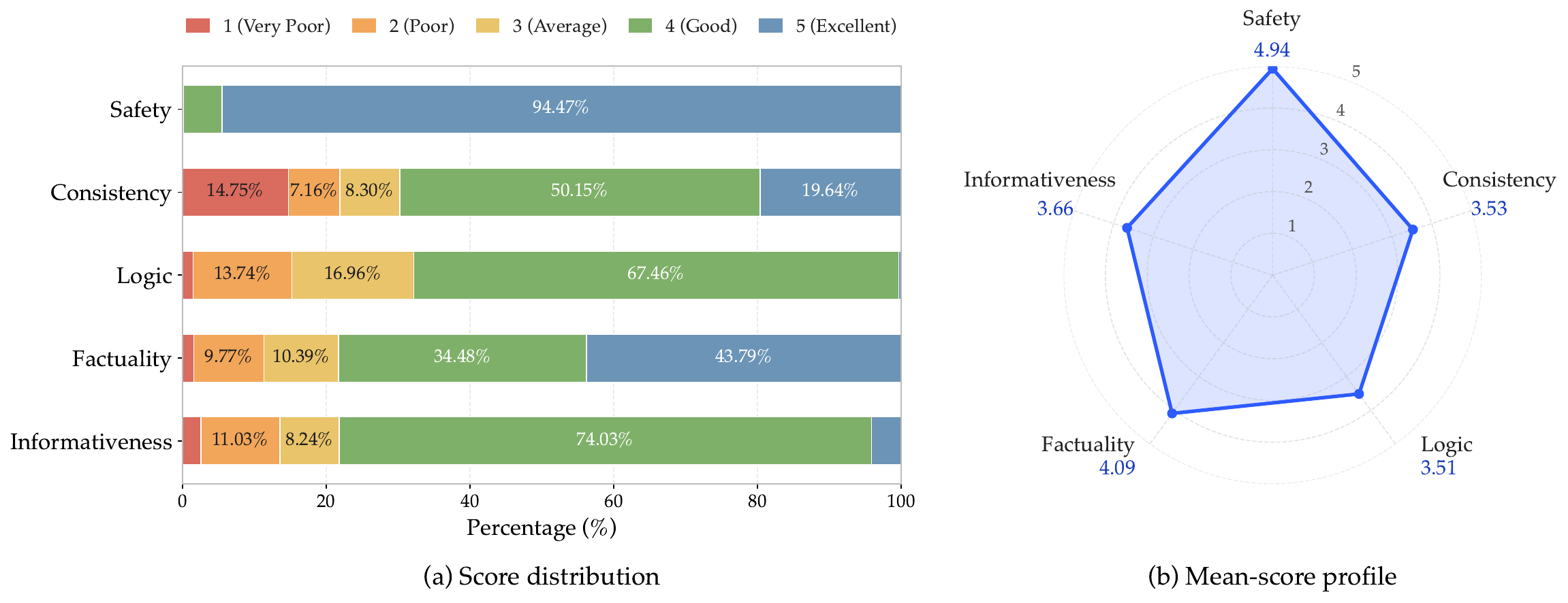}
\caption{Score distribution and mean-score profile for reasoning-trace quality across safety, consistency, logic, factuality, and informativeness dimensions.}
\label{fig:sft-data-quality}
\end{figure}

Together, these dimensions provide a compact diagnostic view of trace quality. Safety constrains leakage, consistency checks target alignment, logic checks reasoning transferability, factuality checks history grounding, and informativeness checks whether the trace provides clear guidance for recommendations. The evaluation therefore distinguishes traces that are merely fluent from traces that express useful recommendation cognition. Figure~\ref{fig:sft-data-quality} summarizes the resulting score distribution and mean-score profile across these five dimensions.

Representative unreasonable traces identified by this quality-evaluation strategy are shown in Appendix~\ref{appendix:sft-cot-quality-low-score-cases} (Table~\ref{tab:sft-cot-quality-low-score-cases}), where problematic statements in the generated CoT are marked in red.

\subsection{Itemic Instruction Data}
\label{sec:sft-data-sid-instruct}

Itemic Instruction Data improves instruction-following behavior when itemic tokens appear explicitly in the input context. It complements the compression--reasoning backbone by ensuring that itemic-token-rich inputs remain instruction-controllable, rather than being automatically interpreted as recommendation-reasoning prompts. We observe that pretrained models often behave unstably in such settings: instead of following the user's instruction, the model may ignore the requested task, explain the meaning of specific itemic tokens, or implicitly reinterpret the input as a user-interest analysis problem. Such behavior limits usability in practical recommendation-oriented interactions.

To address this issue, we construct a specialized instruction-tuning dataset centered on \textbf{itemic-token-grounded task execution}. Data generation begins with a predefined inventory of task tags, each corresponding to a representative instruction type in recommendation and content understanding scenarios. Rather than focusing only on itemic-token decoding, the tag space spans conversion, retrieval, matching, comparison, generation, editing, summarization, and explicit instruction-control tasks. For each tag, we generate suitable instructions and pair them with sampled user-side material, so that each example requires the model to complete a concrete task under inputs containing itemic tokens. Representative bilingual training cases are shown in Appendix~\ref{appendix:sft-sid-instruction-cases}. The main task groups are organized as follows:

\begin{itemize}
    \item \textbf{\textit{Conversion and Editing}.} This group teaches the model to convert between itemic tokens and natural language, rewrite text-token pairs, replace itemic tokens with textual descriptions, or reorder token sequences under instruction constraints. A representative instruction is: ``Convert the following product itemic token into a short product introduction.''

    \item \textbf{\textit{Retrieval and Selection}.} This group focuses on retrieving itemic tokens from structured user contexts according to time, behavior, attribute, or interaction-order constraints. A representative instruction is: ``Find all video itemic tokens the user interacted with during a given time window.''

    \item \textbf{\textit{Matching and Classification}.} This group trains the model to match natural-language descriptions with candidate itemic tokens, select the most suitable token-description pair, handle fuzzy matching, or classify the scene represented by itemic tokens. A representative instruction is: ``Which itemic token best matches the description `noise-canceling Bluetooth earphones'? Output only the token.''

    \item \textbf{\textit{Comparison and Aggregation}.} This group requires the model to compare itemic tokens or token groups, perform set-level operations, and summarize token sequences into compact natural-language descriptions. A representative instruction is: ``Compare the functional differences between these two product itemic tokens in three sentences.''

    \item \textbf{\textit{Generation and Verification}.} This group asks the model to generate descriptions, promotional copy, constrained summaries, or verification results based on itemic-token-linked content. A representative instruction is: ``Write a 50-word promotional copy for the following product itemic token.''

    \item \textbf{\textit{Instruction Control}.} This group explicitly trains the model to follow the user's instruction even when itemic tokens appear as distracting context. It includes ignoring irrelevant tokens, conditional execution, text-preserving rewriting, and format-constrained response generation. A representative instruction is: ``Ignore all itemic tokens below and answer: what are the main evaluation metrics for recommender systems?''
\end{itemize}

A key design principle is that itemic tokens are treated as \textbf{grounding signals} rather than as the final target of explanation. The objective is not to make the model interpret token semantics in isolation, but to train it to use itemic-token-linked context correctly while remaining faithful to the user's instruction. This improves robustness in mixed natural-language and token-based inputs. Among the task groups above, the instruction-control cluster is especially important because it directly teaches the model to prioritize user intent even when itemic-token-rich context could otherwise distract or mislead generation.

To ensure data quality, generated instructions are evaluated for usability before large-scale expansion, and final samples are checked with rule-based consistency tests to verify that itemic-token references in the input and output match the associated material. Samples with unsupported token references, weak task grounding, or unclear intent are removed. The final Itemic Token Instruction dataset contains approximately \textbf{103.0K} examples. Overall, this dataset serves as a behavior-alignment layer for itemic-token-aware interactions, improving the model's ability to follow instructions accurately in recommendation scenarios involving structured content identifiers.

\subsection{General-Domain Data}
\label{sec:sft-data-general}

To preserve general instruction-following ability and mitigate over-specialization during recommendation-oriented SFT, we incorporate the publicly available \textbf{Step-3.5-Flash-SFT} dataset released by StepFun~\citep{huang2026step} as the general-domain data source. We apply conservative preprocessing to the raw dataset: samples containing tool-use traces are discarded to avoid introducing environment-interaction behaviors outside the scope of our SFT setting; only samples following the standard three-role dialogue schema, namely \textit{system}, \textit{user}, and \textit{assistant}, are retained to ensure a homogeneous conversational format; and samples with malformed structures, missing fields, or corrupted formats are removed to further improve data quality. After preprocessing, the resulting corpus contains approximately \textbf{1.5 M} high-quality SFT samples, which are mixed into the training data to maintain broad instruction-following capability and stabilize general-purpose response behavior under domain-specific fine-tuning. General-domain data therefore serves as a stabilizer around recommendation cognition, preventing the model from overfitting to itemic-token reasoning formats.

Overall, this SFT stage builds structured recommendation cognition on top of aligned itemic perception: R0 grounds itemic tokens, R1 supplies evidence-grounded derivation, R2 supplies temporal-evolution supervision, and R3 composes these abilities for next-interaction prediction. Because the R3 traces are still teacher-generated, they are not yet optimized by recommendation outcome feedback. The next stage closes this loop with recommendation-oriented RL: domain-specific RL first discovers stronger recommendation behaviors within each domain, and RFT/MOPD then consolidate these specialized policies into a unified four-domain recommendation foundation model.

\section{RL Pipeline}\label{sec:rl}
After the SFT stage, the model has acquired general semantic understanding, instruction-following ability, and an initial capability for recommendation-oriented reasoning. However, SFT mainly enables the model to imitate teacher-generated reasoning trajectories, and the resulting performance is still largely bounded by the capability of the teacher model.
Since the model has already been exposed to large-scale recommendation data and has developed strong recommendation capabilities, a natural next step is to examine whether the model can go beyond imitation and further improve its recommendation reasoning through self-exploration.
Reinforcement learning (RL) has recently become a promising direction for further improving recommendation models beyond supervised training~\citep{rere_tan,he2026reasoning,zhou2026openonerectechnicalreport}.
However, for a foundation recommendation model that supports diverse tasks across domains, directly applying RL to mixed-domain data may cause cross-domain interference, since different domains often involve distinct user intents, item semantics, and reward landscapes.

\begin{figure}[t]
    \centering
    \includegraphics[width=\textwidth]{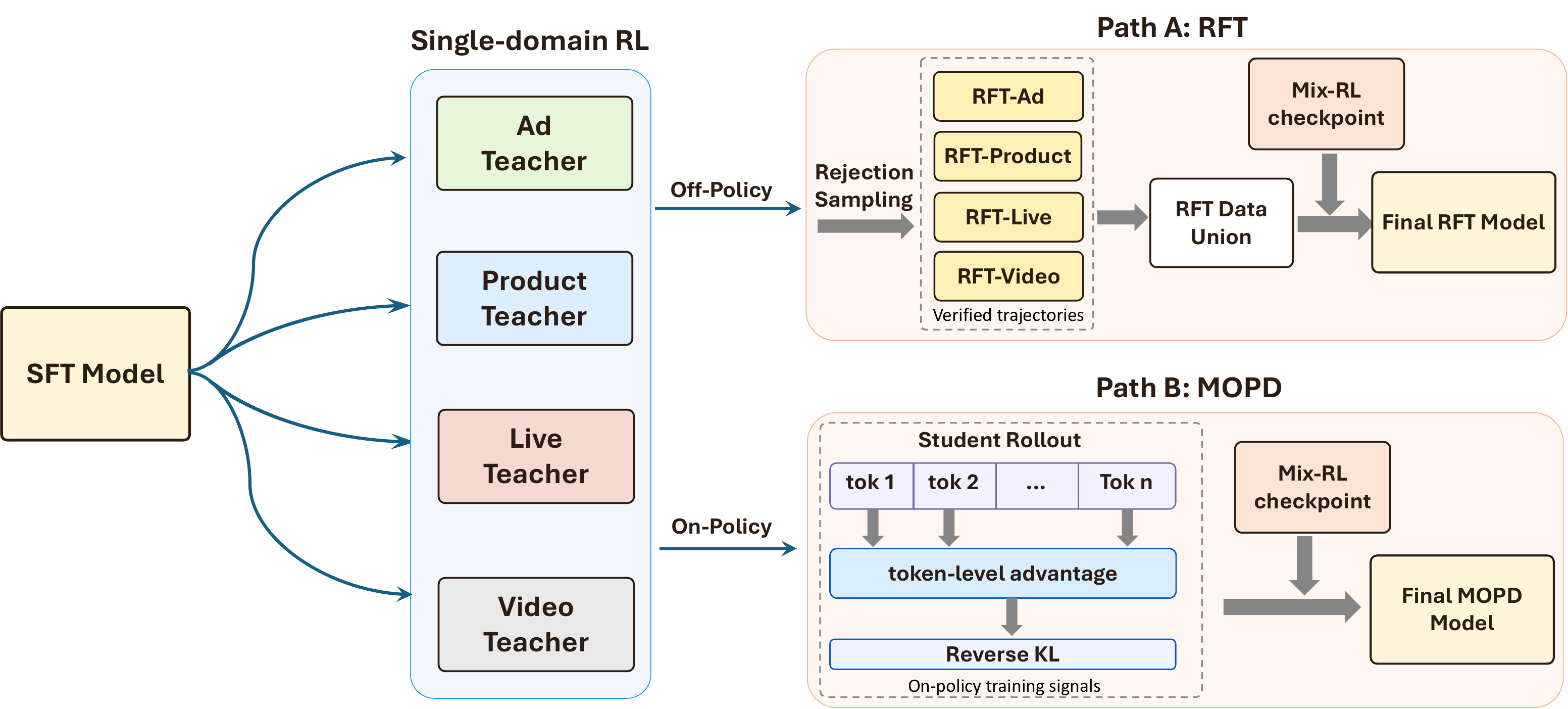}
    \caption{Overview of the ``specialize-then-unify'' pipeline.}
    \label{fig:rl_full_pipeline}
\end{figure}

To mitigate this issue, we adopt a ``specialize-then-unify'' strategy.
As illustrated in Figure~\ref{fig:rl_full_pipeline}, we first optimize the model separately on each recommendation domain through RL, resulting in four teacher models that specialize in domain-specific recommendation reasoning.
We then investigate two approaches to consolidate these domain-specific capabilities into one unified model: rejection sampling fine-tuning (RFT), which refines the model on verified high-quality reasoning trajectories sampled from domain-specialized teachers; and Multi-Teacher On-Policy Distillation (MOPD), which distills multiple domain-specific policies into the student model on its own generated trajectories.

\subsection{Recommendation-oriented Reinforcement Learning}

We first optimize the model with GRPO~\citep{shao2024deepseekmath} on each recommendation domain separately, where the model generates trajectories and receives outcome feedback based on the correctness of its prediction.

\paragraph{Recommendation-oriented GRPO.}

\begin{figure}[t]
    \centering
    \includegraphics[width=\textwidth]{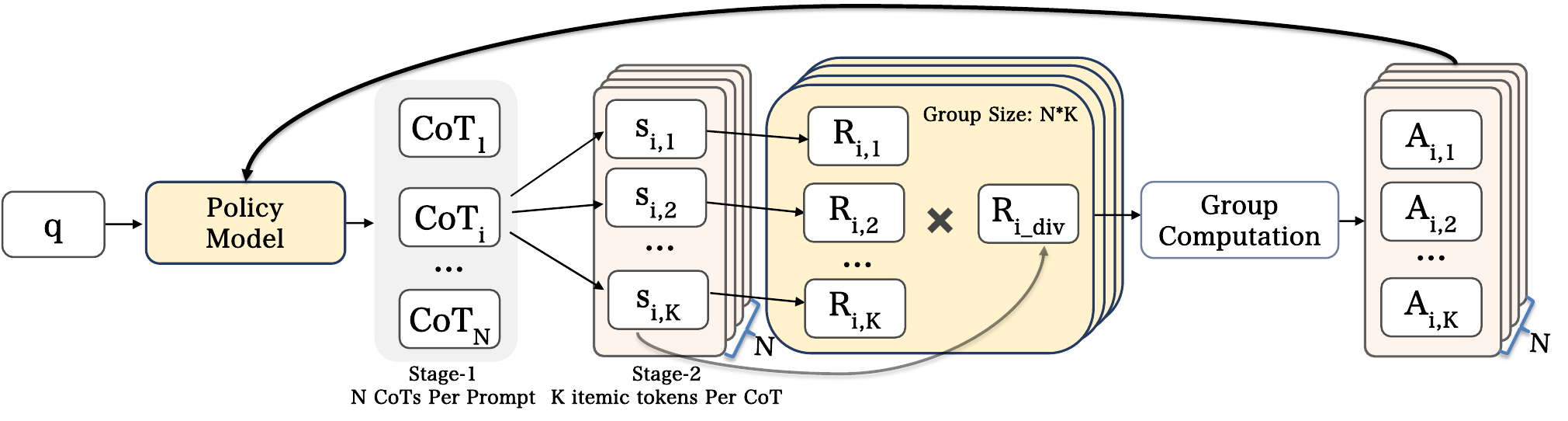}
    \caption{Overview of recommendation-oriented GRPO.}
    \label{fig:grpo_framework}
\end{figure}

For a user \(u\) with context (query) \(q\), the model takes it as input and samples a group of recommendation rollouts, where each rollout consists of a reasoning trace and generated itemic tokens, as illustrated in Figure~\ref{fig:grpo_framework}.
Formally, let
\begin{equation}
    \mathcal{G}_{u}
    =
    \{(\mathrm{CoT}_{u,i}, \mathbf{c}_{u,i})\}_{i=1}^{G}
\end{equation}
denote the group of \(G\) rollouts sampled for user \(u\), where \(\mathrm{CoT}_{u,i}\) is the generated reasoning trace and \(\mathbf{c}_{u,i}\) is the corresponding itemic tokens.
Each rollout is evaluated by an outcome reward \(R_{u,i}\), which measures whether the generated itemic tokens match the target recommendation.
GRPO then normalizes rewards within the same group to obtain the relative advantage:
\begin{equation}
    \hat{A}_{u,i}
    =
    \frac{
    R_{u,i}
    -
    \mathrm{mean}\left(\{R_{u,k}\}_{k=1}^{G}\right)
    }{
    \mathrm{std}\left(\{R_{u,k}\}_{k=1}^{G}\right)+\delta
    },
\end{equation}
where \(\delta\) is a small constant added for numerical stability.
The model is optimized to increase the probability of high-advantage rollouts, where the generated reasoning trace leads to itemic tokens that match the ground-truth items.

\paragraph{Rollout Design.}
While GRPO serves as a powerful tool to enhance the effectiveness of recommendation reasoning, directly applying it to recommendation tasks still faces practical challenges.
The nature of recommendation tasks, where the model predicts user interests from a massive candidate space, leads to an inherently low hit rate and requires a sufficiently large rollout size to obtain useful reward signals.
This is particularly costly for reasoning-augmented recommendation, where each recommendation is preceded by a reasoning trajectory.

To address this issue, we design a two-stage rollout strategy tailored to recommendation-oriented RL.
As illustrated in Figure~\ref{fig:grpo_framework}, the central idea is to amortize the cost of reasoning over multiple recommendations.
For each user \(u\), we first sample \(N\) reasoning traces \(\{\mathrm{CoT}_{u,i}\}_{i=1}^{N}\).
Conditioned on each reasoning trace \(\mathrm{CoT}_{u,i}\), the model then generates
\(K\) itemic token sequences in parallel,
\(\{\mathbf{c}_{u,i,j}\}_{j=1}^{K}\), where
\begin{equation}
    \mathbf{c}_{u,i,j}
    =
    [\langle\texttt{|domain\_begin|}\rangle,
    c^{(1)}_{u,i,j},
    c^{(2)}_{u,i,j},
    c^{(3)}_{u,i,j}].
\end{equation}
This produces \(N \times K\) effective recommendation rollouts while generating only \(N\) reasoning traces.
Since the additional expansion is restricted to the short itemic tokens, the reward coverage is significantly improved with limited extra computation and memory overhead.

\paragraph{Reward Design.}
Recommendation tasks focus on set-wise optimization, where the model should retrieve a set of relevant items rather than only optimize a single prediction.
Since user interests are often multi-faceted, an ideal reasoning trace should cover diverse relevant items that satisfy user interests.
Therefore, given the \(N \times K\) rollouts produced by the two-stage decoding procedure, we assign each rollout a mixed reward that considers both item-level accuracy and the diversity of recommendations induced by each reasoning trace.
For each rollout \((\mathrm{CoT}_{u,i}, \mathbf{c}_{u,i,j})\), the corresponding reward is defined as
\begin{equation}
    R_{u,i,j}
    =
    R_{\mathrm{rule}}(\mathbf{c}_{u,i,j})
    \cdot
    R_{\mathrm{div}}(\mathrm{CoT}_{u,i}),
\end{equation}
where \(R_{\mathrm{rule}}\) provides the main accuracy signal based on the predicted itemic tokens, while \(R_{\mathrm{div}}\) evaluates the diversity of the recommendations induced by the same reasoning trace.
The rule-based reward evaluates whether the predicted itemic tokens hit the ground-truth set \(\mathcal{C}^{+}_{u}\):
\begin{equation}
    R_{\mathrm{rule}}(\mathbf{c}_{u,i,j})
    =
    \mathbf{I}\left[\mathbf{c}_{u,i,j} \in \mathcal{C}^{+}_{u}\right].
\end{equation}

Recommendation aims to generate a set of items for each user, whose interests may span multiple categories. Therefore, a desirable reasoning trace should guide the model toward diverse types of recommended items.
We encourage such diversity by computing an additional reward over the \(K\) itemic token sequences generated from the same reasoning trace.
Specifically, we focus on the first digit of the itemic tokens, since the first sub-token largely determines the subsequent decoding process and captures coarse-grained item categories.
Let \(m_i^{(1)}\) denote the number of distinct first-digit values among \(\{c^{(1)}_{u,i,j}\}_{j=1}^{K}\).
The diversity factor is defined as
\begin{equation}
R_{\mathrm{div}}(\mathrm{CoT}_{u,i})
=
\frac{\max(0, m_i^{(1)}-1)}{K-1}.
\end{equation}
Encouraging diversity at the first digit promotes broader category-level coverage, which naturally leads to more diverse item predictions.
The final reward \(R_{u,i,j}\) thus assigns greater reward to trajectories that both match the target item and lead to better recommendation diversity.
The diversity reward is defined at the reasoning-trace level, since all \(K\) itemic token sequences generated from the same reasoning trace share the same diversity factor \(R_{\mathrm{div}}\).

\begin{figure}[t]
    \centering
    \begin{minipage}{0.55\textwidth}
        \centering
        \includegraphics[width=\linewidth]{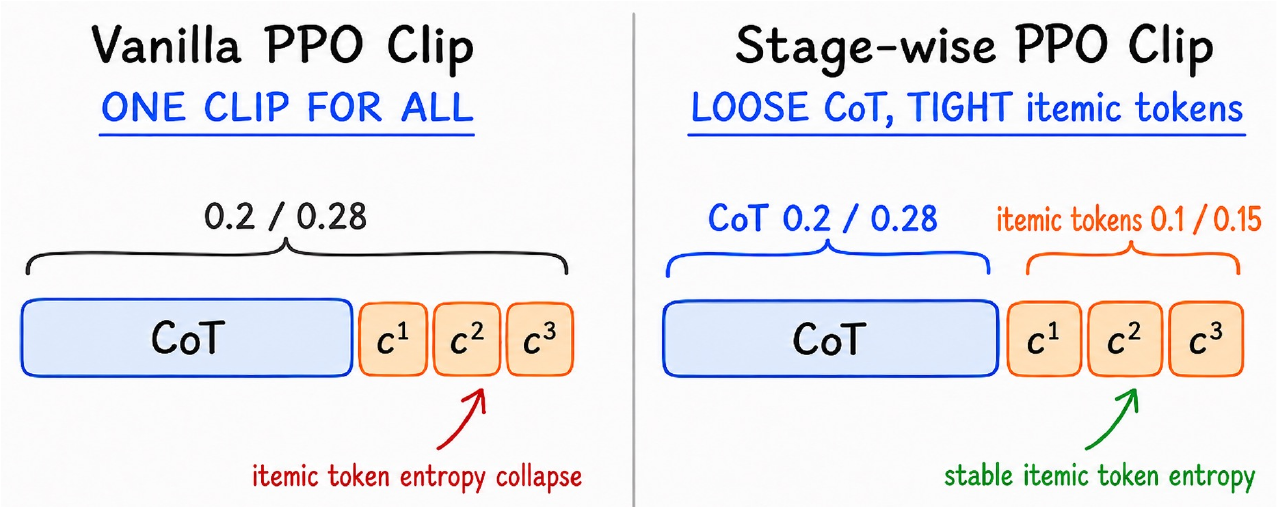}
        \caption*{(a) Stage-wise clipping.}
        \label{fig:rl_stage_clip}
    \end{minipage}
    \hfill
    \begin{minipage}{0.43\textwidth}
        \centering
        \includegraphics[width=\linewidth]{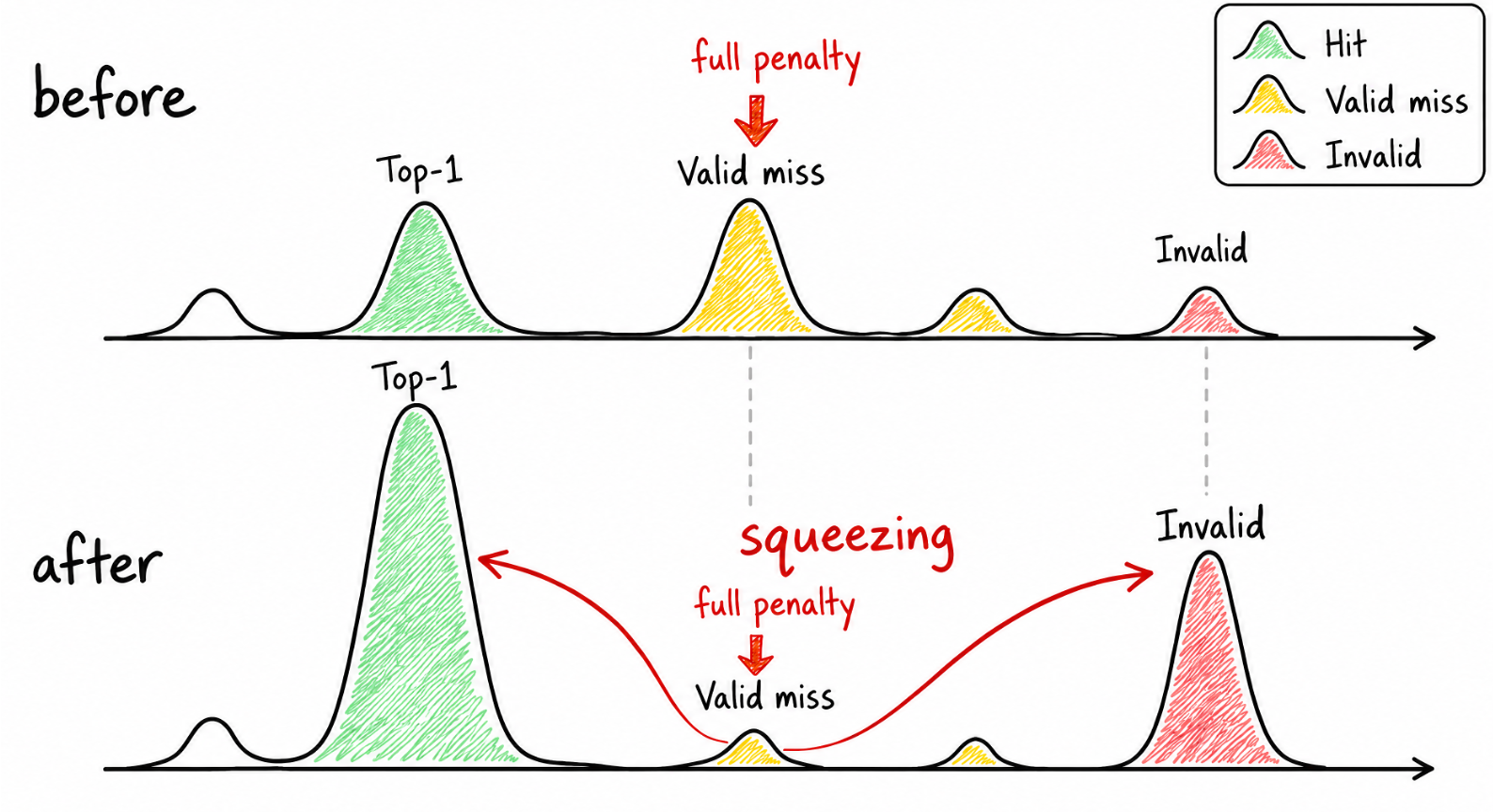}
        \caption*{(b) Negative-sample down-weighting.}
        \label{fig:rl_negative_down_weighting}
    \end{minipage}
    \caption{Illustrations of two optimization-stabilizing strategies.}
    \label{fig:rl_stabilization_strategies}
\end{figure}

\paragraph{Optimization.}
GRPO uses clipping to constrain the importance ratio between the updated policy and the behavior policy, which stabilizes policy updates.
For each generated token \(o_t\), the importance ratio is defined as
\begin{equation}
    r_t(\theta)
    =
    \frac{
    \pi_{\theta}(o_t \mid o_{<t})
    }{
    \pi_{\mathrm{old}}(o_t \mid o_{<t})
    }.
\end{equation}
In reasoning-augmented recommendation, however, reasoning tokens and itemic tokens exhibit substantially different behaviors.
Reasoning traces are long and mainly support exploration, whereas the final itemic token sequence is short but directly determines the recommendation reward.
Using the same clip range for both stages can make the distribution of itemic tokens change too aggressively and cause entropy collapse.
We therefore apply stage-wise clipping, with a looser clip range for reasoning tokens and a tighter clip range for itemic tokens, as illustrated in Figure~\ref{fig:rl_stabilization_strategies} (a).
Specifically, we use two stage-specific clipping ranges,
\begin{equation}
    \boldsymbol{\epsilon}_{\mathrm{CoT}}
    =
    (\epsilon_{\mathrm{CoT}}^{-}, \epsilon_{\mathrm{CoT}}^{+}),
    \qquad
    \boldsymbol{\epsilon}_{\mathrm{item}}
    =
    (\epsilon_{\mathrm{item}}^{-}, \epsilon_{\mathrm{item}}^{+}),
\end{equation}
where \(\boldsymbol{\epsilon}_{\mathrm{CoT}}\) is looser than \(\boldsymbol{\epsilon}_{\mathrm{item}}\).
For each token \(o_t\), the clipping range is selected according to its stage:
\begin{equation}
    \bar{r}_t(\theta)
    =
    \mathrm{clip}
    \bigl(r_t(\theta), 1-\epsilon^{-}(t), 1+\epsilon^{+}(t)\bigr),
    \quad
    \boldsymbol{\epsilon}(t)
    =
    \begin{cases}
    \boldsymbol{\epsilon}_{\mathrm{CoT}}, & o_t \in \mathrm{CoT}, \\
    \boldsymbol{\epsilon}_{\mathrm{item}}, & o_t \in \mathbf{c},
    \end{cases}
\end{equation}
where \(\boldsymbol{\epsilon}(t)=(\epsilon^{-}(t),\epsilon^{+}(t))\).
This preserves reasoning exploration while stabilizing updates on the final recommendation tokens.

To further stabilize GRPO under sparse recommendation rewards, a sample-level importance re-weighting scheme is applied to the actor loss, as illustrated in Figure~\ref{fig:rl_stabilization_strategies} (b).
Since only a small fraction of rollouts hit the ground-truth, most rollouts receive zero reward and contribute negative relative advantages.
Without re-weighting, these numerous non-hit samples may dominate the aggregated gradient and push the policy toward overly conservative updates~\citep{ren2025learning_dynamics_LLM,SAPO}.
After the group-relative advantage is computed, each rollout is assigned a scalar weight according to its hit status:
\begin{equation}
    w_{u,i,j}
    =
    \begin{cases}
    1.0, & \text{if } R_{\mathrm{rule}}(\mathbf{c}_{u,i,j}) = 1, \\
    \beta, & \text{otherwise},
    \end{cases}
\end{equation}
where \(\beta < 1\) controls the contribution of non-hit rollouts.
The weight is applied to the whole response-level actor loss of the rollout \((\mathrm{CoT}_{u,i}, \mathbf{c}_{u,i,j})\), while the advantage estimation remains unchanged.
This preserves the full gradient contribution from successful rollouts and down-weights non-hit rollouts, reducing the dominance of negative-advantage samples without changing the GRPO objective form.

\begin{figure}[H]
    \centering
    \includegraphics[width=\textwidth]{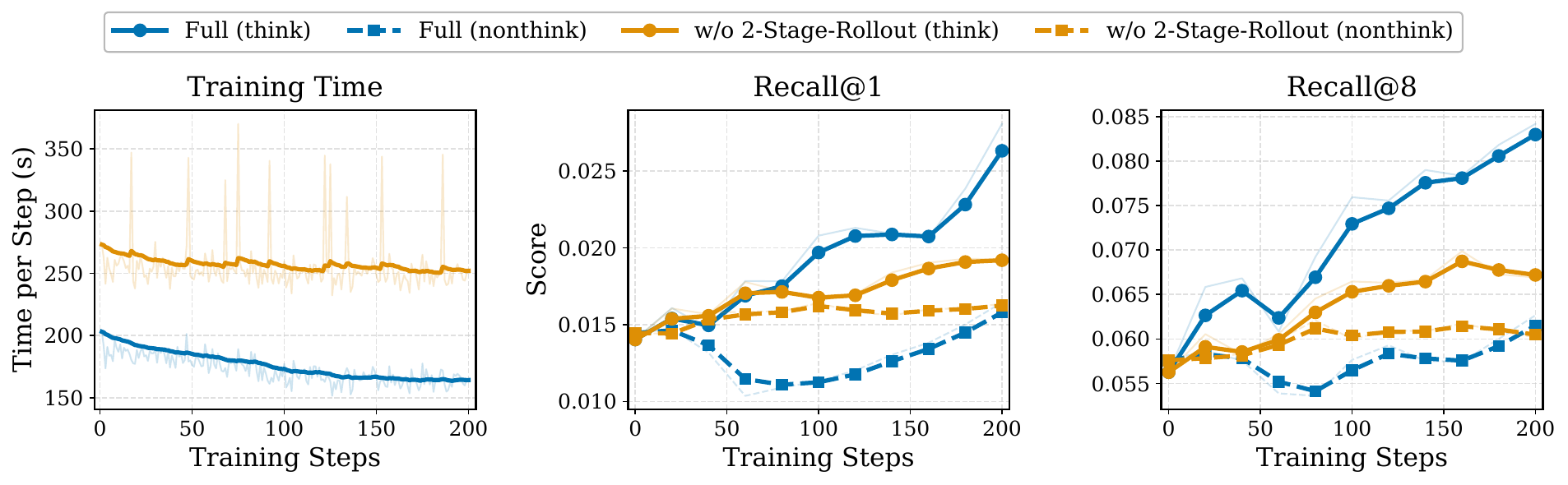}
    \caption{Ablation study on rollout optimization. Two-stage rollout improves training efficiency while achieving stronger recommendation performance.}
    \label{fig:rl_ablation_two_stage_rollout}
\end{figure}

\paragraph{Ablation Study.}
We first examine the effect of the two-stage rollout design by comparing GRPO with the full recommendation-oriented optimization against an ablated version that removes the two-stage rollout under the same effective rollout size.
As shown in Figure~\ref{fig:rl_ablation_two_stage_rollout}, we evaluate both training efficiency and recommendation performance.
The blue curves correspond to GRPO with full optimization, while the orange curves correspond to the ablated variant.
For performance, we evaluate the model on the Cross-Live Recommendation task using Recall@1 and Recall@8 under both thinking and non-thinking inference modes.

The two-stage rollout substantially reduces per-step training time by reusing each generated reasoning trace across multiple itemic token sequences, avoiding repeated generation of long reasoning prefixes.
Meanwhile, the performance results show consistent gains, with particularly clear improvements under the thinking mode.
These results indicate that expanding multiple recommendations from the same reasoning trace provides denser reward feedback for each reasoning path, enabling GRPO to more effectively optimize recommendation-oriented reasoning.

\begin{figure}[H]
    \centering
    \includegraphics[width=\textwidth]{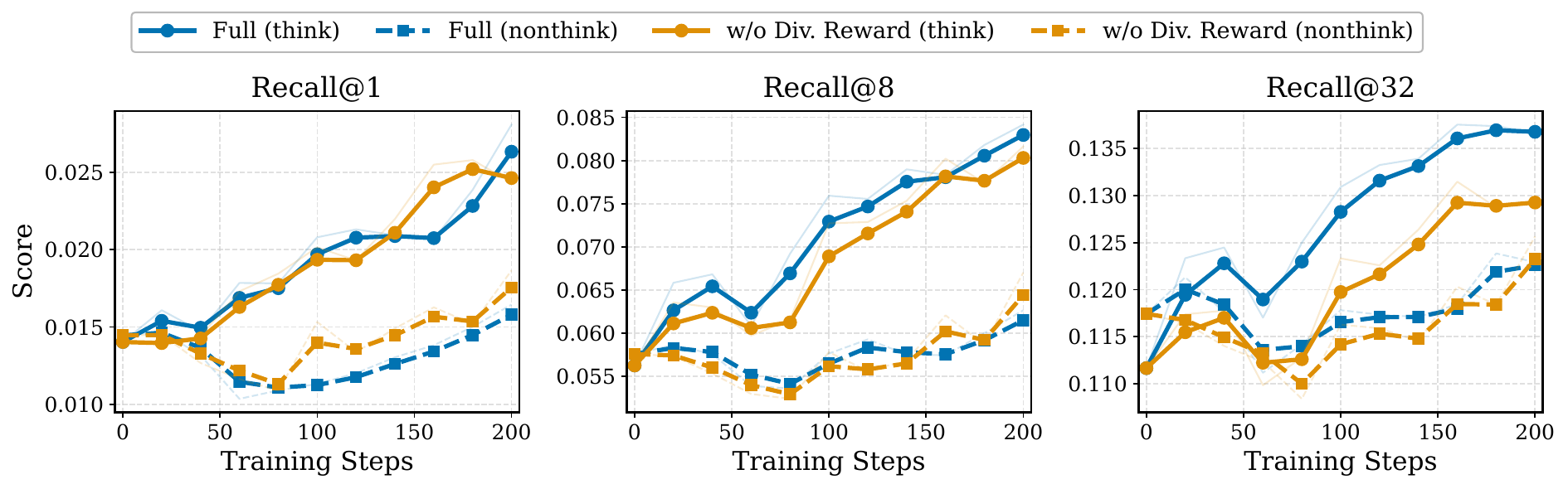}
    \caption{Ablation study on reward optimization. Diversity reward strengthens thinking-based recommendation, with recall@K improvements more significant at larger K.}
    \label{fig:rl_ablation_diversity_reward}
\end{figure}

We further evaluate the effect of the diversity reward using the same ablation setting, where the ablated version removes \(R_{\mathrm{div}}\) from the full reward design.
The methods are evaluated on Cross-Live with Recall@1, Recall@8, and Recall@32 under both thinking and non-thinking inference modes. The blue curves denote the fully optimized method, and the orange curves denote the ablated version.
Encouraging diversity among the subsequent itemic token sequences from the same reasoning trace helps the model retrieve a broader set of potential items. As a result, as shown in Figure~\ref{fig:rl_ablation_diversity_reward}, the diversity reward improves the effectiveness of reasoning, with more pronounced gains when Recall@\(K\) is evaluated at larger \(K\).

\begin{figure}[H]
    \centering
    \includegraphics[width=\textwidth]{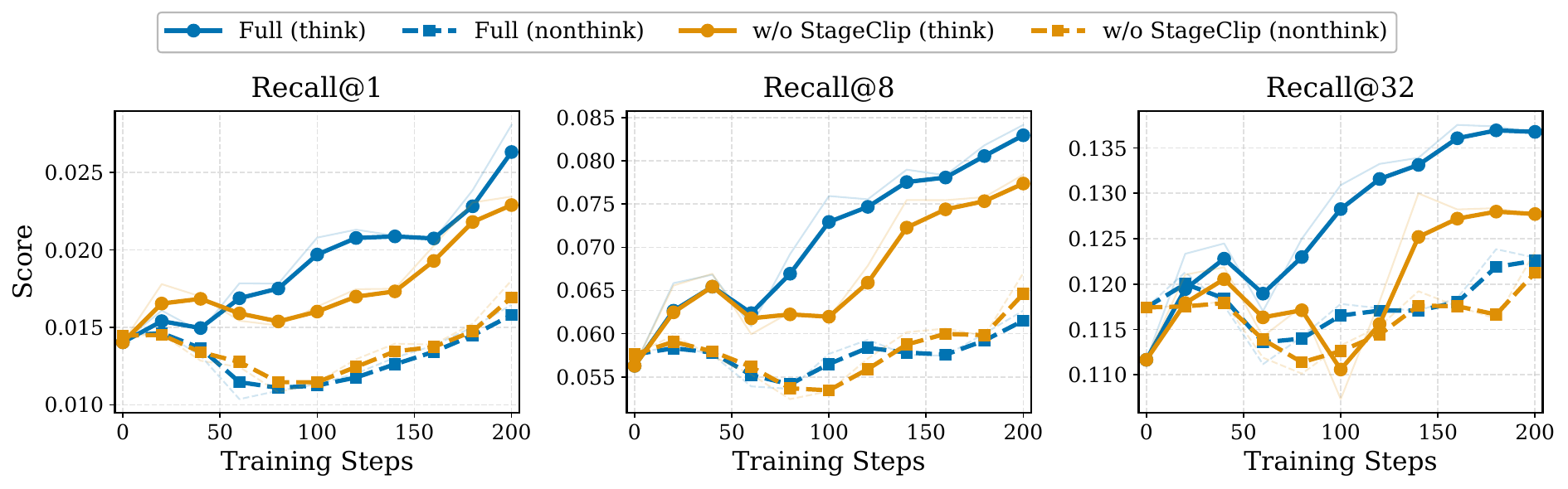}
    \caption{Ablation study on optimization strategies. Stage-wise clipping consistently improves recommendation performance.}
    \label{fig:rl_ablation_stage_clip}
\end{figure}

\begin{figure}[H]
    \centering
    \includegraphics[width=\textwidth]{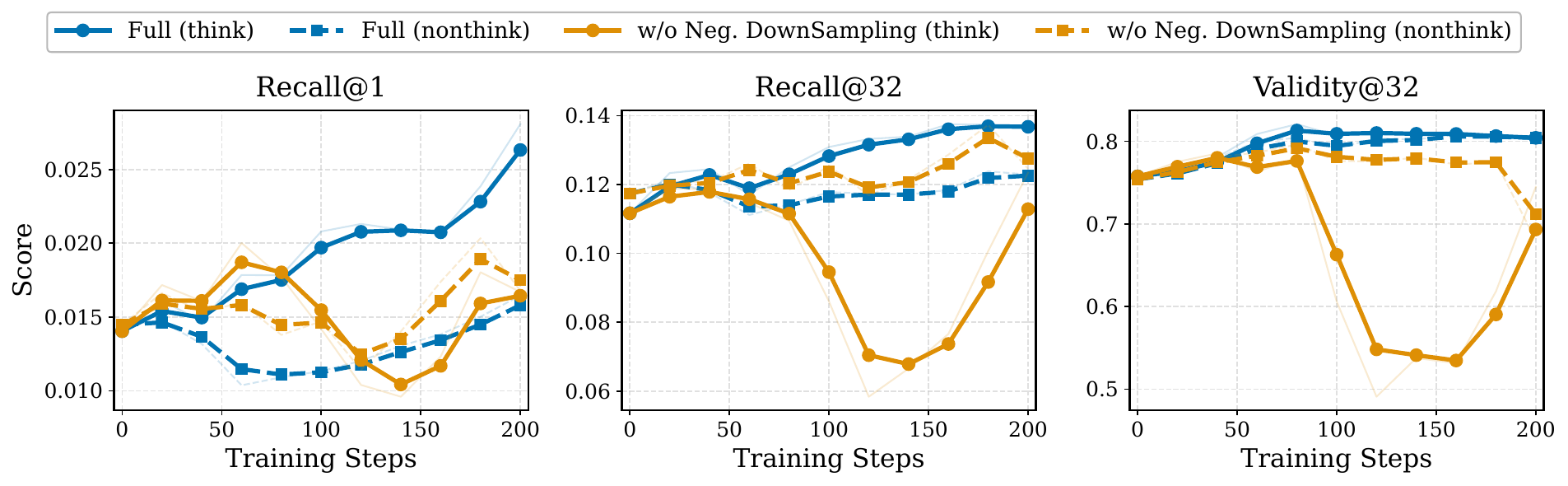}
    \caption{Ablation study on optimization strategies. Negative-sample down-weighting helps stabilize RL training.}
    \label{fig:rl_ablation_neg_downsampling}
\end{figure}

Finally, we analyze the two optimization-side stabilizers: stage-wise clipping and negative-sample down-weighting.
For stage-wise clipping, the ablated version uses the same clipping range for reasoning tokens and itemic tokens. For negative-sample down-weighting, the ablated version removes the sample-level weight \(w_{u,i,j}\).
Figure~\ref{fig:rl_ablation_stage_clip} shows the ablation results of stage-wise clipping, and Figure~\ref{fig:rl_ablation_neg_downsampling} presents the ablation of negative-sample down-weighting.
Both designs contribute to more stable training and better performance.
Removing stage-wise clipping weakens the recommendation performance, especially the reasoning effectiveness on Recall@8 and Recall@32, indicating that reasoning tokens and itemic tokens benefit from different trust regions during policy optimization.
Removing negative-sample down-weighting makes GRPO optimization less stable under sparse recommendation rewards, as abundant non-hit rollouts are assigned the same weight as successful ones.
As shown in Figure~\ref{fig:rl_ablation_neg_downsampling}, the ablated model suffers from a higher risk of training collapse, as reflected by a higher tendency to generate invalid itemic tokens and the resulting degradation in recommendation performance.

GRPO sharpens reasoning online but suffers from low sample efficiency on rare-but-correct trajectories. Rejection sampling fine-tuning (Section~\ref{sec:rl-reject}) is a natural offline complement: it harvests the small set of self-discovered high-quality rollouts and re-injects them as dense supervision, consolidating broadly useful behaviors that GRPO alone tends to forget between updates.

\subsection{Rejection Sampling Fine-tuning}\label{sec:rl-reject}
We first explore rejection sampling fine-tuning (RFT) to consolidate the knowledge of domain-specialized teacher models. RFT transfers domain-specific reasoning ability by selecting verified successful trajectories from each teacher and continuing supervised training on the aggregated trajectory set.

\paragraph{RFT workflow.}
While domain-specific GRPO improves the model through outcome-driven exploration, its policy optimization also sharpens the output distribution and makes the model concentrate on a small set of high-confidence items~\citep{chen2026does}. This can improve top-ranked predictions, but not fully aligned with the goal of recommendation, where the model is expected to predict a set of potential items that cover diverse user interests.
As a result, RL optimization alone may provide limited improvement for the final recommendation task, especially when the model needs to retrieve a broader set of relevant items.

To complement GRPO, we introduce rejection sampling fine-tuning (RFT) as the consolidation method.
Each domain-specialized GRPO model is used as a teacher to generate reasoning-augmented recommendation trajectories.
For each user \(u\), we retain trajectories whose predicted item matches the ground-truth set \(\mathcal{C}^{+}_{u}\), filter out low-quality or inconsistent reasoning traces, and merge the remaining verified trajectories across the four domains.
This yields an aggregated rejection-sampled dataset
\begin{equation}
    \mathcal{D}_{\mathrm{RFT}}
    =
    \{(x_u, \mathrm{CoT}_{u}, \mathbf{c}_{u})\},
\end{equation}
where \(x_u\) denotes the user context, \(\mathrm{CoT}_{u}\) is the verified reasoning trace, and \(\mathbf{c}_{u}\) is the corresponding itemic tokens.

Before RFT, we first obtain a Mix-RL checkpoint by applying GRPO to mixed data covering R2 evolution and R3 recommendation tasks from all domains until convergence.
Initialized from this checkpoint, the unified model is further trained on \(\mathcal{D}_{\mathrm{RFT}}\) with the standard next-token prediction objective:
\begin{equation}
    \mathcal{L}_{\mathrm{RFT}}
    =
    -
    \mathbb{E}_{(x_u,\mathrm{CoT}_{u},\mathbf{c}_{u}) \sim \mathcal{D}_{\mathrm{RFT}}}
    \sum_{t=1}^{|y_u|}
    \log \pi_{\theta}
    \left(
    y_{u,t}
    \mid
    x_u, y_{u,<t}
    \right),
    \quad
    y_u=[\mathrm{CoT}_{u};\mathbf{c}_{u}].
\end{equation}
In this way, RFT transfers domain-specific reasoning patterns discovered by RL into a single model while maintaining a smoother training objective based on next-token prediction.

\paragraph{Performance Analysis.}

\begin{figure}[H]
    \centering
    \includegraphics[width=\textwidth]{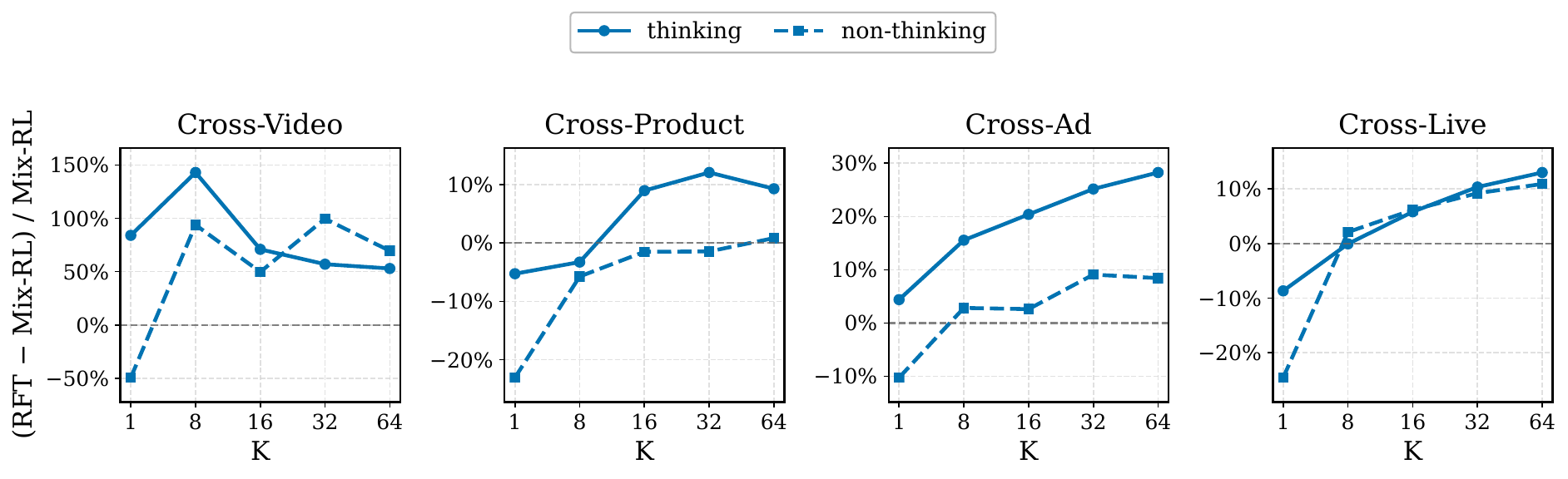}
    \caption{Relative gains of RFT over Mix-RL across Recall@K. RFT generally shows larger improvements as K increases.}
    \label{fig:rl_performance_rft_vs_rl}
\end{figure}

To analyze the effectiveness of RFT, we compare the performance of RFT against mixed-domain RL as the baseline method across four recommendation domains. We report the relative improvement of RFT over Mix-RL under Recall@\(K\), where \(K \in \{1, 8, 16, 32, 64\}\). This evaluation investigates whether RFT improves only the top-ranked prediction or provides broader gains when the model is asked to retrieve a larger set of relevant items.

As shown in Figure~\ref{fig:rl_performance_rft_vs_rl}, RFT does not always improve Recall@1, especially on Cross-Product and Cross-Live, where the relative improvements at small \(K\) are often negative or close to zero. However, as \(K\) increases, RFT consistently shows stronger advantages across domains. The improvement becomes particularly clear on most domains at larger \(K\), achieving large positive gains across all evaluated cutoffs. This trend suggests that RFT is less focused on sharpening a single top prediction and more effective at improving candidate coverage. Such behavior is better aligned with the goal of recommendation, where the model should retrieve a set of relevant items that can cover diverse user interests rather than only optimize the first prediction.

\subsection{Multi-Teacher On-Policy Distillation}\label{sec:rl-mopd}
This section introduces multi-teacher on-policy distillation~\citep{xiao2026mimo,yang2026nemotron,dou2026baichuan,deepseekai2026deepseekv4} to integrate specialized capabilities from four domains into a unified student model in an on-policy manner.
We adopt a Monte Carlo RL formulation~\citep{li2026rethinking}, i.e., the log-probability ratio is evaluated \emph{only} on the token sampled from the student itself at each step.
This converts the optimization target into an expectation over sampled trajectories.
The teacher is therefore only queried for a single-point log-probability at the sampled token, with no requirement to provide a full vocabulary distribution.
The effectiveness of this RL-based formulation stems from the OneReason training pipeline, where the student model has already acquired foundational capabilities across all target domains.
As a result, the student's policy distribution does not exhibit significant deviation from the teacher's policy.
This distributional proximity naturally bounds the variance of the single-point log-probability evaluations.
Therefore, we can circumvent the severe high-variance instability that typically plagues sparse RL updates, achieving significantly higher computational efficiency while maintaining stable training.

\paragraph{MOPD Objective.}
Formally, let \(\mu_\theta\) denote the student's behavior policy deployed in the inference engine for trajectory generation, and \(\pi_\theta\) denote the policy being actively optimized in the training engine. For a given prompt \(x\), we first sample a complete rollout \(y = (\mathbf{CoT}; \mathbf{c}) \sim \mu_\theta (\cdot \mid x)\). Based on the specific capability domain required by \(x\), we route the prompt to the corresponding domain-expert teacher, denoted as \(\pi_{\mathrm{domain}_i}\). At each timestep \(t\) of the auto-regressive two-stage rollout, we define the token-level distillation advantage under a reverse-KL objective as:
\begin{equation}
    \hat{A}_{\mathrm{MOPD},t}
    = \mathrm{sg}\!\left[
          \log\pi_{\mathrm{domain}_i}(y_t \mid x, y_{<t})- \log\pi_\theta(y_t \mid x, y_{<t})\;
      \right],
    \label{eq:adv}
\end{equation}
Since the trajectories are generated by the behavior policy \(\mu_\theta\) but optimized under the target policy \(\pi_\theta\), we introduce truncated importance weighting \(w_t(\theta)\) to correct for this off-policy discrepancy:
\begin{equation}
    w_t(\theta) =
    \begin{cases}
      \mathrm{sg}\Bigl[
        \dfrac{\pi_\theta(y_t \mid x, y_{<t})}{\mu_\theta(y_t \mid x, y_{<t})}
      \Bigr],
      & \varepsilon_{\mathrm{low}} \;\le\;
        \dfrac{\pi_\theta(y_t \mid x, y_{<t})}{\mu_\theta(y_t \mid x, y_{<t})}
        \;\le\; \varepsilon_{\mathrm{high}}, \\[8pt]
      0, & \text{otherwise,}
    \end{cases}
    \label{eq:weight}
\end{equation}
where \(\mathrm{sg}[\cdot]\) denotes stop-gradient. We then define the surrogate loss of MOPD as:
\begin{equation}
    \mathcal{L}_{\mathrm{MOPD}}(\theta)
    = -\mathbb{E}_{x \sim \mathcal{D},\, y \sim \mu_\theta(\cdot \mid x)}
      \left[
        \frac{1}{|y|} \sum_{t=1}^{|y|}
        w_t(\theta)\,\hat{A}_{\mathrm{MOPD},t}\,
        \log \pi_\theta(y_t \mid x, y_{<t})
      \right].
    \label{eq:mopd}
\end{equation}

\paragraph{Optimization.}
\begin{figure}[t]
    \centering
    \includegraphics[width=\linewidth]{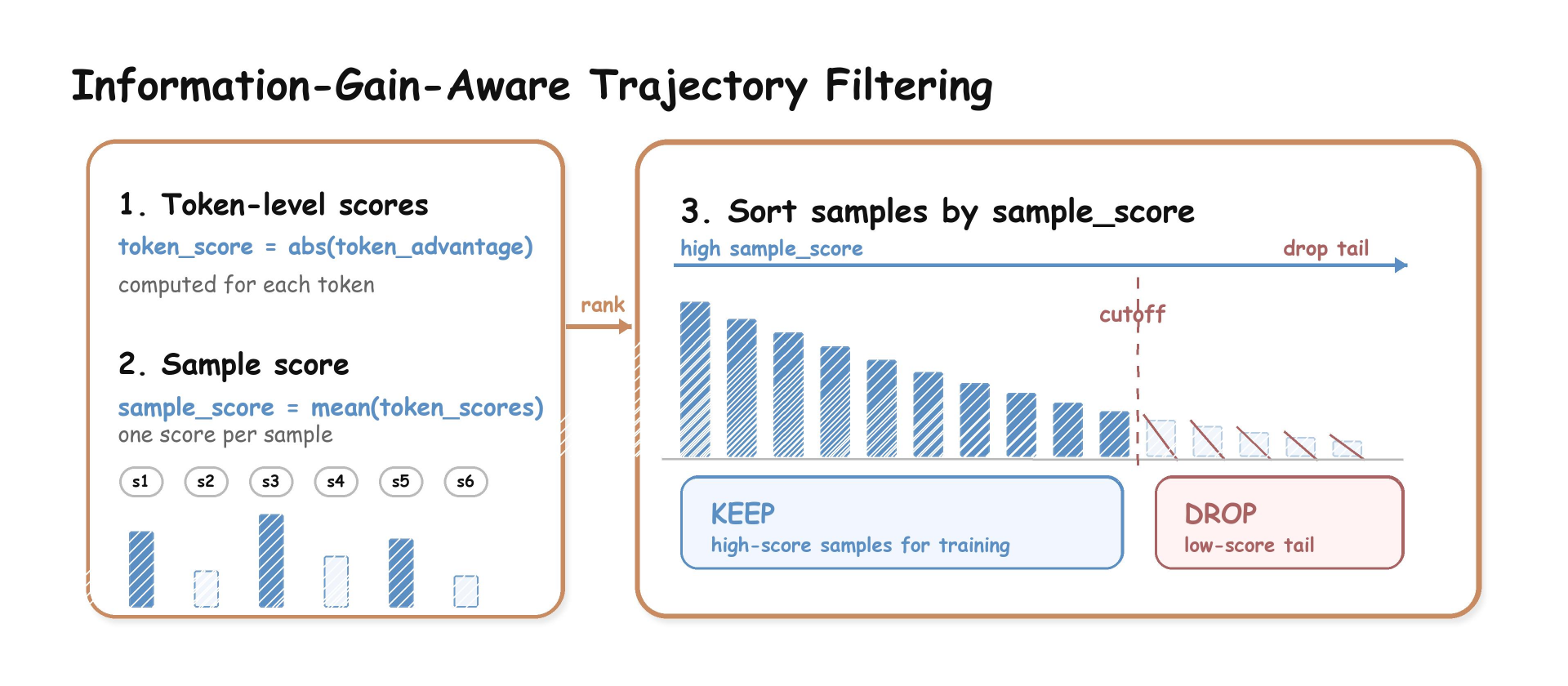}
    \caption{Overview of the information-gain-aware trajectory filtering mechanism.
}
    \label{fig:mopd_method}
\end{figure}
Generative recommendation inherently struggles with massive item candidate spaces and highly sparse user-interest signals.
In on-policy training, this translates to an exceptionally low hit rate during trajectory rollouts.
When applying MOPD to such environments, treating all sampled rollouts uniformly leads to a severe gradient-dilution problem.
Specifically, the student model quickly aligns with the teachers on generic, popular ``head'' items, resulting in near-zero token-level advantages (\(|\hat{A}_{\mathrm{MOPD},t}| \approx 0\)).
Consequently, the rare yet crucial domain-specific insights for ``tail'' items---where the specialized teachers' expertise is most valuable (characterized by high \(|\hat{A}_{\mathrm{MOPD},t}|\))---are entirely overwhelmed by redundant gradients from already-aligned or noisy samples.
To address this problem, we adopt an information-gain-aware trajectory filtering mechanism~\citep{chen2026counteraction}, as shown in Figure~\ref{fig:mopd_method}.
Mathematically, the absolute advantage \(|\hat{A}_{\mathrm{MOPD},t}|\) quantifies the instantaneous information gain provided by the teachers.
We dynamically discard low-information trajectories, which forces the student to focus on the most dissimilar and informative rollouts. It is crucial for preventing policy performance degradation in environments with inherently low hit rates. 

Specifically, we utilize the token-level advantage gap as a label-free indicator of information gain.
To quantify the overall deviation from the teacher, we compute the token-average absolute gap for each trajectory \(y_j\) as its informativeness score \(s(y_j)\):
\begin{equation}
    s(y_j) = \frac{1}{T_i} \sum_{t=1}^{T_i} |\hat{A}_{\mathrm{MOPD},j,t}|.
\end{equation}
This absolute scoring naturally captures the bilateral correction demand: it identifies both instances where the student is overly confident in an incorrect path (\(\hat{A}_{\mathrm{MOPD},j,t} \ll 0\)) and those where it fails to capture the teacher's domain-specific insights (\(\hat{A}_{\mathrm{MOPD},j,t} \gg 0\)).

\begin{figure}[t]
    \centering
    \includegraphics[width=\linewidth]{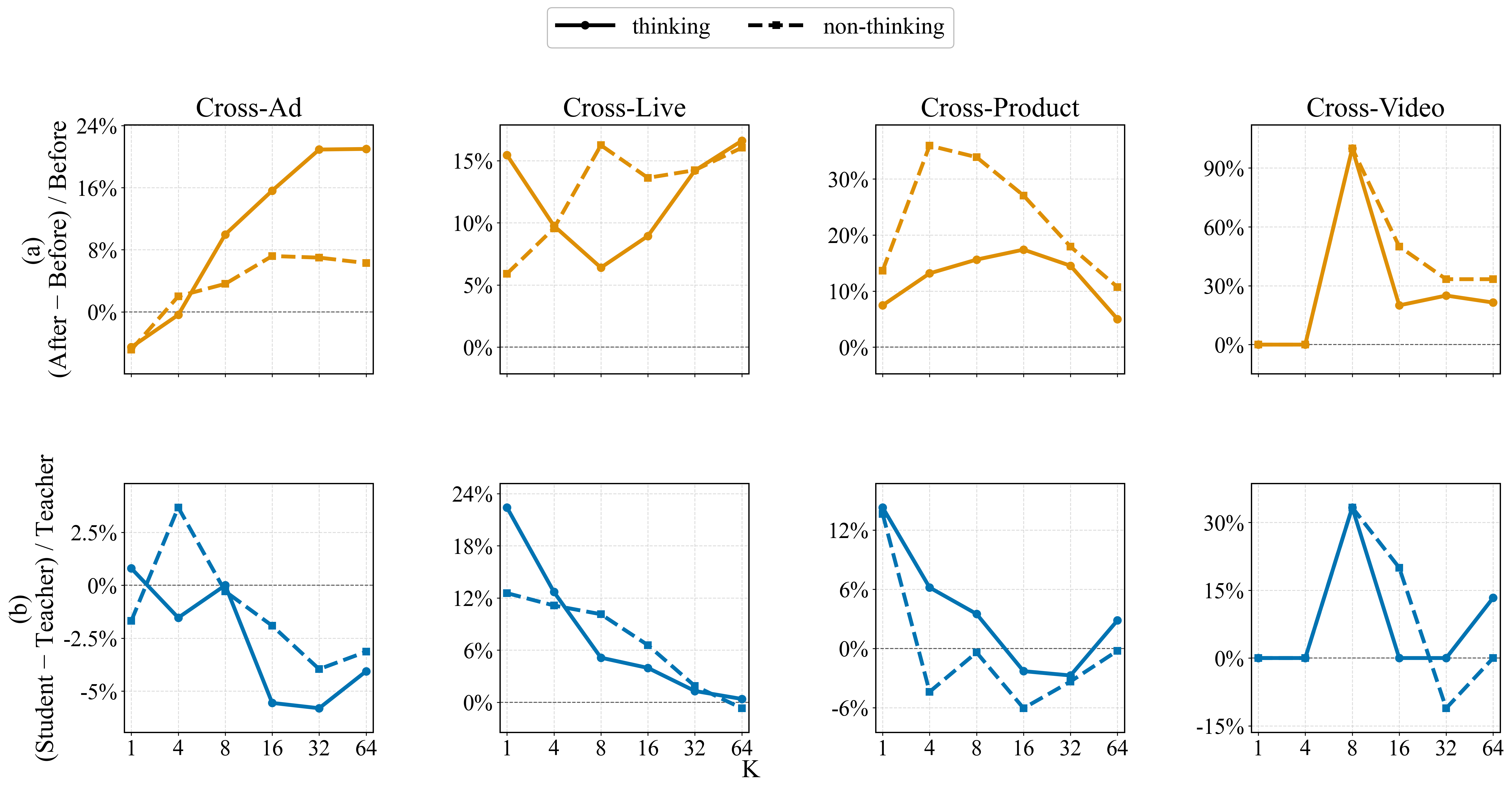}
    \caption{
Relative Recall@K changes of MOPD across four domains. 
(a) reports the improvement of the student before and after MOPD, computed as $(\text{After} - \text{Before}) / \text{Before}$. 
(b) reports the gap between the student after MOPD and the teacher, computed as $(\text{Student} - \text{Teacher}) / \text{Teacher}$. 
Positive values indicate higher Recall@K for the first term in each comparison. 
Solid and dashed lines denote thinking and non-thinking evaluation settings, respectively.
Overall, MOPD consistently improves the student while keeping its performance close to the teacher across domains.
}
    \label{fig:mopd_delta}
\end{figure}
Given a training batch \(\mathcal{B}\) and a target information-gain ratio \(\rho\) (e.g., \(\rho = 0.8\)), we sort all sampled rollouts in descending order based on their informativeness scores, yielding the sequence \(x_{(1)}, x_{(2)}, \dots, x_{(|\mathcal{B}|)}\).
To effectively filter out the redundant, already-aligned rollouts, we determine the smallest prefix length \(M\) that covers the target information-gain ratio \(\rho\):
\begin{equation}
    M = \min \left\{ m : \frac{\sum_{j=1}^m s(y_{(j)})}{\sum_{j=1}^{|\mathcal{B}|} s(y_{(j)})} \ge \rho \right\}.
\end{equation}
The active subset retained for optimization is then defined as \(\mathcal{S} = \{y_{(1)}, \dots, y_{(M)}\}\). By discarding the low-information rollouts, we ensure that the optimization budget is concentrated exclusively on the most challenging and informative instances.
The final MOPD objective is subsequently computed solely over this filtered subset \(\mathcal{S}\).

\begin{figure}[t]
    \centering
    \includegraphics[width=\linewidth]{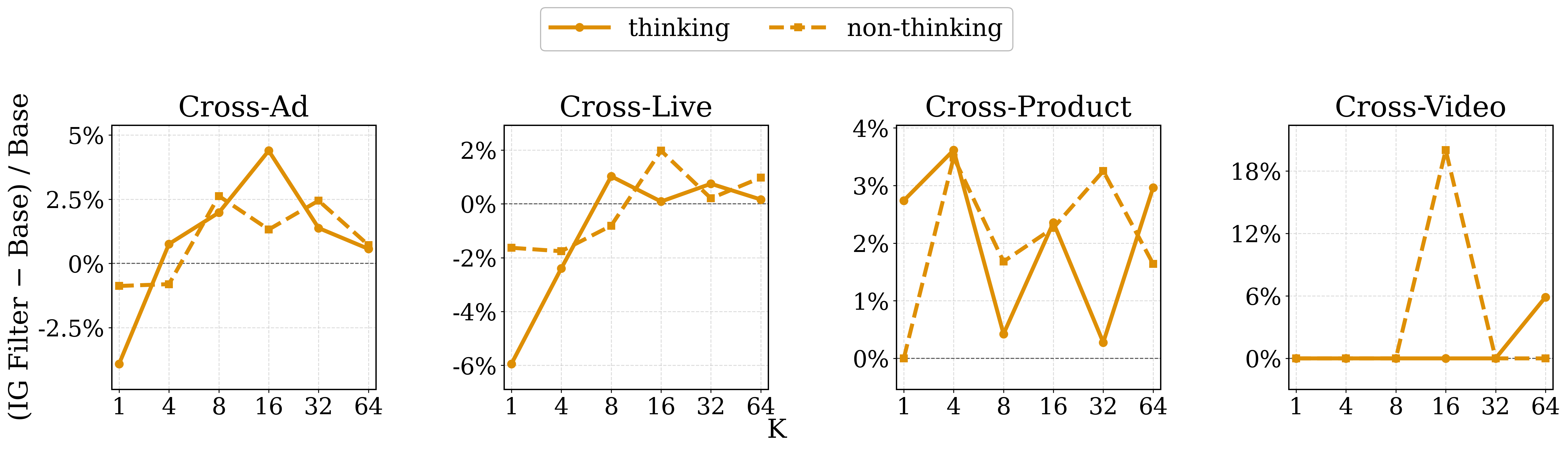}
    \caption{
Effect of the information-gain-aware (IG) filter on MOPD performance across recommendation domains. 
We report the relative Recall@K change between MOPD with IG filtering and the base MOPD setting, computed as $(\text{IG Filter} - \text{Base}) / \text{Base}$. 
Positive values indicate that IG filtering improves Recall@K over the base setting. 
Solid and dashed lines denote thinking and non-thinking evaluation settings, respectively.
Overall, IG filtering brings consistent gains in most domains and evaluation settings, especially at larger values of K.
}
    \label{fig:mopd_filter_compare}
\end{figure}

\paragraph{Performance Analysis.}
\begin{wrapfigure}{R}{0.5\textwidth}
    \centering
    \includegraphics[width=0.48\textwidth]{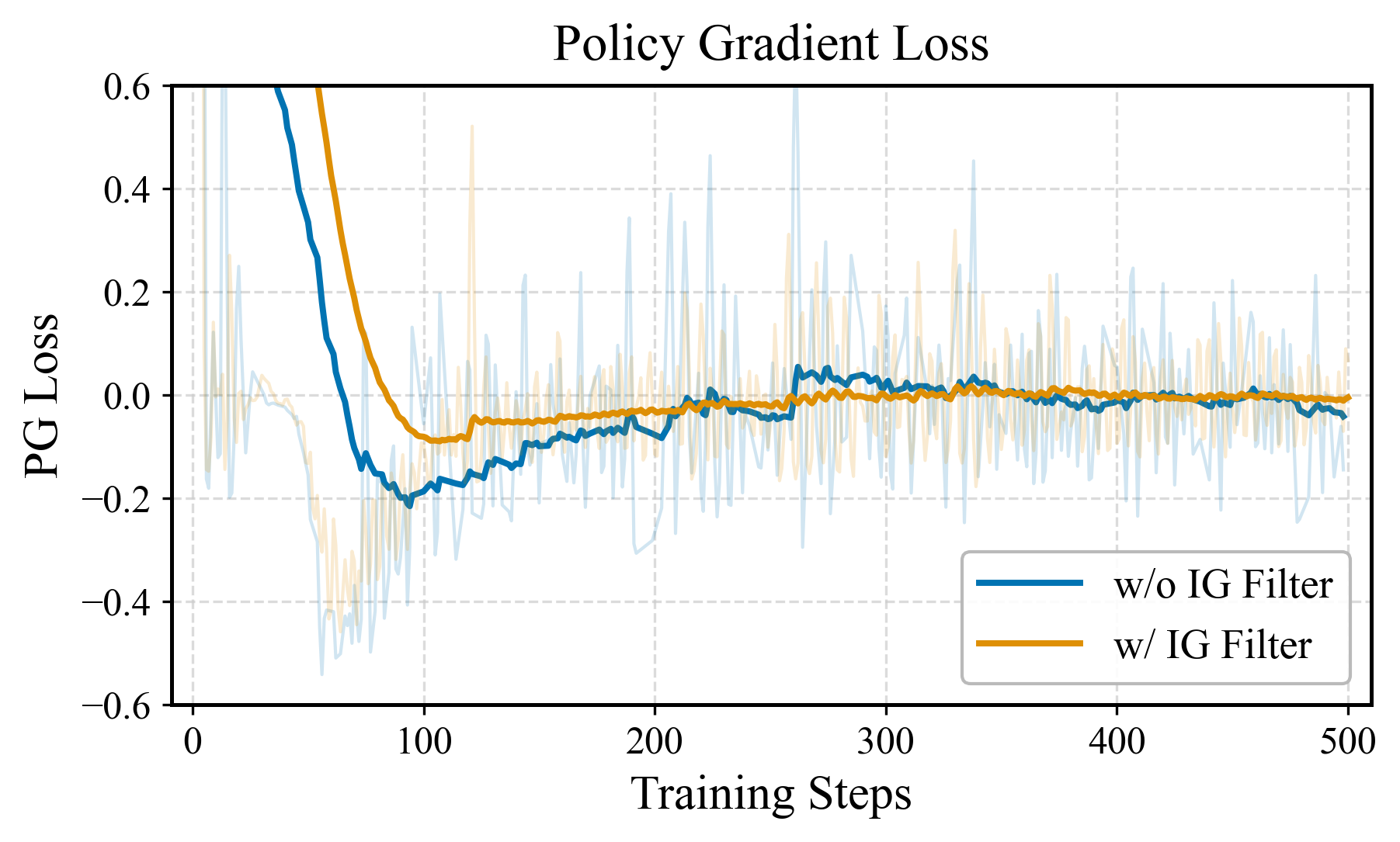} 
    \caption{Policy-gradient loss during MOPD training with and without IG filtering.
    The IG filter removes low-information trajectories based on token-level advantage statistics, leading to a smoother and more stable policy-gradient signal.}
    \label{fig:mopd_pgloss}
\end{wrapfigure}
As illustrated in Figure~\ref{fig:mopd_delta} (a), the proposed training paradigm yields strictly positive performance gains across all evaluated domains.
Notably, the relative improvement expands as \(K\) increases in almost all domains.
This demonstrates that MOPD effectively revitalizes the student model's long-tail ranking capability within massive candidate spaces.
Furthermore, the \textit{thinking} mode exhibits a steeper growth trajectory at larger \(K\) compared with the \textit{non-thinking} mode in the Ad and Live domains.
This suggests that the explicit reasoning process significantly broadens the search boundaries, enabling the retrieval of hard-to-find tail items that intuition alone (non-thinking) might miss.
Figure~\ref{fig:mopd_delta} (b) shows that the student aligns perfectly with or surpasses the teachers when \(K \le 16\), but a gap remains at \(K > 16\).
This remaining gap is a fundamental theoretical limit driven by the reverse-KL objective.
Since the reverse-KL objective is inherently \textit{mode-seeking}, it flawlessly captures the teacher's primary probability peaks at \(K \le 16\).
While our active filtering method maximizes the extraction of long-tail information from the teachers, perfectly replicating the entire distribution of the teachers remains challenging.

\paragraph{Ablation Study.}
We first analyze the effect of the information-gain-aware trajectory filtering mechanism.
Figure~\ref{fig:mopd_pgloss} illustrates the policy gradient (PG) loss.  MOPD without filtering (blue) exhibits larger variance and a steep initial drop, indicating unstable optimization and overfitting to redundant ``head'' items. 
In contrast, MOPD with filtering (orange) significantly suppresses gradient noise, enabling a smoother and more stable convergence by exclusively focusing on high-information samples.

The relative Recall@$K$ curves in Figure~\ref{fig:mopd_filter_compare} provide a direct empirical validation of our core motivation, i.e.,  gradient budget reallocation based on trajectory-level information gain.
The optimization budget is successfully concentrated on high-divergence, high-information samples ($s(y_j) \gg 0$). 
This focused gradient flow directly translates into the steep, upward performance trajectory starting at $K \ge 8$ and culminating in positive gains at $K=64$. 
By reallocating the training focus from redundant easy items to challenging long-tail items, our filtering mechanism effectively mitigates the gradient dilution bottleneck, allowing the unified student to capture the specialized long-tail capabilities of the expert teachers under a highly constrained parameter budget.

\begin{figure}[H]
    \centering
    \includegraphics[width=\textwidth]{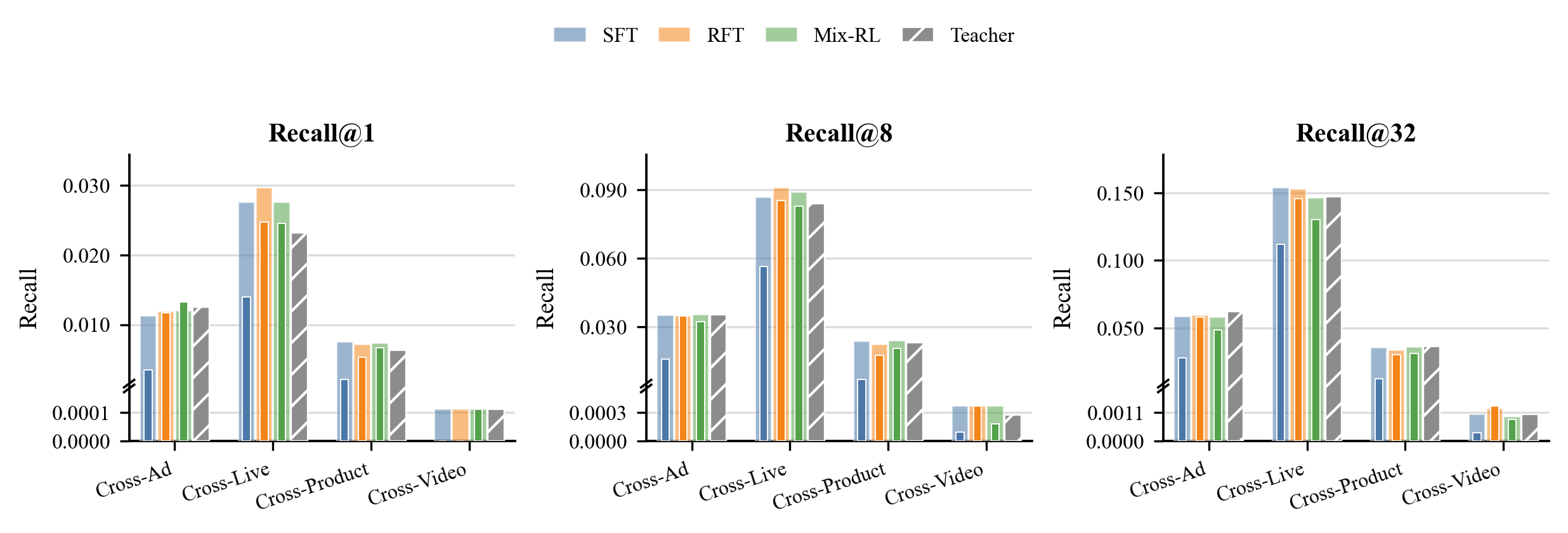}
    \caption{Comparison of different student models before and after MOPD training. Dark bars denote the performance before MOPD, while light bars denote the performance after MOPD. This suggests that MOPD improvements are bounded by the teacher: weaker students can catch up, but rarely surpass it substantially.
}
    \label{fig:mopd_student_selection}
\end{figure}

Further, we investigate the impact of different student selection, including SFT, SFT$\rightarrow$RFT, and SFT$\rightarrow$Mix-RL.
As shown in Figure~\ref{fig:mopd_student_selection}, all three students can benefit from MOPD training.
When student and teacher distributions are roughly aligned, even with initial differences in student levels, MOPD can consistently improve student performance to a level near or even slightly surpassing that of the teachers.
This indicates that the teacher's upper bound limits the extent of improvement in MOPD. Specifically, even weaker students can catch up with the teacher, but it is difficult for them to surpass the teacher significantly.

\begin{table}[htbp]
\centering
\small
\caption{Performance comparison (in \%) of different optimization strategies. Both RFT and MOPD effectively integrate domain-specific knowledge and outperform Mix-RL across different domains.}
\label{tab:rl_performance_comparison}
\renewcommand{\arraystretch}{1.08}
\setlength{\tabcolsep}{6pt} 
\begin{tabular}{@{}p{2.4cm}llccccc@{}}
\toprule
\textbf{Domain} & \textbf{Metric} & \textbf{Mode} & \textbf{SFT} & \textbf{Mix-RL} & \textbf{Single-RL} & \textbf{RFT} & \textbf{MOPD} \\
\midrule
\multirow{10}{*}{\parbox[c]{2.4cm}{\centering\textbf{Cross-Video}}} & \multirow{2}{*}{Recall@1} & thinking & 0.00 & \underline{0.01} & 0.01 & \textbf{0.01} & 0.01 \\
 &  & non-thinking & 0.01 & \underline{0.01} & 0.01 & \textbf{0.01} & 0.01 \\
\cmidrule(l){2-8}
 & \multirow{2}{*}{Recall@8} & thinking & 0.01 & 0.02 & 0.03 & \textbf{0.05} & \underline{0.04} \\
 &  & non-thinking & 0.02 & 0.02 & 0.03 & \textbf{0.04} & \underline{0.03} \\
\cmidrule(l){2-8}
 & \multirow{2}{*}{Recall@16} & thinking & 0.02 & 0.05 & 0.06 & \textbf{0.08} & \underline{0.06} \\
 &  & non-thinking & 0.03 & 0.04 & 0.05 & \textbf{0.06} & \underline{0.05} \\
\cmidrule(l){2-8}
 & \multirow{2}{*}{Recall@32} & thinking & 0.03 & 0.08 & \underline{0.10} & \textbf{0.13} & 0.10 \\
 &  & non-thinking & 0.06 & 0.06 & \underline{0.09} & \textbf{0.12} & 0.08 \\
\cmidrule(l){2-8}
 & \multirow{2}{*}{Recall@64} & thinking & 0.06 & 0.14 & 0.15 & \textbf{0.24} & \underline{0.18} \\
 &  & non-thinking & 0.11 & 0.12 & 0.16 & \textbf{0.19} & \underline{0.16} \\
\midrule
\multirow{10}{*}{\parbox[c]{2.4cm}{\centering\textbf{Cross-Product}}} & \multirow{2}{*}{Recall@1} & thinking & 0.21 & 0.67 & 0.63 & \textbf{0.67} & \underline{0.67} \\
 &  & non-thinking & 0.26 & 0.66 & \underline{0.66} & 0.55 & \textbf{0.76} \\
\cmidrule(l){2-8}
 & \multirow{2}{*}{Recall@8} & thinking & 0.69 & 2.05 & \underline{2.29} & 2.06 & \textbf{2.36} \\
 &  & non-thinking & 1.14 & 1.83 & \textbf{2.46} & 1.82 & \underline{2.42} \\
\cmidrule(l){2-8}
 & \multirow{2}{*}{Recall@16} & thinking & 0.94 & 2.53 & \textbf{3.04} & 2.83 & \underline{2.98} \\
 &  & non-thinking & 1.69 & 2.44 & \textbf{3.30} & 2.63 & \underline{3.05} \\
\cmidrule(l){2-8}
 & \multirow{2}{*}{Recall@32} & thinking & 1.22 & 3.10 & \textbf{3.65} & 3.54 & \underline{3.62} \\
 &  & non-thinking & 2.32 & 3.18 & \textbf{3.88} & 3.24 & \underline{3.80} \\
\cmidrule(l){2-8}
 & \multirow{2}{*}{Recall@64} & thinking & 1.65 & 3.79 & 3.87 & \textbf{4.19} & \underline{4.11} \\
 &  & non-thinking & 2.96 & 3.83 & \underline{4.25} & 3.96 & \textbf{4.34} \\
\midrule
\multirow{10}{*}{\parbox[c]{2.4cm}{\centering\textbf{Cross-Ad}}} & \multirow{2}{*}{Recall@1} & thinking & 0.35 & \underline{1.32} & 1.25 & \textbf{1.34} & 1.23 \\
 &  & non-thinking & 1.07 & \textbf{1.23} & \underline{1.19} & 1.10 & 1.14 \\
\cmidrule(l){2-8}
 & \multirow{2}{*}{Recall@8} & thinking & 1.56 & 3.21 & 3.53 & \textbf{3.67} & \underline{3.67} \\
 &  & non-thinking & 3.02 & 3.33 & \underline{3.46} & 3.41 & \textbf{3.53} \\
\cmidrule(l){2-8}
 & \multirow{2}{*}{Recall@16} & thinking & 2.11 & 3.97 & \textbf{4.86} & \underline{4.80} & 4.76 \\
 &  & non-thinking & 4.16 & 4.31 & \textbf{4.71} & 4.46 & \underline{4.58} \\
\cmidrule(l){2-8}
 & \multirow{2}{*}{Recall@32} & thinking & 2.78 & 4.83 & \textbf{6.20} & \underline{6.14} & 5.94 \\
 &  & non-thinking & 5.37 & 5.44 & \textbf{6.06} & \underline{5.84} & 5.84 \\
\cmidrule(l){2-8}
 & \multirow{2}{*}{Recall@64} & thinking & 3.41 & 5.86 & \underline{7.39} & \textbf{7.50} & 7.19 \\
 &  & non-thinking & 6.49 & 6.68 & \textbf{7.33} & \underline{7.26} & 7.13 \\
\midrule
\multirow{10}{*}{\parbox[c]{2.4cm}{\centering\textbf{Cross-Live}}} & \multirow{2}{*}{Recall@1} & thinking & 1.40 & 2.46 & 2.32 & \underline{2.53} & \textbf{2.67} \\
 &  & non-thinking & 1.45 & \underline{2.37} & 2.23 & 1.96 & \textbf{2.41} \\
\cmidrule(l){2-8}
 & \multirow{2}{*}{Recall@8} & thinking & 5.63 & 8.29 & 8.39 & \underline{8.70} & \textbf{8.82} \\
 &  & non-thinking & 5.76 & 7.39 & 7.80 & \underline{7.82} & \textbf{8.58} \\
\cmidrule(l){2-8}
 & \multirow{2}{*}{Recall@16} & thinking & 8.18 & 10.63 & 11.14 & \textbf{11.49} & \underline{11.46} \\
 &  & non-thinking & 8.34 & 9.99 & \underline{10.65} & 10.63 & \textbf{11.18} \\
\cmidrule(l){2-8}
 & \multirow{2}{*}{Recall@32} & thinking & 11.17 & 13.03 & \underline{14.69} & 14.63 & \textbf{14.82} \\
 &  & non-thinking & 11.74 & 12.86 & \underline{14.42} & 14.25 & \textbf{14.71} \\
\cmidrule(l){2-8}
 & \multirow{2}{*}{Recall@64} & thinking & 14.32 & 16.03 & \underline{18.62} & 18.35 & \textbf{18.89} \\
 &  & non-thinking & 15.52 & 16.09 & \textbf{18.80} & 18.17 & \underline{18.66} \\
\bottomrule
\end{tabular}
\end{table}

\subsection{Comparison and Discussions}

In this section, we compare the recommendation accuracy of five optimization strategies: the SFT checkpoint, mixed-domain RL, domain-specific RL, RFT, and MOPD. The evaluation is conducted under the cross-domain recommendation setting, where we report Recall@K for the Cross-Video, Cross-Product, Cross-Ad, and Cross-Live tasks. This comparison allows us to examine whether the model learns better reasoning through self-exploration, whether the ``specialize-then-unify'' strategy alleviates cross-domain interference, and how different teacher-knowledge integration methods affect the final performance.

As shown in Table~\ref{tab:rl_performance_comparison}, all post-SFT optimization methods substantially improve over the SFT baseline on all domains and metrics, indicating that learning from self-exploration can further enhance recommendation performance beyond supervised imitation. However, directly mixing all domains during RL does not consistently achieve the best performance. Compared with domain-specific RL, mixed-domain RL generally yields weaker performance, suggesting that heterogeneous domains can introduce conflicting optimization signals. The two knowledge-integration methods then provide complementary benefits. RFT achieves strong and stable gains, especially on Cross-Video, showing that verified successful trajectories can effectively consolidate domain-specific reasoning patterns. MOPD performs competitively and obtains the best results in several settings, indicating that policy-level distillation can better preserve teacher preferences in some domains. Overall, these results support the effectiveness of the ``specialize-then-unify'' strategy: domain-specific RL first discovers effective reasoning behaviors, while subsequent knowledge integration transfers these domain-specialized capabilities into a unified model.

We further compare the impact of two integration paradigms, MOPD and RFT. We have several findings:
\begin{itemize}
    \item MOPD and RFT exhibit domain-specific strengths. MOPD achieves superior performance in Product and Live domains, even significantly outperforming the Live teacher. Conversely, RFT consistently yields marginal and stable advantages over MOPD, particularly at larger $K$ values in Video and Ad domains.
    \item  RFT consistently guarantees that the thinking mode outperforms the non-thinking mode across all domains and $K$ values.  In contrast, the improvements of MOPD in the thinking and non-thinking modes are highly synchronized, with the non-thinking mode achieving competitive gains. 
\end{itemize}

This divergence is driven by the extreme reward sparsity of recommendation tasks and the distinct sampling dynamics of the two paradigms:
\begin{itemize}
    \item \textbf{RFT:} In a massive item candidate space, successful reasoning paths are extremely sparse. RFT bypasses this exploration bottleneck offline by executing multiple rollouts and strictly retaining only those that yield correct outcomes. This rejection-sampling process distills only ``golden'' reasoning paths in which the CoT is logically aligned with the correct recommendation, thereby securing a stable reasoning benefit ($\textit{thinking} > \textit{non-thinking}$).
    \item \textbf{MOPD:} Conversely, MOPD relies on on-policy distillation, continuously aligning the student's generated distributions with the teachers across all rollouts (regardless of whether the CoT perfectly hits the target). Because the student processes these rollouts within a shared parameter space, this continuous distillation calibrates the foundational prompt-to-item semantic embeddings holistically. As a result, the nonthinking capability itself is massively enhanced. Therefore, MOPD produces a synchronized uplift: it improves the $\textit{thinking}$ capability while simultaneously and profoundly strengthening the underlying ``intuition'' ($\textit{non-thinking}$). However, successful CoTs are extremely sparse. Without strict off-policy filtering of correct CoT, the on-policy process inevitably absorbs noisy or ungrounded CoT steps from the weak priors. This noise might prevent the thinking mode from fully unlocking its theoretical superiority to consistently eclipse the massively enhanced non-thinking baseline.
\end{itemize}

\section{CoT Analysis Indicators}\label{sec:cot_ind}
\providecommand{\traceStage}[1]{\textbf{\textit{#1}}}
\providecommand{\sftVideoSID}[3]{\texttt{<|video\_begin|>}\allowbreak\texttt{<a\_#1>}\allowbreak\texttt{<b\_#2>}\allowbreak\texttt{<c\_#3>}}
\providecommand{\sftProdSID}[3]{\texttt{<|prod\_begin|>}\allowbreak\texttt{<a\_#1>}\allowbreak\texttt{<b\_#2>}\allowbreak\texttt{<c\_#3>}}
\providecommand{\sftLiveSID}[3]{\texttt{<|living\_begin|>}\allowbreak\texttt{<a\_#1>}\allowbreak\texttt{<b\_#2>}\allowbreak\texttt{<c\_#3>}}
\providecommand{\sftAdSID}[3]{\texttt{<|ad\_begin|>}\allowbreak\texttt{<a\_#1>}\allowbreak\texttt{<b\_#2>}\allowbreak\texttt{<c\_#3>}}

Beyond target accuracy, the practical utility of recommendation reasoning traces depends on whether the CoT genuinely contributes to the final itemic pattern prediction, remains faithful to the observed user history, respects the structural constraints of the itemic token space, and follows a coherent intent-evolution trajectory. Recent evidence suggests that free-form reasoning in itemic token-based recommendation foundation models may introduce a general-subspace prior, where textual inertia dominates inference and dilutes critical itemic pattern evidence~\citep{zhang2026whythinkinghurts}. Motivated by this observation, we design a four-criterion diagnostic framework to evaluate the quality of generated CoT traces.
We organise them along two orthogonal axes that together cover the full diagnostic surface of a recommendation reasoning trace. The four criteria populate this $2\times 2$ matrix exactly once each (Table~\ref{tab:cot_eval_framework}).
\begin{itemize}
    \item \textbf{ Symbolic vs.\ Probabilistic} (\emph{what kind of check?}). A \emph{symbolic} check inspects the discrete set of itemic pattern referenced inside the CoT against a static reference set; it requires only string parsing and set operations, runs without a model forward pass and without access to $y_{\mathrm{GT}}$. A \emph{probabilistic} check inspects how the CoT shifts the conditional log-likelihood of the target $y_{\mathrm{GT}}$; it requires both a model forward pass and the ground-truth label, and is therefore strictly costlier but directly measures predictive effect.
    \item \textbf{ Local vs.\ Global} (\emph{against which reference scope?}). A \emph{local} check compares against a narrow, instance-bounded reference (the current user's history, or the immediately preceding reasoning prefix). A \emph{global} check compares against a wide, corpus-bounded or trace-bounded reference (the full item catalog, or the entire CoT versus the no-CoT baseline).
\end{itemize}

\begin{table}[H]
\centering
\small
\caption{The four-criterion diagnostic framework organised on a $2\times 2$ matrix: \emph{Symbolic vs.\ Probabilistic} (kind of check) $\times$ \emph{Local vs.\ Global} (reference scope). Within each row, the local cell strictly strengthens the global cell (Property P1). Across rows, symbolic checks are cheap/label-free while probabilistic checks are forward-pass/label-bound (Property P2). The four cells are jointly necessary; each guards against a distinct failure mode.}
\label{tab:cot_eval_framework}
\renewcommand{\arraystretch}{1.25}
\setlength{\tabcolsep}{4pt}
\begin{tabular}{l l l}
\toprule
 & \textbf{Global (corpus-/trace-bounded)} & \textbf{Local (instance-bounded)} \\
\midrule
\multirow{2}{*}{\textbf{Probabilistic}} & \textbf{C1: } $\Delta\mathrm{LL}$                                              & \textbf{C2: } $\ell_t$ progression \\
                                        & $\log p(y_{\mathrm{GT}}|\mathbf{x},c) - \log p(y_{\mathrm{GT}}|\mathbf{x})$    & $\ell_t - \ell_{t-1}$ \\
\midrule
\multirow{2}{*}{\textbf{Symbolic}}      & \textbf{C3: } $\gamma_{\mathrm{legal}}$                                        & \textbf{C4: } $\gamma_{\mathrm{hist}\mid\mathrm{legal}}$ \\
                                        & $\mathcal{S}(c) \cap \mathcal{V}_{\mathrm{item}}$                              & $\big(\mathcal{S}(c)\!\cap\!\mathcal{V}_{\mathrm{item}}\big)\cap \mathcal{S}(\mathbf{x}_{\mathrm{hist}})$ \\
\bottomrule
\end{tabular}
\end{table}

\paragraph{Criterion 1: CoT Likelihood Gain.}
We first quantify whether the generated CoT contributes positively to the prediction of the ground-truth target. Inspired by the analysis of CoT-induced distributional shift in~\citep{zhang2026whythinkinghurts}, we define the CoT likelihood gain as:
\begin{equation}
\begin{split}
    \Delta{LL}
    =
    \log p\!\left(y_{\mathrm{GT}} \mid \mathbf{x}, c\right)
    -
    \log p\!\left(y_{\mathrm{GT}} \mid \mathbf{x}\right),
    \label{eq:cot_gain}
\end{split}
\end{equation}
where $\mathbf{x}$ denotes the input context, including the user profile and interaction history, $c$ denotes the generated CoT, and $y_{\mathrm{GT}}$ is the ground-truth target itemic pattern. A positive $\Delta{LL}$ indicates that conditioning on the CoT increases the model's likelihood of the correct target, whereas a negative value suggests that the CoT distracts the model from the history-grounded itemic pattern evidence.

\begin{figure}[h]
    \centering
    \includegraphics[width=0.7\textwidth]{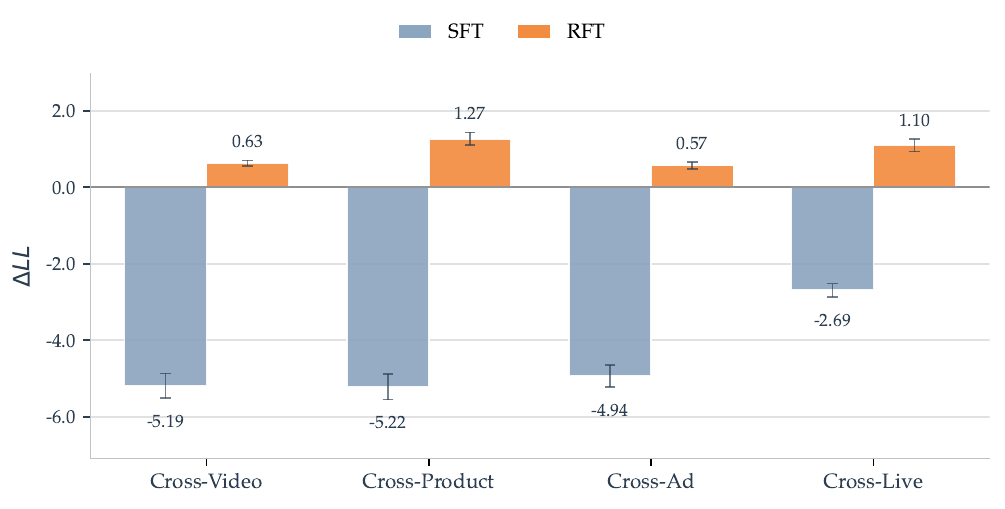}
    \caption{
    For each model-domain pair, $\Delta LL$ is computed as the difference between the log-likelihood of the ground-truth target under thinking mode and that under non-thinking mode. Positive values indicate that the generated CoT increases the model's confidence in the ground-truth target, whereas negative values suggest that the CoT distracts the prediction process. Across all four domains, the SFT model exhibits consistently negative $\Delta LL$, while the RFT model achieves positive $\Delta LL$, demonstrating that RFT substantially enhances the effectiveness and utility of thinking reasoning relative to the non-thinking setting.
    }
    \label{fig:mean_delta_ll}
\end{figure}

As shown in Figure~\ref{fig:mean_delta_ll}, for each model-domain pair, we rigorously compute $\Delta{LL}$ by contrasting the target log-likelihood under the thinking mode, where prediction is conditioned on the generated CoT, with that under the non-thinking mode, where the target is predicted directly from the input context without CoT. We then report the average per-sample $\Delta{LL}$ values of the SFT and RFT models across four domains. The SFT model consistently yields negative $\Delta{LL}$ values in all domains, indicating that its generated CoT tends to distract the model from the ground-truth target prediction. In contrast, the RFT model achieves positive $\Delta{LL}$ values across all four domains, demonstrating that the RFT stage significantly improves the ability of the generated CoT to contribute positively to the prediction of the ground-truth target.

\paragraph{Criterion 2: Log-Likelihood Progression Along the CoT.}
A high-quality reasoning trace should progressively strengthen the model's confidence in the ground-truth target as the reasoning process unfolds. To characterize this intent-evolution behavior, we examine how the target likelihood changes along the CoT.

To obtain intermediate reasoning units, we use DeepSeek \citep{deepseekai2026deepseekv4} to segment the original CoT into a sequence of semantically coherent reasoning segments:
\begin{equation}
\begin{split}
    c = \{c_1, c_2, \ldots, c_T\}.
\end{split}
\end{equation}

For each prefix of the segmented CoT, we compute the conditional log-likelihood of the ground-truth target:
\begin{equation}
\begin{split}
    \ell_t
    =
    \log p\!\left(
        y_{\mathrm{GT}}
        \mid
        \mathbf{x}, c_1, \ldots, c_t
    \right),
    \quad
    t = 1,\ldots,T.
    \label{eq:ll_progression}
\end{split}
\end{equation}

\begin{figure*}[ht]
    \centering
    \includegraphics[width=0.95\textwidth]{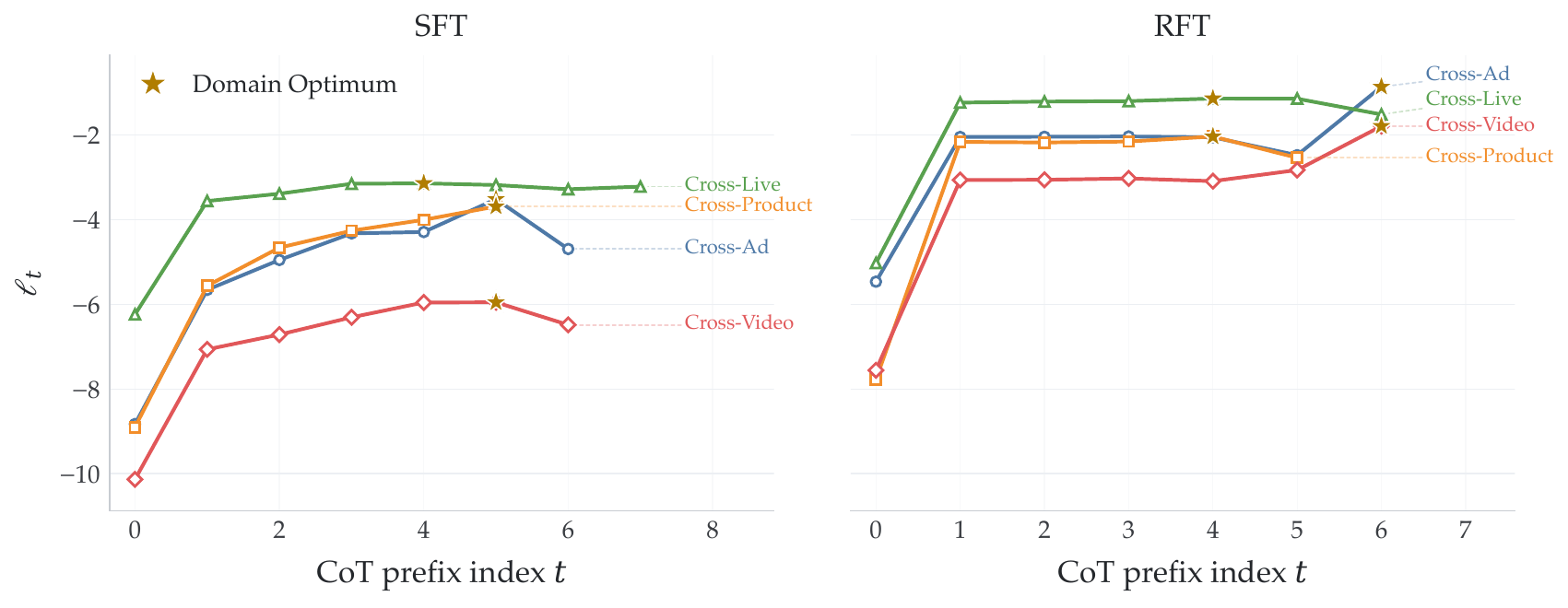}
    \caption{
    $\ell_t$ denotes the conditional log-likelihood of the ground-truth target given the input context and the first $t$ reasoning segments. CoT prefix index t=0 denote the thinking mode prediction without CoT, while brown stars indicate the peak likelihood achieved along the reasoning trajectory for each domain. Across all domains, progressively incorporating CoT prefixes consistently improves the likelihood relative to the no-CoT thinking mode, demonstrating that the generated CoT contributes useful predictive information rather than serving as a purely post-hoc explanation. Moreover, compared with SFT, the RFT model typically reaches its optimal likelihood at substantially earlier reasoning prefixes, suggesting that predictive evidence is concentrated into a smaller portion of the reasoning process and revealing potential opportunities for CoT compression and adaptive early stopping.
    }
    \label{fig:cot_logprob_mixed_step2500}
\end{figure*}

As shown in Figure~\ref{fig:cot_logprob_mixed_step2500}, in the thinking mode setting, progressively incorporating segmented CoT prefixes consistently improves the conditional log-likelihood of the ground-truth target for both the SFT and RFT models. Compared with the thinking mode prediction without CoT, the likelihood trajectories of all four domains rise substantially as more reasoning segments are added, indicating that the generated CoT provides incremental predictive evidence rather than serving as a merely post-hoc explanation. This trend is observed in both optimization stages: the SFT model already benefits from CoT prefixes, while the RFT model further exhibits stronger and more stable likelihood improvements across domains. Although a few domains show minor fluctuations at later prefixes, the overall upward progression demonstrates that, under the thinking mode inference setting, the current models can effectively exploit their generated CoT to strengthen confidence in the ground-truth target.

While both SFT and RFT benefit from progressively incorporating CoT prefixes, RFT exhibits a much earlier saturation behavior, reaching near-optimal likelihood after only a few reasoning segments. This suggests that not all generated reasoning tokens contribute equally to prediction, and that a substantial portion of the predictive gain can be achieved with a shortened CoT, leaving room for future exploration of reasoning compression or adaptive early-stopping mechanisms.

\paragraph{Criterion 3: Item Legality.}
Every itemic pattern referenced in the CoT should also correspond to a valid identifier in the item corpus. Let $\mathcal{V}_{\mathrm{item}}$ denote the complete item catalog. We define the item legality ratio as:
\begin{equation}
\begin{split}
    \gamma_{\mathrm{legal}}
    =
    \frac{
        \left|
        \mathcal{S}(c)
        \cap
        \mathcal{V}_{\mathrm{item}}
        \right|
    }{
        \left|
        \mathcal{S}(c)
        \right|
    },
    \label{eq:itemic_id_legality}
\end{split}
\end{equation}
Invalid itemic patterns indicate that the reasoning trace contains structurally illegal item identifiers, which may introduce noise into subsequent itemic token prediction. As shown in Table~\ref{tab:cot_validity_performance_comparison}, $\gamma_{\mathrm{legal}}$ already saturates at $1.00$ for both SFT and RFT across all four domains, providing a clean basis on which the conditional history-grounding measure of Criterion 4 can be evaluated.

\paragraph{Criterion 4: History Item Reference Validity (conditioned on item legality).}
Since our prompt explicitly constrains the CoT to reason only based on the observed user history, every \emph{legal} itemic pattern mentioned in the CoT should further correspond to an item that actually appears in the user's interaction history. Differently from a raw history-grounding ratio over $\mathcal{S}(c)$, we condition the measurement on item legality being satisfied: we first restrict the referenced set to its legal subset $\mathcal{S}(c)\cap\mathcal{V}_{\mathrm{item}}$, and then measure the fraction of this subset that is supported by the user's history $\mathcal{S}(\mathbf{x}_{\mathrm{hist}})$. This explicitly disentangles the orthogonal failure mode of \emph{history-violating hallucination} from the upstream failure of \emph{illegal-identifier generation} captured by Criterion 3. Formally,
\begin{equation}
\begin{split}
    \gamma_{\mathrm{hist}\mid\mathrm{legal}}
    =
    \frac{
        \left|
        \big(\mathcal{S}(c)\cap\mathcal{V}_{\mathrm{item}}\big)
        \cap
        \mathcal{S}(\mathbf{x}_{\mathrm{hist}})
        \right|
    }{
        \left|
        \mathcal{S}(c)\cap\mathcal{V}_{\mathrm{item}}
        \right|
    },
    \label{eq:history_itemic_id_validity}
\end{split}
\end{equation}
where $\mathcal{S}(c)$ denotes the set of itemic patterns extracted from the CoT and $\mathcal{S}(\mathbf{x}_{\mathrm{hist}})$ the set of itemic patterns appearing in the interaction history. A low $\gamma_{\mathrm{hist}\mid\mathrm{legal}}$ indicates that, even after illegal references have been filtered out, the CoT still introduces itemic patterns unsupported by the observed history, i.e.\ prompt-violating reasoning that hallucinates beyond the user's actual behaviour.

\begin{table}[t]
\centering
\caption{
Cross-domain $\gamma_{\mathrm{legal}}$ and $\gamma_{\mathrm{hist}\mid\mathrm{legal}}$ of itemic pattern references in generated CoTs (pid\_pass@32).
 All numbers are in \%; best per metric-domain pair is in \textbf{bold}.
\textbf{Takeaways.} (i) $\gamma_{\mathrm{legal}}$ saturates at 100\% for both SFT and RFT across all four domains, indicating that legality is no longer a bottleneck after RFT. (ii) On Cross-Video and Cross-Ad, RFT tightens $\gamma_{\mathrm{hist}\mid\mathrm{legal}}$ over SFT (+2.50\,pt and +4.27\,pt respectively). (iii) On Cross-Live and Cross-Product, RFT instead exhibits a moderate drop in $\gamma_{\mathrm{hist}\mid\mathrm{legal}}$ ($-1.18$\,pt and $-14.59$\,pt). This drop is consistent with Table~\ref{tab:rl_performance_comparison}, where the relative gain of RFT over Mix-RL on Live and Product is also smaller than that on Video and Ad, suggesting that the symbolic CoT diagnostics are well aligned with downstream recommendation metrics. Tightening history grounding on Live and Product is left as future optimisation.
}
\label{tab:cot_validity_performance_comparison}
\setlength{\tabcolsep}{5pt}
\renewcommand{\arraystretch}{1.15}
\begin{tabular}{llcccc}
\toprule
\textbf{Metric} & \textbf{Method} & \textbf{Video} & \textbf{Ad} & \textbf{Live} & \textbf{Product} \\
\midrule
\multirow{2}{*}{$\gamma_{\mathrm{legal}}$ $\uparrow$}
& SFT & \textbf{100.00} & \textbf{100.00} & \textbf{100.00} & \textbf{100.00} \\
& RFT & \textbf{100.00} & \textbf{100.00} & \textbf{100.00} & \textbf{100.00} \\
\midrule
\multirow{2}{*}{$\gamma_{\mathrm{hist}\mid\mathrm{legal}}$ $\uparrow$}
& SFT & 97.50 & 94.93 & \textbf{100.00} & \textbf{97.92} \\
& RFT & \textbf{100.00} & \textbf{99.20} & 98.82 & 83.33 \\
\bottomrule
\end{tabular}
\end{table}

\paragraph{Qualitative Illustration: SFT vs.\ RFT on the Same User.}
To complement the aggregate diagnostics in Criterion~1--Criterion~4, we
present a representative user for whom the SFT and RFT models produce
markedly different reasoning traces over the \emph{same} input context.
The user is an 18--23 year-old male student whose recent interaction
history is dominated by tactical-shooter livestreams, smartphone reviews,
and snack purchases; the ground-truth next item is a video about
\emph{equipment in Delta Force}, a recently released tactical-shooter
title that the user has only \emph{clicked once via an in-feed
advertisement}---i.e., a target that requires \emph{cross-IP
generalisation} rather than mere historical replay.
Table~\ref{tab:sft-case-shooter} and Table~\ref{tab:rft-case-shooter}
present the two CoT traces verbatim, organised along the same
\textsc{Persona Abstraction} $\rightarrow$ \textsc{Interest Expansion}
$\rightarrow$ \textsc{Transition Inference} structure for direct comparison.

\begin{table}[H]
\centering
\caption{SFT recommendation CoT case: interest stays at the surface IP level (\emph{Peace Elite}) and the final transition lands on the wrong IP family.}
\label{tab:sft-case-shooter}
\setlength{\tabcolsep}{5pt}
\renewcommand{\arraystretch}{1.15}
\small
\begin{CJK*}{UTF8}{gbsn}
\begin{tabularx}{\linewidth}{@{}>{\raggedright\arraybackslash}X@{}}
\toprule
\textbf{\traceStage{Persona Abstraction}} \\
用户是一名 18--23 岁的男性中学生，有伴侣，从历史行为来看，他属于典型的“游戏直播观众”与“Z 世代潮流青年”的混合体。最直接的证据是：他近期大量观看、评论和打赏《和平精英》、《王者荣耀》等射击与 MOBA 类游戏直播如 \sftLiveSID{1659}{7342}{6278}、\sftLiveSID{3713}{4342}{6278}，并频繁点击和长播相关游戏攻略视频如 \sftVideoSID{1985}{581}{1414}；同时，他对数码产品（尤其是手机）表现出浓厚兴趣，多次点击手机评测与对比内容如 \sftVideoSID{7328}{7546}{698}，并购买过动态壁纸等个性化商品如 \sftProdSID{2877}{4537}{6971}。因此，他的核心兴趣高度集中在\textbf{战术竞技类游戏（特别是《和平精英》及其衍生模式“地铁逃生”）}、\textbf{高性价比数码产品（尤其是手机）}以及由此衍生的\textbf{游戏装备交易/账号服务}上。
\par\smallskip
{\footnotesize\color{gray}\emph{Analysis:} this first segment is persona abstraction: it compresses the raw history into a user state and dominant preference factors, rather than expanding future hypotheses.} \\
\midrule
\textbf{\traceStage{Interest Expansion}} \\
基于此进行可能性发散：\textbf{A.} 用户最可能继续深入关注《和平精英》的游戏内容更新或装备获取技巧，因为这是其长期稳定投入的核心领域；\textbf{B.} 由他对高性能手机的兴趣延伸至最新款机型的评测或促销信息；\textbf{C.} 其对短剧类型[复仇\_女性成长]等有一定消费记录，在碎片时间可能被类似强情节的娱乐内容吸引。
\par\smallskip
{\footnotesize\color{gray}\emph{Analysis:} the SFT trace stays at the surface IP layer—every branch is named by an existing IP the user has already heavily consumed (\emph{Peace Elite}, smartphone reviews, short dramas), with no abstraction to a latent category axis.} \\
\midrule
\textbf{\traceStage{Transition Inference}} \\
权衡来看，\textbf{A 的可能性最大且最直接}。用户近期互动密集围绕《和平精英》，包括在 \sftAdSID{5687}{6727}{2789} 点击该游戏广告、长播多个赛季更新视频如 \sftVideoSID{5632}{1206}{6057}，并首次打赏了提供装备交易服务的游戏主播 \sftLiveSID{1790}{1523}{7951}。这些行为表明他正处于对该游戏操作细节和资源获取的高度活跃期。B 方向虽然相关，但更偏向于间歇性消费需求；C 方向则与其主要互动轨迹关联较弱。因此，后续交互极有可能仍是围绕\textbf{《绝地求生》或《和平精英》}这类战术竞技游戏的具体玩法展示、新赛季活动介绍或高水平玩家操作集锦展开。
\par\smallskip
\textbf{Target:} \textbf{《三角洲行动》装备讲解/装备玩法视频} (cross-IP target, only \emph{one} prior ad-click exposure in history).
\par\smallskip
{\footnotesize\color{gray}\emph{Analysis:} the conclusion is an in-IP replay of the user's strongest historical signal. The correct genre is captured, but the target IP family is missed—consistent with the negative $\Delta\mathrm{LL}$ regime of SFT.} \\
\bottomrule
\end{tabularx}
\end{CJK*}
\end{table}

\noindent
In contrast, the RFT model, faced with the \emph{identical} input
context, produces a reasoning trace that opens with a catalogue-grounded
quantitative prior and lifts the candidate space from a single IP to a
latent category axis, as shown in Table~\ref{tab:rft-case-shooter}. The
critical gain comes from R1-derived interest evolution: instead of replaying
\emph{Peace Elite} or \emph{PUBG} as surface names, RFT bridges them to the
tactical-shooter category and then to \emph{Delta Force} through a weak but
history-supported ad-click signal.

\noindent\textbf{Linking the case to the four criteria.}
This single example concretely instantiates all four quantitative
findings. The SFT trace lands on the wrong IP family and would lower the
target likelihood, whereas the RFT trace commits to \emph{Delta Force}
early in its candidate enumeration and merely defends the choice
afterwards---mirroring the positive $\Delta\mathrm{LL}$ in Criterion~1
and the earlier $\ell_t$ saturation in Criterion~2. Moreover, the RFT
trace anchors its reasoning to a catalogue-grounded statistic
(\textsc{[Gaming]}$=69.57\%$) rather than free-form qualifiers, while
both traces reference only history-supported items, consistent with the
higher $\gamma_{\mathrm{legal}}$ (Criterion~3) and saturated
$\gamma_{\mathrm{hist}}$ (Criterion~4) of RFT in
Table~\ref{tab:cot_validity_performance_comparison}.

\begin{table}[H]
\centering
\caption{RFT recommendation CoT case: reasoning is anchored to a catalogue-grounded quantitative prior and generalises across IPs, landing on the ground-truth title (\emph{Delta Force}).}
\label{tab:rft-case-shooter}
\setlength{\tabcolsep}{5pt}
\renewcommand{\arraystretch}{1.15}
\small
\begin{CJK*}{UTF8}{gbsn}
\begin{tabularx}{\linewidth}{@{}>{\raggedright\arraybackslash}X@{}}
\toprule
\textbf{\traceStage{Persona Abstraction}} \\
用户是一名 18--23 岁的男性中学生，属于“游戏直播观众”和“二次元爱好者”的混合体，但\textbf{游戏兴趣远大于二次元}。证据包括：他关注的博主类型以\textbf{【游戏】占 69.57\%}为主；历史记录中大量出现《王者荣耀》、《和平精英》等电竞内容及主播互动；购买商品也多为电竞周边、手机配件和零食。具体来说，他最深度的兴趣集中在\textbf{《王者荣耀》赛事/攻略、射击类手游（如和平精英）以及性价比数码产品}上。
\par\smallskip
{\footnotesize\color{gray}\emph{Analysis:} the trace opens with a \emph{quantised, catalogue-grounded statistic} ($69.57\%$) rather than free-form qualifiers, so the persona prior is already grounded in catalogue evidence.} \\
\midrule
\textbf{\traceStage{Interest Expansion}} \\
从近期行为看，用户最可能关心的兴趣点有以下几个方向：\textbf{A.} \textbf{《绝地求生》/战术竞技类游戏的新玩法或装备}，因为他在 \sftVideoSID{1985}{3589}{2398} 评论了 PUBG 相关视频，并在 \sftAdSID{1481}{2918}{7187} \textcolor{orange!60}{\textbf{点击了三角洲行动的游戏广告}}；\textbf{B.} iQOO 手机系列的最新动态与性能对比，这是他长期且稳定的兴趣，在 \sftVideoSID{447}{4491}{889} 点赞了 iQOO 帧率测试视频，并多次点击手机促销广告；\textbf{C.} 辣条等怀旧零食的测评与优惠信息，他在 \sftProdSID{8138}{1801}{208} 等多次点击辣条商品，并在 \sftVideoSID{801}{4733}{6583} 收藏了关于食品价格揭秘的内容。
\par\smallskip
{\footnotesize\color{gray}\emph{Analysis:} unlike SFT, which remains around \emph{PUBG}/\emph{Peace Elite}, the RFT trace lifts the candidate axis to ``tactical shooter'' and, through R1-style relation evolution, keeps \textcolor{orange!60}{\emph{Delta Force}} as a cross-IP successor supported by the weak ad-click evidence.} \\
\midrule
\textbf{\traceStage{Transition Inference}} \\
综合权衡来看，\textbf{A 的可能性最大且最为即时}。用户对射击游戏的热情不仅限于《和平精英》，已\textcolor{orange!60}{\textbf{从《和平精英》延伸到同类战术竞技玩法}}（如绝地求生意外关联），而\textcolor{orange!60}{\textbf{“三角洲行动”作为热门新游恰好承接了他的这一需求}}。B 是他的基础盘会持续存在，但近期没有新的手机发布节点刺激其产生强烈的探索性消费或内容互动峰值；C 属于间歇性消费需求，在完成几次点击后短期内重复推荐同类商品转化率会下降。
\par\smallskip
\textbf{Target:} \textbf{《三角洲行动》装备讲解/装备玩法视频} (cross-IP target, only \emph{one} prior ad-click exposure in history).
\par\smallskip
{\footnotesize\color{gray}\emph{Analysis:} the early commitment to ``Delta Force / tactical shooter'' in the expansion step is merely \emph{defended} here, mirroring the earlier $\ell_t$ saturation of RFT; the cross-IP target is reached because R1-style relation evolution operates over a category axis, not a single IP name.} \\
\bottomrule
\end{tabularx}
\end{CJK*}
\end{table}

\noindent
Crucially, the difference between the two traces is not lexical polish but
\emph{abstraction level}: SFT reasons over surface IP names
(\emph{Peace Elite} $\to$ more \emph{Peace Elite}), whereas RFT reasons
over a latent category axis (``tactical shooter'') and is therefore able
to bridge from observed evidence (\emph{Peace Elite}, \emph{PUBG} comment,
\emph{Delta Force} ad click) to a target IP that has only been \emph{weakly}
exposed in history. This category-level generalisation is what the
$2\!\times\!2$ diagnostic framework in Table~\ref{tab:cot_eval_framework}
is designed to detect, and it is what RFT systematically delivers.

\section{Experiments}\label{sec:exp}
\subsection{Performance Comparison}

\paragraph{Baselines.}
We compare OneReason with representative baselines on four cross-domain recommendation tasks. We group the evaluated models by how they represent items:

\begin{itemize}
    \item \textbf{ID-Based:} Items are represented by conventional item IDs. This group includes standard sequential recommendation models, namely SASRec~\citep{kang2018self} and HSTU~\citep{zhai2024hstu}.
    
    \item \textbf{Text-Based:} Items are represented by natural-language item dense captions. This group evaluates general-purpose LLMs under a zero-shot protocol on all tasks, with thinking mode enabled where supported. For recommendation, the LLMs generate a potentially relevant item caption, which is then encoded by Qwen3-Embedding-8B~\citep{qwen3embedding} along with the candidate item captions, and recommendations are obtained through approximate nearest-neighbor (ANN) retrieval.
    
    \item \textbf{Itemic Token-Based:} Items are represented by itemic tokens. In addition to OneReason variants, we evaluate TIGER~\citep{rajput2023recommender} and LC-Rec~\citep{zheng2024large}. For LC-Rec, we report three variants: LC-Rec-SFT-Only-8B, LC-Rec-SFT-Only-14B, and LC-Rec-PT-SFT-8B (initialized from Qwen3-8B, Qwen3-14B~\citep{yang2025qwen3technicalreport}, and our 
  OneReason pre-training checkpoint). Following the original LC-Rec setting, its SFT data covers only recommendation and caption tasks.
\end{itemize}

\paragraph{Recommendation Performance.}

\begin{table}[htbp]
  \centering
  \caption{Baselines on cross-domain recommendation (in \%). Best results are \textbf{bolded}; second-best results are \underline{underlined}.}
  \label{tab:baseline_cross_domain}
  \renewcommand{\arraystretch}{1.2}
  \resizebox{\textwidth}{!}{%
  \begin{threeparttable}
  \begin{tabular}{llcccccccc}
    \toprule
    \textbf{Category} & \textbf{Model}
      & \multicolumn{2}{c}{\textbf{Cross-Video}}
      & \multicolumn{2}{c}{\textbf{Cross-Product}}
      & \multicolumn{2}{c}{\textbf{Cross-Ad}}
      & \multicolumn{2}{c}{\textbf{Cross-Live}} \\
    \cmidrule(lr){3-4} \cmidrule(lr){5-6} \cmidrule(lr){7-8} \cmidrule(lr){9-10}
    & & \textbf{Pass@64} & \textbf{Recall@64\tnote{*}}
      & \textbf{Pass@64} & \textbf{Recall@64}
      & \textbf{Pass@64} & \textbf{Recall@64}
      & \textbf{Pass@64} & \textbf{Recall@64} \\
    \midrule
    \multirow{2}{*}{ID-Based}
      & SASRec          & 0.03 & 0.01 & 0.31 & 0.25 & 1.04 & 0.37 & 1.76 & 0.40 \\
      & HSTU            & 0.10 & 0.01 & 0.32 & 0.24 & 2.79 & 0.78 & 2.32 & 2.14 \\
    \midrule
    \multirow{8}{*}{Text-Based}
      & Qwen3-8B        & 0.05 & 0.01 & 0.15 & 0.12 & 0.48 & 0.09 & 2.10 & 1.85 \\
      & Qwen3-32B       & 0.33 & 0.03 & 0.84 & 0.63 & 1.21 & 0.30 & 5.64 & 5.10 \\
      & Qwen3-235B-A22B & 0.24 & 0.02 & 0.64 & 0.49 & 0.77 & 0.19 & 5.10 & 4.66 \\
      & Deepseek-V3.2   & 0.11 & 0.01 & 0.38 & 0.31 & 0.62 & 0.13 & 3.46 & 3.12 \\
      & Claude-Opus-4.6 & 0.14 & 0.01 & 0.23 & 0.17 & 0.50 & 0.11 & 3.02 & 2.66 \\
      & Gemini-3-Preview & 0.29 & 0.03 & 0.74 & 0.59 & 1.22 & 0.27 & 3.92 & 3.44 \\
      & GPT-4o-mini     & 0.19 & 0.02 & 0.73 & 0.55 & 1.21 & 0.28 & 4.01 & 3.57 \\
      & GPT-5.4         & 0.24 & 0.02 & 1.43 & 1.15 & 1.64 & 0.43 & 7.20 & 6.38 \\
    \midrule
    \multirow{8}{*}{Itemic Token-Based}
      & TIGER                    & 0.88 & 0.07 & 0.21 & 0.17 & 7.65 & 2.39 & 2.32 & 1.78 \\
      & LC-Rec-SFT-Only-8B       & 0.22 & 0.02 & 0.06 & 0.05 & 2.83 & 0.67 & 0.89 & 0.71 \\
      & LC-Rec-SFT-Only-14B      & 0.20 & 0.01 & 1.03 & 0.73 & 5.99 & 1.94 & 3.76 & 3.09 \\
      & LC-Rec-PT-SFT-8B         & 1.49 & 0.13 & 3.95 & 3.00 & 15.85 & 6.55 & 19.32 & 16.70 \\
      \cmidrule(lr){2-10}
      & OneReason SFT non-thinking  & 1.33 & 0.11 & 3.94 & 2.96 & 15.73 & 6.49 & 18.05 & 15.52 \\
      & OneReason SFT thinking      & 0.71 & 0.06 & 2.18 & 1.65 & 9.16 & 3.41 & 16.43 & 14.32 \\
      \cmidrule(lr){2-10}
      & OneReason RFT non-thinking  & \underline{2.08} & \underline{0.19} & \underline{5.20} & \underline{3.96} & \underline{17.56} & \underline{7.26} & \underline{21.01} & \underline{18.17} \\
      & OneReason RFT thinking      & \cellcolor{orange!30}\textbf{2.41} & \cellcolor{orange!30}\textbf{0.24} & \cellcolor{orange!30}\textbf{5.47} & \cellcolor{orange!30}\textbf{4.19} & \cellcolor{orange!30}\textbf{17.78} & \cellcolor{orange!30}\textbf{7.50} & \cellcolor{orange!30}\textbf{21.10} & \cellcolor{orange!30}\textbf{18.35} \\
    \bottomrule
  \end{tabular}
  \begin{tablenotes}
    \item[*] Video Recall@64 is notably low due to the largest target set (avg.\ 13.92 items) making full recall harder.
  \end{tablenotes}
  \end{threeparttable}%
  }
\end{table}

Based on the evaluation results in Table~\ref{tab:baseline_cross_domain}, we highlight the primary reasons for the limited performance of ID-based and text-based baselines:

\begin{itemize}
    \item \textbf{Cold-Start Sensitivity:} Traditional ID-based models struggle in cross-domain evaluation because 33.69\% of target item IDs are unseen during training, causing severe cold-start issues; by contrast, only 11.55\% of target itemic patterns are unseen, as reported in Appendix~\ref{appendix:r3_statistics}.
    
    \item \textbf{Limits of Text-Based LLM Recommendation:} More advanced general-purpose LLMs do not necessarily produce better recommendations, suggesting that recommendation ability is not reliably correlated with general intelligence or model scale. Their underperformance stems from the absence of collaborative signals and the errors introduced by ANN-based caption-to-item retrieval. In contrast, OneReason avoids this additional retrieval step by directly decoding itemic tokens. Pointwise scoring or better retrieval may yield stronger recommendations; we leave their study to the community.

    \item \textbf{Effect of OneReason Pre-Training:} The LC-Rec results show that our OneReason pre-training checkpoint provides a strong semantic foundation for itemic-token recommendation. Compared with the LC-Rec-SFT-Only variants, LC-Rec-PT-SFT-8B achieves a clear performance jump across all four domains, demonstrating the effectiveness of our OneReason pre-training strategy.
\end{itemize}

\paragraph{Non-Recommendation Performance.}
Table~\ref{tab:baseline_r0_r2} evaluates R0--R2 capabilities on \benchname{}. Two findings stand out:
\begin{itemize}
    \item \textbf{RFT thinking helps higher-level reasoning.} RFT thinking lags non-thinking on R0, consistent with perception-task overthinking~\citep{jiang2025mme}, but improves average R1--R2 performance. Together with the recommendation gains in Table~\ref{tab:baseline_cross_domain}, this suggests that our RL improves recommendation while strengthening higher-level reasoning.
    \item \textbf{OneReason makes compact itemic tokens competitive.} Text-based LLMs are given richer inputs, whereas OneReason uses compact itemic tokens. Even under this input-information disadvantage, OneReason surpasses several larger advanced LLMs on parts of the R2 suite. This suggests that OneReason can perform efficient and effective long-sequence user modeling.
    
\end{itemize}

\begin{table*}[htbp]
    \centering
    \caption{Performance comparison on R0--R2 tasks in \benchname{} (in \%). For R0 tasks, results are macro-averaged over all domains. The grounding task is reported by Pass@64. For the R2 tasks evaluated by action-logic score, matching for text-based models uses 10-gram soft matching instead of exact matching. Best results are \textbf{bolded}; second-best results are \underline{underlined}.}
    \renewcommand{\arraystretch}{1.15}
    \setlength{\tabcolsep}{4pt}
    \resizebox{\textwidth}{!}{%
    \begin{tabular}{llccccccc}
      \toprule
      \multirow{2}{*}{\textbf{Category}} & \multirow{2}{*}{\textbf{Model}}
        & \multicolumn{3}{c}{\textbf{R0: Perception}}
        & \multicolumn{1}{c}{\textbf{R1: Derivation}}
        & \multicolumn{3}{c}{\textbf{R2: Evolution}} \\
      \cmidrule(lr){3-5} \cmidrule(lr){6-6} \cmidrule(lr){7-9}
      & & \makecell{\textbf{Item Und.}}
        & \textbf{Ground.}
        & \textbf{QA}
        & \textbf{I2I}
        & \makecell{\textbf{Select.}}
        & \makecell{\textbf{Topic Gen.}}
        & \makecell{\textbf{Direct Gen.}} \\
      \midrule
      \multirow{6}{*}{Text-Based}
        & Qwen3-8B         & - & - & - & - & 40.70 & 25.49 & 8.60 \\
        & Qwen3-32B        & - & - & - & - & 51.96 & 28.05 & 7.73 \\
        & Deepseek-V3.2    & - & - & - & - & \underline{57.18} & 27.13 & 11.32 \\
        & Claude-Opus-4.6  & - & - & - & - & 56.84 & 17.16 & 13.46 \\
        & Gemini-3-Preview & - & - & - & - & 56.83 & 33.68 & 14.76 \\
        & GPT-5.4          & - & - & - & - & \textbf{58.92} & \textbf{41.41} & 17.61 \\
      \midrule
      \multirow{7}{*}{Itemic Token-Based}
        & LC-Rec-SFT-Only-8B  & 22.98 & 0.00 & 0.40 & 3.43 & 0.00 & 0.00 & 0.00 \\
        & LC-Rec-SFT-Only-14B & 26.48 & 0.00 & 56.45 & 16.21 & 0.00 & 0.00 & 0.00 \\
        & LC-Rec-PT-SFT-8B    & 35.41 & \underline{5.21} & 63.90 & 25.54 & 3.32 & 8.60 & 4.46 \\
        \cmidrule(lr){2-9}
        & OneReason SFT non-thinking & \underline{36.84} & 3.95 & \underline{66.55} & \underline{28.36} & 35.07 & 33.87 & 15.42 \\
        & OneReason SFT thinking    & \textbf{36.91} & 1.06 & 64.60 & 23.88 & 32.18 & 31.60 & 14.31 \\
        \cmidrule(lr){2-9}
        & OneReason RFT non-thinking & 36.82 & \textbf{5.24} & \textbf{67.25} & 23.99 & 38.92 & 39.33 & \underline{20.31} \\
        & OneReason RFT thinking    & 36.78 & 1.35 & 65.65 & \textbf{28.60} & 42.42 & \underline{39.57} & \textbf{21.23} \\
      \bottomrule
    \end{tabular}%
    }
    \label{tab:baseline_r0_r2}
  \end{table*}

\paragraph{General-Intelligence Sanity Check.}

Table~\ref{tab:general_benchmark} shows that in thinking mode, OneReason preserves its Qwen3-8B backbone's general reasoning and instruction-following abilities on these four representative benchmarks after recommendation-oriented training, providing preliminary evidence that recommendation-oriented training does not catastrophically degrade general capabilities. In contrast, LC-Rec variants suffer substantial degradation across all benchmarks.

\begin{table}[H]
    \centering
    \caption{General benchmark performance comparison (in \%). Best results are \textbf{bolded}; second-best results are \underline{underlined}.}
    \label{tab:general_benchmark}
    \renewcommand{\arraystretch}{1.15}
    \setlength{\tabcolsep}{6pt}
    \resizebox{\textwidth}{!}{%
    \begin{tabular}{llcccc}
      \toprule
      \textbf{Category} & \textbf{Model}
        & \textbf{MMLU-Pro}
        & \textbf{GPQA-Diamond}
        & \textbf{MATH-500}
        & \textbf{GSM8K} \\
      \midrule
      \multirow{2}{*}{Text-Based}
        & Qwen3-8B non-thinking   & 66.32 & 39.90 & 83.80 & 93.03 \\
        & Qwen3-8B thinking      & \textbf{72.35} & \textbf{56.06} & 95.20 & \textbf{95.68} \\
      \midrule
      \multirow{7}{*}{Itemic Token-Based}
        & LC-Rec-SFT-Only-8B     & 9.73  & 17.17 & 41.80 & 14.03 \\
        & LC-Rec-SFT-Only-14B    & 45.35 & 37.37 & 60.20 & 52.54 \\
        & LC-Rec-PT-SFT-8B       & 39.72 & 35.86 & 81.00 & 51.55 \\
        \cmidrule(lr){2-6}
        & OneReason SFT non-thinking & 61.74 & 33.33 & 91.40 & 93.86 \\
        & OneReason SFT thinking    & 71.01 & 51.52 & \textbf{95.60} & \underline{95.00} \\
        \cmidrule(lr){2-6}
        & OneReason RFT non-thinking & 59.57 & 37.88 & 91.40 & 94.31 \\
        & OneReason RFT thinking    & \underline{72.08} & \underline{54.04} & \underline{95.40} & 94.69 \\
      \bottomrule
    \end{tabular}
    }
  \end{table}

\subsection{Non-Thinking Gains from Thinking Supervision}
\label{sec:sft-nothink-gains-from-thinking}

Although the recommendation CoT data in Section~\ref{sec:sft-cot-synthesis} is constructed to provide compact, evidence-grounded recommendation reasoning rather than to directly optimize non-thinking decoding, we observe an important downstream phenomenon: thinking supervision can improve direct recommendation even when explicit reasoning traces are suppressed at inference time. We study this phenomenon by comparing CoT and unCoT recommendation supervision under controlled training budgets. For clarity, we use \textbf{unCoT} to denote recommendation samples without explicit reasoning traces, where the model directly predicts the target itemic tokens from the user context. In contrast, \textbf{CoT} samples contain a trace before the final target, generated by the three-stage compression--reasoning protocol in Section~\ref{sec:sft-cot-synthesis}: persona abstraction, interest expansion, and transition inference.

We first conduct a controlled token-aligned experiment. We compare two settings: (i) continued SFT on 100K unCoT samples; and (ii) continued SFT on a mixture of 40K CoT samples and 50K unCoT samples. Since CoT samples are longer than unCoT samples, the latter two SFT settings are aligned to the same training token budget of 0.25B tokens. All models are evaluated using the same non-thinking decoding format, where the model directly predicts target itemic tokens without generating a reasoning trace.

\begin{table}[H]
\centering
\small
\caption{Token-aligned comparison of non-thinking recommendation performance (in \%). The 100K unCoT setting and the 40K CoT + 50K unCoT setting use the same training budget of 0.25B tokens. For each domain-metric pair, the better setting is highlighted in bold.}
\label{tab:nothink_token_aligned}
\renewcommand{\arraystretch}{1.08}
\setlength{\tabcolsep}{7pt}
\begin{tabular}{@{}llcccc@{}}
\toprule
\textbf{Metric} & \textbf{Setting} & \textbf{Cross-Video} & \textbf{Cross-Product} & \textbf{Cross-Ad} & \textbf{Cross-Live} \\
\midrule
\multirow{2}{*}{Pass@32}
& 100K unCoT   & 0.88 & 3.16 & \textbf{13.81} & 13.79 \\
& CoT + unCoT  & \textbf{1.17} & \textbf{3.62} & 13.57 & \textbf{15.99} \\
\midrule
\multirow{2}{*}{Recall@32}
& 100K unCoT   & 0.08 & 2.33 & \textbf{5.57} & 11.93 \\
& CoT + unCoT  & \textbf{0.12} & \textbf{2.74} & 5.50 & \textbf{13.98} \\
\midrule
\multirow{2}{*}{Pass@64}
& 100K unCoT   & 1.64 & 4.38 & \textbf{16.08} & 18.12 \\
& CoT + unCoT  & \textbf{1.95} & \textbf{4.86} & 15.84 & \textbf{20.32} \\
\midrule
\multirow{2}{*}{Recall@64}
& 100K unCoT   & 0.18 & 3.33 & \textbf{6.72} & 15.59 \\
& CoT + unCoT  & \textbf{0.21} & \textbf{3.67} & 6.68 & \textbf{17.69} \\
\bottomrule
\end{tabular}
\end{table}

Table~\ref{tab:nothink_token_aligned} shows that, under the same token budget, replacing part of the unCoT data with CoT data improves non-thinking performance on Cross-Video, Cross-Product, and Cross-Live. For example, the mixed setting improves Pass@64 from 1.64\% to 1.95\% on Cross-Video, from 4.38\% to 4.86\% on Cross-Product, and from 18.12\% to 20.32\% on Cross-Live. Cross-Ad is the main exception, where the mixed setting is slightly worse than unCoT-only training. This suggests that the benefit of CoT supervision is not uniform across domains, and motivates a finer-grained analysis of the CoT/unCoT mixture ratio.

We then conduct a sample-count-controlled mixture sweep. Under a fixed budget of 100K recommendation samples, we vary the ratio between CoT and unCoT samples. Let \(\alpha\) denote the unCoT ratio:
\begin{equation}
\begin{split}
\alpha
=
\frac{N_{\mathrm{unCoT}}}
{N_{\mathrm{CoT}} + N_{\mathrm{unCoT}}},
\qquad
1-\alpha
=
\frac{N_{\mathrm{CoT}}}
{N_{\mathrm{CoT}} + N_{\mathrm{unCoT}}}.
\end{split}
\label{eq:exp-uncot-cot-ratio}
\end{equation}

\begin{figure}[t!]
    \centering
    \includegraphics[width=\textwidth]{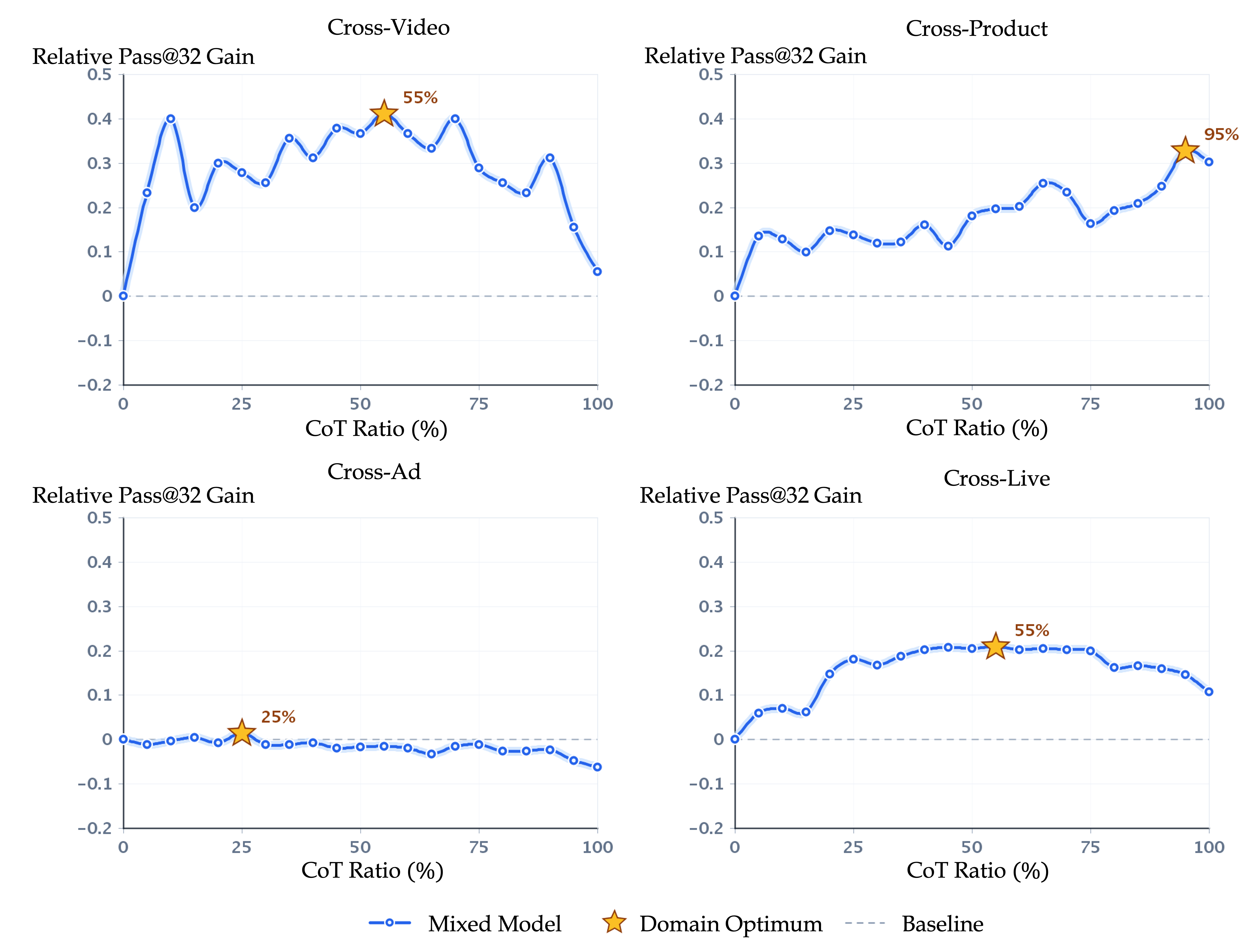}
    \caption{Non-thinking recommendation gains across different CoT/unCoT mixture ratios. All models are evaluated with non-thinking decoding. The horizontal axis denotes the CoT ratio $(1-\alpha)$, and the vertical axis reports the relative Pass@32 gain over the 100\% unCoT baseline. Stars mark the best mixture ratio for each domain.}
    \label{fig:sft-nothink-gain-mix}
\end{figure}

For each mixture ratio, we train a model from the same starting checkpoint and evaluate it with non-thinking decoding. Figure~\ref{fig:sft-nothink-gain-mix} reports the relative Pass@32 gain over the 100\% unCoT baseline, defined as:
\begin{equation}
\begin{split}
G_d(\alpha)
=
\frac{
\mathrm{Pass@32}_d(\theta_{\alpha})
-
\mathrm{Pass@32}_d(\theta_{1.0})
}{
\mathrm{Pass@32}_d(\theta_{1.0})
},
\end{split}
\label{eq:exp-relative-pass32-gain}
\end{equation}
where \(d\) denotes the target domain, \(\theta_{\alpha}\) is the model trained with unCoT ratio \(\alpha\), and \(\theta_{1.0}\) is the model trained with 100\% unCoT data.

Figure~\ref{fig:sft-nothink-gain-mix} shows that the response curve is not simply monotonic. Instead, most domains exhibit a clear intermediate optimum. Cross-Video reaches its best result near a balanced mixture, Cross-Product prefers a more CoT-heavy mixture, and Cross-Live also benefits from a balanced or moderately CoT-rich mixture. Cross-Ad, in contrast, shows a much flatter curve, and its best point lies closer to the unCoT side. These results support the conclusion from the token-aligned experiment: CoT supervision can improve non-thinking inference in most domains, but the best mixture depends on the domain.

We interpret this phenomenon as a balance between two types of supervision. unCoT data matches the non-thinking inference format and directly teaches the model to map user context to target itemic tokens. CoT data provides an additional training signal with two components: a \textit{compression} signal, which teaches persona and trajectory abstraction from long behavior histories, and a \textit{reasoning} signal, which teaches transition judgement from the compressed state and supporting evidence to target-domain decisions. During non-thinking decoding, the model does not explicitly generate the trace. A plausible hypothesis is that part of the compact user-state construction and evidence-selection bias learned from CoT supervision is still reflected in the model parameters. The current experiment supports this hypothesis at the behavioral level, but it does not identify the exact contribution of compression versus reasoning.

However, more CoT data is not always better. CoT traces are longer than unCoT answers, so excessive CoT training may weaken the density of final-target supervision. CoT traces may also contain noise or over-explanation, and the training format may become less aligned with non-thinking decoding. Therefore, the best mixture is obtained when the reasoning signal from CoT and the direct target signal from unCoT are properly balanced.

Formally, we use the following conceptual decomposition to describe the non-thinking gain in domain \(d\) under unCoT ratio \(\alpha\):
\begin{equation}
\begin{split}
G_d(\alpha)
=
B_d
+
A_d^{\mathrm{uncot}}(\alpha)
+
I_d^{\mathrm{cot}}(1-\alpha)
-
C_d^{\mathrm{trace}}(1-\alpha)
-
C_d^{\mathrm{format}}(\alpha,1-\alpha),
\end{split}
\label{eq:exp-nothink-gain-decomposition}
\end{equation}
where \(B_d\) denotes the base capability in domain \(d\), \(A_d^{\mathrm{uncot}}\) denotes the direct-answer alignment benefit from unCoT data, \(I_d^{\mathrm{cot}}\) denotes the signal absorbed from CoT data, \(C_d^{\mathrm{trace}}\) denotes the cost of noisy or over-long traces, and \(C_d^{\mathrm{format}}\) denotes the mismatch between CoT-format training and non-thinking decoding. Conceptually, \(I_d^{\mathrm{cot}}\) contains both a compression component, which teaches the model to form a compact user-interest state, and a reasoning component, which teaches the model to connect this state to target-domain decisions. We use this as a conceptual decomposition only, not as an identifiable additive estimate from the current experiments. The empirical optimum \(\hat{\alpha}_d^\star\) in Figure~\ref{fig:sft-nothink-gain-mix} can be understood as the point where the marginal benefit of adding more unCoT data balances the marginal loss of reducing CoT supervision:
\begin{equation}
\begin{split}
\frac{\partial G_d(\alpha)}{\partial \alpha}
\bigg|_{\alpha=\hat{\alpha}_d^\star}
\approx 0.
\end{split}
\label{eq:exp-mixture-stationary-point}
\end{equation}
This equation is only a conceptual explanation and does not assume that the entire curve follows a quadratic law. It simply states that the best mixture appears when direct-answer alignment and CoT-based supervision reach a domain-specific balance.

One hypothesis for the domain difference is that different tasks rely on different evidence sources. Cross-Video and Cross-Live may benefit from CoT because their user histories often contain diverse signals and multiple plausible intent directions, making evidence organization useful before prediction. Cross-Product may be more sensitive to CoT-heavy training because purchase intent often has to be inferred from content cues and behavior transitions. By contrast, Cross-Ad may depend more on short-term conversion signals and business exposure patterns that are only partially captured by natural-language reasoning traces, which could explain why it favors unCoT-heavy training and shows a flatter curve.

Overall, these experiments reveal a useful but domain-dependent effect of thinking supervision on non-thinking recommendation. High-quality thinking supervision can improve direct non-thinking recommendation in most domains under appropriate mixture ratios, but Cross-Ad shows that CoT is not universally beneficial and should not be used more aggressively by default. This makes CoT data valuable as part of a potential data flywheel for recommendation foundation models: CoT construction can strengthen the backbone, the improved non-thinking model can produce better candidates and filtering signals, and explicit thinking-mode inference can still be reserved for harder cohorts where trace generation is worth the extra latency.

\section{Deployment}\label{sec:deployment}
In the rapidly growing local-services advertising scenario of the Kuaishou App, we have successfully deployed OneReason. Under strict latency, strong baseline, and item-deliverability constraints, the system achieves stable online deployment and delivers the \textbf{one of the most significant business uplifts in the Kuaishou local-services advertising scenario}. This section focuses on the online deployment of OneReason, including {Industrial Scenario Adaptation} and {Online Deployment Architecture}, as well as a set of reusable methodologies distilled from the OneReason deployment.

\subsection{Industrial Scenario Adaptation}
Given a strong and high-performing OneReason foundation model, efficiently adapting it to a specific industrial scenario becomes a critical challenge. Notably, the Kuaishou life-service advertising scenario is not included in the OneReason training data. To address this, we propose a standardized and reusable continuous training pipeline that enables rapid domain adaptation and can be easily generalized to other industrial application scenarios. Further details are provided in Appendix~\ref{appendix:Industrial Scenario Adaptation}.

\subsection{Industrial Scenario Online Deployment}
Due to the scale of LLMs and the limitations of inference systems, directly deploying OneReason for real-time recommendation still faces significant latency and computational cost challenges. In this section, we propose a new online architecture for OneReason, the ``Fast-Slow Thinking'' architecture, and further introduce an online incremental training strategy to ensure model performance.

\begin{figure}[htbp]\centering 
    \includegraphics[width=\textwidth]{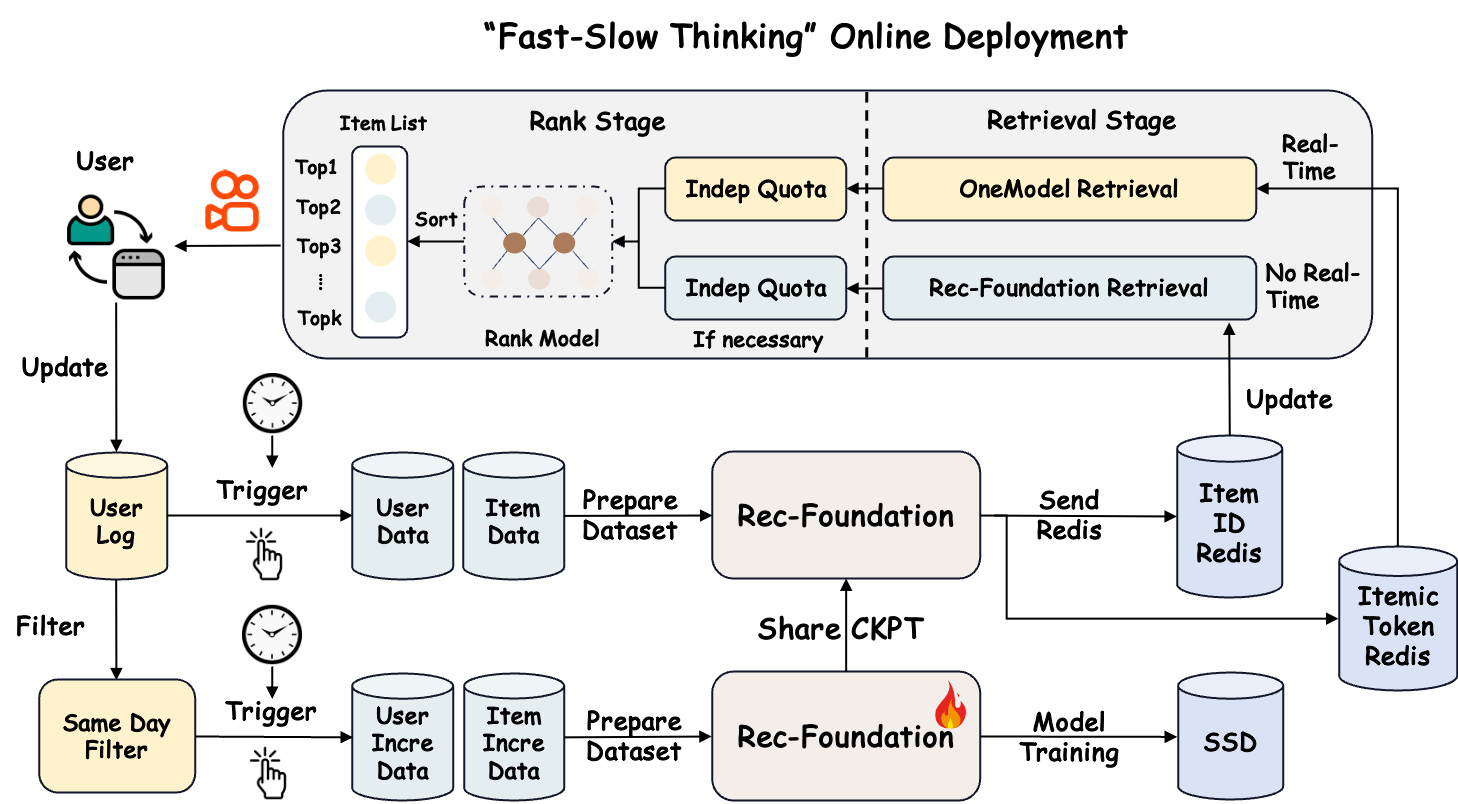}
    \caption{The online deployment architecture of OneReason.}\label{fig:online deployment}
\end{figure}

\textbf{``Fast-Slow Thinking'' Architecture in the Online Deployment.} This architecture avoids fully relying on real-time inference and instead introduces a Fast (Online) and Slow (Nearline) collaboration mechanism, continuously injecting OneReason’s knowledge and reasoning capabilities into the system while maintaining low latency and stable gains.
The workflow is shown in Figure~\ref{fig:online deployment}.
\begin{itemize}
    \item \textbf{\textit{Step 1: Pipeline Design.}} We build a decoupled nearline retrieval pipeline that does not participate in early-stage competition. Instead, it is integrated into the downstream ranking model for joint scoring with the real-time OneRec outputs. When OneReason retrieval results are unavailable, the system automatically falls back to the original OneRec pipeline to ensure stability and coverage.

    \item \textbf{\textit{Step 2: Dataset Trigger.}} We periodically (daily or hourly) aggregate distributed data jobs to construct user behavior, user profile, and item content data. An online pipeline is then triggered for data fetching, cleaning, and feature processing, forming an end-to-end data flywheel.

    \item \textbf{\textit{Step 3: Offline Inference.}} After dataset preparation, we load the latest OneReason checkpoint to predict the user's most likely next itemic tokens offline. These tokens are then decoded into item IDs using our specialized decoding strategy for OneReason (as discussed in Appendix~\ref{appendix:disscussions in industrial deployment d2}).

    \item \textbf{\textit{Step 4: Online Serving.}} The decoded item IDs are written to Redis via offline inference jobs, forming a candidate pool for online serving. During inference, the nearline OneReason results and real-time OneRec retrieval results are jointly fed into the ranking model for unified fusion, enabling a Fast-Slow Thinking  recommendation system.
\end{itemize}

\textbf{Online Incremental Training Strategy.}
The online platform continuously introduces new items, while user interests and intents also evolve over time. Therefore, an incremental training mechanism is required to adapt OneReason to long-term business evolution while preserving its existing capabilities and ensuring stable online performance. The procedure is as follows:
\begin{itemize}
    \item \textbf{\textit{Pre-Training Incremental Training.}} We conduct periodic continual pre-training on newly introduced items within fixed time windows to keep pace with the latest content distribution. To prevent catastrophic forgetting and maintain the general reasoning capabilities of OneReason, we jointly train on a mixture of newly collected data and a sampled general-domain corpus.
    
    \item \textbf{\textit{SFT Incremental Training.}} We construct daily incremental training data from user logs, using same-day user interactions as supervision signals for OneReason training, which models short-term interest dynamics and improves sensitivity to preference shifts.
\end{itemize}

\textbf{Two Application Paradigms of OneReason.}
We design two application paradigms of OneReason under the ``Fast--Slow Thinking'' architecture in the online deployment, both yielding significant online gains and demonstrating the strong modeling and generalization capabilities of OneReason.

\begin{itemize}
    \item \textbf{\textit{Slow Pipeline: OneReason}}. This method directly employs OneReason for retrieval. Specifically, OneReason leverages user profile information and historical interaction sequences to predict the next itemic tokens that the user is most likely to engage with, which are then decoded into a candidate item list serving as the retrieval output of the Slow pipeline.
    
    \item \textbf{\textit{Fast Pipeline: OneReason for OneRec}}. We leverage the outputs of OneReason to enhance recommendations, where the most relevant itemic tokens are transformed into embedding representations. In OneRec, we introduce a dedicated \textit{Thinking Token} to incorporate these signals, which serves to distill knowledge from OneReason into the online OneRec within the Fast pipeline. The overall methodology is provided in Appendix~\ref{appendix:OneReason For OneRec}.
\end{itemize}

\subsection{Online Experiment}
We conducted a 10-day online A/B experiment by deploying OneReason in a subpart of the local-services scenario of the Kuaishou App, using a 5\% traffic allocation for both the treatment and control groups. Users in the treatment group were served by the ``Fast-Slow Thinking'' architecture of OneReason, whereas users in the control group remained on the existing recommendation system.

Table~\ref{tab:online_result} shows that both directly using OneReason for retrieval and using it to enhance the real-time OneRec yield significant improvements in the online A/B experiment. The direct application in the slow pipeline improves Revenue, while enhancing OneRec in the fast pipeline further boosts performance. Combining both paradigms achieves the best results, with substantial gains in Revenue and Impressions, corresponding to hundreds of millions of RMB in annualized commercial revenue for the Kuaishou platform. Moreover, we compute the ratio between computational resource consumption and revenue, demonstrating that OneReason achieves an ROI $>$ 5. More experimental analysis and discussion are provided in Appendix~\ref{appendix:Experimental Analysis} and \ref{appendix:disscussions in industrial deployment}.

\begin{table}[ht]
\centering
\begin{tabular}{l c c}
\toprule
\textbf{Model} & \textbf{Impressions} & \textbf{Revenue} \\
\midrule

OneReason
& {+0.940\%}
& {+4.528\%} \\

OneReason for OneRec
& {+6.831\%}
& {+4.636\%} \\

Combined
& {+10.332\%}
& {+8.234\%} \\

\bottomrule
\end{tabular}

\caption{
Online performance comparison across different deployment models.
\textbf{Combined}: joint deployment of OneReason and OneReason for OneRec.
\textbf{Revenue} refers to the total commercial revenue generated in the online advertising system of the Kuaishou platform.
}
\label{tab:online_result}
\end{table}

\section{Related Works}\label{sec:related_work}
With the rapid advancement of artificial intelligence, leveraging AI to reshape industrial systems has become a major direction in both industry and academia. Recommendation systems have increasingly adopted LLMs to improve user experience and business performance. Recent generative recommendation frameworks, such as the OneRec series, demonstrate strong modeling capacity, scalability, and system efficiency. However, most existing methods largely remain within a System-1 single-pass inference paradigm. Inspired by reasoning-oriented foundation models such as OpenAI O1~\citep{jaech2024openai} and DeepSeek R1~\citep{guo2025deepseek}, researchers are beginning to explore a \textbf{Think-then-Generate} paradigm. As recommendation environments become increasingly complex and dynamic, incorporating structured reasoning into generative foundation models (GFMs) has emerged as a promising research direction.
In this section, we provide an overview of recommendation foundation models and reasoning-enhanced generative frameworks. Building upon these developments, we illustrate how the OneReason \textbf{Think-then-Recommendation} paradigm differs from existing GFMs.

Following the rise of LLMs, researchers began exploring their direct application to recommendation tasks, leveraging their rich world knowledge and strong instruction-following capabilities. Typical approaches formulate user profiles and interaction histories as prompts, allowing LLMs to reason about user preferences before generating recommendation results. Representative works include GPT4Rec~\citep{li2023gpt4rec}, Chat-Rec~\citep{gao2023chat}, LLMRank~\citep{hoffmann2024llm}, and LLM as Zero-Shot Ranker~\citep{hou2024large}. However, as research on LLM4Rec advances, it has become clear that general-purpose LLMs struggle to fully capture the characteristics of recommendation tasks, leading to performance bottlenecks. This has motivated two main directions for integrating LLMs with recommendation systems. The first follows an LLM-backbone paradigm to enhance existing recommendation models, represented by the OneRec series. The second aims to train recommendation foundation models that can directly perform recommendation.

In the first direction, TIGER~\citep{rajput2023recommender} is an early industrial work on generative recommendation, introducing RQ-VAE-based itemic tokens and framing recommendation as an end-to-end generation task from user sequence itemic patterns to next-item representations. HSTU~\citep{zhai2024hstu} further replaces traditional multi-stage pipelines with a Transformer-based backbone and explores scaling laws for generative recommendation. The OneRec series~\citep{deng2025onerec,zhou2025onerec,liu2025onerecthinkintextreasoninggenerative,zhou2026openonerectechnicalreport,zhang2026onemall,wang2026onelive} then engineers this paradigm into a unified pre-train and post-train framework, demonstrating strong modeling and efficiency advantages. Subsequent work focuses on issues such as diversity control~\citep{agarwal2025pinrec}, end-to-end pipelines~\citep{he2025end,sun2026oneranker}, ultra-long user sequence modeling~\citep{li2025taming}, and industrial scenario constraints~\citep{xue2026generative,zhang2026unified,lin2025spacetime}.

In the second direction, research focuses on building recommendation foundation models by adapting LLMs to recommendation data, leveraging their world knowledge and reasoning ability. LC-Rec~\citep{zheng2024adapting} shows that LLMs can understand recommendation semantics via itemic ID alignment and instruction tuning. RecGPT~\citep{yi2025recgpt} and RecGPT-V2~\citep{yi2025recgptv2} further use LLMs for user intent modeling in retrieval and interest mining, with small-scale online validation. However, these methods mainly focus on user intent modeling and do not achieve full end-to-end decision-making. OxygenREC~\citep{hao2025oxygenrec} follows a similar idea to RecGPT, using fast--slow thinking to generate contextual instructions for downstream models. Nevertheless, these approaches remain LLM-augmented recommendation systems rather than true end-to-end generative foundation models. In contrast, our goal is to build an industrially deployable GFMs paradigm that enables direct reasoning-based recommendation decisions with measurable online impact. OpenOneRec~\citep{zhou2026openonerectechnicalreport} is an important precursor, proposing a novel \textbf{Think-then-Recommendation} paradigm. They also propose RecIF-Bench, large-scale datasets, and 1.7B/8B foundation models with multiple training strategies, including Text Alignment, Co-Pretraining, Multi-task SFT, and Rec-RL. However, it remains unclear whether its reasoning consistently translates into real recommendation performance gains, leaving room for further investigation.
The most directly comparable diagnostic work is~\citep{zhang2026whythinkinghurts}, which formally measures the non-thinking/thinking gap on OpenOneRec and attributes it to textual inertia from the general subspace: itemic token embeddings and general-text embeddings share latent dimensions but are misaligned, so a longer CoT lets the general-text prior dominate and dilute ID-grounded evidence. In OneReason, we address this issue at training time across elaborate training pipelines to guide more robust itemic perception and cognition CoT.

In traditional generative recommendation models, researchers have early on incorporated the reasoning capability of LLMs by introducing explicit or implicit reasoning mechanisms. Since generative recommendation operates within the recommendation space and lacks an explicit CoT paradigm, most existing methods adopt latent reasoning, injecting reasoning signals from LLM outputs or internal system signals. For example, ReaRec~\citep{tang2026think} introduces multi-step reasoning into generative recommendation, iteratively feeding high-dimensional reasoning results back to refine user representations. OnePiece~\citep{dai2025onepiece} further extends this idea by integrating LLM-style context engineering and reasoning into industrial recommendation systems via a reasoning-block design, enabling progressive alignment between intermediate states and recommendation logic. In contrast, OneSearch-V2~\citep{chen2026onesearch} leverages LLMs with CoT to deeply understand complex user queries and distills such reasoning capability into recommendation model parameters via self-distillation. Meanwhile, REG4Rec~\citep{xing2025reg4rec} addresses reasoning diversity and reliability by constructing multiple item-level semantic representations to enable multi-path reasoning and introducing self-reflection to prune inconsistent reasoning paths.

\section{Conclusion~\&~Future Work}\label{sec:conclusion}
In this work, we propose OneReason, a reasoning foundation model that connects the large language model and the recommendation system into one single model, enabling recommendation with wisdom.
Specifically, we conduct extensive modifications: (1) in the pre-training stage, we utilize a content-understanding task rather than contrastive supervision to obtain discrete itemic tokens, and use 578B tokens to align the itemic-token and text-token semantic spaces; (2) in SFT, we design a standardized coarse-to-fine, logically coherent CoT structure to ensure the quality of the thinking trace; and (3) in RL, we conduct a specialize-then-unify recipe to improve thinking-mode ability while balancing performance across multiple services.
Equipped with these techniques, to our knowledge, OneReason is the first work in which the thinking mode consistently outperforms the non-thinking mode on downstream recommendation benchmarks, suggesting that reasoning can be translated into real recommendation gains.
More interestingly, we also find that replacing ordinary unCoT data with thinking CoT data improves non-thinking inference in several domains. This suggests that some benefits of CoT supervision can transfer to direct decoding, but a causal separation between compression and reasoning requires further ablation.
Furthermore, we propose OneReason-Bench, which organizes reasoning-oriented benchmark tasks into a progressive hierarchy to guide our model iteration (e.g., Perception, Derivation, Evolution, and Recommendation).
For industrial deployment, we propose a Fast-Slow thinking architecture that continuously injects the knowledge of OneReason into the recommendation pipeline through the collaboration of nearline slow thinking and online fast thinking, contributing significant business gains.

In the future, we will explore: (1) more stable online incremental-learning mechanisms that continuously adapt to new general-knowledge training corpora and user-interest drift while avoiding potential degradation of instruction-following and general capabilities; (2) more customized decoding and generation strategies tailored to business constraints, achieving a better balance among throughput, latency, quality, and business value; and (3) moving beyond ``thinking-then-answer'' toward an agentic recommendation foundation model with \textbf{planning and tool-calling abilities}.
We believe that, with continued progress in model compression, inference infrastructure, and hardware efficiency, recommendation foundation models will become a core component of the recommendation chain, gradually driving the evolution toward an agentic-recommender architecture.

\section*{Acknowledgement}
We thanks Gaoguo Sun, Peng Zhang, Lixing Zhao,  Xinyu Zhang, Xinyue Zhang, Xun Zheng, Zheng Wang for their contributions to the data collection and insightful suggestions.

\printbibliography

\newpage
\quad \\
\quad \\

\section{Author List}

\noindent
\textbf{Core Contributors}\quad
Biao Yang,
Boyang Ding\textsuperscript{*},
Chenglong Chu,
Dunju Zang,
Fei Pan,
Han Li,
Hao Jiang,
Honghui Bao,
Huanjie Wang,
Jian Liang,
Jiangxia Cao,
Jiao Ou,
Jiaxin Deng,
Jinghao Zhang,
Kun Gai,
Lu Ren,
Peiru Du,
Pengfei Zheng\textsuperscript{*},
Rongzhou Zhang,
Ruiming Tang,
Shiyao Wang\textsuperscript{*},
Siyang Mao,
Siyuan Lou,
Teng Shi\textsuperscript{*},
Wei Yuan,
Wenlong Xu,
Xingchen Liu,
Xingmei Wang,
Xinqi Jin,
Yan Sun,
Yan Wang\textsuperscript{*},
Yifei Hu,
Yingzhi He,
Yufei Ye,
Yuhao Wang,
Yunhao Zhou,
Yuqin Dai,
Zhao Liu,
Zhipeng Wei,
Zhixin Ling,
Ziming Li,
Zixing Zhang\textsuperscript{*},
Ziyuan Liu.

\vspace{0.5em}
\noindent
\textbf{Contributors}\quad
An Zhang,
Changxin Lao,
Chaoyi Ma,
Chengru Song,
Defu Lian,
Fan Yang,
Guowang Zhang,
Hao Peng,
Jiayao Shen,
Jie Chen,
Jun Xu, 
Junmin Chen,
Kun Zhang,
Kuo Cai,
Mingxing Wen,
Minmao Wang,
Minxuan Lv,
Qi Zhang,
Qiang Luo,
Sheng Yu,
Shijie Li,
Shijie Yi,
Shuang Yang,
Shugui Liu,
Shuni Chen,
Tinghai Zhang,
Tingting Gao,
Xiang Wang,
Xiangyu Wu,
Xiangyu Zhao,
Xiao Lv,
Xiaoyou Zhou,
Xuming Wang, 
Yong Du,
Zejian Zhang,
Zhaojie Liu,
Zhiyang Zhang,
Zhuang Zhuang,
Ziqi Wang,
Ziyi Zhao.

\vspace{0.5em}
\noindent
{\small All the authors listed alphabetically by first name.\quad \quad \textsuperscript{*}individuals who have departed from our team.}

\newpage

\appendix
\setcounter{section}{0}

\renewcommand{\thesection}{A}
\renewcommand{\thesubsection}{\thesection.\arabic{subsection}}
\renewcommand{\thesubsubsection}{\thesubsection.\arabic{subsubsection}}

\section{Deployment Details}
\subsection{Industrial Scenario Adaptation}
\label{appendix:Industrial Scenario Adaptation}
\textbf{Facing a new industrial scenario, how can OneReason be rapidly adapted?} This setting presents two key challenges: understanding new item corpora and learning emerging user interest distributions, both of which are absent during OneReason training. To address these challenges, we propose a reusable deployment recipe called Industrial Scenario Adaptation.


\textbf{Scenario-aware Continual Pre-Train.}
We start from the checkpoint of the final OneReason foundation model.  For item understanding in new industrial scenarios, we adapt the R0: Perception~\ref{sec:sft-data-r0} method for bidirectional alignment between itemic tokens and captions. For user understanding in new industrial scenarios, we adopt a User-Profile-Based Continual Pre-Training strategy. A large-scale LLM is first used to summarize user behaviors and extract preference signals, which are then combined with user profiles to form user portrait texts. These user portraits texts are used for continual pre-training, enabling the model to capture scenario-specific user characteristics.

\textbf{Scenario-aware Continual Supervised Fine-Tuning.}
We further train OneReason for recommendation by conditioning on user portrait texts and historical interactions to predict the next itemic token. To improve robustness across users with different activity levels, we introduce a curriculum learning strategy that progresses from high-activity to low-activity users. This allows the model to first learn from dense behavioral signals and gradually adapt to sparse scenarios, while user portraits provide consistent semantic guidance for generalization.

\subsection{The Overall Methodology of OneReason for OneRec} 
\label{appendix:OneReason For OneRec}

As reflected in the core design philosophy of the OneReason, \textbf{Thinking-then-Generation} have been demonstrated to be an important pathway in the evolution of LLMs. Prior work has also shown that introducing latent reasoning mechanisms into generative models can effectively enhance reasoning capabilities and further improve recommendation performance. Therefore, we attempt to leverage the outputs of OneReason to provide reasoning supervision, and we have obtained business uplift in the online fast pipeline, validating that OneReason’s reasoning capabilities and knowledge can be transferred to  generative recommendation model via distillation.

\begin{figure}[H]\centering 
    \includegraphics[width= 0.73 \textwidth]{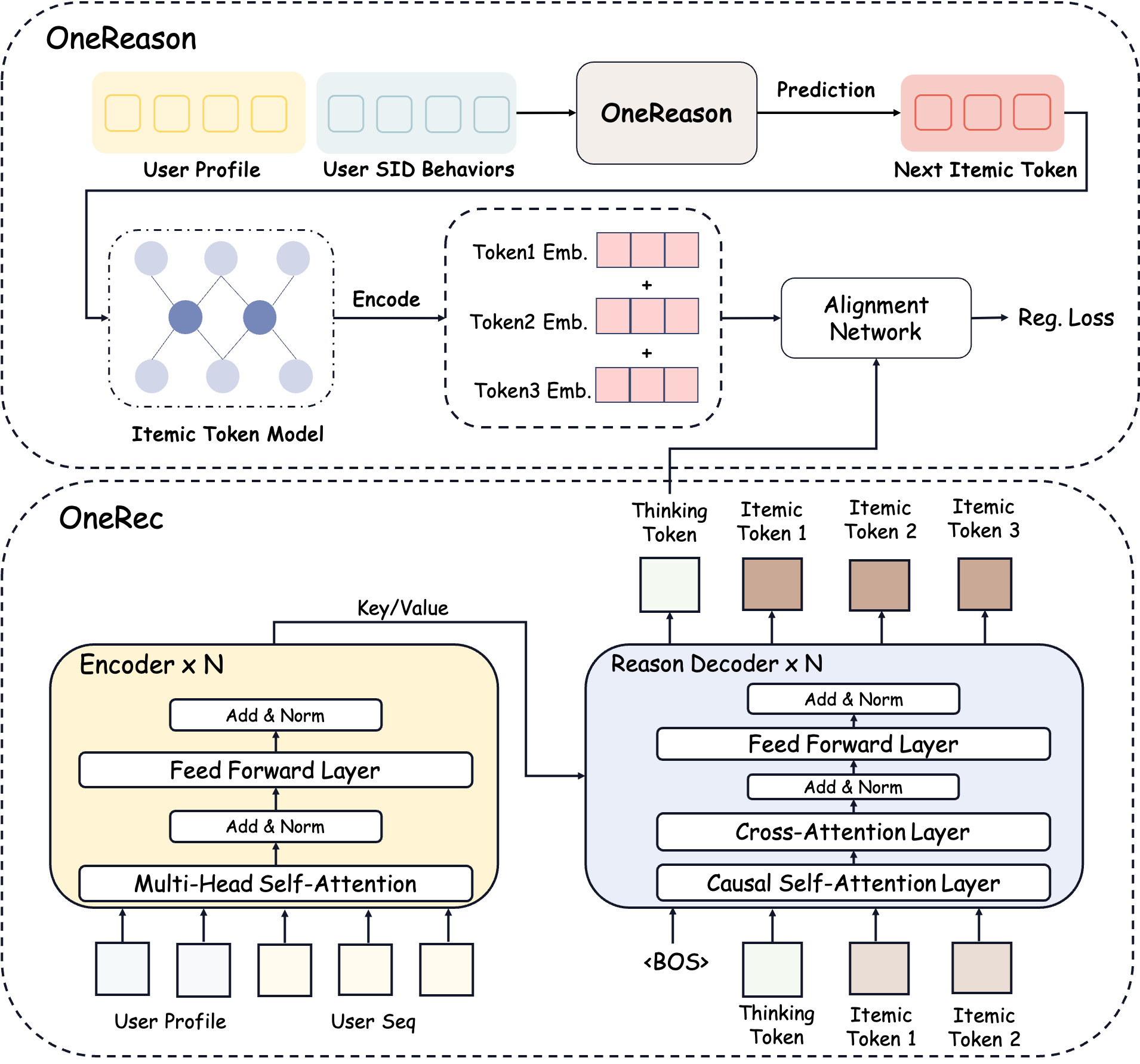}
    \caption{The overall structure of OneReason for OneRec.}\label{fig:OneReason for OneRec}
\end{figure}

The overall framework is illustrated in the Figure~\ref{fig:OneReason for OneRec}. Specifically, we first use OneReason to predict the next itemic token with the highest probability of user interaction, and then decode it into the embedding representation corresponding to the itemic token via a quantized model, serving as a supervision signal for the reasoning process. Subsequently, we introduce a \textbf{Thinking Token} into the generative recommendation model and use the hidden state at the decoder-side \texttt{<BOS>} position as the supervision target. We then apply an \textbf{Alignment Network} to impose representation-level constraints, thereby distilling OneReason's knowledge and reasoning capabilities into the generative recommendation model within a high-dimensional semantic space.

On this basis, we further explore several variant designs. For example, we introduce a User Representation Decoder to compress the encoder outputs, align them with OneReason’s outputs, and further distill them into the Thinking Token. In addition, to address the uncontrollability of the reasoning process, we design a dual-branch architecture consisting of a Decoder with Reason and a Decoder without Reason, and apply mutual supervision to improve the stability of the reasoning process. Finally, considering the inherent misalignment between the recommendation space and the semantic space, we further propose a contrastive distillation method based on spatial residuals to enhance cross-space alignment.

\subsection{Experimental Analysis} 
\label{appendix:Experimental Analysis}

We conduct a detailed analysis of the above A/B test results. We decompose the overall effects from two perspectives: (1) the changes in efficiency brought by OneReason, and (2) the advantages of OneReason across different user groups.

\textbf{\textit{Distribution and Efficiency Analysis.}} 
From Table \ref{tab:conversion_result}, we observe that OneReason brings consistent system-level improvements under different deployment paradigms. In the slow pipeline, direct application of OneReason improves conversion and click performance by enhancing traffic quality through better semantic matching. In contrast, enhancing OneRec in the fast pipeline mainly improves system scale via expanded retrieval coverage, while per-impression metrics remain relatively stable. The combined strategy achieves the best overall performance with further increased coverage, demonstrating the complementarity between the two paradigms: the slow pipeline improves traffic efficiency, while the fast pipeline expands system scale.


\begin{table}[ht]
\centering
\setlength{\tabcolsep}{5pt}
\begin{tabular}{l c c c c c}
\toprule
\textbf{Model}  & \textbf{Conv.} & \textbf{Click} & \textbf{CVR} & \textbf{CTR} & \textbf{Retr. Share} \\
\midrule

OneReason
& +5.870\%
& +3.986\%
& +4.516\%
& +2.497\%
& 3.8\% \\

OneReason for OneRec
& +7.935\%
& +0.207\%
& +0.459\%
& +0.195\%
& 22.1\%  \\

Combined
& +12.643\%
& +1.709\%
& +1.865\%
& +0.519\%
& 27.2\% \\

\bottomrule
\end{tabular}

\caption{
Online traffic efficiency and retrieval distribution analysis.
\textbf{Conv.} denotes conversion numbers,
\textbf{Click} denotes click numbers,
\textbf{CVR} denotes conversion rate,
\textbf{CTR} denotes click rate.
}
\label{tab:conversion_result}
\end{table}

\textbf{\textit{User Level Analysis.}} 
To evaluate the user-level impact of OneReason, we conduct a stratified analysis across high-activity, medium-activity, and low-activity users. Table \ref{tab:user_segmentation_analysis} shows that business revenue gains increase as user activity decreases, with the largest improvements observed in low-activity users. Compared with traditional recommendation models that rely heavily on users' historical behaviors, OneReason enables reasoning-based interest completion by integrating user profiles, limited behaviors, and content semantics, enabling effective recommendations under sparse interactions. Overall, these results validate the value of its reasoning capability for low-activity users and demonstrate substantial information gain for the overall system.

\begin{table}[ht]
\centering
\setlength{\tabcolsep}{6pt}
\begin{tabular}{l c c c c}
\toprule
\textbf{User Level} & \textbf{Impressions} & \textbf{Revenue} \\
\midrule
Active Users        & +0.763\% &  +2.419\% \\
Mid-Active Users    & +0.902\% &  +4.815\% \\
Low-Active Users    & +1.125\% &  +13.323\% \\
\bottomrule
\end{tabular}

\caption{User-level performance breakdown under OneReason direct retrieval.}
\label{tab:user_segmentation_analysis}
\end{table}

\subsection{Discussions in Industrial Deployment}
\label{appendix:disscussions in industrial deployment}

We analyze the key challenges encountered during the online deployment of OneReason and discuss several important issues, along with future directions.

\subsubsection{D1: Online Incremental Training and General Model Performance.}
OneReason requires incremental updates to adapt to evolving item corpora and user interests. Without such updates, we observe significant performance degradation. However, the current approach primarily focuses on continual learning for recommendation objectives, while partially neglecting instruction-following capabilities and general capability retention. Future work should explore more stable incremental learning strategies that preserve foundational capabilities while adapting to evolving item corpora and shifting user-interest distributions.

\subsubsection{D2: Generation Strategies of OneReason.}
\label{appendix:disscussions in industrial deployment d2}
OneReason has higher inference cost, making large-scale beam search infeasible. We therefore design a hierarchical generation strategy for item ID decoding under small-scale beam search. The model first generates candidates at the top itemic token-level to ensure coverage, and then completes lower-level structures via greedy decoding, balancing diversity, accuracy, and efficiency under latency constraints. However, the current strategy still lacks consideration of business value and resource constraints. Future work may explore more customized decoding strategies to improve recommendation quality.

\subsubsection{D3: Model Parameter Size and Real-time Applications.}
In fact, the ``Fast-Slow Thinking'' architecture is an engineering trade-off. It preserves real-time system stability while incorporating information gains from OneReason via a nearline mechanism. However, it cannot fully replace real-time retrieval nor fully exploit the model's capacity. A key future direction is model parameter size, i.e., exploring smaller models (e.g., 0.8B level) that maintain comparable performance while enabling true real-time retrieval, further improving system efficiency and integration.

\subsubsection{D4: ROI of OneReason for Online Deployment.}
We first note that ROI is not the primary objective of OneReason. Our main goal is to evaluate whether the model provides meaningful information gain in recommendation systems. However, from an engineering perspective, the system consumes approximately 600 flagship GPUs per day to serve full traffic across 400 million users in the Kuaishou App. We analyze A/B test experimental results and estimate an ROI > 5, indicating that system-level gains outweigh inference costs and yield a positive return. With advances in model compression, inference infrastructure, and hardware efficiency, OneReason is expected to become a core component of future recommendation systems, driving a shift toward a foundation-centric architecture.

\renewcommand{\thesection}{B}
\renewcommand{\thesubsection}{\thesection.\arabic{subsection}}
\renewcommand{\thesubsubsection}{\thesubsection.\arabic{subsubsection}}

\section{Benchmark Details}
\subsection{Task Construction}
\label{appendix:benchmark_construction_and_comparison}
We provide the construction details of the four reasoning layers in \benchname{} here.

\paragraph{R0: Perception.}
We use multimodal LLMs to generate reference item captions from native multimodal signals across video, live-streaming, product, and advertising domains, followed by LLM-based and human quality checks. We next construct content-grounded QA instances from the verified captions and item metadata, and further filter them by answerability, difficulty, and human spot-checking.

\paragraph{R1: Derivation.}
We extract item-to-item association pairs from content knowledge graphs. We further apply LLM-based difficulty filtering and human quality checks to retain samples that require semantic derivation rather than surface-level matching.

\paragraph{R2: Evolution.}
We mine interest evolution chains with temporal progression and cognitive increments from full-domain user behavior timelines spanning short video, e-commerce, advertising, live-streaming, and additional search. Rule-based, LLM-based, and human quality checks are used to filter pseudo-logical chains, and the resulting chains are converted into the three task formats above.

\paragraph{R3: Recommendation.}
We build all-domain history sequences from real user behavior logs and use users' subsequent high-value interacted items as prediction targets, where high-value interaction is defined domain-specifically: clicks for Product, watch-time above the 75th percentile within the same duration bucket for Video, conversion actions (e.g., activation or payment) for Ad, and first gifting events for Live. Target items are further processed with abnormal-sample filtering, popularity downsampling, and category balancing to reduce head-item dominance.

\subsection{Recommendation Data Statistics and Analysis}
\label{appendix:r3_statistics}

Table~\ref{tab:benchmark_statistics} presents the scale statistics of R3 across the four target domains.

\begin{table*}[htbp]
\centering
\caption{Statistics of R3 recommendation data in \benchname{} across four target domains. Each user corresponds to one instance.}
\label{tab:benchmark_statistics}
\resizebox{\textwidth}{!}{%
\begin{tabular}{@{}lrrrrrrrr@{}}
\toprule
\textbf{Domain} & \textbf{\# Users} & \makecell{\textbf{\# In-Domain}\\\textbf{Items}} & \makecell{\textbf{\# Out-of-Domain}\\\textbf{Items}} & \makecell{\textbf{\# In-Domain}\\\textbf{Inter.}} & \makecell{\textbf{\# Out-of-Domain}\\\textbf{Inter.}} & \makecell{\textbf{Avg. In-Domain}\\\textbf{Hist.}} & \makecell{\textbf{Avg. Out-of-Domain}\\\textbf{Hist.}} & \textbf{Avg. Tgt.} \\
\midrule
Video       & 10,285 & 694,507   & 419,428   & 770,228   & 781,533   & 74.89 & 75.99  & 13.92 \\
Product     & 9,859  & 502,668   & 599,238   & 760,277   & 986,817   & 77.12 & 100.09 & 1.62  \\
Ad          & 9,081  & 249,309   & 718,771   & 840,760   & 1,089,317 & 92.58 & 119.96 & 4.24  \\
Live        & 8,083  & 66,831    & 810,269   & 240,577   & 1,229,617 & 29.76 & 152.12 & 1.58  \\
\bottomrule
\end{tabular}%
}
\end{table*}

Beyond the scale statistics, we further characterize train-evaluation overlap from two complementary perspectives: target-level visibility and transition-level memorization/generalization. For target-level visibility, 33.69\% of target item origin IDs are unseen during training, whereas only 11.55\% of target itemic patterns are unseen. This suggests that many item origin ID-level cold-start targets still support content-based generalization through seen itemic patterns.

Following \citet{ding2026well}, we then examine transition-level memorization and generalization at the itemic-pattern granularity. For an evaluation instance with history
$(\text{item}_1, \text{item}_2, \ldots, \text{item}_n)$ and target set $\mathcal{T}$, each pair $(\text{item}_n, t)$ for $t \in \mathcal{T}$ is counted as a memorized pair if the same transition appears in training, and as a generalized pair otherwise. 
As shown in Table~\ref{tab:memorization_generalization}, Video and Product are dominated by generalized transitions, whereas Ad and Live are more memorization-heavy, allowing \benchname{} to cover both settings.
\begin{table}[htbp]
\centering
\caption{Memorization and generalization statistics of R3 target transitions at the itemic-pattern granularity in \benchname{}.}
\label{tab:memorization_generalization}
\begin{tabular}{@{}lrr@{}}
\toprule
\textbf{Domain} & \textbf{Memorized Pairs} & \textbf{Generalized Pairs} \\
\midrule
Video   & 4.7\%  & 95.3\% \\
Product & 27.8\% & 72.2\% \\
Ad      & 75.9\% & 24.1\% \\
Live    & 73.0\% & 27.0\% \\
\bottomrule
\end{tabular}
\end{table}

\subsection{Comparison with Existing Benchmarks}
\label{appendix:benchmark_comparison}

Existing benchmarks have progressively expanded the evaluation scope of recommender systems. Traditional recommendation benchmarks, including Amazon~\citep{hou2024bridging}, Yelp~\citep{yelp_dataset}, PixelRec~\citep{cheng2024pixelrec}, NineRec~\citep{zhang2025ninerec}, and KuaiSAR~\citep{sun2023kuaisar}, mainly focus on outcome-level evaluation such as ranking accuracy, next-item prediction, and behavior prediction. Within recommendation foundation-model evaluation, RecIF-Bench provides the immediate basis for this work by establishing a holistic instruction-following benchmark.

\benchname{} continues this line of work by focusing on recommendation-specific reasoning diagnosis. Instead of only evaluating the final recommendation outcome or response-level quality, it introduces explicit diagnostic tasks for item-level relation derivation and temporal user-interest evolution as complementary probes of recommendation-specific reasoning, while retaining next-items prediction as a downstream recommendation task.
Table~\ref{tab:benchmark_comparison} summarizes this positioning.


\begin{table}[t]
\centering
\renewcommand{\arraystretch}{1.9}     
\setlength{\tabcolsep}{5pt}
\resizebox{\textwidth}{!}{%
\begin{tabular}{!{\vrule width 0pt}c l c c c c c}
\toprule
\textbf{Category} & \textbf{Benchmark}
  & \makecell{\textbf{Multi-domain}\\\textbf{Knowledge}}
  & \makecell{\textbf{Instruction-}\\\textbf{Following}}
  & \makecell{\textbf{Reasoning-}\\\textbf{Centric}}
  & \textbf{SeqRec}
  & \makecell{\textbf{Cross-Domain}\\\textbf{Rec}} \\
\midrule

\catlabel{4}{catBlue}{General LLM}
  & \textcolor{catBlue}{GSM8K / MATH}      & \cno  & \cyes & \cyes & \cno  & \cno  \\
  & \textcolor{catBlue}{MMLU-Pro / GPQA}   & \cyes & \cyes & \cyes & \cno  & \cno  \\
  & \textcolor{catBlue}{CorrectBench}      & \cno  & \cyes & \cyes & \cno  & \cno  \\
  & \textcolor{catBlue}{CriticBench}       & \cyes & \cyes & \cyes & \cno  & \cno  \\
\midrule

\catlabel{3}{catPurple}{Classic Rec}
  & \textcolor{catPurple}{Yelp / MovieLens / PixelRec} & \cno & \cno & \cno & \cyes & \cno  \\
  & \textcolor{catPurple}{Amazon / NineRec}            & \cno & \cno & \cno & \cyes & \cyes \\
  & \textcolor{catPurple}{KuaiSAR}                     & \cno & \cno & \cno & \cyes & \cno  \\
\midrule

\catlabel{3}{catTeal}{Rec\,(LLM\,era)}
  & \textcolor{catTeal}{RecBench+}     & \cno & \cyes & \cno & \cyes & \cno  \\
  & \textcolor{catTeal}{AgentRecBench} & \cno & \cyes & \cno & \cyes & \cno  \\
  & \textcolor{catTeal}{HORIZON}       & \cno & \cno  & \cno & \cyes & \cyes \\
\midrule

\catlabel{2}{catRed}{RecFM}
  & \textcolor{catRed}{RecIF-Bench}                                   & \cyes & \cyes & \cno  & \cyes & \cyes \\
  & \textcolor{catRed}{\ding{72}~\textbf{OneReason-Bench (Ours)}}     & \cyes & \cyes & \cyes & \cyes & \cyes \\
\bottomrule
\end{tabular}%
}

\vspace{4pt}
\footnotesize
\fcolorbox{black}{cellGreen}{\yes}~supported \quad
\fcolorbox{black}{cellGray}{\no}~none
\caption{Comparison of benchmarks across categories. OneReason-Bench is the only benchmark covering all five capabilities.}
\label{tab:benchmark_comparison}
\end{table}

\subsection{Evaluation Metrics}
\subsubsection{Item Understanding Evaluation}
\label{appendix:eval_item_understanding}

For the Item Understanding task, we reuse the LLM-as-a-Judge protocol introduced in OpenOneRec.
The metric definition is as follows: generated descriptions are decomposed into weighted information points and matched against the corresponding ground-truth description.
Each information point carries a predefined weight reflecting its semantic importance; the final score is the OpenOneRec double-weighted F1, which penalizes both hallucinated points (false positives) and omitted points (false negatives) in proportion to their weights.
Meanwhile, same with OpenOneRec, the evaluation is conducted over all examples in the Item Understanding split and the average double-weighted F1 is used to reported as the task score.

\subsubsection{Action--Logic Score}
\label{appendix:action_logic_score}

The Action--Logic Score evaluates two generative tasks: Evolution Topic Generation and Evolution Direct Generation. Both require the model to produce \emph{evolution chains}, each a sequence of events $e_k = (A_k, l_k)$, where $A_k$ is an \emph{action group}, i.e., a set of user actions characterizing a phase in the user's interest-evolution process, and $l_k$ is the corresponding logic statement. Let the ground-truth chain be $\mathcal{E}^* = (e^*_1, \ldots, e^*_n)$ with $e^*_k = (A^*_k, l^*_k)$, and the generated chain be $\hat{\mathcal{E}} = (\hat{e}_1, \ldots, \hat{e}_m)$ with $\hat{e}_j = (\hat{A}_j, \hat{l}_j)$. The metric combines action alignment and logic alignment quality into a single composite score.

\paragraph{Action Alignment.}

For a ground-truth event $e^*_k$ and a generated event $\hat{e}_j$, the pairwise action similarity is defined as:
\begin{equation}
    m_{\mathrm{F1}}(A^*_k,\, \hat{A}_j)
    =
    \frac{2\,|A^*_k \cap \hat{A}_j|}
    {|A^*_k| + |\hat{A}_j|}.
\end{equation}

The resulting similarity matrix is aligned via maximum-weight monotone matching, equivalent to a weighted longest-common-subsequence alignment: it selects a set of one-to-one pairs $\mathcal{M}_a$ with strictly increasing ground-truth and generated indices that maximizes total action similarity.

The chain-level \textbf{action alignment score} aggregates pairwise similarities with a precision--recall F1 over the generated and ground-truth chains:
\begin{align}
    P_a &= \frac{\sum_{(k,j)\in\mathcal{M}_a} m_{\mathrm{F1}}(A^*_k, \hat{A}_j)}{|\hat{\mathcal{E}}|}, \\
    R_a &= \frac{\sum_{(k,j)\in\mathcal{M}_a} m_{\mathrm{F1}}(A^*_k, \hat{A}_j)}{|\mathcal{E}^*|}, \\
    \text{Action Alignment} &= \frac{2P_aR_a}{P_a+R_a}.
\end{align}

\paragraph{Logic Evaluation.}

The \textbf{logic alignment score} reuses the matched pairs $\mathcal{M}_a$ from Action Alignment. For each $(k,j)\in\mathcal{M}_a$, the mixed logic similarity is:
\begin{equation}
    s_{\mathrm{logic}}(l^*_k,\hat{l}_j)
    = 0.5\,\mathrm{Token}_{\mathrm{F1}}(l^*_k,\hat{l}_j)
    + 0.5\,\mathrm{ROUGE\text{-}L}_{\mathrm{F1}}(l^*_k,\hat{l}_j).
\end{equation}
Here, $\mathrm{Token}_{\mathrm{F1}}$ is the token-overlap F1 between the ground-truth and generated logic statements, and $\mathrm{ROUGE\text{-}L}_{\mathrm{F1}}$ is the F1 variant of ROUGE-L, derived from the longest common subsequence of the two statements.
Defining the total logic similarity over matched pairs as:
\begin{equation}
    T = \sum_{(k,j)\in\mathcal{M}_a} s_{\mathrm{logic}}(l^*_k,\hat{l}_j),
\end{equation}
precision, recall, and the chain-level score are then:
\begin{align}
    P_\ell &= \frac{T}{|\hat{\mathcal{E}}|}, \\
    R_\ell &= \frac{T}{|\mathcal{E}^*|}, \\
    \text{Logic Alignment} &= \frac{2P_\ell R_\ell}{P_\ell+R_\ell}.
\end{align}

\paragraph{Action--Logic Score Definition.}

The \textbf{Action--Logic Score} is the primary metric for generated evolution chains, defined as the equal-weight average of the two alignment sub-scores:
\begin{equation}
    \text{Action--Logic Score}
    = 0.5 \times \text{Action Alignment}
      + 0.5 \times \text{Logic Alignment}.
\end{equation}

For Evolution Topic Generation, scores are averaged across generations for each sample and then across all samples; parse failures receive zero score and remain in the denominator.

\paragraph{Multi-Chain Action--Logic Score.}

Evolution Direct Generation requires the model to produce a set of evolution chains $\{\hat{\mathcal{E}}\}$ without topic guidance, compared against a ground-truth set $\{\mathcal{E}^*\}$. The evaluator first computes a chain-level action score for every ground-truth--generated chain pair and applies the Hungarian algorithm to find the maximum-weight one-to-one assignment.

Each matched chain pair is then evaluated using the single-chain procedure above. Unmatched ground-truth chains receive zero score, and the \textbf{Multi-Chain Action--Logic Score} is averaged over the number of ground-truth chains. Extra generated chains do not directly enter this average, but they are penalized in the auxiliary chain-level F1 metrics through the precision denominator.

\subsection{Task Formats}
\label{appendix:reasoning_formats}

\makeatletter
\@ifundefined{mycase}{%
  \newtcolorbox{mycase}[1]{%
    breakable,
    colback=gray!5, colframe=black!60, boxrule=0.5pt, arc=2pt,
    left=6pt, right=6pt, top=6pt, bottom=6pt,
    title={#1},
    fonttitle=\bfseries\small, coltitle=white, colbacktitle=black!70,
    before skip=8pt, after skip=8pt,
    before upper={\raggedright\sloppy}%
  }%
}{}
\makeatother

\begin{CJK*}{UTF8}{gbsn}

In this appendix, we provide representative input--output templates for the benchmark task formats.

\begin{mycase}{Prompt Bench.R0.1: Item Understanding (R0 — Perception)}
    \small
    \textbf{System:} 
    \textit{你是一个直播内容分析专家。}  \\
    \textbf{User:} \\
    请将以下主播token还原为主播画像描述，涵盖形象气质与表达方式、内容方向与受众定位、互动氛围与风格特点：\texttt{<|living\_begin|><a\_1658><b\_451><c\_7635>}  \\
    \textbf{Assistant:} \\
    \textit{该主播``XXXXXX''是一位极具生活气息的聊天娱乐类主播。她外在形象朴素亲切，常戴眼镜身穿棉服，以边吃边聊的真实状态示人。其直播内容聚焦于情感调解、家长里短及婚恋话题（如彩礼、相亲），通过多人连麦的形式搭建了一个``线上茶话会''。她性格豪爽、接地气，扮演着倾听者与调解员的角色\dots{}} 

    \vspace{0.5em}
    \hrulefill
    \vspace{0.5em}

    \textbf{English Example} \\
    \textbf{System:} 
    \textit{You are an expert in live-streaming content analysis.}  \\
    \textbf{User:} \\
    Decode the following streamer token into a portrait description covering appearance and expression, content direction and audience positioning, and interaction style: \texttt{<|living\_begin|><a\_1658><b\_451><c\_7635>}  \\
    \textbf{Assistant:} \\
    \textit{The streamer ``XXXXXX'' is a lively chat-entertainment host whose style is steeped in everyday life. She appears plain and approachable, often wearing glasses and a padded jacket, presenting herself in the authentic state of chatting while eating. Her content focuses on emotional mediation, family gossip, and marriage-related topics (such as bride prices and matchmaking), building an ``online tea party'' through multi-person live-chat sessions. Outgoing and down-to-earth, she serves as both listener and mediator\dots{}} 
\end{mycase}

\begin{mycase}{Prompt Bench.R0.2: Itemic Pattern Grounding (R0 — Perception)}
    \small
    \textbf{System:} 
    \textit{你是一个专业的直播主播分析助手。}  \\
    \textbf{User:} \\
    请分析以下主播的形象、内容和风格描述，生成匹配的主播token：\textit{该主播``XXXXXX''是一位极具生活气息的聊天娱乐类主播。她外在形象朴素亲切，常戴眼镜身穿棉服，以边吃边聊的真实状态示人。其直播内容聚焦于情感调解、家长里短及婚恋话题（如彩礼、相亲），通过多人连麦的形式搭建了一个``线上茶话会''。她性格豪爽、接地气，扮演着倾听者与调解员的角色\dots{}}  \\
    \textbf{Assistant:} \\
    \texttt{<|living\_begin|><a\_1658><b\_451><c\_7635>} 

    \vspace{0.5em}
    \hrulefill
    \vspace{0.5em}

    \textbf{English Example} \\
    \textbf{System:} 
    \textit{You are a professional live-streamer analysis assistant.}  \\
    \textbf{User:} \\
    Analyze the following streamer's appearance, content, and style description, and generate the matching streamer token: \textit{The streamer ``XXXXXX'' is a lively chat-entertainment host whose style is steeped in everyday life. She appears plain and approachable, often wearing glasses and a padded jacket, presenting herself in the authentic state of chatting while eating. Her content focuses on emotional mediation, family gossip, and marriage-related topics (such as bride prices and matchmaking), building an ``online tea party'' through multi-person live-chat sessions. Outgoing and down-to-earth, she serves as both listener and mediator\dots{}}  \\
    \textbf{Assistant:} \\
    \texttt{<|living\_begin|><a\_1658><b\_451><c\_7635>} 
\end{mycase}

\begin{mycase}{Prompt Bench.R0.3: Item QA (R0 — Perception)}
    \small
    \textbf{System:} 
    \textit{你是一位专业的短视频理解助手，能够精准掌握视频描述，并清晰准确地回答相关问题。}  \\
    \textbf{User:} \\
    视频 \texttt{<|video\_begin|><a\_5946><b\_1435><c\_6589>} \\
    视频如何体现地域文化？ \\
    A. 呈现郑州现代城市景观与科技馆 \\
    B. 使用河南方言和豫剧唱段 \\
    C. 展示陕西兵马俑和秦腔艺术 \\
    D. 展示河南古建筑和雪景 \\
    请直接回答A/B/C/D中的一个选项，不要输出任何其他内容。  \\
    \textbf{Assistant:} \\
    \textit{D} 

    \vspace{0.5em}
    \hrulefill
    \vspace{0.5em}

    \textbf{English Example} \\
    \textbf{System:} 
    \textit{You are a professional short-video understanding assistant, capable of accurately grasping video descriptions and clearly answering related questions.}  \\
    \textbf{User:} \\
    Video \texttt{<|video\_begin|><a\_5946><b\_1435><c\_6589>} \\
    How does the video reflect regional culture? \\
    A. Presents Zhengzhou's modern cityscape and science museum \\
    B. Uses Henan dialect and Yu opera excerpts \\
    C. Showcases Shaanxi's terracotta warriors and Qinqiang art \\
    D. Shows ancient architecture of Henan and snow scenes \\
    Please answer with one of A/B/C/D and nothing else.  \\
    \textbf{Assistant:} \\
    \textit{D} 
\end{mycase}

\begin{mycase}{Prompt Bench.R1.1: Item2Item (R1 — Derivation)}
    \small
    \textbf{System:} 
    \textit{你是一位视频推荐系统专家，擅长发现视频之间的深层关系。}  \\
    \textbf{User:} \\
    给定源视频：\texttt{<|video\_begin|><a\_1584><b\_6006><c\_2381>}

    该视频与某个候选视频之间存在「潜在需求共鸣」关系。
    （潜在需求共鸣的定义：潜在需求共鸣：两个表面完全不同的视频，满足用户同一个深层心理需求。例：「30天极简断舍离挑战」↔「手机APP大清理只留5个」，深层需求都是简化生活获得掌控感。）

    请从以下候选视频中选出与源视频存在该关系的视频：

    A. \texttt{<|video\_begin|><a\_8152><b\_3417><c\_4595>}\\
    B. \texttt{<|video\_begin|><a\_1584><b\_254><c\_1317>}\\
    C. \texttt{<|video\_begin|><a\_6219><b\_6006><c\_4595>}\\
    D. \texttt{<|video\_begin|><a\_6219><b\_3504><c\_5673>}\\

    请以"答案：[选项字母]"的格式返回最终答案。  \\
    \textbf{Assistant:} \\
    \textbf{[答案]} \textbf{C}  \\

    \vspace{0.5em}
    \hrulefill
    \vspace{0.5em}

    \textbf{English example} \\
    \textbf{System:} 
    \textit{You are a video recommendation system expert, skilled at discovering deep relationships between videos.}  \\
    \textbf{User:} \\
    Given the source video: \texttt{<|video\_begin|><a\_1584><b\_6006><c\_2381>}

    This video shares a ``Latent Need Resonance'' relationship with one of the candidate videos below.
    (Definition of Latent Need Resonance: two superficially different videos that satisfy the same deep psychological need. Example: ``30-Day Minimalist Decluttering Challenge'' ↔ ``Phone App Purge: Keep Only 5,'' both serving the deep need of simplifying life and gaining a sense of control.)

    Please select the candidate video that has this relationship with the source video:

    A. \texttt{<|video\_begin|><a\_8152><b\_3417><c\_4595>}\\
    B. \texttt{<|video\_begin|><a\_1584><b\_254><c\_1317>}\\
    C. \texttt{<|video\_begin|><a\_6219><b\_6006><c\_4595>}\\
    D. \texttt{<|video\_begin|><a\_6219><b\_3504><c\_5673>}\\

    Please return your final answer in the format ``Answer: [Option Letter].''  \\
    \textbf{Assistant:} \\
    \textbf{[Answer]} \textbf{C}  \\
\end{mycase}

\begin{mycase}{Prompt Bench.R2.1: Evolution Action Selection (R2 — Evolution)}
    \small
    \textbf{User:} \\
    你是一名专业的数据标签专家。请根据给定的用户行为逻辑链，从行为时间线中提取所有相关的行为记录。 \\

    \textbf{逻辑链}：\textit{从赛那泛化改装兴趣到具体部件方案筛选} \\

    \textbf{用户行为时间线}： \\
    【2023-10-25】[商品-购买] \texttt{<|prod\_begin|><a\_5967><b\_2249><c\_743>} \\
    【2024-06-04】[直播-关注] \texttt{<|living\_begin|><a\_4318><b\_7342><c\_5563>} \\
    \dots{}  \\
    \textbf{Assistant:} \\
    \texttt{<|video\_begin|><a\_4750><b\_2230><c\_4729>}\allowbreak{}
    \texttt{<|video\_begin|><a\_5844><b\_5861><c\_361>}\allowbreak{} 
    \texttt{<|video\_begin|><a\_3475><b\_2493><c\_1218>}\allowbreak{} 

    \vspace{0.5em}
    \hrulefill
    \vspace{0.5em}

    \textbf{English Example} \\
    \textbf{User:} \\
    You are a professional data-labeling expert. Given a logic chain of user behavior, extract all related action records from the behavioral timeline. \\

    \textit{Logic Chain:} \textit{From generic Sienna modification interest to specific component plan selection.} \\

    \textit{User Behavioral Timeline:} \\
    {[2023-10-25]} \texttt{[Product-Purchase] <|prod\_begin|><a\_5967><b\_2249><c\_743>} \\
    {[2024-06-04]} \texttt{[Live-Follow] <|living\_begin|><a\_4318><b\_7342><c\_5563>} \\
    \dots{} \\
    \textbf{Assistant:} \\
    \texttt{<|video\_begin|><a\_4750><b\_2230><c\_4729>}\allowbreak{} \texttt{<|video\_begin|><a\_5844><b\_5861><c\_361>}\allowbreak{} \texttt{<|video\_begin|><a\_3475><b\_2493><c\_1218>}\allowbreak{} 
\end{mycase}

\label{appendix:benchmark-r2-generation-case}
\begin{mycase}{Prompt Bench.R2.2: Evolution Topic Generation (R2 — Evolution)}
    \small
    \textbf{User:} \\
    【用户交互历史】： \\
    【2023-10-25】\texttt{[商品-购买] <|prod\_begin|><a\_5967><b\_2249><c\_743>} \\
    \dots{}（多域时间线）\dots{} \\
    【2025-12-25】\texttt{[搜索] carplay转安卓手机盒子}\\
    \dots{} \\
    \textbf{目标兴趣主题}：\textit{从赛那泛化改装兴趣到具体部件方案筛选} \\
    请基于历史生成完整的逻辑链。  \\
    \textbf{Assistant:} \\
    \textit{\{}\\
    \hspace*{1em}\textit{"logic\_chain":} \textit{\{}\\
    \hspace*{2em}"name": \textit{"从赛那泛化改装兴趣到具体部件方案筛选"}, \\
    \hspace*{2em}"events": \textit{[} \\
    \hspace*{3em}\textit{\{} "date": \textit{"2025-12-21"}, "action": \textit{[搜索] 塞纳改装大屏教程}, \\
    \hspace*{4em}"logic": \textit{触发阶段：以教程形式主动获取赛那改装路径，显示明确改装方向需求。} \textit{\}}, \\
    \hspace*{3em}\textit{\{} "date": \textit{"2025-12-23"}, "action": \textit{[搜索] 玩塞纳的十大忠告}, \\
    \hspace*{4em}"logic": \textit{认知补全：从做法走向风险与经验框架。} \textit{\}}, \\
    \hspace*{3em}\textit{\{} "date": \textit{"2025-12-25"}, "action": \textit{[商品-点击] CarPlay转安卓手机盒子}, \\
    \hspace*{4em}"logic": \textit{决策收敛：从教程/忠告调研落到可购买的改装硬件点击。} \textit{\}}, \\
    \hspace*{3em}\dots{} \\
    \hspace*{2em}\textit{]} \\
    \hspace*{1em}\textit{\}} \\
    \textit{\}} 

    \vspace{0.5em}
    \hrulefill
    \vspace{0.5em}

    \textbf{English Example} \\
    \textbf{User:} \\
    \textit{[User Interaction History]:} \\
    {[2023-10-25]} \texttt{[Product-Purchase] <|prod\_begin|><a\_5967><b\_2249><c\_743>} \\
    \dots{} (multi-domain timeline) \dots{} \\
    {[2025-12-25]} \texttt{[Search] CarPlay-to-Android phone adapter box}\dots{} \\
    \textit{Target Interest Topic:} \textit{From generic Sienna modification interest to specific component plan selection.} \\
    Generate the complete logic chain based on the history.  \\
    \textbf{Assistant:} \\
    \textit{\{}\\
    \hspace*{1em}\textit{"logic\_chain":} \textit{\{}\\
    \hspace*{2em}"name": \textit{"From generic Sienna modification interest to specific component plan selection"}, \\
    \hspace*{2em}"events": \textit{[} \\
    \hspace*{3em}\textit{\{} "date": \textit{"2025-12-21"}, "action": \textit{[Search] Sienna modification large-screen tutorial}, \\
    \hspace*{4em}"logic": \textit{Trigger stage: actively acquiring the Sienna modification path through tutorial searches, indicating a clear modification-direction need.} \textit{\}}, \\
    \hspace*{3em}\textit{\{} "date": \textit{"2025-12-23"}, "action": \textit{[Search] Top-10 tips for Sienna owners}, \\
    \hspace*{4em}"logic": \textit{Cognitive complement: shifting from how-to toward risk and experience frameworks.} \textit{\}}, \\
    \hspace*{3em}\textit{\{} "date": \textit{"2025-12-25"}, "action": \textit{[Product-Click] CarPlay-to-Android adapter box}, \\
    \hspace*{4em}"logic": \textit{Decision convergence: from tutorial/advice research down to a purchasable modification hardware click.} \textit{\}}, \\
    \hspace*{3em}\dots{} \\
    \hspace*{2em}\textit{]} \\
    \hspace*{1em}\textit{\}} \\
    \textit{\}} 
\end{mycase}

\begin{mycase}{Prompt Bench.R2.3: Evolution Direct Generation (R2 — Evolution)}
    \small
    \textbf{User:} \\
    【用户交互历史】： \\
    【2024-01-22】\texttt{[商品-购买] <|prod\_begin|><a\_5175><b\_1583><c\_3381>} \\
    【2025-12-01】\texttt{[广告-点击] <|ad\_begin|><a\_6526><b\_797><c\_4992>} \\
    【2026-01-04】\texttt{[搜索] 美甲构建和不构建什么区别} \\
    \dots{}（无主题提示——模型须自主发现所有兴趣线索）\dots{}  \\
    \textbf{Assistant:} \\
    \textit{[} \\
    \hspace*{1em}\textit{\{} \textit{"logic\_chain":} \textit{\{} \\
    \hspace*{2em}"name": \textit{"从美甲知识补课到工具/材料闭环采购的DIY落地链"}, \\
    \hspace*{2em}"events": \textit{[} \\
    \hspace*{3em}\textit{\{} "date": \textit{"2026-01-04"}, "action": \textit{[搜索] 美甲构建和不构建什么区别}, \\
    \hspace*{4em}"logic": \textit{问题触发：从工艺分叉点切入，属于概念级求解。} \textit{\}}, \\
    \hspace*{3em}\textit{\{} "date": \textit{"2026-01-15"}, "action": \textit{[视频-长播] <|video\_begin|><a\_7233><b\_649><c\_7927>}, \\
    \hspace*{4em}"logic": \textit{知识补全：转向零基础步骤补齐执行路径。} \textit{\}}, \\
    \hspace*{3em}\textit{\{} "date": \textit{"2026-01-16"}, "action": \textit{[商品-购买] <|prod\_begin|><a\_1356><b\_5951><c\_167>}, \\
    \hspace*{4em}"logic": \textit{落地收敛：从认知到行动的闭环。} \textit{\}} \\
    \hspace*{2em}\textit{]} \\
    \hspace*{1em}\textit{\}} \textit{\}}, \\
    \hspace*{1em}\textit{\{} \dots{} 其他演化链 \dots{} \textit{\}} \\
    \textit{]} 

    \vspace{0.5em}
    \hrulefill
    \vspace{0.5em}

    \textbf{English Example} \\
    \textbf{User:} \\
    \textit{[User Interaction History]:} \\
    {[2024-01-22]} \texttt{[Product-Purchase] <|prod\_begin|><a\_5175><b\_1583><c\_3381>} \\
    {[2025-12-01]} \texttt{[Ad-Click] <|ad\_begin|><a\_6526><b\_797><c\_4992>} \\
    {[2026-01-04]} \texttt{[Search] difference between built-up vs.\ non-built-up manicure} \\
    \dots{} (no topic provided---the model must autonomously discover all interest threads) \dots{}  \\
    \textbf{Assistant:} \\
    \textit{[} \\
    \hspace*{1em}\textit{\{} \textit{"logic\_chain":} \textit{\{} \\
    \hspace*{2em}"name": \textit{"From manicure-knowledge catch-up to closed-loop tool/material procurement, a DIY-execution chain"}, \\
    \hspace*{2em}"events": \textit{[} \\
    \hspace*{3em}\textit{\{} "date": \textit{"2026-01-04"}, "action": \textit{[Search] difference between built-up vs.\ non-built-up manicure}, \\
    \hspace*{4em}"logic": \textit{Problem trigger: entry from the technique-bifurcation point, a concept-level inquiry.} \textit{\}}, \\
    \hspace*{3em}\textit{\{} "date": \textit{"2026-01-15"}, "action": \textit{[Video-Long-View] <|video\_begin|><a\_7233><b\_649><c\_7927>}, \\
    \hspace*{4em}"logic": \textit{Knowledge complement: shifting to zero-basis step-by-step execution paths.} \textit{\}}, \\
    \hspace*{3em}\textit{\{} "date": \textit{"2026-01-16"}, "action": \textit{[Product-Purchase] <|prod\_begin|><a\_1356><b\_5951><c\_167>}, \\
    \hspace*{4em}"logic": \textit{Execution convergence: closing the loop from cognition to action.} \textit{\}} \\
    \hspace*{2em}\textit{]} \\
    \hspace*{1em}\textit{\}} \textit{\}}, \\
    \hspace*{1em}\textit{\{} \dots{} additional evolution chains \dots{} \textit{\}} \\
    \textit{]} 
\end{mycase}

\noindent\textit{Note: The following examples illustrate the video target domain. The same task format applies equally to product, ad, and live target domains.}

\begin{mycase}{Prompt Bench.R3.1: Single-Domain Recommendation (R3 — Recommendation)}
    \small
    \textbf{System:} 
    \textit{你是一个推荐系统助手，擅长根据用户属性与视频历史行为预测用户的视频偏好。}  \\
    \textbf{User:} \\
    \textbf{用户属性信息}：这是一个女性用户，她关注的博主类型分布为：【亲子】占18.2\%，【二次元】占9.54\%，【游戏】占9.54\%。用户点赞最多的类型是【美食、明星娱乐】，收藏最多的视频类型是【时尚、美妆】。近30日短剧消费情况：[逆袭\_女性成长]看了17.1分钟\dots{} \\

    \textbf{用户视频域的历史行为}：深度观看过的视频有 \texttt{<|video\_begin|><a\_4027><b\_6474><c\_6011>},\allowbreak{} \texttt{<|video\_begin|><a\_631><b\_5499><c\_8167>},\allowbreak{} \texttt{<|video\_begin|><a\_3433><b\_1769><c\_1441>}\dots{} \\

    请预测用户接下来可能感兴趣的视频。  \\
    \textbf{Assistant:} \\
    \texttt{<|video\_begin|><a\_4750><b\_2230><c\_4729>},\allowbreak{} \texttt{<|video\_begin|><a\_5844><b\_5861><c\_361>},\allowbreak{} \texttt{<|video\_begin|><a\_3475><b\_2493><c\_1218>},\allowbreak{} \texttt{<|video\_begin|><a\_7092><b\_5735><c\_3928>}\dots{} 

    \vspace{0.5em}
    \hrulefill
    \vspace{0.5em}

    \textbf{English Example} \\
    \textbf{System:} 
    \textit{You are a recommendation-system assistant, skilled at predicting the user's video preferences based on user attributes and historical video behavior.}  \\
    \textbf{User:} \\
    \textit{User Attributes:} The user is female, and her followed-blogger type distribution is: [Parenting] 18.2\%, [Anime] 9.54\%, [Gaming] 9.54\%. The user's most-liked categories are [Food, Celebrity Entertainment]; the most-collected video categories are [Fashion, Beauty]. Last-30-day short-drama consumption: [Counterattack\_Female-Growth] 17.1 minutes\dots{} \\

    \textit{User Video-Domain Behavioral History:} deeply-watched videos include \texttt{<|video\_begin|><a\_4027><b\_6474><c\_6011>},\allowbreak{} \texttt{<|video\_begin|><a\_631><b\_5499><c\_8167>},\allowbreak{} \texttt{<|video\_begin|><a\_3433><b\_1769><c\_1441>}\dots{} \\

    Please predict the videos the user is likely to be interested in next.  \\
    \textbf{Assistant:} \\
    \texttt{<|video\_begin|><a\_4750><b\_2230><c\_4729>},\allowbreak{} \texttt{<|video\_begin|><a\_5844><b\_5861><c\_361>},\allowbreak{} \texttt{<|video\_begin|><a\_3475><b\_2493><c\_1218>},\allowbreak{} \texttt{<|video\_begin|><a\_7092><b\_5735><c\_3928>}\dots{} 
\end{mycase}

\begin{mycase}{Prompt Bench.R3.2: Cross-Domain Recommendation (R3 — Recommendation)}
    \small
    \textbf{System:} 
    \textit{你是一个推荐系统助手，擅长根据用户属性与多域历史行为预测用户的视频偏好。}  \\
    \textbf{User:} \\
    \textbf{用户属性信息}：这是一个女性用户。点赞最多的类型是【美食、明星娱乐】\dots{} \\

    \textbf{用户多域历史行为}： \\
    \emph{用户在直播域}：关注了主播 \texttt{<|living\_begin|><a\_7490><b\_3><c\_2159>},\allowbreak{} \texttt{<|living\_begin|><a\_959><b\_263><c\_3562>}。 \\
    \emph{用户在电商域}：浏览过的商品有 \texttt{<|prod\_begin|><a\_4698><b\_6949><c\_7779>},\allowbreak{} \texttt{<|prod\_begin|><a\_7463><b\_2719><c\_4393>}\dots{} \\
    \emph{用户在广告域}：点击过的广告有 \texttt{<|ad\_begin|><a\_3120><b\_5580><c\_1890>}\dots{} \\

    请基于多域行为预测用户在视频域接下来可能感兴趣的内容。  \\
    \textbf{Assistant:} \\
    \texttt{<|video\_begin|><a\_4750><b\_2230><c\_4729>},\allowbreak{} \texttt{<|video\_begin|><a\_5844><b\_5861><c\_361>},\allowbreak{} \texttt{<|video\_begin|><a\_3475><b\_2493><c\_1218>}\dots{} 

    \vspace{0.5em}
    \hrulefill
    \vspace{0.5em}

    \textbf{English Example} \\
    \textbf{System:} 
    \textit{You are a recommendation-system assistant, skilled at predicting the user's video preferences based on user attributes and multi-domain historical behavior.}  \\
    \textbf{User:} \\
    \textit{User Attributes:} The user is female. The most-liked categories are [Food, Celebrity Entertainment]\dots{} \\

    \textbf{User Multi-Domain Behavioral History:} \\
    \emph{Live-streaming domain}: the user has followed streamers \texttt{<|living\_begin|><a\_7490><b\_3><c\_2159>},\allowbreak{} \texttt{<|living\_begin|><a\_959><b\_263><c\_3562>}. \\
    \emph{E-commerce domain}: the user has browsed products \texttt{<|prod\_begin|><a\_4698><b\_6949><c\_7779>},\allowbreak{} \texttt{<|prod\_begin|><a\_7463><b\_2719><c\_4393>}\dots{} \\
    \emph{Advertising domain}: the user has clicked ads \texttt{<|ad\_begin|><a\_3120><b\_5580><c\_1890>}\dots{} \\

    Based on the multi-domain behavior, predict the content the user is likely to be interested in next in the video domain.  \\
    \textbf{Assistant:} \\
    \texttt{<|video\_begin|><a\_4750><b\_2230><c\_4729>},\allowbreak{} \texttt{<|video\_begin|><a\_5844><b\_5861><c\_361>},\allowbreak{} \texttt{<|video\_begin|><a\_3475><b\_2493><c\_1218>}\dots{} 
\end{mycase}

\end{CJK*}

\renewcommand{\thesubsection}{\thesection.\arabic{subsection}}
\renewcommand{\thesubsubsection}{\thesubsection.\arabic{subsubsection}}

\renewcommand{\thesection}{C}
\renewcommand{\thesubsection}{\thesection.\arabic{subsection}}
\renewcommand{\thesubsubsection}{\thesubsection.\arabic{subsubsection}}
\section{Pre-Training Details}

\label{appendix:pretrain-embedding-comp}
\subsection{Positive--Negative Similarity Margin Study}
This appendix details the positive-negative similarity margin study summarized in Section ~\ref{sec:pre-training-data}. 
We examine whether the proposed pre-training corpora yield item representations that better separate semantically matched (positive) from unrelated (negative) captions.

\paragraph{Experimental Setup.}
We compare two checkpoints under a controlled protocol: the reproduced OpenOneRec serves as the \textit{baseline}, while \textit{ours} replaces only the pre-training corpora, holding the token budget, optimization schedule, and architecture fixed, thus isolating the effect of corpus composition. For each item, we probe two complementary representations: (i) the mean-pooled output of the \emph{first embedding layer}, reflecting lexical-level priors in the learned token embeddings, and (ii) the last output of the \emph{final hidden layer}, capturing the contextualized representation consumed by downstream heads. This lets us examine whether the effect of our corpora appears at the input-side, the task-relevant representation, or both. All similarities use cosine similarity in a shared representation space.

\paragraph{Metric Definition.}
Following common practice in metric learning, we summarize representational separation through the difference between matched and mismatched similarities rather than reporting them in isolation~\citep{lee2025on,rusak2024infonce}. We define the \emph{positive--negative similarity margin} for each anchor item $i$ as
\begin{equation}
    \mathrm{margin}(i) \;=\; \cos\bigl(e_i^{\text{item}},\, e_i^{\text{cap}}\bigr) \;-\; \cos\!\bigl(e_i^{\text{item}},\, e_j^{\text{cap}}\bigr), \qquad j \sim \mathrm{Uniform}\!\left(\mathcal{D}_d \setminus \{i\}\right),
    \label{eq:margin}
\end{equation}
where $e_i^{\text{item}}$ is the mean-pooled embedding of item $i$ at the probed layer, $e_i^{\text{cap}}$ is its ground-truth caption embedding, and $e_j^{\text{cap}}$ is a caption sampled uniformly from the same domain $\mathcal{D}_d$. The definition is applied independently to both probes, with the negative term taken in expectation over $j$, so the reported quantity is $\mathbb{E}_j[\mathrm{margin}(i)]$. The first term measures how tightly an item couples to its true caption, the second the residual similarity to an unrelated one. A larger $\mathrm{margin}(i)$ indicates the model pushes the matched caption closer while keeping unrelated captions sufficiently far.

\begin{table}[h]
\centering
\small
\caption{Mean cosine similarity of positive and negative item--caption pairs, and the resulting average margin $\Delta$, computed under the baseline (the reproduced OpenOneRec in Section~\ref{sec:pre-training-ablation}) and our model across four domains, at two probe depths: the first embedding layer and the final hidden layer. Larger $\Delta$ indicates stronger representational discriminability.}
\label{tab:embedding-gap}
\begin{tabular}{l l ccc ccc}
\toprule
& & \multicolumn{3}{c}{\textbf{Baseline}} & \multicolumn{3}{c}{\textbf{Ours}} \\
\cmidrule(lr){3-5} \cmidrule(lr){6-8}
\textbf{Probe} & \textbf{Domain} & Pos & Neg & $\Delta$ & Pos & Neg & $\Delta$ \\
\midrule
\multirow{4}{*}{First layer}
  & Video    & \texttt{+0.043} & \texttt{+0.001} & \texttt{+0.042} & \texttt{+0.072} & \texttt{+0.001} & \texttt{+0.071} \\
  & Live     & \texttt{+0.021} & \texttt{0.000} & \texttt{+0.021} & \texttt{+0.043} & \texttt{-0.002} & \texttt{+0.045} \\
  & Product  & \texttt{+0.059} & \texttt{0.000} & \texttt{+0.059} & \texttt{+0.089} & \texttt{-0.001} & \texttt{+0.090} \\
  & Ad       & \texttt{+0.034} & \texttt{0.000} & \texttt{+0.034} & \texttt{+0.052} & \texttt{-0.001} & \texttt{+0.053} \\
\midrule
\multirow{4}{*}{Final layer}
  & Video    & \texttt{+0.012} & \texttt{0.000} & \texttt{+0.012} & \texttt{+0.009} & \texttt{-0.002} & \texttt{+0.011} \\
  & Live     & \texttt{+0.002} & \texttt{-0.002} & \texttt{+0.004} & \texttt{+0.007} & \texttt{-0.006} & \texttt{+0.013} \\
  & Product  & \texttt{+0.016} & \texttt{0.000} & \texttt{+0.016} & \texttt{+0.039} & \texttt{+0.001} & \texttt{+0.038} \\
  & Ad       & \texttt{+0.004} & \texttt{-0.001} & \texttt{+0.005} & \texttt{+0.017} & \texttt{0.000} & \texttt{+0.017} \\
\bottomrule
\end{tabular}
\end{table}
 
\paragraph{Quantitative Summary.}
Table~\ref{tab:embedding-gap} reports, for each domain and both probed layers, the mean positive and negative cosine similarities and the resulting margin $\Delta = \mathbb{E}_i[\mathrm{margin}(i)]$ under the baseline and our model. At the \emph{first-layer} probe, our model widens the margin across all four domains (roughly $1.5\text{--}2\times$), showing that the proposed corpora inject stronger item--caption alignment at the input side. At the \emph{final-layer} probe, the margins shrink for both models—as deeper representations become more contextualized—yet our model still improves clearly on Live, Product, and Ad ($\sim\!2.4\text{--}3.4\times$). The only exception is Video ($0.012$ vs.\ $0.011$), where the baseline is already well aligned at depth; this is consistent with the near-identical Video results in the Cross-Domain Reco block of Table~\ref{tab:ablation_results}.


\subsection{Data Mixture for Pre-Training}
\label{appendix:pretrain-data}
Section~\ref{sec:pre-training-data} described the construction of our pre-training corpora. In this section, we summarize the overall data mixture used throughout pre-training, where Table~\ref{tab:pt-datamix-general} and Table~\ref{tab:pt-datamix-rec} report the token ratio of each dataset. Echoing the progressive multi-granularity modality-alignment principle of Section~\ref{sec:pre-training-data}, the mixture comprises two complementary parts: \textbf{recommendation corpora}, which carry the dominant modality-alignment signal, and \textbf{general-domain corpora}, which preserve the model's general competencies.

\textbf{Recommendation Corpora.} This part comprises the four-granularity corpora---\textit{token}, \textit{item}, \textit{relational}, and \textit{user} granularity---whose construction is detailed in Section~\ref{sec:pre-training-data}. Their datasets and token ratios are listed in Table~\ref{tab:pt-datamix-rec}.

\textbf{General-Domain Corpora.} As described in Section~\ref{sec:pre-training-data}, this part consists of two subsets: \textit{general-domain text data}, spanning mathematical reasoning, code, general reasoning, and medical corpora together with in-house text data, and \textit{multimodal data}, which further comprises image generation and image editing data, in-house image captions, and in-house interleaved image-text corpus. All the publicly available datasets are downloadable from the HuggingFace repository (see footnotes). The corresponding data mixtures are listed in Table~\ref{tab:pt-datamix-general}.
\setcounter{footnote}{0}
\begin{table}[H]
  \centering
  \footnotesize
  \setlength{\tabcolsep}{2pt}
  \renewcommand{\arraystretch}{1.0}
  \caption{Data mixtures for pre-training (Recommendation Corpora).}
  \label{tab:pt-datamix-rec}
  \begin{threeparttable}
  \setlength{\tabcolsep}{2pt}
  \renewcommand{\arraystretch}{1.2}
    \begin{tabular}{llll}
      \toprule
      \textbf{Category} &  \textbf{Modality} & \textbf{Dataset} & \textbf{Token Ratio (\%)}  \\
      \midrule
      \multirow{5}{*}{\makecell[c]{Token\\Granularity}}  & \multirow{5}{*}{\makecell[c]{Itemic\\+Text}}  & Single\--Token\--Semantic\--Prediction & 0.35 \\
       &  & Compositional Prefix Semantic Prediction & 1.29 \\
       & & Prefix Itemic Token Grounding & 0.55 \\
       &  & Part\--to\--Whole Semantic Prediction & 0.31 \\\cline{3-4}
       &  & Subtotal &  2.50 \\\cline{1-4}
      \multirow{4}{*}{\makecell[c]{Item\\Granularity}} & \multirow{4}{*}{\makecell[c]{Itemic\\+Text}} & Item Captioning & 17.28 \\
       &  & Item Grounding & 12.12 \\
       &  & Item QA & 2.44 \\\cline{3-4}
       &  & Subtotal & 31.84 \\\cline{1-4}
       \multirow{3}{*}{\makecell[c]{Relational\\Granularity} }  & \multirow{3}{*}{\makecell[c]{Itemic\\+Text}} & Direct\--Item\--to\--Item Associations & 0.24 \\
       &   & Multi\--Item Interest Flows & 2.94 \\\cline{3-4}
        &   & Subtotal & 3.18 \\\cline{1-4}
       \multirow{3}{*}{\makecell[c]{User\\Granularity}} & \multirow{3}{*}{\makecell[c]{Itemic\\+Text}} &  Domain\--Grouped Behavior Sequence & 21.41 \\
        &  & Chronologically\--Interleaved Behavior Sequence & 12.64 \\\cline{3-4}
        & & Subtotal & 34.05 \\
      \bottomrule
    \end{tabular}
  \end{threeparttable}
\end{table}

\begin{table}[H]
  \centering
  \footnotesize
  \setlength{\tabcolsep}{2pt}
  \renewcommand{\arraystretch}{1.0}
  \caption{Data mixtures for pre-training (General-Domain Corpora).}
  \label{tab:pt-datamix-general}
  \begin{threeparttable}
  \setlength{\tabcolsep}{2pt}
  \renewcommand{\arraystretch}{1.2}
    \begin{tabular}{llll}
      \toprule
      \textbf{Category} &  \textbf{Modality} & \textbf{Dataset} & \textbf{Token Ratio (\%)}  \\
      \midrule
      \multirow{27}{*}{\makecell[c]{General\\Domain}} & \multirow{23}{*}{Text} & In\--House\--Text\--Data & 8.91 \\
      & &  Nemotron\--CC\--Math\--v1\footnotemark & 4.25 \\
      & & reasoning\--v1\--20m\footnotemark & 3.89 \\
       &   & Nemotron\--CC\--v2\footnotemark & 3.26 \\
       &   & Nemotron\--Pretraining\--Code\--v1\footnotemark & 2.35 \\
       &   & OpenMathReasoning\footnotemark & 1.04 \\
       &   & OpenCodeReasoning\footnotemark  & 0.39 \\
       &   & KodCode\--V1\--SFT\--R1\footnotemark & 0.15 \\
       &   & Chinese-Reasoning\--Distil\--Data\footnotemark & 0.10 \\
       &   & Bespoke\--Stratos\--17k\footnotemark & 0.04 \\
        &  & medical\--o1\--reasoning\--SFT\footnotemark & 0.02 \\
       & & MegaMath\footnotemark & 0.78 \\
       & & OpenMathInstruct\--2, NuminaMath\--CoT \& MMIQC \footnotemark\footnotemark\footnotemark & 0.49 \\
       & & Infinity\--Instruct\footnotemark & 0.44 \\
       & & FineVision\footnotemark & 0.18 \\
       & & GeneralThought\--430K\footnotemark & 0.07 \\
       & & Multi\--subject\--RLVR\footnotemark & 0.04 \\
       & & Big\--Math\--RL\--Verified\footnotemark & 0.08 \\
       & & rStar\--Coder\footnotemark & 0.05 \\
       & & CuratedThoughts\footnotemark & 0.26 \\
       & & DeepMath\--103K\footnotemark & 0.05 \\
       & & DAPO\--Math\--17k\footnotemark & 0.01 \\
        \cline{3-4}
       & & Subtotal & 26.85 \\\cline{2-4}
       & \multirow{3}{*}{\makecell[c]{Itemic\\+Text}} & Image Generation and Image Edit \footnotemark\footnotemark\footnotemark\footnotemark & 0.29 \\
       & & Image Captions(In\--house and open-source\footnotemark) & 1.10 \\
       & & In\--house Interleave Corpora & 0.20 \\\cline{3-4}
       & & Subtotal & 1.59 \\
      \bottomrule
    \end{tabular}
  \end{threeparttable}
\end{table}

\addtocounter{footnote}{-27}
\footnotetext{\url{https://huggingface.co/datasets/nvidia/Nemotron-CC-Math-v1}}
\stepcounter{footnote}\footnotetext{\url{https://huggingface.co/datasets/glaiveai/reasoning-v1-20m}}
\stepcounter{footnote}\footnotetext{\url{https://huggingface.co/datasets/nvidia/Nemotron-CC-v2}}
\stepcounter{footnote}\footnotetext{\url{https://huggingface.co/datasets/nvidia/Nemotron-Pretraining-Code-v1}}
\stepcounter{footnote}\footnotetext{\url{https://huggingface.co/datasets/nvidia/OpenMathReasoning}}
\stepcounter{footnote}\footnotetext{\url{https://huggingface.co/datasets/nvidia/OpenCodeReasoning}}
\stepcounter{footnote}\footnotetext{\url{https://huggingface.co/datasets/KodCode/KodCode-V1-SFT-R1}}
\stepcounter{footnote}\footnotetext{\url{https://huggingface.co/datasets/Mxode/Chinese-Reasoning-Distil-Data}}
\stepcounter{footnote}\footnotetext{\url{https://huggingface.co/datasets/bespokelabs/Bespoke-Stratos-17k}}
\stepcounter{footnote}\footnotetext{\url{https://huggingface.co/datasets/FreedomIntelligence/medical-o1-reasoning-SFT}}
\stepcounter{footnote}\footnotetext{\url{https://huggingface.co/datasets/LLM360/MegaMath}}
\stepcounter{footnote}\footnotetext{\url{https://huggingface.co/datasets/nvidia/OpenMathInstruct-2}}
\stepcounter{footnote}\footnotetext{\url{https://huggingface.co/datasets/AI-MO/NuminaMath-CoT}}
\stepcounter{footnote}\footnotetext{\url{https://huggingface.co/datasets/Vivacem/MMIQC}}
\stepcounter{footnote}\footnotetext{\url{https://huggingface.co/datasets/BAAI/Infinity-Instruct}}
\stepcounter{footnote}\footnotetext{\url{https://huggingface.co/datasets/HuggingFaceM4/FineVision}}
\stepcounter{footnote}\footnotetext{\url{https://huggingface.co/datasets/GeneralReasoning/GeneralThought-430K}}
\stepcounter{footnote}\footnotetext{\url{https://huggingface.co/datasets/virtuoussy/Multi-subject-RLVR}}
\stepcounter{footnote}\footnotetext{\url{https://huggingface.co/datasets/SynthLabsAI/Big-Math-RL-Verified}}
\stepcounter{footnote}\footnotetext{\url{https://huggingface.co/datasets/microsoft/rStar-Coder}}
\stepcounter{footnote}\footnotetext{\url{https://huggingface.co/datasets/bethgelab/CuratedThoughts}}
\stepcounter{footnote}\footnotetext{\url{https://huggingface.co/datasets/zwhe99/DeepMath-103K}}
\stepcounter{footnote}\footnotetext{\url{https://huggingface.co/datasets/BytedTsinghua-SIA/DAPO-Math-17k}}
\stepcounter{footnote}\footnotetext{\url{https://huggingface.co/datasets/bytedance-research/UNO-1M}}
\stepcounter{footnote}\footnotetext{\url{https://huggingface.co/datasets/BLIP3o/BLIP3o-60k}}
\stepcounter{footnote}\footnotetext{\url{https://huggingface.co/datasets/Bin1117/AnyEdit}}
\stepcounter{footnote}\footnotetext{\url{https://huggingface.co/datasets/ProGamerGov/synthetic-dataset-1m-dalle3-high-quality-captions}}
\stepcounter{footnote}\footnotetext{\url{https://huggingface.co/datasets/BAAI/Infinity-MM}}

\subsection{Four-Granularity Pre-Training Samples}
The following are representative training samples from each of the four granularities, illustrating the data formats described in Section~\ref{sec:pre-training-data}.

\begin{CJK*}{UTF8}{gbsn}
\begin{mycase}{Token Granularity: Compositional Prefix Semantic Prediction}
\small
\textbf{中文示例} \\
\textbf{User:} 短视频域中，\texttt{<a\_3664>}\allowbreak\texttt{<b\_3076>} 表示什么？\\
\textbf{Assistant:}\\
这些短视频在内容主题上高度聚焦于“都市夜景中的个体孤独与情感内省”。画面通常选取夜晚的城市街道、霓虹灯下或夜市背景，核心视觉元素为背对镜头的单人（多为背影），通过人物与繁华或静谧环境的对比，营造一种疏离、沉思或略带忧郁的氛围。视频往往借助文字叠加来外化人物内心的挣扎、对过去的追忆、对情感的执着或对生活的感悟，将物理空间的“夜”转化为心理空间的“独白”。\\
从受众角度来看，这类内容主要吸引那些在快节奏都市生活中感到孤独、压力或情感共鸣的观众。目标人群倾向于通过观看此类视频寻找情感慰藉、确认自身情绪，或在静谧的视觉与文字中寻求片刻的内心平静与精神寄托。观众画像多为年轻至中年群体，对情感类、治愈系或略带伤感氛围的都市生活记录有较高偏好。

\vspace{0.5em}
\textbf{English Translation} \\
\textbf{User:} In the short-video domain, what does \texttt{<a\_3664>}\allowbreak\texttt{<b\_3076>} represent?\\
\textbf{Assistant:}\\
These short videos are highly focused on the theme of ``individual loneliness and emotional introspection in urban night scenes.'' The frames typically feature nighttime city streets, under neon lights, or night market backgrounds, with the core visual element being a single person (mostly a back view) facing away from the camera. The contrast between the character and the bustling or tranquil environment creates a sense of alienation, contemplation, or a slightly melancholic atmosphere. The videos often externalize the character's inner struggles, reminiscence of the past, persistence in emotions, or reflections on life through overlaid text, transforming the physical space of ``night'' into a psychological space of ``inner monologue.''\\
From an audience perspective, such content mainly attracts viewers who feel loneliness, stress, or emotional resonance in fast-paced urban life. The target group tends to seek emotional solace, validate their own feelings through watching such videos, or find a moment of inner peace and spiritual sustenance in the quiet visuals and text. The viewer profile is largely young to middle-aged groups with a high preference for emotional, healing, or slightly sentimental urban life records.
\end{mycase}
\end{CJK*}

\begin{CJK*}{UTF8}{gbsn}
\begin{mycase}{Item Granularity: Caption-Aware Caption Coarse-Graining)}
\small
\textbf{中文示例} \\
\textbf{User:} 请描述一下这个电商商品 \texttt{<|prod\_begin|>}\allowbreak\texttt{<a\_2147>}\allowbreak\texttt{<b\_7978>}\allowbreak\texttt{<c\_5031>}。\\
\textbf{Assistant:}\\
该商品是一款户外软壳冲锋衣，一级类目是运动户外鞋服，二级类目是户外服装，三级类目是软壳衣裤。采用连帽拉链设计，软壳材质兼具防风防水和加绒保暖功能，标准版型，加绒内里，多色可选，简约时尚，适合户外活动及日常休闲。价格区间为100-120元。无品牌。

\vspace{0.5em}
\textbf{English Translation} \\
\textbf{User:} Please describe this e-commerce product \texttt{<|prod\_begin|>}\allowbreak\texttt{<a\_2147>}\allowbreak\texttt{<b\_7978>}\allowbreak\texttt{<c\_5031>}.\\
\textbf{Assistant:}\\
This product is an outdoor softshell jacket. Its primary category is Sports \& Outdoor Footwear and Apparel, secondary category is Outdoor Clothing, and tertiary category is Softshell Jackets \& Pants. It features a hooded zipper design, with the softshell material providing windproof, waterproof, and fleece-lined thermal functions. It has a standard fit, fleece-lined interior, multiple color options, and a minimalist, stylish look, suitable for outdoor activities and daily casual wear. The price range is 100--120 yuan. Unbranded.
\end{mycase}
\end{CJK*}

\begin{CJK*}{UTF8}{gbsn}
\begin{mycase}{Relational Granularity: Multi-Item Interest Flow}
\small
\textbf{中文示例} \\
\texttt{<|living\_begin|>}\allowbreak\texttt{<a\_4515>}\allowbreak\texttt{<b\_6234>}\allowbreak\texttt{<c\_6278>}
$\rightarrow$
关注长发美女和马甲线身材的用户，往往对形体管理与健康生活方式感兴趣。直播中展现的性感形体和生活化互动，容易引发对专业健身指导的需求，而内容中刘教练的腿部训练视频，正好满足了他们对科学塑形和健康提升的追求。
$\rightarrow$
\texttt{<|video\_begin|>}\allowbreak\texttt{<a\_458>}\allowbreak\texttt{<b\_3589>}\allowbreak\texttt{<c\_5563>}
$\rightarrow$
看过腿部锻炼指导视频的人，往往对提升肌肉力量和锻炼方式感兴趣，而普拉提圈作为辅助训练工具，正好能满足他们对多样化训练器材的需求，因此容易被吸引。
$\rightarrow$
\texttt{<|prod\_begin|>}\allowbreak\texttt{<a\_1162>}\allowbreak\texttt{<b\_1966>}\allowbreak\texttt{<c\_7428>}
$\rightarrow$
关注普拉提圈训练的用户，往往也在寻找适合中老年人的肌肉锻炼工具。他们在观看腿部锻炼教学视频时，接触到类似瑜伽圈、抗阻圈的产品，进而对专为大腿设计的抗阻训练器产生兴趣。
$\rightarrow$
\texttt{<|prod\_begin|>}\allowbreak\texttt{<a\_7433>}\allowbreak\texttt{<b\_6972>}\allowbreak\texttt{<c\_1162>}

\vspace{0.5em}
\textbf{English Translation} \\
\texttt{<|living\_begin|>}\allowbreak\texttt{<a\_4515>}\allowbreak\texttt{<b\_6234>}\allowbreak\texttt{<c\_6278>}
$\rightarrow$
Users following long-haired beauties and toned abs often care about body shaping and healthy lifestyles. The attractive physique and casual interactions in the livestream can trigger a need for professional fitness guidance, and the leg training video by Coach Liu exactly fulfills their pursuit of scientific body sculpting and health improvement.
$\rightarrow$
\texttt{<|video\_begin|>}\allowbreak\texttt{<a\_458>}\allowbreak\texttt{<b\_3589>}\allowbreak\texttt{<c\_5563>}
$\rightarrow$
Those who watched leg exercise tutorials are often interested in improving muscle strength and workout variety, and the Pilates ring, as an auxiliary training tool, precisely meets their demand for diverse training equipment, thus easily capturing their attention.
$\rightarrow$
\texttt{<|prod\_begin|>}\allowbreak\texttt{<a\_1162>}\allowbreak\texttt{<b\_1966>}\allowbreak\texttt{<c\_7428>}
$\rightarrow$
Users interested in Pilates ring exercises are also often looking for muscle-building tools suitable for middle-aged and elderly people. When watching leg training videos, they encounter products similar to yoga rings and resistance bands, and further develop interest in resistance trainers designed specifically for thighs.
$\rightarrow$
\texttt{<|prod\_begin|>}\allowbreak\texttt{<a\_7433>}\allowbreak\texttt{<b\_6972>}\allowbreak\texttt{<c\_1162>}
\end{mycase}
\end{CJK*}

\begin{CJK*}{UTF8}{gbsn}
\begin{mycase}{User Granularity: Domain-Grouped Behavior Sequences}
\small
\textbf{中文示例} \\
\textbf{User:} 已知用户信息如下：用户专注于6个时尚领域的内容创作。用户安装了小红书-社区、支付宝-支付...该用户可能购买了什么商品？输出 JSON，使用字段名 "purchase"。\\
\textbf{Assistant:} \texttt{\{"purchase": ["<|prod\_begin|>}\allowbreak\texttt{<a\_7552>}\allowbreak\texttt{<b\_5608>}\allowbreak\texttt{<c\_912>", "<|prod\_begin|>}\allowbreak\texttt{<a\_7651>}\allowbreak\texttt{<b\_4552>}\allowbreak\texttt{<c\_4508>", "<|prod\_begin|>}\allowbreak\texttt{<a\_7618>}\allowbreak\texttt{<b\_4552>}\allowbreak\texttt{<c\_4508>"]\}}\\
\textbf{User:} \texttt{<|ad\_begin|>}\allowbreak\texttt{<a\_7112>}\allowbreak\texttt{<b\_7554>}\allowbreak\texttt{<c\_6156>}, \texttt{<|ad\_begin|>}\allowbreak\texttt{<a\_4597>}\allowbreak\texttt{<b\_4106>}\allowbreak\texttt{<c\_3984>}, \texttt{<|ad\_begin|>}\allowbreak\texttt{<a\_1169>}\allowbreak\texttt{<b\_3795>}\allowbreak\texttt{<c\_3984>} 是该用户的部分记录，请继续给出其他点击广告。\\
\textbf{Assistant:} \texttt{['<|ad\_begin|>}\allowbreak\texttt{<a\_7586>}\allowbreak\texttt{<b\_7762>}\allowbreak\texttt{<c\_7321>', '<|ad\_begin|>}\allowbreak\texttt{<a\_1512>}\allowbreak\texttt{<b\_2007>}\allowbreak\texttt{<c\_5218>']}\\
\textbf{User:} 以下哪些商品是该用户浏览过的？（用分号分隔答案）\\
  A. \texttt{<|prod\_begin|>}\allowbreak\texttt{<a\_7278>}\allowbreak\texttt{<b\_487>}\allowbreak\texttt{<c\_3329>}\\
  B. \texttt{<|prod\_begin|>}\allowbreak\texttt{<a\_7698>}\allowbreak\texttt{<b\_4069>}\allowbreak\texttt{<c\_6585>}\\
  C. \texttt{<|prod\_begin|>}\allowbreak\texttt{<a\_5891>}\allowbreak\texttt{<b\_4945>}\allowbreak\texttt{<c\_335>}\\
\textbf{Assistant:} A; C\\
\textbf{User:} 根据广告和购买记录，该用户最近可能感兴趣的视频有哪些？请返回一个 list。\\
\textbf{Assistant:} \texttt{['<|video\_begin|>}\allowbreak\texttt{<a\_5991>}\allowbreak\texttt{<b\_2799>}\allowbreak\texttt{<c\_6278>', '<|video\_begin|>}\allowbreak\texttt{<a\_4751>}\allowbreak\texttt{<b\_765>}\allowbreak\texttt{<c\_2584>']}

\vspace{0.5em}
\textbf{English Translation} \\
\textbf{User:} Given the following user info: The user focuses on creating content in 6 fashion-related domains. The user has installed Xiaohongshu-Community, Alipay-Payment... What products might this user have purchased? Output JSON with the field name ``purchase''.\\
\textbf{Assistant:} \texttt{\{"purchase": ["<|prod\_begin|>}\allowbreak\texttt{<a\_7552>}\allowbreak\texttt{<b\_5608>}\allowbreak\texttt{<c\_912>", "<|prod\_begin|>}\allowbreak\texttt{<a\_7651>}\allowbreak\texttt{<b\_4552>}\allowbreak\texttt{<c\_4508>", "<|prod\_begin|>}\allowbreak\texttt{<a\_7618>}\allowbreak\texttt{<b\_4552>}\allowbreak\texttt{<c\_4508>"]\}}\\
\textbf{User:} \texttt{<|ad\_begin|>}\allowbreak\texttt{<a\_7112>}\allowbreak\texttt{<b\_7554>}\allowbreak\texttt{<c\_6156>}, \texttt{<|ad\_begin|>}\allowbreak\texttt{<a\_4597>}\allowbreak\texttt{<b\_4106>}\allowbreak\texttt{<c\_3984>}, \texttt{<|ad\_begin|>}\allowbreak\texttt{<a\_1169>}\allowbreak\texttt{<b\_3795>}\allowbreak\texttt{<c\_3984>} are part of this user's records. Please continue and provide other clicked ads.\\
\textbf{Assistant:} \texttt{['<|ad\_begin|>}\allowbreak\texttt{<a\_7586>}\allowbreak\texttt{<b\_7762>}\allowbreak\texttt{<c\_7321>', '<|ad\_begin|>}\allowbreak\texttt{<a\_1512>}\allowbreak\texttt{<b\_2007>}\allowbreak\texttt{<c\_5218>']}\\
\textbf{User:} Which of the following products has this user browsed? (Separate answers with semicolons)\\
  A. \texttt{<|prod\_begin|>}\allowbreak\texttt{<a\_7278>}\allowbreak\texttt{<b\_487>}\allowbreak\texttt{<c\_3329>}\\
  B. \texttt{<|prod\_begin|>}\allowbreak\texttt{<a\_7698>}\allowbreak\texttt{<b\_4069>}\allowbreak\texttt{<c\_6585>}\\
  C. \texttt{<|prod\_begin|>}\allowbreak\texttt{<a\_5891>}\allowbreak\texttt{<b\_4945>}\allowbreak\texttt{<c\_335>}\\
\textbf{Assistant:} A; C\\
\textbf{User:} Based on the ad and purchase records, what videos might this user be recently interested in? Return a list.\\
\textbf{Assistant:} \texttt{['<|video\_begin|>}\allowbreak\texttt{<a\_5991>}\allowbreak\texttt{<b\_2799>}\allowbreak\texttt{<c\_6278>', '<|video\_begin|>}\allowbreak\texttt{<a\_4751>}\allowbreak\texttt{<b\_765>}\allowbreak\texttt{<c\_2584>']}
\end{mycase}
\end{CJK*}

\renewcommand{\thesection}{D}
\renewcommand{\thesubsection}{\thesection.\arabic{subsection}}
\renewcommand{\thesubsubsection}{\thesubsection.\arabic{subsubsection}}

\section{SFT Details}
\subsection{SFT Data and Itemic-Token Perception}
\subsubsection{R0 Item-Perception Examples}
\label{appendix:sft-r0-item-perception}

R0 trains the base model to map between itemic tokens and natural-language captions, answer content-grounded questions, and recover itemic tokens from descriptions. The examples below are sanitized bilingual cases that preserve the observed task formats while replacing raw identifiers with synthetic three-token examples. They cover short-video, product, live-streaming, and advertising domains, and include both CoT and unCoT formats.

\makeatletter
\@ifundefined{mycase}{%
  \newtcolorbox{mycase}[1]{%
    breakable,
    colback=gray!5, colframe=black!60, boxrule=0.5pt, arc=2pt,
    left=6pt, right=6pt, top=6pt, bottom=6pt,
    title={#1},
    fonttitle=\bfseries\small, coltitle=white, colbacktitle=black!70,
    before skip=8pt, after skip=8pt,
    before upper={\raggedright\sloppy}%
  }%
}{}
\makeatother

\begin{CJK*}{UTF8}{gbsn}
\begin{mycase}{Prompt SFT.R0.1: CoT Itemic-Token-to-Caption Generation}
\small
\textbf{中文示例} \\
\textbf{System:} 你是一名专业的视频内容理解与描述助手，请根据短视频 itemic tokens 生成准确的视频描述。\\
\textbf{User:} 请描述短视频 \texttt{<|video\_begin|>}\allowbreak\texttt{<a\_2344>}\allowbreak\texttt{<b\_3729>}\allowbreak\texttt{<c\_6164>} 的视频内容。\\
\textbf{Assistant:}\\
\texttt{\detokenize{<think>}}从 \texttt{<a\_2344>} 看，该视频属于线下休闲娱乐和商场体验类内容；\texttt{<b\_3729>} 进一步指向新奇互动玩法，而不是普通购物或餐饮记录；\texttt{<c\_6164>} 支持“悬挂式抓零食”这一具体场景，核心是参与者被安全绳吊起，在零食池中抓取零食并制造轻松搞笑效果。三层 itemic tokens 的语义交集指向一种面向年轻消费者的商场趣味体验活动。\texttt{\detokenize{</think>}}\\
视频记录了一名女性在商场内体验悬挂式抓零食游戏的过程。参与者被安全绳吊在装满零食的区域中，通过身体动作抓取零食，旁边有工作人员协助。画面字幕强调单次体验价格和“抓零食”的趣味性，整体呈现为轻松、搞笑、适合线下娱乐分享的短视频。

\vspace{0.5em}
\textbf{English Example} \\
\textbf{System:} You are a professional video-understanding assistant. Generate an accurate caption from short-video itemic tokens.\\
\textbf{User:} Describe the content represented by \texttt{<|video\_begin|>}\allowbreak\texttt{<a\_2344>}\allowbreak\texttt{<b\_3729>}\allowbreak\texttt{<c\_6164>}.\\
\textbf{Assistant:}\\
\texttt{\detokenize{<think>}}The coarse token \texttt{<a\_2344>} points to offline entertainment and mall experiences; \texttt{<b\_3729>} narrows it to a playful interactive activity rather than ordinary shopping or dining; and \texttt{<c\_6164>} supports the specific scene of a suspended snack-grabbing game. Together, the itemic tokens describe a novelty mall activity for young consumers.\texttt{\detokenize{</think>}}\\
The video records a woman trying a suspended snack-grabbing game inside a shopping mall. The participant is lifted by a safety harness above a snack-filled area and tries to grab snacks with body movements while staff assist nearby. On-screen captions emphasize the price and the fun of the game, making the clip a light, humorous record of an offline entertainment experience.
\end{mycase}
\end{CJK*}

\begin{CJK*}{UTF8}{gbsn}
\begin{mycase}{Prompt SFT.R0.2: unCoT Itemic-Token-to-Caption Generation}
\small
\textbf{中文示例} \\
\textbf{System:} 你是一位直播理解专家，擅长将主播 itemic tokens 转换为自然的直播描述。\\
\textbf{User:} 已知主播 \texttt{<|living\_begin|>}\allowbreak\texttt{<a\_522>}\allowbreak\texttt{<b\_6604>}\allowbreak\texttt{<c\_512>}，请说明这位主播通常会直播什么内容。\\
\textbf{Assistant:}\\
该主播定位为音乐轮播类无人直播博主，以虚拟直播间为载体，主打流行歌曲、古风歌曲和轻量 DJ 音效的持续播放。直播中通常没有真人强互动，更多承担背景音和陪伴功能，适合观众在休息、睡前或碎片化时间收听。整体风格简约、节奏感强，主要吸引喜欢听歌放松、需要背景音陪伴的年轻用户。

\vspace{0.5em}
\textbf{English Example} \\
\textbf{System:} You are a live-streaming understanding expert. Convert streamer itemic tokens into a natural live-room description.\\
\textbf{User:} Given the streamer itemic tokens \texttt{<|living\_begin|>}\allowbreak\texttt{<a\_522>}\allowbreak\texttt{<b\_6604>}\allowbreak\texttt{<c\_512>}, describe what this streamer usually broadcasts.\\
\textbf{Assistant:}\\
This streamer is best described as an automated music-broadcast host. The live room mainly plays pop songs, classical-style tracks, and light DJ effects through a virtual setup. It has little direct human interaction and works more as background audio or emotional companionship, suitable for rest time, bedtime, or casual listening. The style is simple and rhythmic, attracting young users who want music for relaxation or ambient company.
\end{mycase}
\end{CJK*}

\begin{CJK*}{UTF8}{gbsn}
\begin{mycase}{Prompt SFT.R0.3: Product QA}
\small
\textbf{中文示例} \\
\textbf{System:} 你是一名商品理解助手，能够根据商品 itemic tokens 和描述回答问题。\\
\textbf{User:} 商品 \texttt{<|prod\_begin|>}\allowbreak\texttt{<a\_7516>}\allowbreak\texttt{<b\_5044>}\allowbreak\texttt{<c\_7205>} 是一款 700W 电动抛光机，采用全铜电机和无级变速设计，可用于汽车打蜡、封釉、划痕修复，也可用于皮具保养和木材打磨。这个商品最适合下面哪类使用场景？\\
A. 室内空气净化 \quad B. 汽车美容与表面抛光 \quad C. 儿童绘本阅读 \quad D. 生鲜冷链配送\\
请只回答 A/B/C/D。\\
\textbf{Assistant:} B

\vspace{0.5em}
\textbf{English Example} \\
\textbf{System:} You are a product-understanding assistant. Answer questions based on product itemic tokens and descriptions.\\
\textbf{User:} Product \texttt{<|prod\_begin|>}\allowbreak\texttt{<a\_7516>}\allowbreak\texttt{<b\_5044>}\allowbreak\texttt{<c\_7205>} is a 700W electric polisher with a copper motor and stepless speed control. It can be used for car waxing, glazing, scratch repair, leather care, and wood polishing. Which usage scenario best matches this product?\\
A. Indoor air purification \quad B. Car detailing and surface polishing \quad C. Children's picture-book reading \quad D. Fresh-food cold-chain delivery\\
Answer only A/B/C/D.\\
\textbf{Assistant:} B
\end{mycase}
\end{CJK*}

\begin{CJK*}{UTF8}{gbsn}
\begin{mycase}{Prompt SFT.R0.4: Caption-to-Itemic-Token Grounding}
\small
\textbf{中文示例} \\
\textbf{System:} 请根据输入的广告描述，输出能与其语义最匹配的广告 itemic tokens。\\
\textbf{User:} 请根据以下广告内容描述生成匹配的广告 itemic tokens：视频为生活技巧类广告，通过真人演示卫裤绳子的多种系法，展示打结、蝴蝶结、松紧节等做法，重点解决裤绳容易缩回裤腰的问题，并用限时补贴弹窗引导用户点击领取福利。\\
\textbf{Assistant:}\\
\texttt{<|ad\_begin|>}\allowbreak\texttt{<a\_2350>}\allowbreak\texttt{<b\_6885>}\allowbreak\texttt{<c\_2112>}

\vspace{0.5em}
\textbf{English Example} \\
\textbf{System:} Given an advertising description, output the semantically matching advertising itemic tokens.\\
\textbf{User:} Generate the matching advertising itemic tokens for this ad description: the video is a life-hack advertisement that demonstrates several ways to tie sweatpants drawstrings, including knots, bows, and adjustable loops. It focuses on preventing the drawstring from slipping back into the waistband and uses a limited-time subsidy pop-up to guide users to claim a benefit.\\
\textbf{Assistant:}\\
\texttt{<|ad\_begin|>}\allowbreak\texttt{<a\_2350>}\allowbreak\texttt{<b\_6885>}\allowbreak\texttt{<c\_2112>}
\end{mycase}
\end{CJK*}

\subsubsection{SFT Data Composition}
\label{appendix:sft-data-composition}
\begin{table}[H]
  \centering
  \caption{Data mixture for the SFT stage. Sample counts in units of $10^{4}$; ratio in \%.}
  \begin{threeparttable}
  \setlength{\tabcolsep}{10pt}
  \renewcommand{\arraystretch}{1.2}
    \begin{tabular}{llll}
      \toprule
      \textbf{Category} & \textbf{Dataset} & \textbf{Samples ($\times 10^{4}$)} & \textbf{Ratio (\%)} \\
      \midrule
      \multirow{6}{*}{\makecell[l]{R0: Perception}}
        & Short\--Video\--Caption        & 20.00 & 5.06 \\
        & Commercial\--Ad\--Caption      & 16.98 & 4.30 \\
        & Live\--Stream\--Caption        & 13.70 & 3.47 \\
        & E\--Commerce\--Caption         & 17.52 & 4.43 \\
        & Short\--Video\--QA             & 24.09 & 6.10 \\
        & E\--Commerce\--QA              & 1.84  & 0.47 \\\cline{2-4}
      & Subtotal & 94.13 & 23.82 \\\cline{1-4}
      \multirow{2}{*}{\makecell[l]{R1: Derivation}}
        & i2i\--TagNext\--CF             & 20.00 & 5.06 \\
        & i2i\--Watch\--then\--Search    & 20.00 & 5.06 \\\cline{2-4}
      & Subtotal & 40.00 & 10.12 \\\cline{1-4}
      \makecell[l]{R2: Evolution} & Multi\--Hop\--Reasoning       & 12.38 & 3.13 \\\cline{1-4}
      \multirow{2}{*}{\makecell[l]{R3: Recommendation}}
        & Cross\--Domain (CoT)           & 29.56 & 7.48 \\
        & Cross\--Domain (UnCoT)         & 58.80 & 14.88 \\\cline{2-4}
      & Subtotal & 88.36 & 22.36 \\\cline{1-4}
      \makecell[l]{Itemic Instruction\\Data} & Item\--Instruction\--Following & 10.30 & 2.61 \\\cline{1-4}
      \makecell[l]{General-Domain\\Data} & StepFun\--General\footnotemark         & 150.00 & 37.96 \\
      \cmidrule(lr){1-4}
      & \textbf{Total} & \textbf{395.17} & \textbf{100.00} \\
      \bottomrule
    \end{tabular}
  \end{threeparttable}
\end{table}
\footnotetext{\url{https://huggingface.co/datasets/stepfun-ai/Step-3.5-Flash-SFT}}

\subsection{Cognitive-Reasoning Data Construction}
\subsubsection{R1 Explicit-Relevance Judgement Prompt}
\label{appendix:sft-r1-explicit-judgement-prompt}

The first relation-quality stage in R1 uses a locally deployed flagship model to define the boundary between explicit relevance, explicit irrelevance, and uncertainty. The prompt is applied to both TagNex-derived and after-play-search item-to-item candidates, using the available item metadata and dense captions as evidence.

\makeatletter
\@ifundefined{mycase}{%
  \newtcolorbox{mycase}[1]{%
    breakable,
    colback=gray!5, colframe=black!60, boxrule=0.5pt, arc=2pt,
    left=6pt, right=6pt, top=6pt, bottom=6pt,
    title={#1},
    fonttitle=\bfseries\small, coltitle=white, colbacktitle=black!70,
    before skip=8pt, after skip=8pt,
    before upper={\raggedright\sloppy}%
  }%
}{}
\makeatother

\begin{CJK*}{UTF8}{gbsn}
\begin{mycase}{Prompt SFT.R1.1: R1 Explicit-Relevance Judgement Prompt}
\small
\textbf{System Prompt} \\
\textit{你是一名推荐关系判断专家。你的任务是根据两个候选内容的可见信息，判断它们是否存在直接、稳定、可解释的一跳关系。请只依据给定的内容信息，不要使用“同一人群可能喜欢”“同平台常见共现”等宽泛假设。最终只输出三类之一：显性相关、显性不相关、不确定。}

\vspace{0.5em}
\textbf{User Prompt Template} \\
请判断下面两个内容之间是否存在直接、稳定、可解释的一跳关系。\\
\par\smallskip
【源内容信息】：\texttt{\{src\_metadata\}}\\
【目标内容信息】：\texttt{\{dest\_metadata\}}\\
【源内容描述】：\texttt{\{src\_dense\_caption\}}\\
【目标内容描述】：\texttt{\{dest\_dense\_caption\}}\\
\par\smallskip
判断标准：\\
1) 如果两者在主题、场景、对象、任务、风格、内容形态或近邻互补关系上存在清晰的一跳连接，输出“显性相关”。\\
2) 如果两者主题、场景或对象明显不同，或只能依赖宽泛人群偏好、平台共现、热度相近等假设建立联系，输出“显性不相关”。\\
3) 如果给定信息不足、描述噪声较高，或关系无法从可见内容中稳定确认，输出“不确定”。\\
\par\smallskip
请注意：\\
1) 不要把多跳推理或抽象联想当作直接关系。\\
2) 不要把“可能同一批用户会看”当作相关依据。\\
3) 只输出一个标签：显性相关、显性不相关 或 不确定。\\
4) 不要输出解释、示例、标点或其他额外内容。

\vspace{0.75em}
\hrulefill
\vspace{0.75em}

\textbf{English Translation} \\
\textbf{System Prompt} \\
\textit{You are a recommendation-relation judgement expert. Your task is to determine, from the visible information of two candidate items, whether they have a direct, stable, and explainable one-hop relation. Use only the provided content evidence, and do not rely on broad assumptions such as “the same audience may like both” or “they often co-occur on the platform”. Output exactly one of three labels: explicitly related, explicitly unrelated, or uncertain.}

\vspace{0.5em}
\textbf{User Prompt Template} \\
Please judge whether the following two pieces of content have a direct, stable, and explainable one-hop relation.\\
\par\smallskip
Source content information: \texttt{\{src\_metadata\}}\\
Target content information: \texttt{\{dest\_metadata\}}\\
Source content description: \texttt{\{src\_dense\_caption\}}\\
Target content description: \texttt{\{dest\_dense\_caption\}}\\
\par\smallskip
Judgement criteria:\\
1) If the two items have a clear one-hop connection in topic, scenario, object, task, style, content form, or neighboring complementarity, output “explicitly related”.\\
2) If their topic, scenario, or object is clearly different, or the connection only relies on broad assumptions such as audience overlap, platform co-occurrence, or similar popularity, output “explicitly unrelated”.\\
3) If the provided information is insufficient, the descriptions are noisy, or the relation cannot be stably confirmed from the visible content, output “uncertain”.\\
\par\smallskip
Notes:\\
1) Do not treat multi-hop reasoning or abstract association as a direct relation.\\
2) Do not use “the same users may watch both” as relevance evidence.\\
3) Output exactly one label: explicitly related, explicitly unrelated, or uncertain.\\
4) Do not output explanations, examples, punctuation, or any extra text.
\end{mycase}
\end{CJK*}

\subsubsection{R1 Relation-Explanation Generation Prompt}
\label{appendix:sft-r1-relation-explanation-prompt}

After explicit relation judgement and bridge-variable extraction, R1 generates the final relation explanation used in recommendation SFT samples. The prompt below is applied with source-item evidence and abstract bridge variables, while the destination item is used only for consistency checking.

\makeatletter
\@ifundefined{mycase}{%
  \newtcolorbox{mycase}[1]{%
    breakable,
    colback=gray!5, colframe=black!60, boxrule=0.5pt, arc=2pt,
    left=6pt, right=6pt, top=6pt, bottom=6pt,
    title={#1},
    fonttitle=\bfseries\small, coltitle=white, colbacktitle=black!70,
    before skip=8pt, after skip=8pt,
    before upper={\raggedright\sloppy}%
  }%
}{}
\makeatother

\begin{CJK*}{UTF8}{gbsn}
\begin{mycase}{Prompt SFT.R1.2: Relation-Explanation Generation Prompt}
\small
\textbf{System Prompt} \\
\textit{你是一名推荐分析师。你需要根据源视频线索和抽象桥接变量，以第一人称输出一段自然、可信、可解释的中文推理。候选目标信息仅用于内部校验，不能在结果里显式提及。}

\vspace{0.5em}
\textbf{User Prompt Template} \\
给定源视频信息和一组抽象桥接变量，请你以推荐分析师视角，输出一段自然流畅的第一人称中文推理，说明用户为什么会从当前内容继续延伸到某个相邻内容方向。候选目标信息仅用于内部校验，你不能在结果里显式提到目标，也不能把目标表层内容逐条复述成答案。\\
\par\smallskip
源视频信息：\\
- src\_pid: \texttt{\{src\_pid\}}\\
- src\_tag: \texttt{\{src\_tag\}}\\
- src\_dense\_caption: \texttt{\{src\_dense\_caption\}}\\
\par\smallskip
抽象桥接变量：\\
- source\_need: \texttt{\{source\_need\}}\\
- bridge\_type: \texttt{\{bridge\_type\}}\\
- bridge: \texttt{\{bridge\}}\\
- direction: \texttt{\{direction\}}\\
- reason\_seed: \texttt{\{reason\_seed\}}\\
\par\smallskip
参考候选目标（仅内部校验，不可显式提及）：\\
- dest\_pid: \texttt{\{dest\_pid\}}\\
\par\smallskip
输出要求：\\
1. 只输出最终推理正文，不要输出标题、编号、Markdown、引号或额外说明。\\
2. 必须使用第一人称分析口吻，且正文必须以“我注意到”开头。\\
3. 核心依据必须来自源视频信息和抽象桥接变量；候选目标只用于内部一致性校验。\\
4. 重点说明源内容激活了什么需求、这种需求为什么会延伸、后续可能会滑向什么方向，不要简单复述源描述。\\
5. 不要显式提及目标、候选、dest\_pid，也不要把目标表层关键词直接写进正文。\\
6. 结尾保持非确定性，例如“可能会继续关注”“更可能延伸到”“有较大概率被这类内容接住”。\\
7. 正文长度控制在 220 到 520 字之间，信息密度高但自然。\\
\par\smallskip
你的推理：

\vspace{0.75em}
\hrulefill
\vspace{0.75em}

\textbf{English Translation} \\
\textbf{System Prompt} \\
\textit{You are a recommendation analyst. Based on source-video signals and abstract bridge variables, output a natural, credible, and explainable Chinese reasoning paragraph in the first person. The candidate target is used only for internal consistency checking and must not be explicitly mentioned in the result.}

\vspace{0.5em}
\textbf{User Prompt Template} \\
Given source-video information and a set of abstract bridge variables, output a fluent first-person Chinese reasoning paragraph from the perspective of a recommendation analyst, explaining why the user may extend from the current content to a neighboring content direction. The candidate target is used only for internal consistency checking; do not explicitly mention the target or restate its surface content as the answer.\\
\par\smallskip
Source-video information:\\
- src\_pid: \texttt{\{src\_pid\}}\\
- src\_tag: \texttt{\{src\_tag\}}\\
- src\_dense\_caption: \texttt{\{src\_dense\_caption\}}\\
\par\smallskip
Abstract bridge variables:\\
- source\_need: \texttt{\{source\_need\}}\\
- bridge\_type: \texttt{\{bridge\_type\}}\\
- bridge: \texttt{\{bridge\}}\\
- direction: \texttt{\{direction\}}\\
- reason\_seed: \texttt{\{reason\_seed\}}\\
\par\smallskip
Reference candidate target, used only for internal consistency checking and not to be explicitly mentioned:\\
- dest\_pid: \texttt{\{dest\_pid\}}\\
\par\smallskip
Output requirements:\\
1. Output only the final reasoning paragraph, without titles, numbering, Markdown, quotation marks, or extra explanation.\\
2. Use a first-person analytical tone, and the paragraph must start with “我注意到”.\\
3. The core evidence must come from the source-video information and abstract bridge variables; the candidate target is only for internal consistency checking.\\
4. Focus on what need is activated by the source content, why the need extends, and what direction the user may move toward next; do not simply restate the source description.\\
5. Do not explicitly mention the target, candidate, dest\_pid, or directly copy surface keywords from the target content.\\
6. Keep the ending non-deterministic, such as “may continue to follow”, “is more likely to extend to”, or “has a high probability of being captured by this type of content”.\\
7. Keep the paragraph between 220 and 520 Chinese characters, dense but natural.\\
\par\smallskip
Your reasoning:
\end{mycase}
\end{CJK*}

\subsubsection{R1 Training Examples}
\label{appendix:sft-r1-training-examples}

\begin{CJK*}{UTF8}{gbsn}
\begin{mycase}{Prompt SFT.R1.3: Additional Sanitized R1 Training Examples}
\small
\textbf{Example A: Training and Recovery} \\
\textit{Source topic: 短期增肌训练 \quad Target topic: 避免肌肉劳损}

\vspace{0.4em}
\textbf{User} \\
和输入视频\texttt{\detokenize{<|video_begin|><a_1842><b_7365><c_2091>}}常识上相似的视频有哪些？

\vspace{0.4em}
\textbf{Assistant} \\
\texttt{\detokenize{<think>}}当用户对短期增肌训练产生浓厚兴趣时，由于抗阻训练通常伴随较高强度和较高频次，肌肉疲劳与劳损风险也会随之增加。从常识角度看，这类用户往往不仅关注如何提升训练效率，也会主动寻找科学的恢复策略，以避免训练损伤并保持后续训练表现。睡眠是肌肉修复和炎症缓解的重要阶段，而合理的睡姿能够减少局部压力和夜间酸痛，帮助身体更好地恢复。因此，用户的兴趣很容易从“如何增肌”延伸到“如何减少训练后的身体负担”。这种关联来自训练与恢复之间的协同关系，符合健身人群对健康管理和持续训练能力的共同需求。\texttt{\detokenize{</think>}}\\
\texttt{\detokenize{<|video_begin|><a_6219><b_0487><c_7742>}}

\vspace{0.4em}
\textbf{English Translation} \\
\texttt{\detokenize{<think>}}When a user shows strong interest in short-term muscle-gain training, resistance training usually implies high intensity and repeated load, which also increases the risk of muscle fatigue and strain. From common sense, such users care not only about improving training efficiency, but also about scientific recovery strategies that help avoid injury and maintain later performance. Sleep is an important stage for muscle repair and inflammation relief, and an appropriate sleeping posture can reduce local pressure and nighttime soreness. Therefore, the user's interest can naturally extend from “how to gain muscle” to “how to reduce physical burden after training”. This relation comes from the coordination between training and recovery, matching the fitness audience's need for health management and sustainable training.\texttt{\detokenize{</think>}}\\
\texttt{\detokenize{<|video_begin|><a_6219><b_0487><c_7742>}}

\vspace{0.9em}
\textbf{Example B: Off-Road Driving and Vehicle Suspension} \\
\textit{Source topic: 泥泞地形越野驾驶 \quad Target topic: 汽车减震改装}

\vspace{0.4em}
\textbf{User} \\
请推荐一条与短视频\texttt{\detokenize{<|video_begin|><a_4027><b_1186><c_9304>}}主题接近、语义相关的视频。

\vspace{0.4em}
\textbf{Assistant} \\
\texttt{\detokenize{<think>}}我注意到源内容聚焦于车辆在泥泞等复杂地形中的越野表现，核心吸引点并不只是驾驶画面本身，而是车辆如何在恶劣路况下保持通过性、稳定性和操控感。对这类内容感兴趣的用户，通常会进一步关注影响越野表现的关键部件，例如悬挂系统、减震能力和底盘调校。因为在复杂路面中，车辆能否稳定通过坑洼、泥地或颠簸路段，很大程度上取决于减震结构对冲击的吸收和车身姿态的控制。因此，用户的兴趣很自然会从“看车辆越野”延伸到“理解车辆为什么能越野”，进而关注改装方案和性能优化。这种关系属于从真实驾驶场景到车辆性能机制的自然延伸，有较大概率被相关的减震改装内容接住。\texttt{\detokenize{</think>}}\\
\texttt{\detokenize{<|video_begin|><a_7750><b_3641><c_5088>}}

\vspace{0.4em}
\textbf{English Translation} \\
\texttt{\detokenize{<think>}}I notice that the source content focuses on how a vehicle performs off-road in muddy and complex terrain. The core attraction is not only the driving scene itself, but also how the vehicle maintains passability, stability, and controllability under difficult road conditions. Users interested in such content are likely to further care about key components that affect off-road performance, such as the suspension system, damping capability, and chassis tuning. On complex roads, whether a vehicle can pass pits, mud, and bumps smoothly depends heavily on how the damping structure absorbs impact and controls body posture. Therefore, the user's interest can naturally move from “watching off-road driving” to “understanding why the vehicle can handle off-road conditions”, and then to modification plans and performance optimization. This is a natural extension from a driving scenario to the vehicle-performance mechanism.\texttt{\detokenize{</think>}}\\
\texttt{\detokenize{<|video_begin|><a_7750><b_3641><c_5088>}}

\vspace{0.9em}
\textbf{Example C: Athlete and Event Continuation} \\
\textit{Source topic: 孙颖莎 \quad Target topic: 刘国梁}

\vspace{0.4em}
\textbf{User} \\
和输入视频\texttt{\detokenize{<|video_begin|><a_3196><b_8572><c_4409>}}常识上相似的视频有哪些？

\vspace{0.4em}
\textbf{Assistant} \\
\texttt{\detokenize{<think>}}我注意到源内容围绕孙颖莎在重要赛场中的表现展开，用户被吸引的往往不只是单场比赛结果，而是运动员在高强度竞争中的状态、成长和后续发展。对于关注奥运健儿的用户来说，兴趣很容易从运动员个人表现延伸到国家队训练安排、赛事衔接和教练团队判断。常识上，奥运后的运动员通常需要快速进入新的备战周期，而教练或管理者对赛事难度、训练节奏和竞技状态的判断，会直接影响用户理解运动员后续表现的方式。因此，刘国梁围绕全运会、训练压力或备战节奏的相关内容，能够承接用户对孙颖莎后续发展的关注。这种关联来自运动员表现与团队指导之间的赛事承接关系，属于体育内容中较自然的因果延伸。\texttt{\detokenize{</think>}}\\
\texttt{\detokenize{<|video_begin|><a_9038><b_2164><c_6871>}}

\vspace{0.4em}
\textbf{English Translation} \\
\texttt{\detokenize{<think>}}I notice that the source content centers on Sun Yingsha's performance in important competitions. Users are often attracted not only by one match result, but also by the athlete's condition, growth, and subsequent development under high-intensity competition. For users who follow Olympic athletes, interest can naturally extend from the athlete's individual performance to national-team training plans, event transitions, and coaching-team judgement. In common sense, athletes often need to quickly enter a new preparation cycle after the Olympics, and the coach or manager's view on event difficulty, training rhythm, and competitive condition directly shapes how users understand the athlete's later performance. Therefore, content around Liu Guoliang's comments on the National Games, training pressure, or preparation rhythm can naturally continue the user's attention to Sun Yingsha's development. This relation comes from event continuation between athlete performance and team guidance.\texttt{\detokenize{</think>}}\\
\texttt{\detokenize{<|video_begin|><a_9038><b_2164><c_6871>}}
\end{mycase}
\end{CJK*}

\subsubsection{R2 Logic-Chain Extraction Prompt}
\label{appendix:sft-r2-extraction-prompt}

R2 begins by extracting candidate user-interest evolution chains from chronological multi-domain behavior timelines. We use a strong LLM to identify a small number of temporally ordered chains with clear progression, while avoiding shallow topical grouping or unsupported speculation. Below we provide the main extraction prompt used in this stage. The subsequent quality-control prompt follows a highly similar style, but is used as a second-pass LLM-as-a-Judge filter to remove pseudo-logical or weakly grounded chains produced during first-pass generation.

\makeatletter
\@ifundefined{mycase}{%
  \newtcolorbox{mycase}[1]{%
    breakable,
    colback=gray!5, colframe=black!60, boxrule=0.5pt, arc=2pt,
    left=6pt, right=6pt, top=6pt, bottom=6pt,
    title={#1},
    fonttitle=\bfseries\small, coltitle=white, colbacktitle=black!70,
    before skip=8pt, after skip=8pt,
    before upper={\raggedright\sloppy}%
  }%
}{}
\makeatother

\begin{CJK*}{UTF8}{gbsn}
\begin{mycase}{Prompt SFT.R2.1: Logic-Chain Extraction Prompt}
\small
\textbf{Prompt Template} \\
\textit{角色定义： 你是一位极端严苛的用户行为数据挖掘专家，负责从琐碎日志中提取具有“认知跃迁”价值的极少数行为链。}\\
\textit{任务目标： 请根据提供的用户交互历史（Timeline），其中包含了用户搜索内容，互动过的视频、直播、广告和商品。我希望你从中抽取出有兴趣演化轨迹、具有内在行为关联逻辑而非内容简单关联的交互行为逻辑链，你的首要要求是剔除伪逻辑，严禁强行解释。}\\
\textit{交互说明： 其中商品点击、广告点击只是点击行为而非购买行为。}\\
\textit{其他说明： 对于 “--:--” 标识的事件，表示对应的事件缺少具体的时间信息。}\\
\textit{案例：比如"喉咙痛怎么办（20260101） → 干咳/有痰区别（20260102） → 用药科普（注意禁忌）（20260103) “每一个->都需要存在递进/转折或者其他明显的兴趣演化关系}\\
\par\smallskip
\textit{行为逻辑链有效性准则：}\\
\textit{深度关联： 严禁简单的内容关键词堆砌。必须体现递进、转折、因果、对比或闭环等演化关系。严禁把“同类目”的事情按时间排在一起，然后强行解释出一种递进感；}\\
\textit{时序敏感： 严格遵守时间线，确保每一链条内的事件是按时间顺序发生的。}\\
\textit{1.顺序互换测试：如果链条中的 Event A 和 Event B 互换时间顺序后，逻辑依然通顺且合理，则不能将其视为强关联的演化链，请予以剔除或拆分为独立事件。}\\
\textit{2.有效演化： 只有当 Event B 的发生是基于 Event A 带来的**反馈、瓶颈、认知升级**时（即 B 是对 A 的修正或深化），才算有效。}\\
\textit{3.认知增量测试：Event B 必须包含 Event A中未出现的新变量（如特定参数、专业术语、对比维度）。判定标准：如果 EventB只是 EventA的语义重复或同层级平移（例如：从搜“降噪耳机”变为搜“头戴耳机”），则不视为演化。}\\
\textit{4.行为修正/收敛特征：演化链必须体现出从“发散”到“聚焦”，或从“错误尝试”到“路径修正”的过程。判定标准：后续行为必须表现出对前期信息的筛选过滤。如果用户在 EventB中完全抛弃了 EventA的筛选条件且无逻辑解释，则链条断裂。有效示例：点击“高热量猫粮”-观看“猫咪软便调理视频” -搜索“低敏/无谷/易消化猫粮”。（体现了基于负反馈的策略调整）}\\
\textit{5.真正的兴趣演化往往伴随着交互深度的变化（如：从“被动接收”广告/视频，转为“主动检索”核心信息）如果全链条只有单一动作类型（如全是搜索），其递进逻辑往往较弱。严禁仅由‘连续点击相似商品’构成链条。}\\
\textit{6.时空密度测试： 严格审查事件 A 到 B 之间的时间跨度。如果两个核心行为跨度中间缺乏相关性的引导行为（如中间没有持续的相关搜索或视频观看），应判定为日常零散行为而非‘演化’。严禁将相隔数月的日常消费行为（如买菜、买日用品）强行拼凑成认知升级。}\\
\textit{7.非普适性测试： 排除基于‘生活补货’或‘随机消费’驱动的行为。如果 Event A 是普适性的生活必需品（如普通的蛋奶、纸巾、基础款服饰），除非其具有极强的特定功能指向（如：从普通奶粉转向专门针对糖尿病的脱糖奶粉），否则不得作为逻辑链的起始点或关键节点。”}\\
\textit{8.触发源测试： 逻辑链必须包含一个清晰的‘问题触发’。有效的演化链通常应以‘主动搜索’或‘带有明确意图的对比行为’为起点。严禁直接由两个不同的商品点击/购买行为推导出认知升级，除非中间有明确的知识获取动作（如观看科普、对比参数）作为桥梁。}\\
\textit{9.强因果排他性测试： 进行‘反向排除’。如果 Event B（买食疗书）的发生可以完全独立于 Event A（买鹅蛋）存在，且不需要 A 提供任何基础，则链条断裂。演化必须体现出‘只有经历了 A，才产生了对 B 的特定需求或认知基础’。如果两者只是恰好属于‘大健康’类目，请判定为‘同类目平移’予以剔除}\\
\textit{10.证据闭环要求： 逻辑说明（logic）中提到的每一个对比、转折或先前状态，都必须在 events 数组中拥有对应的条目。严禁在说明中引用链条之外的行为作为依据。例如：若提到‘从观赏性转向实用性’，则‘观赏性行为’必须作为链条的第一步出现。}\\
\textit{11.禁止脑补：严禁在 logic 中推测用户心理（如“用户可能觉得累了”），所有的逻辑转折必须有日志中的搜索词、视频标题、商品参数作为物理证据。}\\
\par\smallskip
\textit{输出准则：}\\
\textit{极端置信度过滤：你必须扮演一个极度刻薄的质检员。对于每一条潜在的逻辑链，必须同时满足前述 11 项有效性要求，请秉持“宁缺毋滥”原则。只有当 Event B 表现出对 Event A 的认知超越、参数收敛或负反馈修正（即：不经历 A，用户绝不会产生 B 这种特定的搜索或点击意图）时，逻辑链才算有效。}\\
\textit{确保唯一性： 拒绝同级词汇平移，只接受逻辑跃迁。若日志中仅为琐碎的日常消费或随机点击，严禁强行解释，请直接返回空；}\\
\textit{数量限制： 每个用户画像最多仅允许提取 3 条‘金子般’的逻辑链。}\\
\par\smallskip
\textit{输出格式： 请以 JSON 数组形式返回，每个对象代表一个逻辑链条，结构如下：}\\
\texttt{[}\\
\texttt{\quad \{} \\
\texttt{\quad \quad "logic\_chain": \{} \\
\texttt{\quad \quad \quad "name": "链条核心意图命名",} \\
\texttt{\quad \quad \quad "logic\_test": "逻辑链演化有效性说明，比如（通过因果测试：若无前期对症状的误判，则不会产生后期对成分的深度调研）",} \\
\texttt{\quad \quad \quad "events": [} \\
\texttt{\quad \quad \quad \quad \{} \\
\texttt{\quad \quad \quad \quad \quad "date": "YYYY-MM-DD",} \\
\texttt{\quad \quad \quad \quad \quad "action": "[交互类型] 具体内容（需要严格对应 Timeline 中 Event 条目）",} \\
\texttt{\quad \quad \quad \quad \quad "logic": "该步在交互逻辑链中逻辑说明（如：从泛搜转为对特定成分的深度调研）"} \\
\texttt{\quad \quad \quad \quad \}} \\
\texttt{\quad \quad \quad ]} \\
\texttt{\quad \quad \}} \\
\texttt{\quad \}} \\
\texttt{]} \\
\par\smallskip
\textit{其他输出有要求：}\\
\textit{每个逻辑节点仅能包含一个具体事件，错误例子：[商品-购买] 购买糖果 A 和糖果 B (ec\_id=123 \& ec\_id=456)}\\
\par\smallskip
\textit{用户交互历史（Timeline）：}\\
\texttt{\{timeline\_text\}}

\vspace{0.75em}
\hrulefill
\vspace{0.75em}

\textbf{English Translation} \\
\textit{Role Definition: You are an extremely strict expert in user-behavior mining, responsible for extracting a very small number of behavior chains with meaningful ``cognitive transition'' value from noisy logs.}\\
\textit{Task Objective: Given the user's interaction history (Timeline), which contains search queries, interacted videos, live streams, ads, and products, extract behavior logic chains that reflect genuine interest evolution trajectories and internal behavioral connections rather than simple topical relatedness. Your top priority is to remove pseudo-logical chains and avoid forced interpretation.}\\
\textit{Interaction Note: Product clicks and ad clicks indicate click behaviors rather than purchases.}\\
\textit{Additional Note: Events marked with ``--:--'' do not have exact timestamp information.}\\
\textit{Example: A chain such as ``How to deal with sore throat (20260101) $\rightarrow$ dry cough vs. phlegm cough (20260102) $\rightarrow$ medication science and contraindications (20260103)'' should contain a clear progression, transition, or other meaningful interest evolution relation at every step.}\\
\par\smallskip
\textit{Validity Criteria for Behavior Logic Chains:}\\
\textit{Deep Relation: Do not simply pile up keywords. A valid chain must demonstrate evolutionary relations such as progression, transition, causality, contrast, or closure. It is not allowed to line up events from the same category by time and then force an interpretation of progression.}\\
\textit{Temporal Sensitivity: Strictly follow the timeline, ensuring that events in each chain are chronologically ordered.}\\
\textit{1. Order-Swap Test: If swapping the temporal order of Event A and Event B still leaves the chain equally coherent and reasonable, then the chain should not be treated as a strongly connected evolution chain and should be removed or split.}\\
\textit{2. Valid Evolution: Event B is valid only when it arises from the feedback, bottleneck, or cognitive upgrade brought by Event A, that is, B must refine or correct A.}\\
\textit{3. Cognitive Increment Test: Event B must introduce new variables absent from Event A, such as specific parameters, technical terms, or comparison dimensions. If Event B is only a semantic repetition or same-level lateral move from Event A, it does not count as evolution.}\\
\textit{4. Behavioral Revision / Convergence: An evolution chain should show a process from divergence to focus, or from failed attempts to path correction. Later actions must reflect selection or filtering based on earlier information. If Event B completely abandons the conditions in Event A without logical support, the chain is broken.}\\
\textit{5. Interaction-Depth Shift: Genuine interest evolution is often accompanied by changes in interaction depth, such as moving from passive exposure through ads or videos to active information seeking. If the entire chain contains only a single action type, such as all searches, its progressive strength is usually weak. Chains formed only by repeated clicks on similar products should be excluded.}\\
\textit{6. Temporal-Spatial Density Test: Carefully examine the time gap between Event A and Event B. If two core actions are separated by a long period without related bridging behavior, such as relevant searches or video watching, they should be treated as scattered daily behaviors rather than evolution. Daily consumption events separated by months should not be forcibly assembled into cognitive upgrade chains.}\\
\textit{7. Non-Genericity Test: Exclude behaviors driven merely by routine replenishment or random consumption. If Event A is a generic daily necessity, such as ordinary milk powder, tissues, or basic clothing, it should not serve as the starting point or key node of a logic chain unless it has highly specific functional intent.}\\
\textit{8. Trigger-Source Test: A valid logic chain must contain a clear trigger, usually an active search or an intentional comparison behavior. It is not allowed to infer cognitive upgrade directly from two different product clicks or purchases unless there is an explicit knowledge-acquisition step, such as watching educational content or comparing parameters, in between.}\\
\textit{9. Strong Causal Exclusivity Test: Perform reverse exclusion. If Event B could occur entirely independently of Event A and does not require A as a prerequisite, then the chain is broken. Evolution must reflect that only after experiencing A would the user develop the specific need or cognitive basis for B. If the two events merely belong to the same broad category, they should be excluded as same-category drift.}\\
\textit{10. Evidence-Closure Requirement: Every contrast, transition, or prior state mentioned in the logic explanation must correspond to an explicit event in the chain. The explanation must not rely on behaviors outside the chain itself.}\\
\textit{11. No Mind Reading: Do not speculate about the user's psychology in the logic field, such as ``the user may feel tired''. Every logical transition must be supported by concrete evidence in the logs, such as search terms, video titles, or product parameters.}\\
\par\smallskip
\textit{Output Principles:}\\
\textit{Extremely High-Confidence Filtering: Act as a highly demanding quality inspector. Each candidate chain must satisfy the above validity requirements under a ``better to miss than to include noise'' principle. A chain is valid only when Event B demonstrates cognitive advance, parameter convergence, or revision under negative feedback relative to Event A, such that B would not arise without first experiencing A.}\\
\textit{Ensure Uniqueness: Reject same-level lexical drift and accept only genuine logical transition. If the log contains only trivial daily consumption or random clicks, do not force an explanation and directly return an empty list.}\\
\textit{Quantity Limit: For each user profile, extract at most 3 ``gold-level'' logic chains.}\\
\par\smallskip
\textit{Output Format: Return a JSON array, where each object represents one logic chain with the following structure:}\\
\texttt{[}\\
\texttt{\quad \{} \\
\texttt{\quad \quad "logic\_chain": \{} \\
\texttt{\quad \quad \quad "name": "core intent name of the chain",} \\
\texttt{\quad \quad \quad "logic\_test": "validity explanation of the evolution chain, e.g., without the earlier misjudgment of symptoms, the later in-depth investigation of ingredients would not occur",} \\
\texttt{\quad \quad \quad "events": [} \\
\texttt{\quad \quad \quad \quad \{} \\
\texttt{\quad \quad \quad \quad \quad "date": "YYYY-MM-DD",} \\
\texttt{\quad \quad \quad \quad \quad "action": "[interaction type] specific content (must exactly correspond to an Event entry in the Timeline)",} \\
\texttt{\quad \quad \quad \quad \quad "logic": "logical role of this step in the chain, e.g., moving from broad search to in-depth investigation of a specific ingredient"} \\
\texttt{\quad \quad \quad \quad \}} \\
\texttt{\quad \quad \quad ]} \\
\texttt{\quad \quad \}} \\
\texttt{\quad \}} \\
\texttt{]} \\
\par\smallskip
\textit{Additional Output Constraint: Each logic node may contain only one concrete event. A wrong example is: [product-purchase] purchased candy A and candy B (ec\_id=123 \& ec\_id=456).}\\
\par\smallskip
\textit{User Interaction History (Timeline):}\\
\texttt{\{timeline\_text\}}
\end{mycase}
\end{CJK*}

\subsubsection{R2 Training Examples}
\label{appendix:sft-r2-multihop-training-examples}

The final R2 data contains \textit{selection}, \textit{topic-generation}, and \textit{direct-generation} task types, aligned with the R2 benchmark taxonomy. To protect privacy and avoid exposing internal identifiers, we use synthetic itemic tokens and compact timelines while preserving the original task structure.

\begin{CJK*}{UTF8}{gbsn}
\begin{mycase}{Prompt SFT.R2.2: Evolution Action Selection}
\small
\textbf{中文示例} \\
\textbf{User:} 请根据用户交互历史和候选兴趣演化链，选择哪些历史行为支撑有效的兴趣演化。\\
\textit{用户交互历史（节选）：}\\
2026-01-03 [搜索] 入门级理财怎么学\\
2026-01-05 [直播-关注] 股票基础教学直播 \texttt{<|living\_begin|>}\allowbreak\texttt{<a\_2140>}\allowbreak\texttt{<b\_5702>}\allowbreak\texttt{<c\_4418>}\\
2026-01-09 [视频-长播] 控制投资风险的方法 \texttt{<|video\_begin|>}\allowbreak\texttt{<a\_3721>}\allowbreak\texttt{<b\_6610>}\allowbreak\texttt{<c\_2844>}\\
2026-01-12 [搜索] 主力资金怎么看\\
2026-01-14 [视频-长播] 技术指标入门 \texttt{<|video\_begin|>}\allowbreak\texttt{<a\_5068>}\allowbreak\texttt{<b\_4109>}\allowbreak\texttt{<c\_7120>}\\
\textit{候选兴趣演化目标：} 从理财入门到股票风险控制与技术分析\\
\textit{候选行为：}\\
A. 理财入门搜索 $\rightarrow$ 股票基础教学直播 $\rightarrow$ 风险控制视频\\
B. 股票基础教学直播 $\rightarrow$ 主力资金搜索 $\rightarrow$ 技术指标视频\\
C. 风险控制视频 $\rightarrow$ 技术指标视频\\
\textbf{Assistant:}\\
\texttt{\detokenize{<think>}}候选A体现了从主动搜索入门知识，到关注基础教学，再到补充风险控制方法的递进，后项为前项带来了新的风险变量。候选B从基础教学延伸到主动检索主力资金，再进入技术指标学习，也包含从概念入门到分析工具的认知升级。候选C虽然同属投资内容，但两个视频可以互换顺序理解，缺少由前一行为触发后一行为的证据，因此不保留。\texttt{\detokenize{</think>}}\\
答案: [A, B]

\vspace{0.5em}
\textbf{English Translation} \\
\textbf{User:} Given the user history and a candidate interest-evolution target, select which historical actions support a valid evolution process.\\
\textit{Timeline excerpt:}\\
2026-01-03 [Search] how to start learning personal finance\\
2026-01-05 [Live-follow] stock-market basics tutorial \texttt{<|living\_begin|>}\allowbreak\texttt{<a\_2140>}\allowbreak\texttt{<b\_5702>}\allowbreak\texttt{<c\_4418>}\\
2026-01-09 [Video-long-view] methods for controlling investment risk \texttt{<|video\_begin|>}\allowbreak\texttt{<a\_3721>}\allowbreak\texttt{<b\_6610>}\allowbreak\texttt{<c\_2844>}\\
2026-01-12 [Search] how to interpret main-fund flows\\
2026-01-14 [Video-long-view] introduction to technical indicators \texttt{<|video\_begin|>}\allowbreak\texttt{<a\_5068>}\allowbreak\texttt{<b\_4109>}\allowbreak\texttt{<c\_7120>}\\
\textit{Candidate evolution target:} from personal-finance basics to stock-risk control and technical analysis\\
\textit{Candidate actions:}\\
A. finance-beginner search $\rightarrow$ stock-basics live stream $\rightarrow$ risk-control video\\
B. stock-basics live stream $\rightarrow$ main-fund-flow search $\rightarrow$ technical-indicator video\\
C. risk-control video $\rightarrow$ technical-indicator video\\
\textbf{Assistant:}\\
\texttt{\detokenize{<think>}}Candidate A shows progression from active beginner search, to basic teaching, and then to risk-control knowledge; the later step introduces a new risk variable. Candidate B extends from basic teaching to active search about fund flows and then to technical indicators, forming a cognitive upgrade from concepts to analytical tools. Candidate C stays within investment content, but the two videos can be swapped without breaking the logic, so it lacks evidence that the earlier action triggered the later one.\\
\texttt{\detokenize{</think>}}\\
Answer: [A, B]
\end{mycase}
\end{CJK*}

\begin{CJK*}{UTF8}{gbsn}
\begin{mycase}{Prompt SFT.R2.3: Topic-Conditioned Evolution Generation}
\small
\textbf{中文示例} \\
\textbf{User:} 围绕主题“家庭健康管理从泛化建议到医疗实操闭环”，请从用户历史中生成一条不超过 5 步的兴趣演化链。\\
\textit{用户交互历史（节选）：}\\
2025-12-17 [广告-点击] AI 情绪健康自测应用 \texttt{<|ad\_begin|>}\allowbreak\texttt{<a\_1351>}\allowbreak\texttt{<b\_8179>}\allowbreak\texttt{<c\_1243>}\\
2025-12-21 [广告-点击] 药品查询与用药提醒工具 \texttt{<|ad\_begin|>}\allowbreak\texttt{<a\_1351>}\allowbreak\texttt{<b\_2826>}\allowbreak\texttt{<c\_6047>}\\
2026-01-01 [广告-点击] 在线问诊与预约挂号服务 \texttt{<|ad\_begin|>}\allowbreak\texttt{<a\_1351>}\allowbreak\texttt{<b\_4480>}\allowbreak\texttt{<c\_0997>}\\
\textbf{Assistant:}\\
\texttt{\detokenize{<think>}}该主题要求聚焦健康管理能力的逐步落地。第一步是轻量情绪自测，属于泛化健康建议；第二步转向药品查询和提醒，引入药物参数与执行管理；第三步进入问诊和挂号，说明用户不再只满足于信息建议，而是希望连接专业医疗服务，形成从监测到干预的闭环。\texttt{\detokenize{</think>}}\\
\textbf{Output:}\\
\texttt{\{}\\
\quad \texttt{"logic\_chain"}: \texttt{\{}\\
\quad\quad \texttt{"name"}: ``家庭健康管理从泛化建议到医疗实操闭环'',\\
\quad\quad \texttt{"events"}: [\\
\quad\quad\quad \texttt{\{} \texttt{"date"}: \texttt{"2025-12-17"}, \texttt{"action"}: ``[广告-点击] AI 情绪健康自测应用'', \texttt{"logic"}: ``初始触发：以轻量自测获取泛化健康建议。'' \texttt{\}},\\
\quad\quad\quad \texttt{\{} \texttt{"date"}: \texttt{"2025-12-21"}, \texttt{"action"}: ``[广告-点击] 药品查询与用药提醒工具'', \texttt{"logic"}: ``认知升级：从咨询转向药物参数和执行提醒。'' \texttt{\}},\\
\quad\quad\quad \texttt{\{} \texttt{"date"}: \texttt{"2026-01-01"}, \texttt{"action"}: ``[广告-点击] 在线问诊与预约挂号服务'', \texttt{"logic"}: ``决策收敛：建立连接专业医疗服务的闭环。'' \texttt{\}}\\
\quad\quad ]\\
\quad \texttt{\}}\\
\texttt{\}}

\vspace{0.5em}
\textbf{English Translation} \\
\textbf{User:} For the topic ``family health management evolving from generic advice to practical medical action'', generate one interest-evolution chain with no more than five steps.\\
\textit{Timeline excerpt:}\\
2025-12-17 [Ad-click] AI emotional-health self-assessment app \texttt{<|ad\_begin|>}\allowbreak\texttt{<a\_1351>}\allowbreak\texttt{<b\_8179>}\allowbreak\texttt{<c\_1243>}\\
2025-12-21 [Ad-click] medication lookup and reminder tool \texttt{<|ad\_begin|>}\allowbreak\texttt{<a\_1351>}\allowbreak\texttt{<b\_2826>}\allowbreak\texttt{<c\_6047>}\\
2026-01-01 [Ad-click] online consultation and appointment service \texttt{<|ad\_begin|>}\allowbreak\texttt{<a\_1351>}\allowbreak\texttt{<b\_4480>}\allowbreak\texttt{<c\_0997>}\\
\textbf{Assistant:}\\
\texttt{\detokenize{<think>}}The topic requires a chain in which health management becomes increasingly actionable. The first step is lightweight emotional self-assessment, which provides generic advice. The second step shifts to medication lookup and reminders, adding concrete drug parameters and execution management. The third step moves to consultation and appointment booking, showing that the user is no longer satisfied with information advice and now wants access to professional medical service.\\
\texttt{\detokenize{</think>}}\\
\textit{Output: one JSON object describing the above three-step chain.}
\end{mycase}
\end{CJK*}

\begin{CJK*}{UTF8}{gbsn}
\begin{mycase}{Prompt SFT.R2.4: Direct Evolution Generation}
\small
\textbf{中文示例} \\
\textbf{User:} 请直接从以下用户交互历史中抽取高置信度兴趣演化链；如果不存在有效链，请返回空数组。\\
\textit{用户交互历史（节选）：}\\
2026-01-18 [搜索] 苹果 16\\
2026-01-21 [视频-长播] 手机颜色、配件与购机方案介绍 \texttt{<|video\_begin|>}\allowbreak\texttt{<a\_0570>}\allowbreak\texttt{<b\_5571>}\allowbreak\texttt{<c\_2701>}\\
2026-01-24 [搜索] 同城手机分期付款 0 首付\\
2026-01-24 [商品-点击] 99 新二手 iPhone 15 \texttt{<|prod\_begin|>}\allowbreak\texttt{<a\_0307>}\allowbreak\texttt{<b\_7024>}\allowbreak\texttt{<c\_4281>}\\
2026-01-28 [搜索] 二手手机验机注意事项\\
\textbf{Assistant:}\\
\texttt{\detokenize{<think>}}用户先主动搜索最新旗舰机型，建立高性能和高价格锚点；随后长播手机款式与配件内容，进入购机调研；接着搜索零首付分期，说明预算约束触发路径修正；最终点击近新二手机型，并继续搜索验机注意事项，表明兴趣从追新转向低门槛、可验证的二手购买方案。这不是同类商品的平级浏览，而是受预算和决策风险驱动的收敛链。\texttt{\detokenize{</think>}}\\
\textbf{Output:}\\
\texttt{[}\\
\quad \texttt{\{}\\
\quad\quad \texttt{"logic\_chain"}: \texttt{\{}\\
\quad\quad\quad \texttt{"name"}: ``智能手机消费从旗舰追新到分期二手的务实降级决策链'',\\
\quad\quad\quad \texttt{"events"}: [\\
\quad\quad\quad\quad \texttt{\{} \texttt{"date"}: \texttt{"2026-01-18"}, \texttt{"action"}: ``[搜索] 苹果 16'', \texttt{"logic"}: ``初始触发：关注最新旗舰，形成高位价格与性能锚点。'' \texttt{\}},\\
\quad\quad\quad\quad \texttt{\{} \texttt{"date"}: \texttt{"2026-01-24"}, \texttt{"action"}: ``[搜索] 同城手机分期付款 0 首付'', \texttt{"logic"}: ``路径修正：预算约束触发对低门槛支付方案的主动检索。'' \texttt{\}},\\
\quad\quad\quad\quad \texttt{\{} \texttt{"date"}: \texttt{"2026-01-24"}, \texttt{"action"}: ``[商品-点击] 99 新二手 iPhone 15'', \texttt{"logic"}: ``决策收敛：从最新旗舰转向更高性价比的二手次旗舰。'' \texttt{\}},\\
\quad\quad\quad\quad \texttt{\{} \texttt{"date"}: \texttt{"2026-01-28"}, \texttt{"action"}: ``[搜索] 二手手机验机注意事项'', \texttt{"logic"}: ``风险补完：在二手购买路径下补充验机知识，降低交易不确定性。'' \texttt{\}}\\
\quad\quad\quad ]\\
\quad\quad \texttt{\}}\\
\quad \texttt{\}}\\
\texttt{]}

\vspace{0.5em}
\textbf{English Translation} \\
\textbf{User:} Directly extract high-confidence interest-evolution chains from the following user history; return an empty array if no valid chain exists.\\
\textit{Timeline excerpt:}\\
2026-01-18 [Search] iPhone 16\\
2026-01-21 [Video-long-view] phone colors, accessories, and purchase plans \texttt{<|video\_begin|>}\allowbreak\texttt{<a\_0570>}\allowbreak\texttt{<b\_5571>}\allowbreak\texttt{<c\_2701>}\\
2026-01-24 [Search] local phone installment plan with zero down payment\\
2026-01-24 [Product-click] 99\%-new used iPhone 15 \texttt{<|prod\_begin|>}\allowbreak\texttt{<a\_0307>}\allowbreak\texttt{<b\_7024>}\allowbreak\texttt{<c\_4281>}\\
2026-01-28 [Search] checklist for inspecting a used phone\\
\textbf{Assistant:}\\
\texttt{\detokenize{<think>}}The user first searches for the newest flagship phone, forming a high-performance and high-price anchor. The later search for zero-down-payment installment shows a budget-driven path correction. The product click on a nearly new used phone indicates convergence toward a lower-entry-cost option, and the later inspection search supplements risk-control knowledge for the used-phone path. This is not same-level browsing of similar products, but a convergent decision chain driven by budget and transaction risk.\\
\texttt{\detokenize{</think>}}\\
\textit{Output: a JSON array containing the above smartphone decision chain.}
\end{mycase}
\end{CJK*}

\subsection{Recommendation CoT Construction}
\subsubsection{R3 Recommendation Reasoning Trace Construction Prompt}
\label{appendix:sft-recommendation-trace-prompt}

R3 uses the following prompt template, implemented as
\texttt{build\_i2i\_chain\_prompt\_cn}, to synthesize recommendation reasoning traces.
The prompt implements the three-stage protocol described in
Section~\ref{sec:sft-cot-synthesis}: persona abstraction, interest expansion,
and transition inference. During construction, the target interaction is visible
to the teacher for consistency checking, but the generated rationale must be
written as if the answer were unknown. The trace should therefore enumerate
possible interest directions, weigh them against the user profile and behavior
history, and avoid direct leakage of the target interaction. For readability,
the appendix shows a shortened version of the prompt with representative
persona examples.

\makeatletter
\@ifundefined{mycase}{%
  \newtcolorbox{mycase}[1]{%
    breakable,
    colback=gray!5, colframe=black!60, boxrule=0.5pt, arc=2pt,
    left=6pt, right=6pt, top=6pt, bottom=6pt,
    title={#1},
    fonttitle=\bfseries\small, coltitle=white, colbacktitle=black!70,
    before skip=8pt, after skip=8pt,
    before upper={\raggedright\sloppy}%
  }%
}{}
\makeatother

\begin{CJK*}{UTF8}{gbsn}
\begin{mycase}{Prompt SFT.R3.1: Recommendation Reasoning Trace Construction Prompt}
\small
\textbf{Prompt Template}

你是一个短视频平台的用户的推荐逻辑专家, 请你用简洁有条理的逻辑说清楚，我们应该如何根据下面这位用户的用户画像和之前的交互历史，通过发散思维推出所有的可能性，然后进行筛选，最终推导得到后续交互：

\textbf{\# 已知信息}

\textbf{\#\# 用户画像}\\
\texttt{\{user\_context\}}

\textbf{\#\# 用户在平台的历史记录}\\
\texttt{\{input\_data\}}

\textbf{\#\# 用户的后续交互(答案, 不可以显式出现在推理中)}\\
\texttt{\{target\_data\}}

\textbf{\# 分析技巧参考}
\begin{itemize}
    \item 关注每一个行为背后的动机：认真关注并深入分析用户每一条搜索词条，因为它们直观反映了用户的当前兴趣。你需要从搜索中解释出用户的真实需求。
    \item 时序敏感：你会注重事件发生的先后关系，关注后发生的事件是如何从先前的事件中演化得到的。
    \item 注重用户兴趣切换的成因：你会确保后续行为表现出对前期信息的关联筛选与聚焦过程。如果用户在 EventB 中完全抛弃了 EventA 的筛选条件且无逻辑解释，你会判定为无关联。有效关联示例：点击``高热量猫粮'' - 观看``猫咪软便调理视频易消化猫粮''。（分析它们的成因）
    \item 干扰项排除：你会排除基本不会产生前因后果的行为。例如：生活补货（买了就不会再买了，因此不会产生后续行为）或随机消费驱动的消费行为。
\end{itemize}

\textbf{\# 常见用户画像}

为你提供一些常见用户画像，用于快速捕捉群体特征，后续你可以参考该特征进行进一步分析。注意，你最后推荐的内容一定不要局限于推荐的大类，要具体情况具体分析，例如用户是二次元，喜欢看王者荣耀，那你不能仅仅分析到他喜欢打游戏，而要分析到他喜欢王者荣耀这一款游戏（因为他可能不喜欢玩别的，只喜欢这一款），然后针对性推荐：

\begin{itemize}
    \item 游戏直播观众：18-30岁男性为主，关注电竞赛事、主播日常、游戏攻略及外设开箱。高度依附于特定游戏或主播，跨游戏迁移意愿低。若用户频繁点击某款游戏相关内容或直播切片，可归为此类。重点关注：他具体追的游戏与主播（不做跨品类推荐），赛事战报、角色攻略、外设装备内容。
    \item Z世代潮流青年：16-25岁，一二线城市为主，对潮流文化极度敏感。追热门梗、流行音乐、街舞、说唱、潮牌穿搭，更新迭代快。若用户频繁互动挑战赛、踩点视频、潮牌开箱，可归为此类。重点关注：当前热门BGM与挑战赛、具体潮牌与款式、他具体追的说唱歌手或舞蹈风格内容。
    \item 二次元爱好者：16-28岁男女均有，对日本动漫文化有浓厚兴趣，圈层归属感强。若用户频繁点击动漫相关、虚拟偶像、二次元梗视频，可归为此类。重点关注：他看过的具体番剧、他追的具体Vtuber、特定IP手办与周边内容。
    \item 职场新人：22-28岁，应届或工作1-3年，一二线城市居多。关注面试技巧、职场潜规则、副业变现、城市租房攻略，对内容真实性要求高。若用户点击职场经验、简历优化、升职加薪相关内容，可归为此类。重点关注：具体行业职场经验（如互联网/金融）、副业类型（如自媒体/接单）、城市租房与生活成本攻略内容。
    \item 健身爱好者：20-35岁男女均有，集中在一二线城市，注重外形管理。关注减脂增肌教程、训练计划、蛋白粉测评、运动打卡。若用户高频点击健身教程、体脂分析、运动装备内容，可归为此类。重点关注：具体训练部位与方法（如背部/臀腿）、具体营养补剂品牌、他关注的具体健身博主内容。
    \item 美食爱好者：18-45岁男女均有，分布广泛，活跃时段集中在饭点前后（11点、17-19点）。喜欢家常菜教程、地方特色探店、吃播、食材选购。若用户集中在饭点点击菜谱、探店视频，可归为此类。重点关注：他喜欢的具体菜系（如川菜/粤菜）、具体菜品教程、所在城市或感兴趣城市的探店内容。
    \item 共用手机用户：无固定年龄，设备被多人（如夫妻、父母子女）轮流使用。行为数据中存在明显的风格跳变——如同一账号在家常菜、少儿动画、体育赛事之间无规律切换，且不同时段内容偏好截然不同。需优先识别并过滤非当前使用者的历史行为，聚焦当前活跃时段的近期稳定偏好。若画像标签与行为严重矛盾（如标注60岁但点赞内容全为电竞），或同一账号多个人群特征明显共存，应触发共用手机判断。重点关注：按时段拆分行为，以最近活跃行为为准，警惕被``他人历史''带偏。
\end{itemize}

【重要说明】以上画像仅供参考，不可机械套用。现实中用户往往横跨多个画像特征，或属于上述类别未覆盖的特殊群体。分析时应以用户的实际行为数据为第一依据，若用户特征与已知画像不符，请自行归纳总结其核心需求与偏好，切勿强行套用已有标签。

\textbf{\# 推理步骤}
\begin{enumerate}
    \item 用户画像归纳: 若证据充分，可以给出候选用户画像类型，并简短补充他相较于该类人群的特点，如: 由【证据, 列三条以内最直接的证据就行，不要超过30字】推测用户偏向【类型】，并且具体来说最喜欢的是【三个以内，写最明显的就行】。如果证据不足或不属于常见用户类型，请直接总结行为簇和核心需求，不要强制归类，例如: 用户近期集中关注 xx，最明显的内容需求推测为 xx...
    \item 可能性枚举： 从哪些关键信息可以得到, 用户最近最可能关心什么, 并列出所有可能的的兴趣点, 用 A, B, C ... 分点列出. 例如: A. 用户最近最关心的是 xx, 具体表现为 xxx, 另外, B. 由 xx 引发的兴趣也可能是用户发生该类行为的原因..
    \item 权衡后得到最终决定：对所有可能性进行权衡，分析每一个可能性有多大，是否是当前最大, 最终得到最终决定. 例如: 综合来看, A 是最有可能的, 因为 xxx, B 也有一定道理, 但 xxx 也可能由 A 导致, 因此...
\end{enumerate}

\textbf{\# 要求}
\begin{enumerate}
    \item[0.] 绝对不能暴露你早就知道你的后续交互是什么。要装作还不知道后续交互是什么的情况下进行推理。
    \item 输出一段连贯、简洁的分析 不要使用 JSON、列表格式或分主题的小标题。
    \item 仅包含你的推理即可. 简洁明了
    \item 最重要的一点, 你可以省略任何你觉得不需要的步骤。 例如你觉得推理过程过于简单或者不完整，那么你可以直接跳过某些步骤，只给出最终结论。
    \item 当提及交互历史中的具体视频、商品或广告时, \textbf{必须使用原始ID并用单独括号标注, 因为我们后续要使用他进行匹配}, 如(video\_id=XXXXX)、(ec\_id=YYYYY)、(ad\_id=ZZZZZ)。不要在同一括号中包含多个条目，不要使用标题或名称。
\end{enumerate}

你的推理：

\vspace{0.75em}
\hrulefill
\vspace{0.75em}

\textbf{Prompt Template}

You are a recommendation-logic expert for users on a short-video platform. Please use concise and well-structured logic to explain how, given the following user profile and previous interaction history, we should use divergent reasoning to derive all possible directions, filter them, and finally infer the user's subsequent interaction.

\textbf{\# Known Information}

\textbf{\#\# User Profile}\\
\texttt{\{user\_context\}}

\textbf{\#\# User History on the Platform}\\
\texttt{\{input\_data\}}

\textbf{\#\# User's Subsequent Interaction (answer; must not appear explicitly in the reasoning)}\\
\texttt{\{target\_data\}}

\textbf{\# Analysis Heuristics}
\begin{itemize}
    \item \textbf{Behavior motivation:} Pay close attention to and deeply analyze each search query, because search terms directly reflect the user's current interest. You need to infer the user's real need from these searches.
    \item \textbf{Temporal sensitivity:} Pay attention to the chronological order of events and explain how later events evolve from earlier ones.
    \item \textbf{Interest-transition cause:} Ensure that later behaviors show filtering and focusing based on earlier information. If Event B completely abandons the filtering conditions from Event A without a logical explanation, treat the two as unrelated. A valid relation example is clicking ``high-calorie cat food'' and then watching ``easy-to-digest cat food for cats with soft stool''; analyze the cause of this relation.
    \item \textbf{Noise filtering:} Exclude behaviors that usually do not create cause-and-effect follow-ups, such as routine replenishment or random consumption.
\end{itemize}

\textbf{\# Common User Personas}

The following common personas are provided to help quickly capture group-level characteristics. They can be used as references for later analysis. However, the final recommendation must not stop at a broad category. It should be analyzed case by case. For example, if the user is an ACG fan who likes Honor of Kings, do not merely say the user likes games; infer that the user likes Honor of Kings specifically, because the user may not like other games.

\begin{itemize}
    \item \textbf{Game live-streaming viewer:} Mostly men aged 18--30 who follow esports, streamer clips, game guides, and gear unboxing. They are often attached to a specific game or streamer and have low willingness to transfer across games. If the user frequently clicks content related to a specific game or livestream clips, this persona may apply. Focus on the exact game and streamer they follow, and recommend tournament recaps, character guides, and gaming gear content without making cross-category recommendations.
    \item \textbf{Gen-Z trend follower:} Young users aged 16--25 in major cities, highly sensitive to memes, pop music, street dance, rap, fashion, and challenges. Focus on current BGM, challenges, fashion brands, singers, and dance styles.
    \item \textbf{Anime enthusiast:} Users aged 16--28 of any gender with strong interest in Japanese anime culture and a strong sense of community belonging. If the user frequently clicks anime-related content, virtual idols, or ACG memes, this persona may apply. Focus on the exact shows they have watched, the Vtubers they follow, and specific IP figures and merchandise.
    \item \textbf{Early-career worker:} Users aged 22--28, often recent graduates or with 1--3 years of work experience, mostly in major cities. They pay attention to interview skills, workplace norms, side-income opportunities, and city rental guides, and they expect content to feel credible. If the user clicks workplace experience, resume advice, or promotion-related content, this persona may apply. Focus on concrete industry-specific workplace experience, side-job types, and city renting or living-cost guidance.
    \item \textbf{Fitness enthusiast:} Users aged 20--35 of any gender, often in major cities, who care about appearance management. They follow fat-loss and muscle-gain tutorials, training plans, protein supplement reviews, and workout check-ins. If the user frequently clicks fitness tutorials, body-fat analysis, or sports gear content, this persona may apply. Focus on specific training body parts and methods, supplement brands, and the fitness creators they follow.
    \item \textbf{Food lover:} Users aged 18--45 of any gender, broadly distributed across regions, and especially active around meal times (11:00 and 17:00--19:00). They enjoy home-cooking tutorials, local restaurant exploration, mukbang, and ingredient selection. If the user tends to click recipes or restaurant videos around meal times, this persona may apply. Focus on the exact cuisine, dish tutorials, and the cities they live in or want to explore.
    \item \textbf{Shared-device user:} There is no fixed age range, and the device may be shared by multiple people, such as spouses or parents and children. The behavior log may show obvious style jumps, for example abrupt switching among home cooking, children's animation, and sports content, with very different preferences across time slots. Prioritize identifying and filtering out the history of non-current users, and focus on the recent stable preferences during the currently active time period. If the persona label and behavior strongly contradict each other, or multiple user-type signals coexist in one account, shared-device use should be considered. Focus on splitting behavior by time and avoiding being misled by other users' history.
\end{itemize}

\textbf{Important note:} These personas are references only and must not be applied mechanically. Real users may span multiple persona features or fall outside these categories. Always prioritize actual behavior data. If the user does not fit an existing persona, summarize the user's core needs and preferences directly.

\textbf{\# Reasoning Steps}
\begin{enumerate}
    \item \textbf{Persona abstraction:} If the evidence is sufficient, provide a candidate persona type and briefly describe what distinguishes this user from that group. If the evidence is insufficient, mixed, or outside the listed personas, summarize the observed behavior clusters and core needs directly instead of forcing a category label.
    \item \textbf{Possibility enumeration:} Identify key information indicating what the user is most likely to care about recently, and list all possible interest points as A, B, C, and so on.
    \item \textbf{Final weighted decision:} Weigh all possibilities, analyze how likely each one is, determine which is currently strongest, and reach the final decision.
\end{enumerate}

\textbf{\# Requirements}
\begin{enumerate}
    \item[0.] Never reveal that you already know the subsequent interaction. Reason as if the subsequent interaction is unknown.
    \item Output one coherent and concise analysis paragraph. Do not use JSON, list format, or section headings.
    \item Include only the reasoning. Keep it concise and clear.
    \item You may skip any step that you find unnecessary. If the reasoning process is simple or incomplete, directly provide the final conclusion.
    \item When mentioning specific videos, products, or ads from the history, use the original ID in a separate parenthesis because it will be used for later matching, e.g., (video\_id=XXXXX), (ec\_id=YYYYY), or (ad\_id=ZZZZZ). Do not put multiple entries in one parenthesis, and do not use titles or names.
\end{enumerate}

Your reasoning:
\end{mycase}
\end{CJK*}

\subsubsection{Low-Score Reasoning-Trace Cases}
\label{appendix:sft-cot-quality-low-score-cases}

\begin{table}[H]
\centering
\caption{Representative low-quality recommendation CoT cases. Problematic CoT text is marked in red; target fragments are reference evidence.}
\label{tab:sft-cot-quality-low-score-cases}
\setlength{\tabcolsep}{3pt}
\renewcommand{\arraystretch}{1.12}
\scriptsize
\begin{CJK*}{UTF8}{gbsn}
\begin{tabularx}{\linewidth}{@{}>{\raggedright\arraybackslash}p{0.17\linewidth}>{\raggedright\arraybackslash}X@{}}
\toprule
\textbf{Low dimension} & \textbf{Behavior / target / CoT fragment with analysis} \\
\midrule
\textbf{Consistency}
&
\textbf{Input:} 用户点赞最多的是房产家居和宠物，完播最多的是明星娱乐和房产家居；近期评论多次怀念虎皮鹦鹉，也搜索“附近2公里出售二手鹦鹉”“亚克力鸟笼”“虎皮鹦鹉”。商业正反馈同时出现“建华大街小高层现房，首付1万月供两千多”“一楼带院真便宜”等本地房产线索。\par\smallskip
\textbf{Target:} 石家庄建华大街旁低价小高层现房，两居室、交通便利，面向关注经济实惠住房的购房者。\par\smallskip
\textbf{CoT:} 认为宠物鹦鹉是最强即时兴趣，房产只是间歇关注，后续推荐应 {\color{red}紧密围绕鹦鹉饲养}，例如虎皮鹦鹉选购、手养经验、{\color{red}鸟笼用品测评} 或宠物内容创作技巧。
\par\smallskip
{\footnotesize\color{gray}\emph{Analysis:} the target-facing housing signals are present in the behavior evidence, but the trace suppresses them and ends on pet-care recommendations. The conclusion may fit part of the history, yet it does not support the intended local-housing target.} \\
\midrule
\textbf{Factuality}
&
\textbf{Input:} 用户评论常出现“马到成功，我接住”“谢谢佛祖保佑”“好运连连”等祈福表达；短剧消费集中在家庭伦理、复仇和寻亲题材；购物与加购包含铝箔胶带、调味品、扒鸡、鞋油、墙纸等实用家居或食品。\par\smallskip
\textbf{Target:} 一位中年女性主播的生活互动和戏曲才艺直播，标签包括情感抒发、室内明亮氛围、戏曲文化和生活化才艺展示。\par\smallskip
\textbf{CoT:} 解释祈福需求时写道“临近春节，{\color{red}购物中有年货窗花}，这种心理需求可能强化”，并据此推断用户会对团圆、转运主题内容产生兴趣。
\par\smallskip
{\footnotesize\color{gray}\emph{Analysis:} the highlighted purchase evidence is not present in the displayed behavior fragment. It makes the seasonal argument look grounded, but the support comes from a fabricated detail, so this is a factuality failure rather than merely weak target alignment.} \\
\midrule
\textbf{Informativeness}
&
\textbf{Input:} 用户是31--40岁男性，近期搜索“燕窝八宝粥”“洗浴筐”“穿越秦朝赵龙短剧”；关注博主中颜值占比较高，商业正反馈包含金属加工、焊接、美缝、越野车改装等实用技巧，购买侧有食品调料、工装、钓鱼用品和话费等信号。\par\smallskip
\textbf{Target:} 观赏鱼喂养指导视频，讲解喂食频率、喂食量、上浮性鱼粮、避免水质变浑等养鱼技巧。\par\smallskip
\textbf{CoT:} 最终认为应围绕家庭育儿或生活技巧展开，尤其是 {\color{red}婴幼儿护理、喂养知识}，并补充说 {\color{red}他近期没有相关内容的互动记录，因此这是一个潜在的兴趣缺口}。
\par\smallskip
{\footnotesize\color{gray}\emph{Analysis:} the trace lists several user interests, but the final guidance does not help predict or explain the fish-feeding target. It invents a broad ``interest gap'' and gives no actionable attributes about aquarium care, fish food, or feeding routines.} \\
\midrule
\textbf{Logic}
&
\textbf{Input:} 用户近期搜索“婚期已定邀请函”“瑶山大剿匪第40集”“母画眉求爱扇翅视频”“母鸟扇翅求爱大叫”和“坦克摩托车”；短剧消费偏向逆袭、复仇、悬疑，商业正反馈含榫卯结构、农村自建房、摩托车和小游戏。\par\smallskip
\textbf{Target:} 年轻男子在室内吹奏长笛并展示音乐才华，背景为沙发和室内布置，整体是音乐表演内容。\par\smallskip
\textbf{CoT:} 把“母画眉求爱”作为主线，最终判断用户下一步最可能继续探索 {\color{red}画眉鸟求偶、饲养或训练} 相关内容。
\par\smallskip
{\footnotesize\color{gray}\emph{Analysis:} the trace over-commits to one recent search and does not build a bridge to the music-performance target. A logical trace should either explain an evidence-supported transition toward instrumental music or reject the weak relation; here the final conclusion follows a disconnected branch.} \\
\midrule
\textbf{Safety}
&
\textbf{Input:} 用户是50岁以上城镇退休男性，近期多次点击健康管理应用广告、体检饮食提醒、睡眠健康和本地中医男士调理服务；同时也有交友类应用活跃、女性展示/同城聊天广告点击、汽车养护和实用购物行为。历史中出现过“男士功能调理”类本地服务广告，但后续答案仍应被隐藏。\par\smallskip
\textbf{Target:} 一条男性性健康检查服务广告，画面含“早泄/阳痿检查”海报、低价检查、地址和联系方式等信息，目标是关注性健康和本地调理服务的人群。\par\smallskip
\textbf{CoT:} 在权衡健康与社交需求后，直接写道用户后续最可能交互的是 {\color{red}围绕男性健康，如性功能、前列腺等的本地化解决方案广告}，并在结尾进一步点明 {\color{red}一个关于男性性健康检查或调理的广告，如早泄阳痿检查} 是高度可能的后续方向。
\par\smallskip
{\footnotesize\color{gray}\emph{Analysis:} this is a direct target-leakage failure. The trace should reason from prior health-service and local-service evidence without naming the hidden answer, but it exposes the target category and even repeats the target-specific sensitive phrase, making the answer recoverable from the CoT itself.} \\
\bottomrule
\end{tabularx}
\end{CJK*}
\end{table}

\subsubsection{Recommendation CoT Training Cases}
\label{appendix:sft-recommendation-cot-cases}

The final recommendation CoT data uses a three-turn message format: a short
system instruction, a user message containing the user profile and multi-domain
behavior history, and an assistant message that first produces a
\texttt{\detokenize{<think>}} reasoning trace and then emits the target itemic
token. The assistant trace follows the same three-stage pattern used in the main
R3 construction: it first abstracts a persona prior, then expands several
evidence-backed interest hypotheses, and finally compares them to infer the most
plausible transition direction. During construction, the teacher trace marks
concrete historical evidence with original domain ids, and the final SFT sample
replaces matched ids with itemic tokens. The following cases therefore show the
post-processed training format: natural-language reasoning interleaved with
synthetic itemic-token references inside the CoT, followed by the target itemic-token answer. We
omit raw user identifiers, data paths, timestamps, and original item ids.

\makeatletter
\@ifundefined{mycase}{%
  \newtcolorbox{mycase}[1]{%
    breakable,
    colback=gray!5, colframe=black!60, boxrule=0.5pt, arc=2pt,
    left=6pt, right=6pt, top=6pt, bottom=6pt,
    title={#1},
    fonttitle=\bfseries\small, coltitle=white, colbacktitle=black!70,
    before skip=8pt, after skip=8pt,
    before upper={\raggedright\sloppy}%
  }%
}{}
\makeatother

\begin{CJK*}{UTF8}{gbsn}
\begin{mycase}{Prompt SFT.R3.2: Cross-Domain Recommendation with Text--Itemic-Token CoT}
\small
\textbf{中文示例} \\
\textbf{System:} 你擅长理解用户画像、跨场景行为和 itemic tokens，请根据输入信息归纳该用户的目标内容。\\
\textbf{User:} 以下是一个用户的画像与多域历史行为信息：\\
该用户是一名年轻男性学生，长期活跃在短视频、直播、电商和广告场景中。用户安装并活跃使用射击类游戏和短视频应用；近期搜索多次围绕“和平精英攻略”“主播操作”“游戏准星设置”等内容展开。\\
用户在视频域多次长播和转发游戏技巧视频，例如 \texttt{<|video\_begin|>}\allowbreak\texttt{<a\_3062>}\allowbreak\texttt{<b\_4733>}\allowbreak\texttt{<c\_4451>} 和 \texttt{<|video\_begin|>}\allowbreak\texttt{<a\_5708>}\allowbreak\texttt{<b\_2219>}\allowbreak\texttt{<c\_7034>}；在直播域关注了多位游戏主播，如 \texttt{<|living\_begin|>}\allowbreak\texttt{<a\_6815>}\allowbreak\texttt{<b\_2789>}\allowbreak\texttt{<c\_3026>} 和 \texttt{<|living\_begin|>}\allowbreak\texttt{<a\_6263>}\allowbreak\texttt{<b\_5727>}\allowbreak\texttt{<c\_2164>}；\\
在电商域浏览并购买电竞指套、手机散热夹等游戏外设，如 \texttt{<|prod\_begin|>}\allowbreak\texttt{<a\_4180>}\allowbreak\texttt{<b\_6532>}\allowbreak\texttt{<c\_0907>}。同时，用户也消费复仇爽剧、颜值自拍和少量怀旧零食内容。\\
\textbf{Assistant:}\\
\texttt{\detokenize{<think>}}由用户画像中的年轻男性学生身份，以及近期围绕射击游戏的搜索、直播关注和游戏外设购买，可以判断该用户最稳定且最强的兴趣是《和平精英》相关内容。历史中的游戏技巧视频 \texttt{<|video\_begin|>}\allowbreak\texttt{<a\_3062>}\allowbreak\texttt{<b\_4733>}\allowbreak\texttt{<c\_4451>} 和高手操作合集 \texttt{<|video\_begin|>}\allowbreak\texttt{<a\_5708>}\allowbreak\texttt{<b\_2219>}\allowbreak\texttt{<c\_7034>} 不是孤立观看，它们和直播关注 \texttt{<|living\_begin|>}\allowbreak\texttt{<a\_6815>}\allowbreak\texttt{<b\_2789>}\allowbreak\texttt{<c\_3026>}、\texttt{<|living\_begin|>}\allowbreak\texttt{<a\_6263>}\allowbreak\texttt{<b\_5727>}\allowbreak\texttt{<c\_2164>} 一起指向“看主播操作并学习打法”的需求。电商侧的电竞指套和手机散热夹 \texttt{<|prod\_begin|>}\allowbreak\texttt{<a\_4180>}\allowbreak\texttt{<b\_6532>}\allowbreak\texttt{<c\_0907>} 进一步说明用户已经从泛娱乐观看升级到提升实际操作表现。\\
在可能的兴趣方向中，A 是《和平精英》的高阶战术、主播精彩操作和设备配置；B 是复仇爽剧与打脸剧情，这是用户长期的休闲内容；C 是颜值自拍类视频，属于稳定但较浅的基础偏好；D 是怀旧零食和乡土生活片段，证据较零散。综合近期密度、跨域一致性和行为投入度，A 的权重最高，因为视频、直播和商品 itemic tokens 共同支持游戏能力提升这条转移路径。\\
因此，后续推荐应优先落到《和平精英》的实战技巧、高手操作合集、灵敏度和准星配置教学，同时可少量保留复仇爽剧或颜值内容作为辅助兴趣。\texttt{\detokenize{</think>}}\\
该用户最近喜欢的视频有: \texttt{<|video\_begin|>}\allowbreak\texttt{<a\_0016>}\allowbreak\texttt{<b\_1906>}\allowbreak\texttt{<c\_0378>}

\vspace{0.8em}
\textbf{English Translation} \\
\textbf{System:} You are good at understanding user profiles, cross-scenario behavior, and itemic tokens. Infer the user's target content from the input information.\\
\textbf{User:} The following is a user's profile and multi-domain behavior history:\\
The user is a young male student who is active across short-video, live-streaming, e-commerce, and advertising scenarios. He installs and actively uses shooting games and short-video apps. Recent searches repeatedly involve ``Peace Elite guides'', ``streamer operations'', and ``crosshair settings''.\\
In the video domain, the user repeatedly long-viewed and shared game-skill videos such as \texttt{<|video\_begin|>}\allowbreak\texttt{<a\_3062>}\allowbreak\texttt{<b\_4733>}\allowbreak\texttt{<c\_4451>} and \texttt{<|video\_begin|>}\allowbreak\texttt{<a\_5708>}\allowbreak\texttt{<b\_2219>}\allowbreak\texttt{<c\_7034>}. In the live domain, he followed several game streamers such as \texttt{<|living\_begin|>}\allowbreak\texttt{<a\_6815>}\allowbreak\texttt{<b\_2789>}\allowbreak\texttt{<c\_3026>} and \texttt{<|living\_begin|>}\allowbreak\texttt{<a\_6263>}\allowbreak\texttt{<b\_5727>}\allowbreak\texttt{<c\_2164>}.\\
In the e-commerce domain, he browsed and purchased gaming accessories such as finger sleeves and phone cooling clips, represented by \texttt{<|prod\_begin|>}\allowbreak\texttt{<a\_4180>}\allowbreak\texttt{<b\_6532>}\allowbreak\texttt{<c\_0907>}. The user also consumes revenge-oriented short dramas, attractiveness/selfie videos, and a small amount of nostalgic-snack content.\\
\textbf{Assistant:}\\
\texttt{\detokenize{<think>}}Based on the young male student profile, together with recent searches, live follows, and gaming-accessory purchases around shooting games, the strongest and most stable interest is Peace Elite-related content. The game-skill video \texttt{<|video\_begin|>}\allowbreak\texttt{<a\_3062>}\allowbreak\texttt{<b\_4733>}\allowbreak\texttt{<c\_4451>} and the expert-operation compilation \texttt{<|video\_begin|>}\allowbreak\texttt{<a\_5708>}\allowbreak\texttt{<b\_2219>}\allowbreak\texttt{<c\_7034>} are not isolated views; together with game-streamer follows \texttt{<|living\_begin|>}\allowbreak\texttt{<a\_6815>}\allowbreak\texttt{<b\_2789>}\allowbreak\texttt{<c\_3026>} and \texttt{<|living\_begin|>}\allowbreak\texttt{<a\_6263>}\allowbreak\texttt{<b\_5727>}\allowbreak\texttt{<c\_2164>}, they indicate a need to learn streamer operations and gameplay tactics. The e-commerce item \texttt{<|prod\_begin|>}\allowbreak\texttt{<a\_4180>}\allowbreak\texttt{<b\_6532>}\allowbreak\texttt{<c\_0907>} further shows that the user has moved from watching game content to improving actual gameplay performance.\\
Among possible directions, A is advanced Peace Elite tactics, streamer highlights, and device configuration; B is revenge-oriented short drama, which is a stable leisure preference; C is attractiveness/selfie content, a shallow but persistent baseline preference; and D is nostalgic snacks or rural-life clips, supported only by scattered evidence. Considering recent density, cross-domain consistency, and investment level, A receives the highest weight because the video, live, and product itemic tokens jointly support the game-skill-improvement path.\\
Therefore, the recommendation should prioritize Peace Elite practical tutorials, expert-operation compilations, sensitivity and crosshair configuration guides, while keeping revenge drama or attractiveness content as auxiliary interests.\texttt{\detokenize{</think>}}\\
The user may recently like the following video: \texttt{<|video\_begin|>}\allowbreak\texttt{<a\_0016>}\allowbreak\texttt{<b\_1906>}\allowbreak\texttt{<c\_0378>}
\end{mycase}
\end{CJK*}
\begin{CJK*}{UTF8}{gbsn}
\begin{mycase}{Prompt SFT.R3.3: Purchase-Oriented Recommendation with Text--Itemic-Token CoT}
\small
\textbf{中文示例} \\
\textbf{System:} 你是推荐理解助手。你需要根据用户画像与多域历史行为，输出该用户在各推荐场景中的目标内容。\\
\textbf{User:} 下面给出用户画像及其历史行为线索：\\
该用户是一名中年女性，关注生活、穿搭、美食、美妆和短剧内容。她近期多次搜索“高档小衫”“同款连衣裙”“睡衣男”“东北特色菜”等关键词，关注的主播中包含服饰带货、美妆和生活类账号。\\
用户在电商域频繁浏览和购买女装、男士打底衫、睡衣、羽绒服和食品，例如 \texttt{<|prod\_begin|>}\allowbreak\texttt{<a\_1927>}\allowbreak\texttt{<b\_6270>}\allowbreak\texttt{<c\_1880>}、\texttt{<|prod\_begin|>}\allowbreak\texttt{<a\_2388>}\allowbreak\texttt{<b\_4708>}\allowbreak\texttt{<c\_0107>} 和 \texttt{<|prod\_begin|>}\allowbreak\texttt{<a\_6821>}\allowbreak\texttt{<b\_3569>}\allowbreak\texttt{<c\_6806>}；在广告域深度点击家庭伦理与复仇短剧广告，如 \texttt{<|ad\_begin|>}\allowbreak\texttt{<a\_3578>}\allowbreak\texttt{<b\_6234>}\allowbreak\texttt{<c\_3026>}；在视频域点赞和收藏美食、情感、穿搭内容。\\
\textbf{Assistant:}\\
\texttt{\detokenize{<think>}}该用户的画像和行为更接近中年家庭主妇，同时带有明显的直播购物偏好。她的长期内容基本盘是家庭伦理、复仇逆袭短剧和生活化美食，但近期最密集、最主动的行为集中在服装穿搭与家庭采购上：女装浏览 \texttt{<|prod\_begin|>}\allowbreak\texttt{<a\_1927>}\allowbreak\texttt{<b\_6270>}\allowbreak\texttt{<c\_1880>}、男士打底衫 \texttt{<|prod\_begin|>}\allowbreak\texttt{<a\_2388>}\allowbreak\texttt{<b\_4708>}\allowbreak\texttt{<c\_0107>} 和食品补货 \texttt{<|prod\_begin|>}\allowbreak\texttt{<a\_6821>}\allowbreak\texttt{<b\_3569>}\allowbreak\texttt{<c\_6806>} 共同说明她在为自己和家人做实际购买决策。广告侧的家庭伦理与复仇短剧 \texttt{<|ad\_begin|>}\allowbreak\texttt{<a\_3578>}\allowbreak\texttt{<b\_6234>}\allowbreak\texttt{<c\_3026>} 则提供情绪内容背景，但不如服饰采购链路直接。\\
可能方向包括 A. 大码或中年女性穿搭，尤其是显瘦、丝绒、国风和春季外套；B. 家庭伦理与女性逆袭短剧，承接其高粘性的短剧兴趣；C. 东北食品和家常菜，服务于家庭生活补货；D. 美妆小物，属于个人形象管理的辅助需求。综合近期搜索、商品点击和购买闭环，A 的证据最强，因为多个商品 itemic tokens 指向服装和家庭采购，而广告 itemic tokens 更像辅助兴趣。\\
因此，后续推荐更适合推送高品质、性价比明确、适合中年女性或家庭场景的穿搭视频或带货内容，并可结合家庭剧情或生活化表达增强吸引力。\texttt{\detokenize{</think>}}\\
该用户最近喜欢的视频有: \texttt{<|video\_begin|>}\allowbreak\texttt{<a\_2713>}\allowbreak\texttt{<b\_0084>}\allowbreak\texttt{<c\_5586>}

\vspace{0.8em}
\textbf{English Translation} \\
\textbf{System:} You are a recommendation-understanding assistant. Given the user profile and multi-domain behavior, output the user's target content in recommendation scenarios.\\
\textbf{User:} The following are the user's profile and historical behavior signals:\\
The user is a middle-aged woman interested in daily life, outfits, food, beauty, and short dramas. Recent searches include ``high-end blouse'', ``same-style dress'', ``men's sleepwear'', and ``Northeastern home dishes''. Followed creators include apparel-selling, beauty, and lifestyle accounts.\\
In e-commerce, the user frequently browsed and purchased women's clothing, men's thermal tops, sleepwear, down jackets, and food items, such as \texttt{<|prod\_begin|>}\allowbreak\texttt{<a\_1927>}\allowbreak\texttt{<b\_6270>}\allowbreak\texttt{<c\_1880>}, \texttt{<|prod\_begin|>}\allowbreak\texttt{<a\_2388>}\allowbreak\texttt{<b\_4708>}\allowbreak\texttt{<c\_0107>}, and \texttt{<|prod\_begin|>}\allowbreak\texttt{<a\_6821>}\allowbreak\texttt{<b\_3569>}\allowbreak\texttt{<c\_6806>}.\\
In ads, she deeply engaged with family-ethics and revenge-drama advertisements such as \texttt{<|ad\_begin|>}\allowbreak\texttt{<a\_3578>}\allowbreak\texttt{<b\_6234>}\allowbreak\texttt{<c\_3026>}. In videos, she liked and saved food, emotion, and outfit content.\\
\textbf{Assistant:}\\
\texttt{\detokenize{<think>}}The profile and behavior fit a middle-aged homemaker with a strong preference for live shopping. Her long-term baseline is family-ethics and revenge-oriented short dramas, together with daily-life food content, but the densest recent behavior centers on outfits and household purchasing. The women's-clothing item \texttt{<|prod\_begin|>}\allowbreak\texttt{<a\_1927>}\allowbreak\texttt{<b\_6270>}\allowbreak\texttt{<c\_1880>}, the men's thermal top \texttt{<|prod\_begin|>}\allowbreak\texttt{<a\_2388>}\allowbreak\texttt{<b\_4708>}\allowbreak\texttt{<c\_0107>}, and the food-replenishment item \texttt{<|prod\_begin|>}\allowbreak\texttt{<a\_6821>}\allowbreak\texttt{<b\_3569>}\allowbreak\texttt{<c\_6806>} jointly show practical purchasing for herself and her family. The family-ethics ad \texttt{<|ad\_begin|>}\allowbreak\texttt{<a\_3578>}\allowbreak\texttt{<b\_6234>}\allowbreak\texttt{<c\_3026>} supplies an emotional-content background, but it is less direct than the outfit-purchase chain.\\
Possible directions include A. midlife women's outfits, especially slimming, velvet, Chinese-style, and spring outerwear; B. family-ethics and female-comeback dramas; C. Northeastern food and home cooking for household replenishment; and D. small beauty items for personal image management. Considering recent search, product-click, and purchase signals, A has the strongest evidence because multiple product itemic tokens point to outfit and household purchase decisions, while the ad itemic tokens are a secondary preference.\\
Therefore, suitable recommendations should emphasize high-quality, cost-effective outfit videos or shopping content for midlife women and family scenarios, optionally expressed through family-story or daily-life contexts.\texttt{\detokenize{</think>}}\\
The user may recently like the following video: \texttt{<|video\_begin|>}\allowbreak\texttt{<a\_2713>}\allowbreak\texttt{<b\_0084>}\allowbreak\texttt{<c\_5586>}
\end{mycase}
\end{CJK*}

\subsection{Itemic Instruction Details}
\subsubsection{Itemic Token Instruction-Following Cases}
\label{appendix:sft-sid-instruction-cases}

Itemic Token Instruction Data covers task execution under mixed natural-language and itemic-token inputs. The cases below are representative sanitized examples drawn from the observed task patterns in the instruction data. We use synthetic itemic tokens and omit raw user identifiers, data paths, timestamps, and internal metadata.

\makeatletter
\@ifundefined{mycase}{%
  \newtcolorbox{mycase}[1]{%
    breakable,
    colback=gray!5, colframe=black!60, boxrule=0.5pt, arc=2pt,
    left=6pt, right=6pt, top=6pt, bottom=6pt,
    title={#1},
    fonttitle=\bfseries\small, coltitle=white, colbacktitle=black!70,
    before skip=8pt, after skip=8pt,
    before upper={\raggedright\sloppy}%
  }%
}{}
\makeatother

\begin{CJK*}{UTF8}{gbsn}
\begin{mycase}{Prompt SFT.IIF.1: Comparing Two Product Itemic Tokens}
\small
\textbf{中文示例} \\
\textbf{User:} 请对比以下两款商品的适用场景，并总结它们各自的主打功能。\\
1. \texttt{<|prod\_begin|>}\allowbreak\texttt{<a\_1201>}\allowbreak\texttt{<b\_3402>}\allowbreak\texttt{<c\_5603>}\\
2. \texttt{<|prod\_begin|>}\allowbreak\texttt{<a\_7704>}\allowbreak\texttt{<b\_2250>}\allowbreak\texttt{<c\_8191>}\\
\textbf{Assistant:}\\
1. \texttt{<|prod\_begin|>}\allowbreak\texttt{<a\_1201>}\allowbreak\texttt{<b\_3402>}\allowbreak\texttt{<c\_5603>} 更适合日常面部护理场景，主打肤色提亮、淡化瑕疵和基础养护。\\
2. \texttt{<|prod\_begin|>}\allowbreak\texttt{<a\_7704>}\allowbreak\texttt{<b\_2250>}\allowbreak\texttt{<c\_8191>} 更适合营养补充场景，主打肠道调节和免疫支持。\\
两者都服务于健康与生活质量提升，但前者面向外用护肤，后者面向内服调理，使用方式和需求入口明显不同。

\vspace{0.5em}
\textbf{English Example} \\
\textbf{User:} Compare the following two products in terms of usage scenarios, and summarize the main function of each product.\\
1. \texttt{<|prod\_begin|>}\allowbreak\texttt{<a\_1201>}\allowbreak\texttt{<b\_3402>}\allowbreak\texttt{<c\_5603>}\\
2. \texttt{<|prod\_begin|>}\allowbreak\texttt{<a\_7704>}\allowbreak\texttt{<b\_2250>}\allowbreak\texttt{<c\_8191>}\\
\textbf{Assistant:}\\
1. \texttt{<|prod\_begin|>}\allowbreak\texttt{<a\_1201>}\allowbreak\texttt{<b\_3402>}\allowbreak\texttt{<c\_5603>} is better suited for daily facial-care scenarios, focusing on brightening, blemish reduction, and basic skin maintenance.\\
2. \texttt{<|prod\_begin|>}\allowbreak\texttt{<a\_7704>}\allowbreak\texttt{<b\_2250>}\allowbreak\texttt{<c\_8191>} is better suited for nutritional supplementation, focusing on gut regulation and immune support.\\
Both improve everyday well-being, but the former is an external skincare product, while the latter is an ingestible health supplement.
\end{mycase}
\end{CJK*}

\begin{CJK*}{UTF8}{gbsn}
\begin{mycase}{Prompt SFT.IIF.2: Set Operation over Itemic Token Lists}
\small
\textbf{中文示例} \\
\textbf{User:} 请分析以下两组视频列表，并输出两组共同包含的视频 Token。\\
第一组：\\
1. \texttt{<|video\_begin|>}\allowbreak\texttt{<a\_3108>}\allowbreak\texttt{<b\_6412>}\allowbreak\texttt{<c\_2046>}\\
2. \texttt{<|video\_begin|>}\allowbreak\texttt{<a\_5520>}\allowbreak\texttt{<b\_1183>}\allowbreak\texttt{<c\_6905>}\\
3. \texttt{<|video\_begin|>}\allowbreak\texttt{<a\_4807>}\allowbreak\texttt{<b\_2701>}\allowbreak\texttt{<c\_7338>}\\
第二组：\\
1. \texttt{<|video\_begin|>}\allowbreak\texttt{<a\_5520>}\allowbreak\texttt{<b\_1183>}\allowbreak\texttt{<c\_6905>}\\
2. \texttt{<|video\_begin|>}\allowbreak\texttt{<a\_9056>}\allowbreak\texttt{<b\_4420>}\allowbreak\texttt{<c\_1374>}\\
3. \texttt{<|video\_begin|>}\allowbreak\texttt{<a\_3108>}\allowbreak\texttt{<b\_6412>}\allowbreak\texttt{<c\_2046>}\\
\textbf{Assistant:} 两组列表中共同包含的视频 Token 为：\\
1. \texttt{<|video\_begin|>}\allowbreak\texttt{<a\_3108>}\allowbreak\texttt{<b\_6412>}\allowbreak\texttt{<c\_2046>}\\
2. \texttt{<|video\_begin|>}\allowbreak\texttt{<a\_5520>}\allowbreak\texttt{<b\_1183>}\allowbreak\texttt{<c\_6905>}

\vspace{0.5em}
\textbf{English Example} \\
\textbf{User:} Compare the following two video-token lists and output all video tokens that appear in both lists.\\
List 1:\\
1. \texttt{<|video\_begin|>}\allowbreak\texttt{<a\_3108>}\allowbreak\texttt{<b\_6412>}\allowbreak\texttt{<c\_2046>}\\
2. \texttt{<|video\_begin|>}\allowbreak\texttt{<a\_5520>}\allowbreak\texttt{<b\_1183>}\allowbreak\texttt{<c\_6905>}\\
3. \texttt{<|video\_begin|>}\allowbreak\texttt{<a\_4807>}\allowbreak\texttt{<b\_2701>}\allowbreak\texttt{<c\_7338>}\\
List 2:\\
1. \texttt{<|video\_begin|>}\allowbreak\texttt{<a\_5520>}\allowbreak\texttt{<b\_1183>}\allowbreak\texttt{<c\_6905>}\\
2. \texttt{<|video\_begin|>}\allowbreak\texttt{<a\_9056>}\allowbreak\texttt{<b\_4420>}\allowbreak\texttt{<c\_1374>}\\
3. \texttt{<|video\_begin|>}\allowbreak\texttt{<a\_3108>}\allowbreak\texttt{<b\_6412>}\allowbreak\texttt{<c\_2046>}\\
\textbf{Assistant:} The video tokens appearing in both lists are:\\
1. \texttt{<|video\_begin|>}\allowbreak\texttt{<a\_3108>}\allowbreak\texttt{<b\_6412>}\allowbreak\texttt{<c\_2046>}\\
2. \texttt{<|video\_begin|>}\allowbreak\texttt{<a\_5520>}\allowbreak\texttt{<b\_1183>}\allowbreak\texttt{<c\_6905>}
\end{mycase}
\end{CJK*}

\begin{CJK*}{UTF8}{gbsn}
\begin{mycase}{Prompt SFT.IIF.3: Instruction Control under Distracting Itemic Tokens}
\small
\textbf{中文示例} \\
\textbf{User:} 不用理会这些视频标记：\\
1. \texttt{<|video\_begin|>}\allowbreak\texttt{<a\_5061>}\allowbreak\texttt{<b\_7220>}\allowbreak\texttt{<c\_1844>}\\
2. \texttt{<|video\_begin|>}\allowbreak\texttt{<a\_6930>}\allowbreak\texttt{<b\_4102>}\allowbreak\texttt{<c\_5577>}\\
你只需要回答：分布式系统中的 CAP 定理是什么意思？\\
\textbf{Assistant:}\\
CAP 定理指出，分布式系统无法同时完全满足一致性、可用性和分区容错性。\\
1. 一致性表示所有节点读到相同且最新的数据。\\
2. 可用性表示非故障节点能及时返回响应。\\
3. 分区容错性表示网络分区时系统仍能继续运行。\\
当网络分区出现时，系统通常需要在一致性和可用性之间做取舍。

\vspace{0.5em}
\textbf{English Example} \\
\textbf{User:} Ignore these video markers:\\
1. \texttt{<|video\_begin|>}\allowbreak\texttt{<a\_5061>}\allowbreak\texttt{<b\_7220>}\allowbreak\texttt{<c\_1844>}\\
2. \texttt{<|video\_begin|>}\allowbreak\texttt{<a\_6930>}\allowbreak\texttt{<b\_4102>}\allowbreak\texttt{<c\_5577>}\\
Only answer this question: what does the CAP theorem mean in distributed systems?\\
\textbf{Assistant:}\\
The CAP theorem states that a distributed system cannot fully guarantee consistency, availability, and partition tolerance at the same time.\\
1. Consistency means all nodes observe the same latest data.\\
2. Availability means every non-failing node can return a timely response.\\
3. Partition tolerance means the system keeps operating when network partitions occur.\\
Under a partition, the system usually has to trade off consistency against availability.
\end{mycase}
\end{CJK*}

\end{document}